\documentclass[twocolumn]{aastex63}
\usepackage{newtxtext,newtxmath}
\usepackage{enumitem}
\usepackage{ae,aecompl}
\usepackage{longtable}
\usepackage{graphicx}    
\usepackage{amsmath}    
\usepackage{amssymb}    
\usepackage{verbatim}

\received{}
\revised{}
\accepted{}
\submitjournal{ApJS}

\shorttitle{X-ray catalogs in W-CDF-S and ELAIS-S1}
\shortauthors{Ni et al.}

\newcommand{\xmm}{\hbox{\textit{XMM-Newton}}}
\newcommand{\xray}{\hbox{X-ray}}
\newcommand{\chandra}{\hbox{\textit{Chandra}}}
\newcommand{\spitzer}{\hbox{\textit{Spitzer}}}
\newcommand{\flux}{{erg~cm$^{-2}$~s$^{-1}$}}

\newcommand{\pany}{{$p_{\rm any}$}}
\newcommand{\wcdfs}{\hbox{W-CDF-S}}
\newcommand{\es}{\hbox{ELAIS-S1}}

\begin{document}

\title{The XMM-SERVS survey: {\it XMM-Newton} point-source catalogs for the W-CDF-S and ELAIS-S1 fields}

\author{Qingling Ni}
\altaffiliation{qingling1001@gmail.com}
\affiliation{Department of Astronomy and Astrophysics, 525 Davey Lab, The Pennsylvania State University, University Park, PA 16802, USA}
\affiliation{Institute for Gravitation and the Cosmos, The Pennsylvania State University, University Park, PA 16802, USA}

\author{W. N. Brandt}
\affiliation{Department of Astronomy and Astrophysics, 525 Davey Lab, The Pennsylvania State University, University Park, PA 16802, USA}
\affiliation{Institute for Gravitation and the Cosmos, The Pennsylvania State University, University Park, PA 16802, USA}
\affiliation{Department of Physics, 104 Davey Laboratory, The Pennsylvania State University, University Park, PA 16802, USA}

\author{Chien-Ting Chen}
\affiliation{Marshall Space Flight Center, Huntsville, AL 35811, USA}

\author{Bin Luo}
\affiliation{School of Astronomy and Space Science, Nanjing University,
Nanjing, Jiangsu 210093, China}
\affiliation{Key Laboratory of Modern Astronomy and Astrophysics
(Nanjing University), Ministry of Education, Nanjing, Jiangsu 210093, China}

\author{Kristina Nyland}
\affiliation{National Research Council, resident at the U.S. Naval Research Laboratory, 4555 Overlook Ave. SW, Washington, DC 20375, USA}

\author{Guang Yang}
\affiliation{Department of Physics and Astronomy, Texas A\&M University, College Station, TX 77843-4242, USA}
\affiliation{George P. and Cynthia Woods Mitchell Institute for Fundamental Physics and Astronomy, Texas A\&M University, College Station, TX 77843-4242, USA}

\author{Fan Zou}
\affiliation{Department of Astronomy and Astrophysics, 525 Davey Lab, The Pennsylvania State University, University Park, PA 16802, USA}
\affiliation{Institute for Gravitation and the Cosmos, The Pennsylvania State University, University Park, PA 16802, USA}

\author{James Aird}
\affiliation{Institute for Astronomy, University of Edinburgh, Royal Observatory, Edinburgh EH9 3HJ, UK}

\author{David M. Alexander}
\affiliation{Centre for Extragalactic Astronomy, Department of Physics, Durham University, Durham DH1 3LE, UK}

\author{Franz Erik Bauer}
\affiliation{Instituto de Astrof{\'{\i}}sica and Centro de Astroingenier{\'{\i}}a, Facultad de F{\'{i}}sica, Pontificia Universidad Cat{\'{o}}lica de Chile, Casilla 306, Santiago 22, Chile} 
\affiliation{Millennium Institute of Astrophysics, Nuncio Monse{\~{n}}or S{\'{o}}tero Sanz 100, Of 104, Providencia, Santiago, Chile} 
\affiliation{Space Science Institute, 4750 Walnut Street, Suite 205, Boulder, Colorado 80301} 

\author{Mark Lacy}
\affiliation{National Radio Astronomy Observatory, 520 Edgemont Road, Charlottesville, VA 22903, USA}

\author{Bret D. Lehmer}
\affiliation{Department of Physics, University of Arkansas, 226 Physics Building, 825 West Dickson Street, Fayetteville, AR 72701, USA}

\author{Labani Mallick}
\affiliation{Indian Institute of Astrophysics, Block II, Koramangala, Bangalore 560034, India}

\author{Mara Salvato}
\affiliation{MPE, Giessenbachstrasse 1, Garching 85748, Germany}

\author{Donald P. Schneider}
\affiliation{Department of Astronomy and Astrophysics, 525 Davey Lab, The Pennsylvania State University, University Park, PA 16802, USA}
\affiliation{Institute for Gravitation and the Cosmos, The Pennsylvania State University, University Park, PA 16802, USA}

\author{Paolo Tozzi}
\affiliation{INAF, Osservatorio Astrofisico di Firenze, Largo Enrico Fermi 5, I-50125, Firenze, Italy}

\author{Iris Traulsen}
\affiliation{Leibniz-Institut fuer Astrophysik Potsdam (AIP), An der Sternwarte 16, 14482 Potsdam, Germany}

\author{Mattia Vaccari}
\affiliation{Inter-university Institute for Data Intensive Astronomy, Department of Physics and Astronomy, University of the Western Cape, Robert Sobukwe Road, 7535, Bellville, Cape Town, South Africa}
\affiliation{INAF - Istituto di Radioastronomia, via Gobetti 101, 40129 Bologna, Italy}

\author{Cristian Vignali}
\affiliation{Dipartimento di Fisica e Astronomia, Universit\`{a} degli Studi di Bologna, via Gobetti 93/2, 40129 Bologna, Italy}
\affiliation{INAF - Osservatorio di Astrofisica e Scienza dello Spazio di Bologna - via Gobetti 93/3, 40129 Bologna, Italy}

\author{Fabio Vito}
\affiliation{Scuola Normale Superiore, Piazza dei Cavalieri 7, 56126, Pisa (Italy)}

\author{Yongquan Xue}
\affiliation{CAS Key Laboratory for Research in Galaxies and Cosmology, Department of Astronomy, University of Science and Technology of China, Hefei 230026, China}
\affiliation{School of Astronomy and Space Sciences, University of Science and Technology of China, Hefei 230026, China}

\author{Manda Banerji}
\affiliation{School of Physics \& Astronomy, University of Southampton, Highfield Campus, Southampton S017 1BJ, UK}

\author{Kate Chow}
\affiliation{CSIRO Astronomy and Space Science, PO Box 76, Epping, NSW, 1710, Australia}

\author{Andrea Comastri}
\affiliation{INAF - Osservatorio di Astrofisica e Scienza dello Spazio di Bologna - via Gobetti 93/3, 40129 Bologna, Italy}

\author{Agnese Del Moro}
\affiliation{German Aerospace Center (DLR), Space Operation and Astronaut Training, Oberpfaffenhofen, 82234 We$\beta$ling, Germany}

\author{Roberto Gilli}
\affiliation{INAF - Osservatorio di Astrofisica e Scienza dello Spazio di Bologna - via Gobetti 93/3, 40129 Bologna, Italy}

\author{James Mullaney}
\affiliation{Department of Physics and Astronomy, The University of Sheffield, Hounsfield Road, Sheffield, S3 7RH, UK}

\author{Maurizio Paolillo}
\affiliation{Dipartimento di Fisica, Universit\`{a}di Napoli ``Federico II'', via Cinthia 9, 80126 Napoli, Italy}
\affiliation{INAF - Osservatorio Astronomico di Capodimonte, Salita Moiariello 16, I-80131, Napoli, Italy}
\affiliation{INFN - Sezione di Napoli, via Cinthia 9, 80126 Napoli, Italy}

\author{Axel Schwope}
\affiliation{Leibniz-Institut fuer Astrophysik Potsdam (AIP), An der Sternwarte 16, 14482 Potsdam, Germany}

\author{Ohad Shemmer}
\affiliation{Department of Physics, University of North Texas, Denton, TX 76203, USA}

\author{Mouyuan Sun}
\affiliation{Department of Astronomy, Xiamen University, Xiamen, Fujian 361005, China}

\author{John D. Timlin III}
\affiliation{Department of Astronomy and Astrophysics, 525 Davey Lab, The Pennsylvania State University, University Park, PA 16802, USA}
\affiliation{Institute for Gravitation and the Cosmos, The Pennsylvania State University, University Park, PA 16802, USA}

\author{Jonathan R. Trump}
\affiliation{Department of Physics, University of Connecticut, 2152 Hillside Rd Unit 3046, Storrs, CT 06269, USA}

\begin{abstract}
We present the X-ray point-source catalogs in two of the XMM-\spitzer\ Extragalactic Representative Volume Survey (XMM-SERVS) fields, W-CDF-S (4.6 deg$^2$) and ELAIS-S1 (3.2 deg$^2$), aiming to fill the gap between deep pencil-beam \xray\ surveys and shallow \xray\ surveys over large areas.
The W-CDF-S and ELAIS-S1 regions were targeted with 2.3 Ms and 1.0 Ms of \xmm\ observations, respectively; 1.8 Ms and 0.9 Ms exposures remain after flare filtering.
The survey in W-CDF-S has a flux limit of $1.0\times10^{-14}$ \flux\ over 90\% of its area in the 0.5--10 keV band; 4053 sources are detected in total. The survey in \es\ has a flux limit of $1.3\times10^{-14}$ \flux\ over 90\% of its area in the 0.5--10 keV band; 2630 sources are detected in total. Reliable optical-to-IR multiwavelength counterpart candidates are identified for $\approx$ 89\% of the sources in \wcdfs\ and $\approx$~87\% of the sources in \es.
3186 sources in \wcdfs\ and 1985 sources in \es\ are classified as AGNs.
We also provide photometric redshifts for \xray\ sources; $\approx$~84\% of the 3319/2001 sources in \wcdfs/\es\ with optical-to-NIR forced photometry available have either spectroscopic redshifts or high-quality photometric redshifts. The completion of the \xmm\ observations in the \wcdfs\ and \es\ fields marks the end of the XMM-SERVS survey data gathering. The $\approx$ 12,000 point-like \xray\ sources detected in the whole $\approx 13$ deg$^2$ XMM-SERVS survey will benefit future large-sample AGN studies.
\end{abstract}


\section{Introduction}

Owing to the penetrating nature of X-rays and their reduced dilution by host-galaxy starlight, \xray\ surveys have been effectively utilized to identify reliable and nearly complete samples of active galactic nuclei (AGNs), which provide essential insights into the demographics, ecology, and physics of growing supermassive black holes (SMBHs) over most of cosmic history \citep[e.g.,][]{Brandt2015,Xue2017}.

\textit{XMM-Newton} and \chandra\ surveys have provided the most efficient method in assembling reliable and quite complete samples of distant AGNs, including obscured systems otherwise difficult to find. 
The currently publicly available wide-field X-ray surveys such as the 8–10 ks \textit{XMM-Newton} depth Stripe 82X \citep{LaMassa2016} and XMM-XXL \citep[e.g.,][]{Liu2016} have made excellent progress sampling the luminous AGN populations and their environments. At the same time, they lack the sensitivity to detect the bulk of SMBH growth as they only probe $\approx2$--6 times below the knee of the X-ray luminosity function at $z=0.5$--2.5, and the AGNs detected produce less than half of cosmic accretion power \citep[e.g.,][]{Ueda2014,Aird2015}.
The narrow-field deep X-ray surveys (0.1–1.1 deg$^2$), such as the CDF-S \citep{Luo2017}, CDF-N \citep{Xue2016}, E-CDF-S \citep{Xue2016}, AEGIS-X \citep{Nandra2015}, and SXDS \citep{Ueda2008}, are able to sample AGNs that produce the bulk ($>$ 70\%) of cosmic accretion power at $z \lesssim 3$--5 well \citep[e.g.,][]{Ueda2014,Aird2015,Vito2018}.
However, they do not have the contiguous volume needed to explore AGN activity over a wide dynamic range of cosmic environments and to sample substantially the high-luminosity tail of the AGN population.
Simulations indicate that at $z \approx 1$, the largest structures (e.g., superclusters) extend up to 2--3 deg$^2$ on the sky (e.g., \citealt{Klypin2016}). Thus, even the $\approx 2$ deg$^2$ COSMOS field \citep[e.g.,][]{Cappelluti2009,Civano2016} is not able to sample the full range of cosmic environments. 
Therefore, it is necessary to obtain several distinct medium-deep X-ray surveys, each over several deg$^2$, in addition to COSMOS for investigating SMBH growth across the full range of cosmic environments and minimizing cosmic variance \citep[e.g.,][]{Driver2010,Moster2011}.

To this end, we designed an \textit{XMM-Newton} survey, XMM-SERVS, to provide medium-deep \xray\ coverage in the SERVS \citep{Mauduit2012} regions of the W-CDF-S ($\approx$ 4.6 deg$^2$), ELAIS-S1 ($\approx$ 3.2 deg$^2$), and XMM-LSS ($\approx$ 5.3 deg$^2$) fields, all of which have superb multiwavelength coverage.
The point-source catalog for the XMM-LSS field has been published in \citet{Chen2018}.
In this work, we provide point-source catalogs for the two remaining fields, W-CDF-S and ELAIS-S1.
Data products from the full XMM-SERVS survey (including all the three fields) are available online.\footnote{\href{https://personal.psu.edu/wnb3/xmmservs/xmmservs.html}{https://personal.psu.edu/wnb3/xmmservs/xmmservs.html}.}

The paper is structured as follows.
Section~\ref{s-xo} describes the new \xmm\ observations in the \wcdfs\
 and \es\ fields as well as overlapping archival multiwavelength data in these areas.
Section~\ref{s-xcat} presents the \xray\ source detection process and the properties of the derived \xray\ sources.
Section~\ref{s-mc} describes the multiwavelength counterpart identification process for the \xray\ sources.
Section~\ref{s-z} presents spectroscopic redshifts and photometric redshifts of \xray\ sources.
In Section~\ref{sc-sp}, basic AGN classification is presented.
Section~\ref{s-sum} gives the summary of the work.
Appendix~\ref{a-column} describes the columns included in our \xray\ source catalogs.
Appendix~\ref{a-blagn} describes the identification of broad-line (BL) AGNs among the \xray\ sources detected. Appendix~\ref{a-star} describes the classification of \xray\ sources that are not AGNs.
A $\Lambda$CDM cosmology with $H_0 = 70$ km s$^{-1}$ Mpc$^{-1}$, $\Omega_m = 0.3$, and $\Omega_\Lambda = 0.7$ is assumed throughout the paper.
A Galactic column density $N_{\rm H} = 8.4 \times 10^{19}~\rm cm^{-2}$ is adopted for the \wcdfs\ field, and $N_{\rm H} = 3.4 \times 10^{20}~\rm cm^{-2}$ is adopted for the \es\ field \citep{Willingale2013}.

\section{XMM-Newton observations in the W-CDF-S and ELAIS-S1 regions} \label{s-xo}

\subsection{Multiwavelength data coverage and archival \xmm\ and \chandra\ observations in the W-CDF-S region}

\begin{figure*}
\centering
\includegraphics[width=0.525\textwidth]{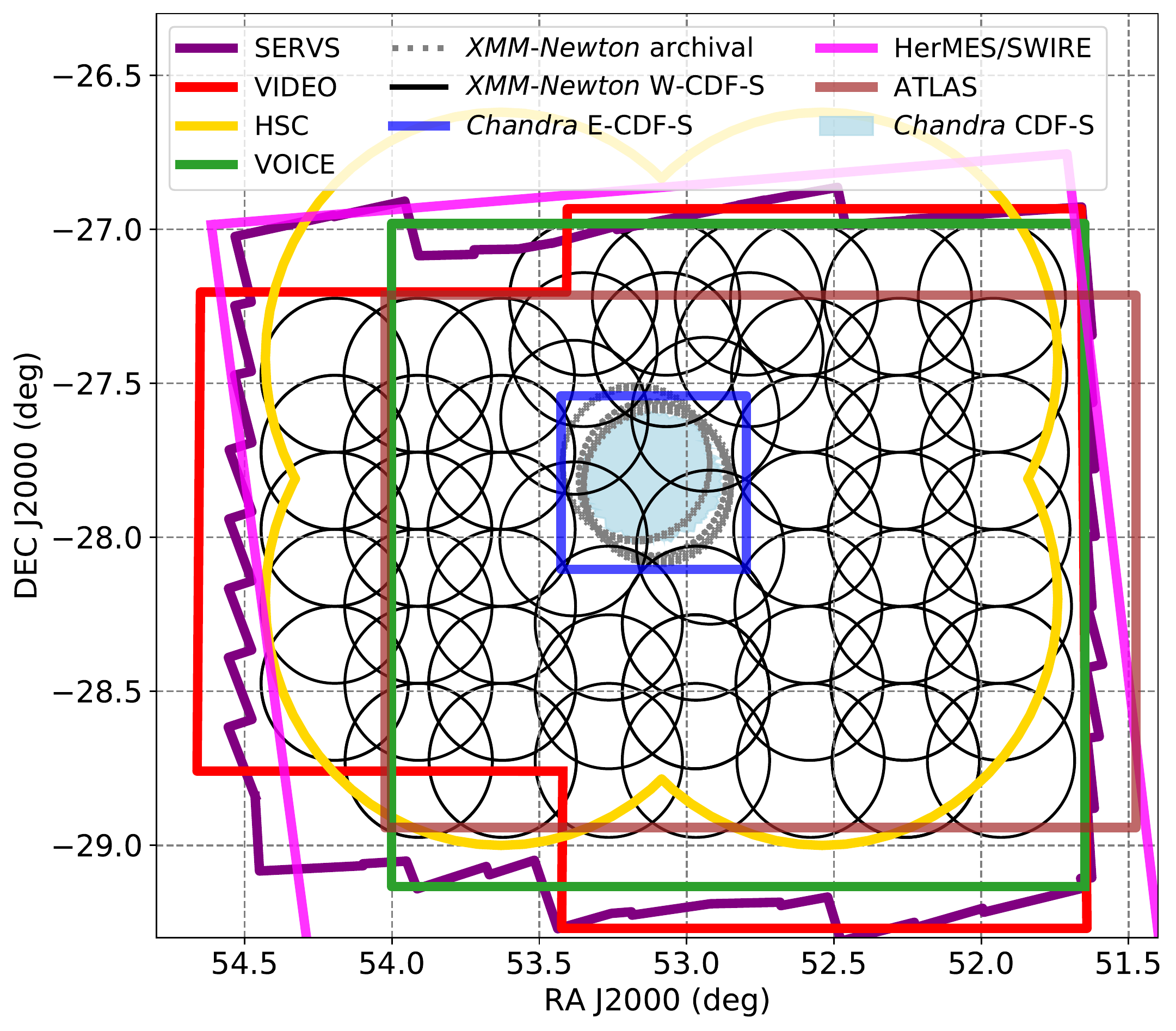}
~
\includegraphics[width=0.455\textwidth]{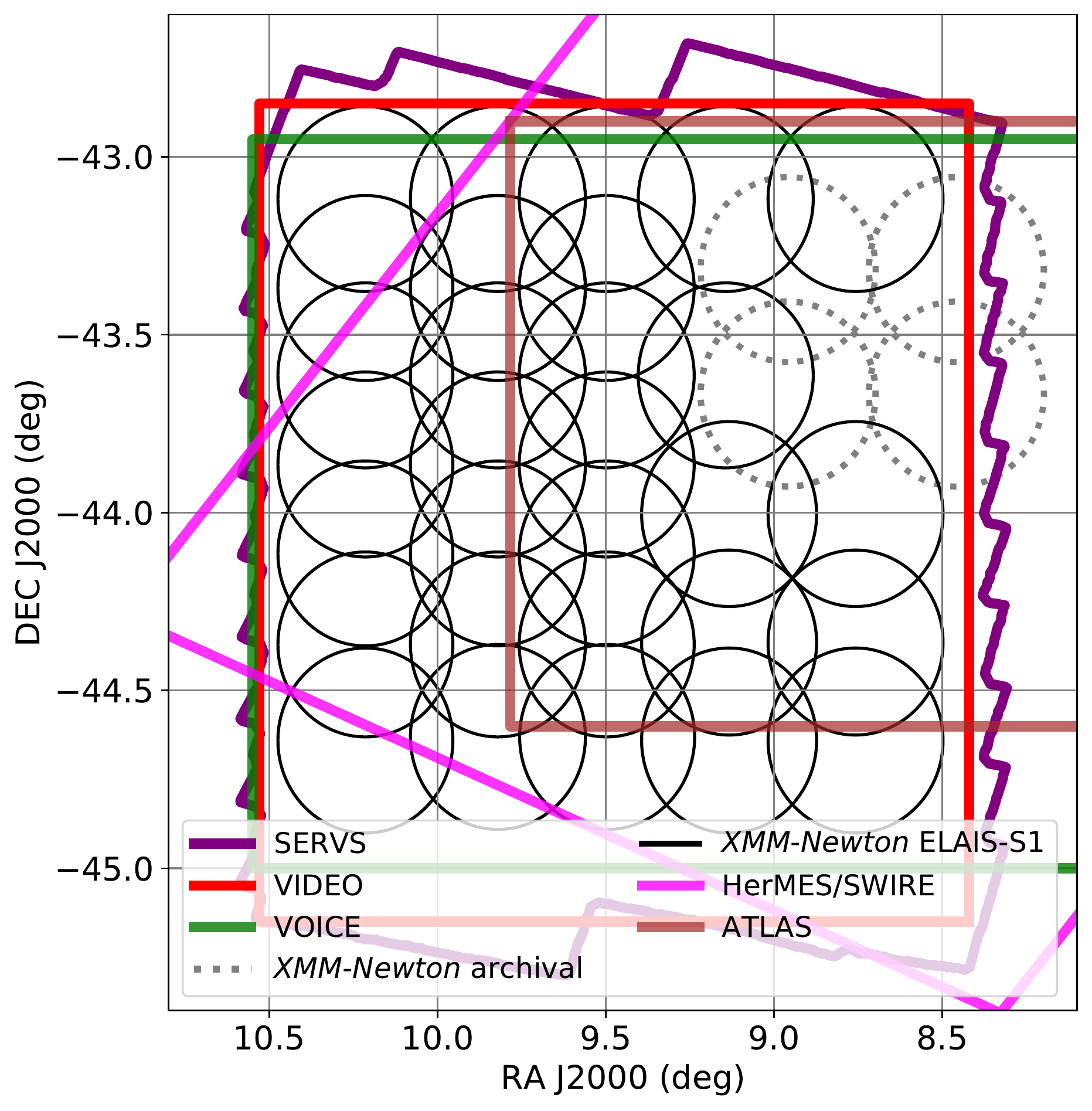}
\vspace{-0.1in}
\caption{\textit{Left:} Locations of the {\it XMM-Newton} observations in the W-CDF-S field (black circles), presented together with the multiwavelength coverage from selected surveys and the primary archival \xray\ observations in this area (as labeled in the figure key). DES wide-field survey (see Table~\ref{tab:wcdfsmw}) in the optical cover the whole area, and thus are not plotted in the figure. \textit{Right:} Locations of the {\it XMM-Newton} observations in the ELAIS-S1 field (black circles), presented together with the selected multiwavelength coverage and the primary archival \xray\ observations in this area (as labeled in the figure key). DES wide-field survey and ESIS (see Table~\ref{tab:wcdfsmw}) in the optical cover the whole area, and thus are not plotted in the figure.}    
\label{fig:loc}
\end{figure*}

There are deep archival \chandra\ and \xmm\ observations in the center of the  $\approx 4.5$ deg$^2$ W-CDF-S field, covering a relatively small area (see Figure~\ref{fig:loc}). 
The \chandra\ Deep Field-South (CDF-S) survey has now reached a 7~Ms depth, covering 482 arcmin$^2$ \citep[e.g.,][]{Xue2011,Luo2017}; \xmm\ has also observed this field for 3.3 Ms (covering $\approx$ 790 arcmin$^2$; e.g., \citealt{Comastri2011,Ranalli2013}). 
The 250 ks Extended \chandra\ Deep Field-South (E-CDF-S) survey further increases the \xray\ coverage to 1128.6 arcmin$^2$ \citep[e.g.,][]{Xue2016}.
There are also several additional $\approx 5$--15 ks \chandra\ observations in the \wcdfs\ region (including four observations just to the south of E-CDF-S; PI: W. N. Brandt). These \chandra\ data are utilized in our study to help the multiwavelength counterpart matching of \xmm\ sources (see Section~\ref{s-mc}).
All of the above \xray\ observations along with the multiwavelength data have enabled many AGN studies.

The W-CDF-S region, which is $\approx 30$ times larger in solid angle than the original CDF-S, also has extensive multiwavelength coverage (see Table~\ref{tab:wcdfsmw} for a list of the key multiwavelength photometric data).
It aligns with one of the well-studied \textit{Spitzer} Extragalactic Representative
Volume Survey (SERVS; \citealt{Mauduit2012}) fields, and it is also one of the deep-drilling fields of the upcoming Legacy Survey of Space and Time (LSST) to be conducted by the Vera C. Rubin Observatory \citep[e.g.,][]{Brandt2018,Scolnic2018}.
With the XMM-SERVS survey covering the \wcdfs\ region, multiwavelength data in this area can be utilized together with the X-ray data, enabling large-sample studies of AGNs and other \xray\ sources.

\begin{table*}
\footnotesize
\caption{\label{tab:wcdfsmw}
Key multiwavelength imaging coverage of the W-CDF-S and ELAIS-S1 fields.}
\vspace{-0.5 cm}
\begin{center}
\begin{tabular}{lllll}
\noalign{\smallskip}\hline\noalign{\smallskip}
{Band}                   &
Field(s)$^*$ & 
{Survey Name} &
{Coverage}; Notes         & 
Example Reference \\
\noalign{\smallskip}\hline \hline\noalign{\smallskip}
Radio    & C/E & Australia Telescope Large Area Survey ({\bf ATLAS}) & 3.6/2.7 deg$^2$; 14/17~$\mu$Jy rms depth at 1.4~GHz     &  \cite{Franzen2015} \\
         & C/E & {\bf MIGHTEE} Survey (in progress) & 3/4.5 deg$^2$; 1~$\mu$Jy rms depth at 1.4~GHz  & \cite{Jarvis2016b} \\
\noalign{\smallskip}\hline\noalign{\smallskip}

MIR--FIR      & C/E &  {\it Herschel} Multi-tiered Extragal.\ Surv.\ ({\bf HerMES}) & 11.4/3.7 deg$^2$;  5--60~mJy depth at 100--500~$\mu$m  & \cite{Oliver2012} \\

     & C/E &  {\it Spitzer} Wide-area IR Extragal.\ Survey ({\bf SWIRE}) & 7.1/14.3 deg$^2$;  0.01--200~mJy depth at 3.6--160~$\mu$m & \cite{Vaccari2015}\\
\noalign{\smallskip}\hline\noalign{\smallskip}

NIR     & C/E &  {\it Spitzer} survey of Deep Drilling Fields ({\bf DeepDrill}) & 9/9 deg$^2$; 2~$\mu$Jy depth at 3.6 and 4.5~$\mu$m  & \cite{Lacy2021}\\
        & C/E &  {\it Spitzer} Extragal.\ Rep.\ Vol.\ Survey ({\bf SERVS}) & 4.5/3 deg$^2$; 2~$\mu$Jy depth at 3.6 and 4.5~$\mu$m  & \cite{Mauduit2012}\\
         & C/E & VISTA Deep Extragal.\ Obs.\ Survey ({\bf VIDEO}) & 4.5/3 deg$^2$; $ZYJHK_s$ to $m_{\rm AB}\approx23.8$--25.7 & \cite{Jarvis2013}\\
\noalign{\smallskip}\hline\noalign{\smallskip}

Optical        & C/E & Dark Energy Survey ({\bf DES}) Data Release 2 & 9/6 deg$^2$;  $grizy$ to $m_{\rm AB}\approx23$--25.5 & \cite{Abbott2021} \\
             &  C & Hyper Suprime Cam (HSC) optical imaging & 5.7 deg$^2$; $griz$ to $m_{\rm AB}\approx 25$--26 & \cite{Ni2019} \\
              & C  & VST Opt. Imaging of CDF-S and ES1 ({\bf VOICE}) & 4 deg$^2$; $ugri$ to $m_{\rm AB}\approx26$& \cite{Vaccari2016} \\
               & E & VST Opt. Imaging of CDF-S and ES1 ({\bf VOICE}) & 4 deg$^2$; $u$ to $m_{\rm AB}\approx26$, $gri$ observations planned & \cite{Vaccari2016} \\ 
              & C &  SWIRE optical imaging & 7.1 deg$^2$; $u'g'r'i'z'$ to $m_{\rm AB}\approx23$--25 & \cite{Lonsdale2003} \\
              & E &  ESO-Spitzer Imaging Extragalactic Survey ({\bf ESIS}) & 4.5 deg$^2$; $BVR$ to $m_{\rm AB}\approx24$--25 & \cite{Berta2006}\\    
              & C/E & {\bf LSST} deep-drilling field (Planned) & 10/10 deg$^2$; $ugrizy$ to $m_{\rm AB} \approx 26$--28 & \cite{Brandt2018} \\
\noalign{\smallskip}\hline\noalign{\smallskip}
UV & C/E  & {\it GALEX} Deep Imaging Survey &  7/15~deg$^2$; NUV, FUV to $m_{\rm AB}\approx24$--24.5 & \cite{Martin2005}\\
\noalign{\smallskip}\hline\noalign{\smallskip}
\end{tabular}
\end{center}
\vspace{-0.2 cm}
	\raggedright
$^*$``C'' stands for \wcdfs; ``E'' stands for \es.
\end{table*}

\subsection{Multiwavelength data coverage and archival \xmm\ and \chandra\ observations in the ELAIS-S1 region}
About 0.6 deg$^2$ of the $\approx 3$ deg$^2$ ELAIS-S1 region has been targeted with both \xmm\ ($\approx 50$ ks depth) and \chandra\ ($\approx 30$ ks depth) \citep[e.g.,][]{Puccetti2006,Feruglio2008}.
There are also several additional \chandra\ observations in the \es\ region.
The multiwavelength data coverage of the \es\ field is listed in Table~\ref{tab:wcdfsmw}. Similar to the \wcdfs\ field, \es\ aligns with one of the SERVS fields, and will be one of the LSST deep-drilling fields.
As can be seen in Table~\ref{tab:wcdfsmw}, the optical data in \es\ are not yet as deep as that in \wcdfs\ (also see \citealt{Zou2021a} for further details).

\subsection{New \xmm\ observations and data reduction} \label{ss-dr}
\textit{XMM-Newton} observations in the \wcdfs\ field were obtained between July 2018 to January 2021 (see the left panel of Figure~\ref{fig:loc} for the pointing layout) with a total of 2.3 Ms exposure time, including 80 successful observations. 
For the \es\ field, \xmm\ observations were performed between May 2019 to December 2020 (see the right panel of Figure~\ref{fig:loc} for the pointing layout), with a total of 1.0 Ms exposure time, including 31 successful observations.
All the observations were performed with a THIN filter for the EPIC cameras, and Optical Monitor data were taken in parallel as well (we do not include these data in our catalogs, due to the exisiting optical/UV coverage listed in Table~\ref{tab:wcdfsmw}).
As these fields are far from the Galactic plane, the numbers of very bright stars in these fields are small, and the optical loading effects for the \xray\ CCDs are negligible.
The details of each observation are listed in Table~\ref{tab:xmmdata}.
As described in \citet{Chen2018}, we first observed the desired pointings with 33 ks exposures, and then re-observed the sky regions strongly affected by \xmm\ background flaring to achieve better uniformity.
For the \wcdfs\ field, we do not re-analyze all of the archival \xmm\ observations of the CDF-S proper (that cover $\approx 0.25$ deg$^2$) in this work; instead, we selected one observation (ObsID: 0604960501) from the archival data to reach a uniform depth across the \wcdfs\ field and process this consistently in the same manner as the rest of our data.
For the \es\ field, all the archival \xmm\ observations are included in the analyses.

\begin{table*}[htbp]
\begin{center}
\footnotesize
\caption{\label{tab:xmmdata}
\xmm\ observations in the W-CDF-S and ELAIS-S1 fields. This table is available in its entirety online.}
\begin{tabular}{ccccccccccc} 
        \hline
        Field                    & Revolution & ObsID        & UT Date & R.A.   & Decl.   & GTI (PN) & GTI (MOS1) & GTI (MOS2)  & Expo \\
                                    &                   &                    &         & (deg) & (deg)  & (ks)          & (ks)              & (ks)    &  (ks)  \\
        \hline
W-CDF-S & 3403 & 0827210101 & 2018-07-08 23:34:26 & 52.579042 & $-$28.723972 & 27.9 & 30.5 & 29.7 & 33\\
W-CDF-S & 3403 & 0827210201 & 2018-07-09 09:04:26 & 52.582875 & $-$28.473972 &  27.9 & 29.4 & 29.2 & 33\\
W-CDF-S & 3406 & 0827210301 & 2018-07-15 05:20:04 & 52.586667 & $-$28.223972  & 28.9 & 30.7 & 30.6 & 33\\
ELAIS-S1             &  3561         &  0827251301 & 2019-05-20 07:26:00 & 9.143708 & $-$43.614139 & 28.8 & 30.5 &  30.6 & 33\\
ELAIS-S1             &  3568         &  0827240101 & 2019-06-03 05:48:52 & 8.757958 & $-$44.004000 & 29.4 & 31.6 & 31.3  & 34 \\
\hline
\end{tabular}
\end{center}
Columns from left to right: target field, \xmm\ revolution, \xmm\ Observation ID, observation starting date/time, Right Ascension and Declination of the pointing center (J2000, degrees), cleaned exposure time (included in the ``good time intervals''; GTIs) for PN, MOS1, and MOS2 in each pointing, total EPIC exposure time (during which PN, MOS1, and MOS2 take exposures simultaneously).
\end{table*}

The \textit{XMM-Newton} Science Analysis System (SAS) 19.0.0\footnote{\url{https://www.cosmos.esa.int/web/xmm-newton/sas-release-notes-1900}.} and HEASOFT 6.26\footnote{\url{https://heasarc.gsfc.nasa.gov/FTP/software/ftools/release/archive/Release_Notes_6.26}.} are utilized for our data analysis.
We use the SAS tasks \texttt{epproc} and \texttt{emproc} to process the \xmm\ Observation Data Files (ODFs), creating MOS1, MOS2, PN, and PN out-of-time (OOT) event files for each observation ID. 
Following Section~2.2 of \cite{Chen2018}, single-event light curves are created for each event file in time bins of 100~s at high (10--12~keV) and low (\hbox{0.3--10}~keV) energies to select time intervals without significant background flares (the ``good time intervals''; GTIs); we note that real sources provide minimal contributions to these total event file light curves.
For the \hbox{10--12~keV} light curve, we remove time intervals with count rates $
> 3\sigma$ above the mean count rate.
The same procedure is also performed for the \hbox{0.3--10}~keV light curves.
For a small number of event files with intense background flares, the event files are filtered using the nominal count-rate thresholds suggested by the {\it XMM-Newton} Science Operations Centre.\footnote{\url{https://www.cosmos.esa.int/web/xmm-newton/sas-thread-epic-filterbackground}}

For the \wcdfs\ field, a total of 1.8~Ms (1.5~Ms) of MOS (PN) exposure remains after flare filtering; for the \es\ field, a total of 0.9~Ms (0.8~Ms) of MOS (PN) exposure remains.
We do not exclude events in energy ranges that overlap with the instrumental background lines (Al K$\alpha$ lines at 1.39--1.55~keV for MOS and PN; Cu lines at 7.35--7.60~keV and 7.84--8.28~keV for PN; Si lines at 1.69–1.80 keV for MOS), as keeping these events improves the positional accuracy of the detected sources due to the higher number of counts detected.

\begin{figure*}
\centering
    \includegraphics[width=0.55\textwidth]{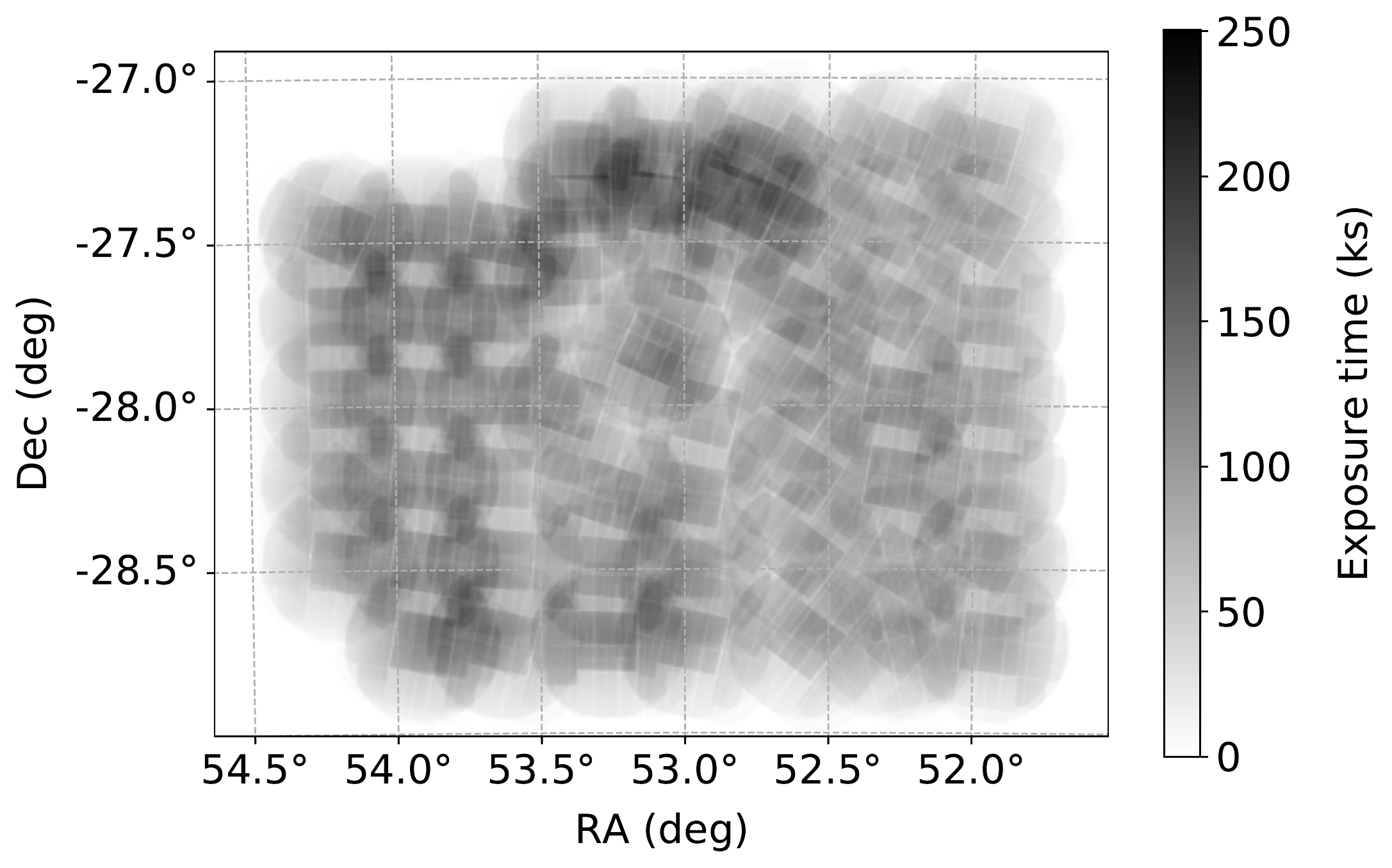}
    ~~~~
    \includegraphics[width=0.42\textwidth]{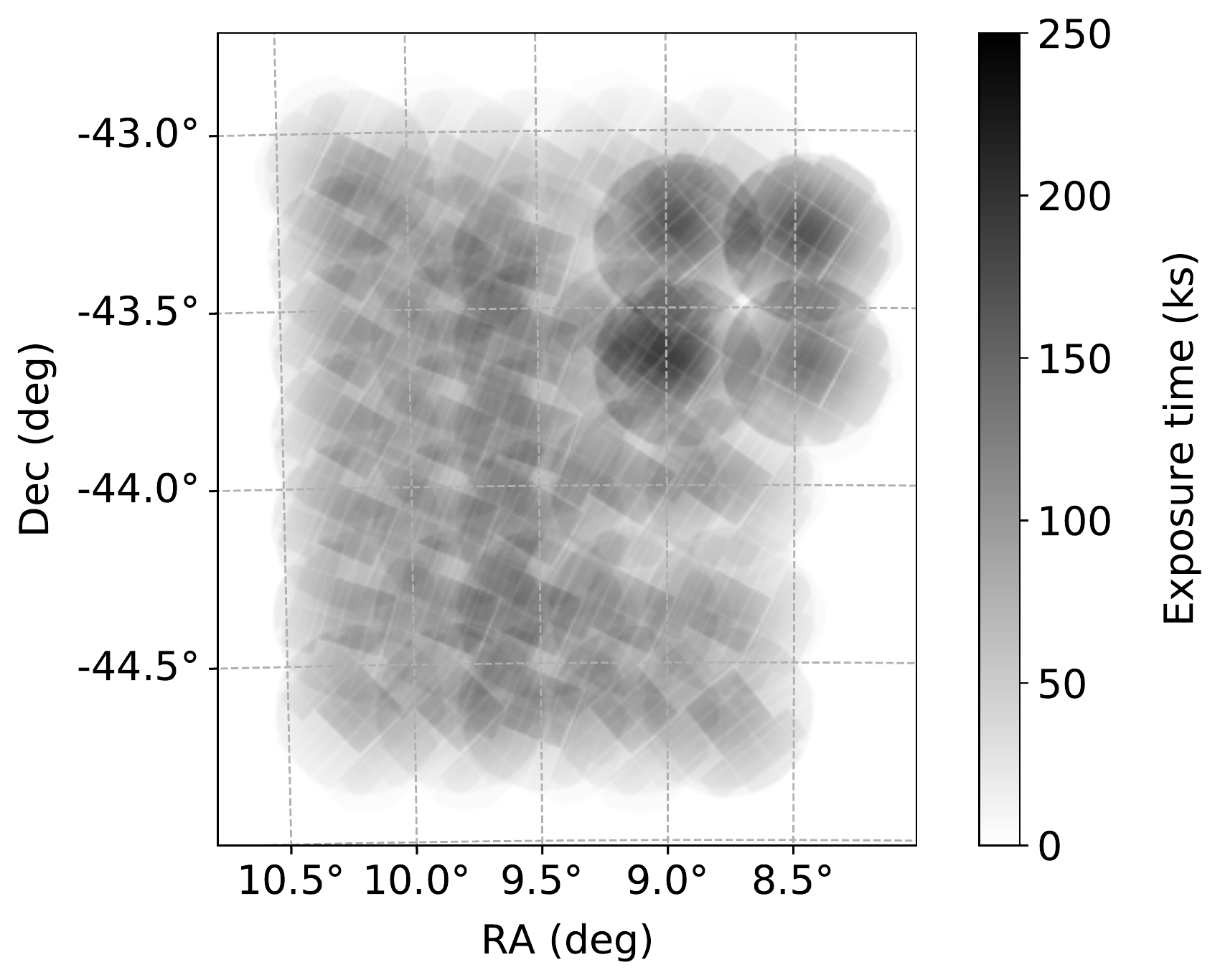}
    \caption{
    {\it Left:} Effective exposure map (PN + MOS) in the full band for W-CDF-S. The {\it XMM-Newton} coverage in the survey region is generally uniform.
      {\it Right:} Similar to the left panel, but for \es.}
    \label{fig:exposure}
\end{figure*}

\begin{figure*}
\centering
    \includegraphics[width=0.287\textwidth]{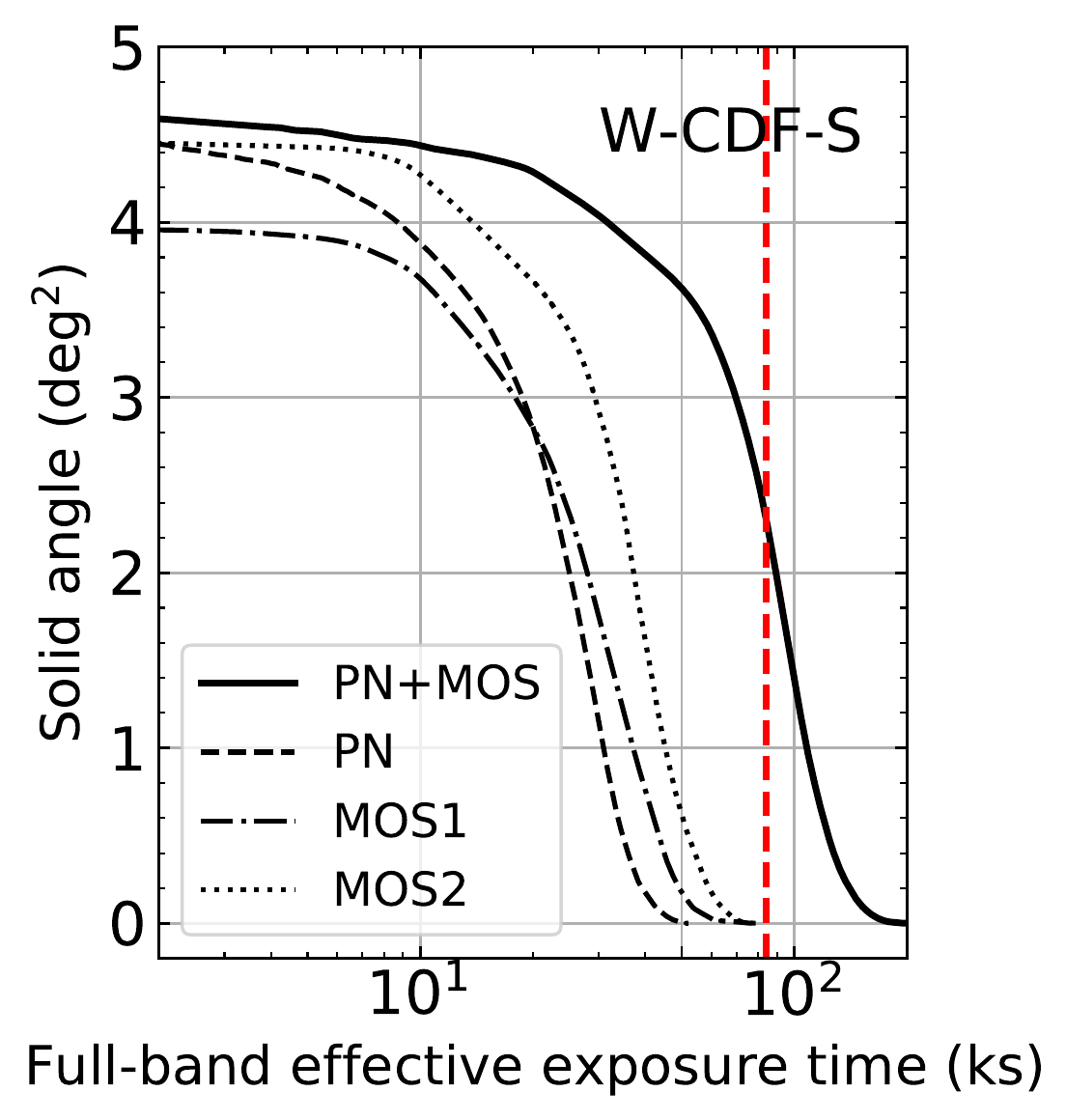}
    \includegraphics[width=0.3\textwidth]{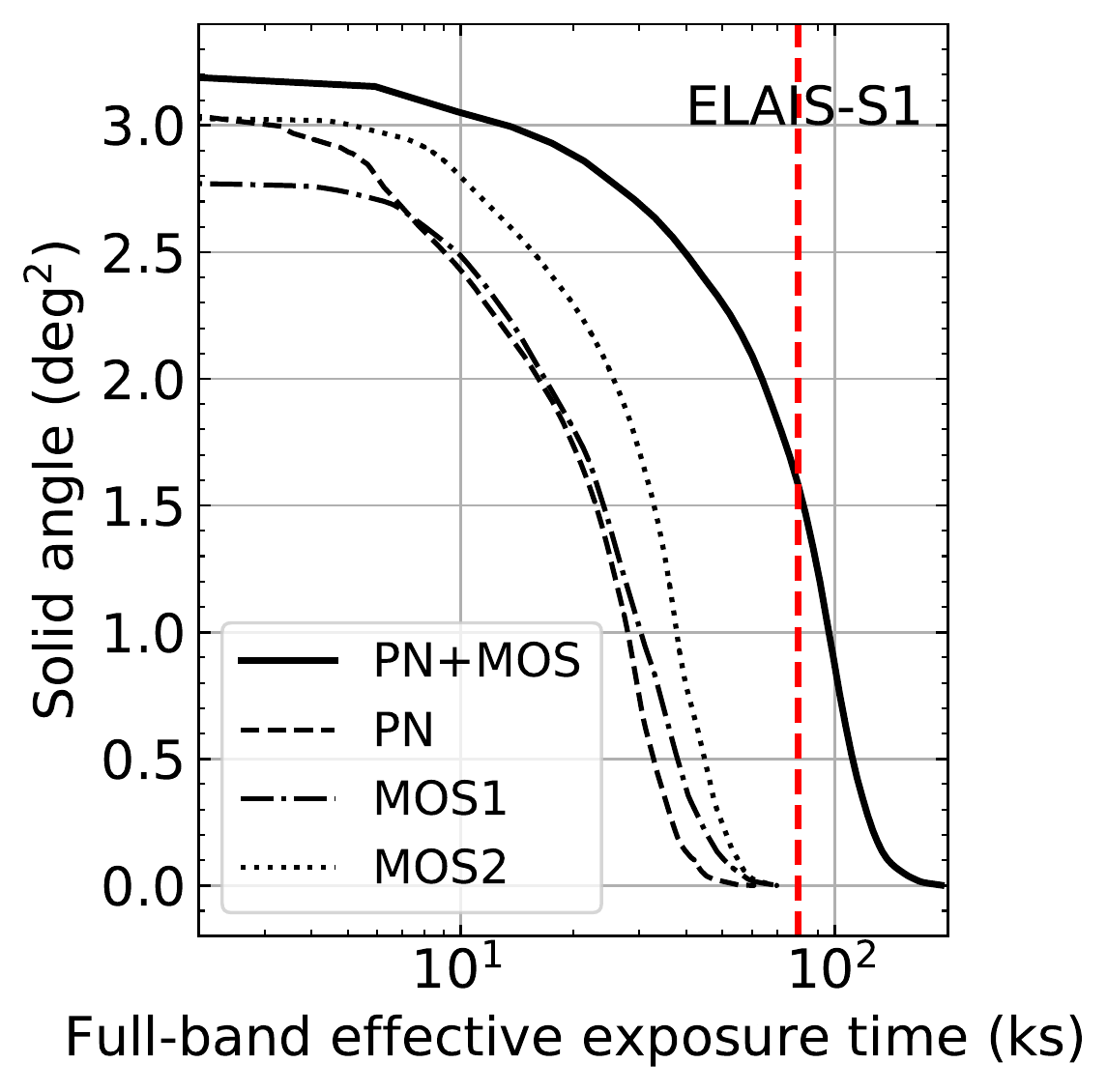}
    \includegraphics[width=0.375\textwidth]{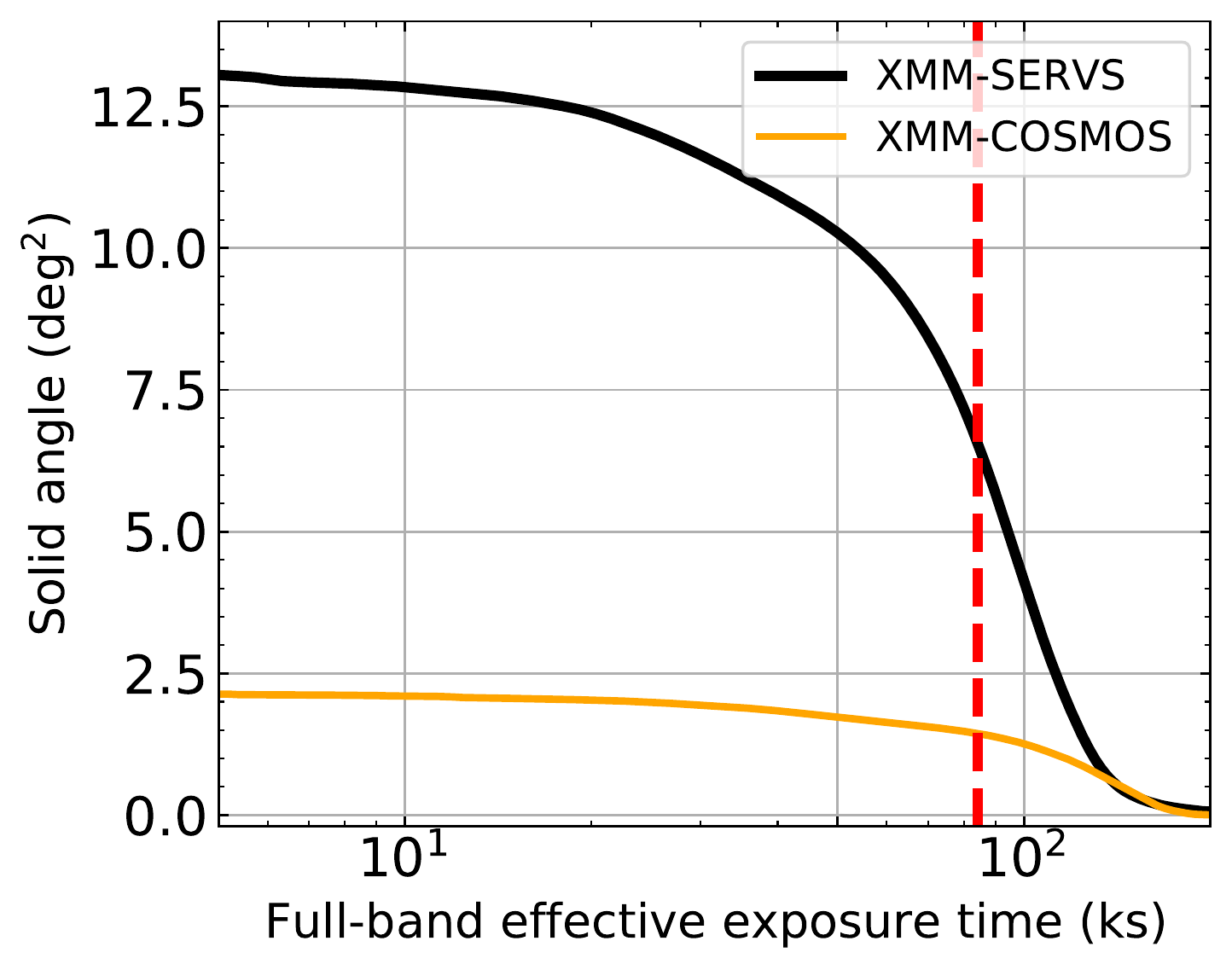}
    \caption{Cumulative survey solid angle as a function of full-band effective (i.e., vignetted) exposure in W-CDF-S (\textit{Left}), ELAIS-S1 (\textit{middle}), and the entire XMM-SERVS survey  (\textit{right}). The black solid/dashed/dash-dotted/dotted line is for the PN+MOS/PN/MOS1/MOS2 exposure. The relatively small solid angle of MOS1 coverage is due to the lost CCDs for MOS1. The solid orange line in the right panel is for the PN+MOS exposure in XMM-COSMOS \citep{Cappelluti2009}. The red vertical line marks the median exposure.}
    \label{fig:allexposure}
\end{figure*}

We use \texttt{evselect} to construct images with a $4^{\prime\prime}$ pixel size from the flare-filtered event file in the full band (\hbox{0.2--12}~keV).
We use \texttt{eexpmap} to generate exposure maps with {\sc usefastpixelization=0} and {\sc attrebin=0.5}, both with and without vignetting corrections. Detector masks were constructed with \texttt{emask}. 
The mosaicked vignetting-corrected PN+MOS1+MOS2 exposure map in the \wcdfs/\es\ field and the distribution of the exposure time across the survey field are presented in Figures~\ref{fig:exposure} and \ref{fig:allexposure}. 
As can be seen in Figure~\ref{fig:allexposure}, $\approx$ 4.6 deg$^2$ of the \wcdfs\ field is covered by \xmm; $\approx$~3.2~deg$^2$ of the \es\ field is covered by \xmm.
The median PN+MOS1+MOS2 exposure time across the \wcdfs/\es\ field is $\approx$ 84/80~ks.
More than 80\% of the \wcdfs/\es\ footprints have PN+MOS1+MOS2 exposure time $\gtrsim$ 47/37 ks. 
Figure~\ref{fig:allexposure} shows the cumulative survey solid angle as a function of full-band effective exposure for the full three-field XMM-SERVS survey.
The median PN+MOS1+MOS2 exposure time across the full XMM-SERVS survey field is $\approx$ 85~ks.

\section{Source detection and the main X-ray source catalogs} \label{s-xcat}

\subsection{First-pass source detection and astrometric correction}

Following \citet{Chen2018}, we run a first-pass source detection in the full band to register the \xmm\ observations onto a common World Coordinate System (WCS) frame with the following steps:

\begin{enumerate}
    
\item For each observation, \texttt{eboxdetect} is used to generate a temporary source list with {\sc likemin=8} for each of the three instruments.

\item This temporary source list is utilized to generate background images for each instrument (with the input sources removed), using \texttt{esplinemap} with {\sc method=asmooth}. This adaptive-smoothing method has been widely adopted in recent \xmm\ catalogs \citep[e.g.,][]{Traulsen2019,Traulsen2020,Webb2020}, as it can well-characterize the local \xray\ background level.

\item We run \texttt{eboxdetect} again in the map mode (with {\sc likemin=8}), combining images, exposure maps, and background maps from all the instruments for each observation. 

\item With this new source list generated by \texttt{eboxdetect} as the input, the PSF-fitting tool \texttt{emldetect} is used to determine the \hbox{X-ray} positions and detection likelihoods utilizing all the instruments of each observation. We only keep the point-like sources, and a stringent likelihood threshold ({\sc likemin} $=10.8$) is adopted to ensure that astrometric corrections are calculated based on significant detections that are unlikely to be spurious.

\end{enumerate}

For each observation, we use {\sc catcorr} to match the output source list from \texttt{emldetect} with an optical/IR reference catalog (available from the relevant \xmm\ Processing Pipeline Subsystem products; \citealt{Rosen2016}) created from the Sloan Digital Sky Survey (SDSS; \citealt{Abazajian2009}), 2MASS \citep{Skrutskie2006}, and USNO-B1.0 \citep{Monet2003} catalogs.
By matching the \xray\ sources to the reference catalogs (the median number of matched sources is 18 among all the observations), the needed astrometric offsets and rotation corrections are calculated.
The RA/DEC offsets are typically $\lesssim 3''$. The rotation corrections are less than $\approx 0.17$~deg. 
The event files and the attitude file for each observation are then projected onto the new frame.

\subsection{Second-pass source detection} \label{ss-ssd}

\begin{figure*}  
\includegraphics[width = 0.576\textwidth]{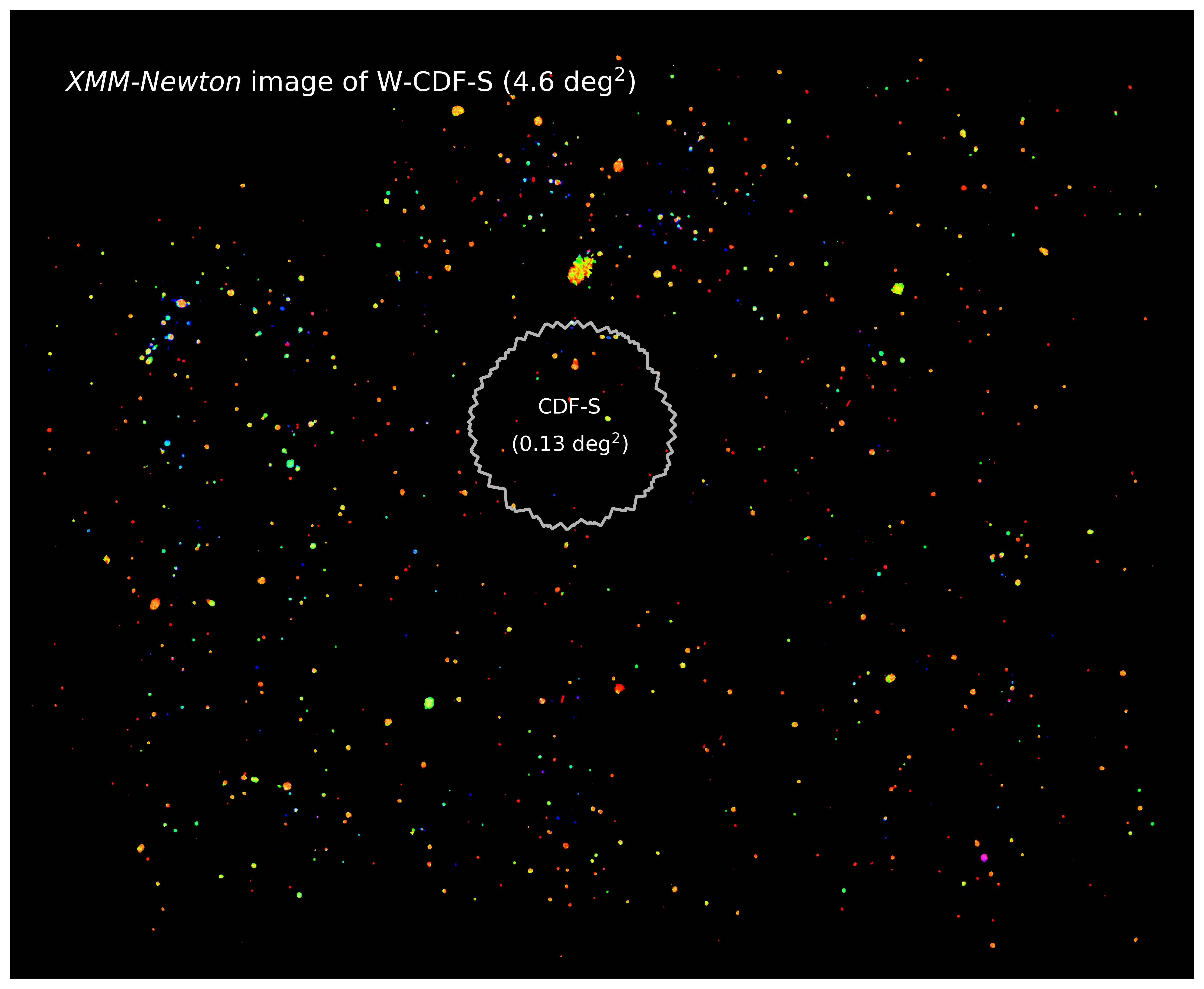}
~
\includegraphics[width = 0.422\textwidth]{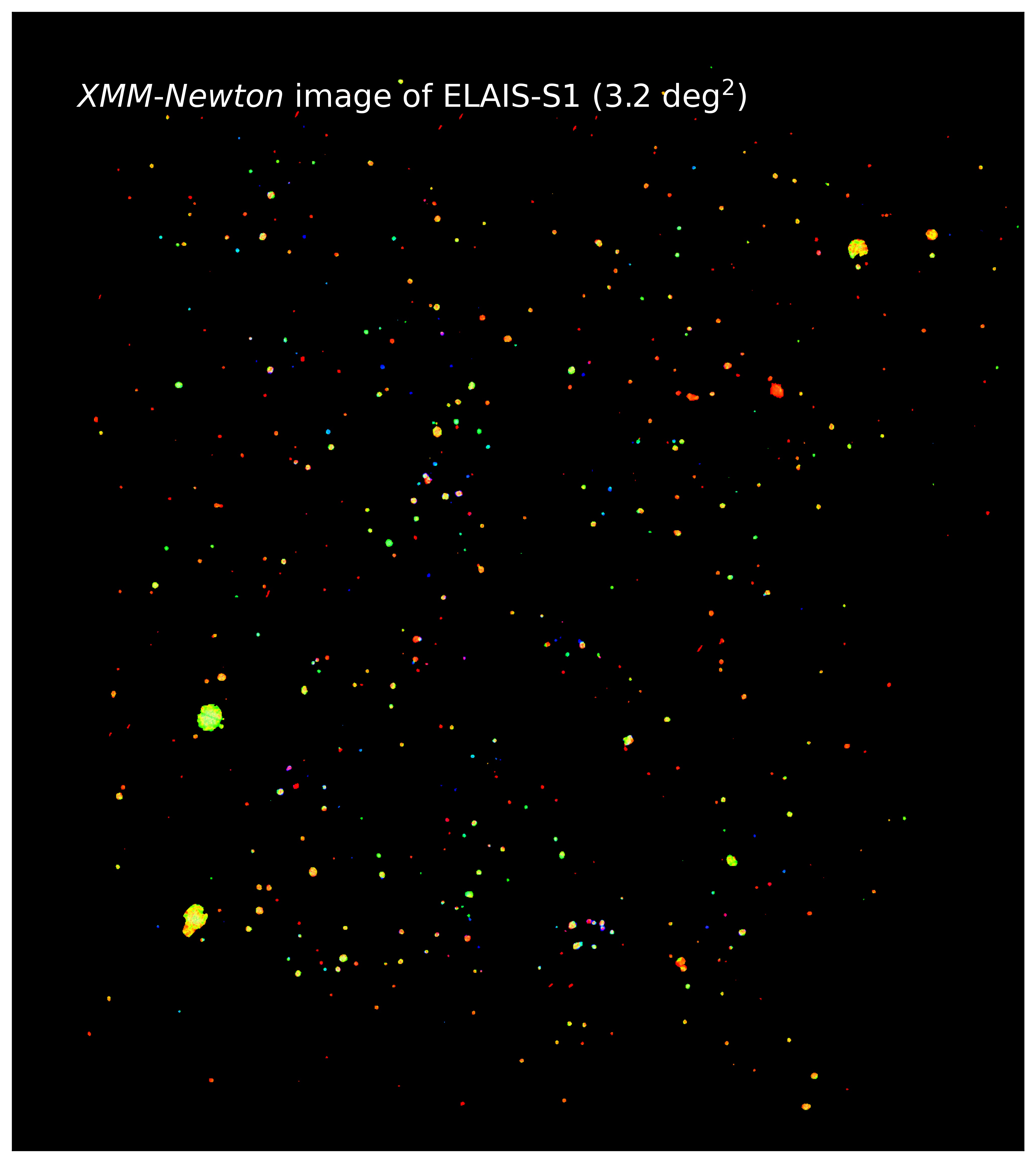}
\caption{\textit{Left:} ``False-color'' smoothed \xray\ image of W-CDF-S.  Band 1+2, band 3, and band 4+5 are represented by the colors red, green, and blue, respectively. Redder sources are softer; bluer sources are harder. An asinh stretch is utilized. The white solid curve indicates the footprint of the 7~Ms CDF-S \citep{Luo2017}. 
\textit{Right:} Similar to the left panel, but for the \es\ field.}
\label{fig:smoimagefull}
\end{figure*}

\begin{figure*}  
\includegraphics[width = 0.49\textwidth]{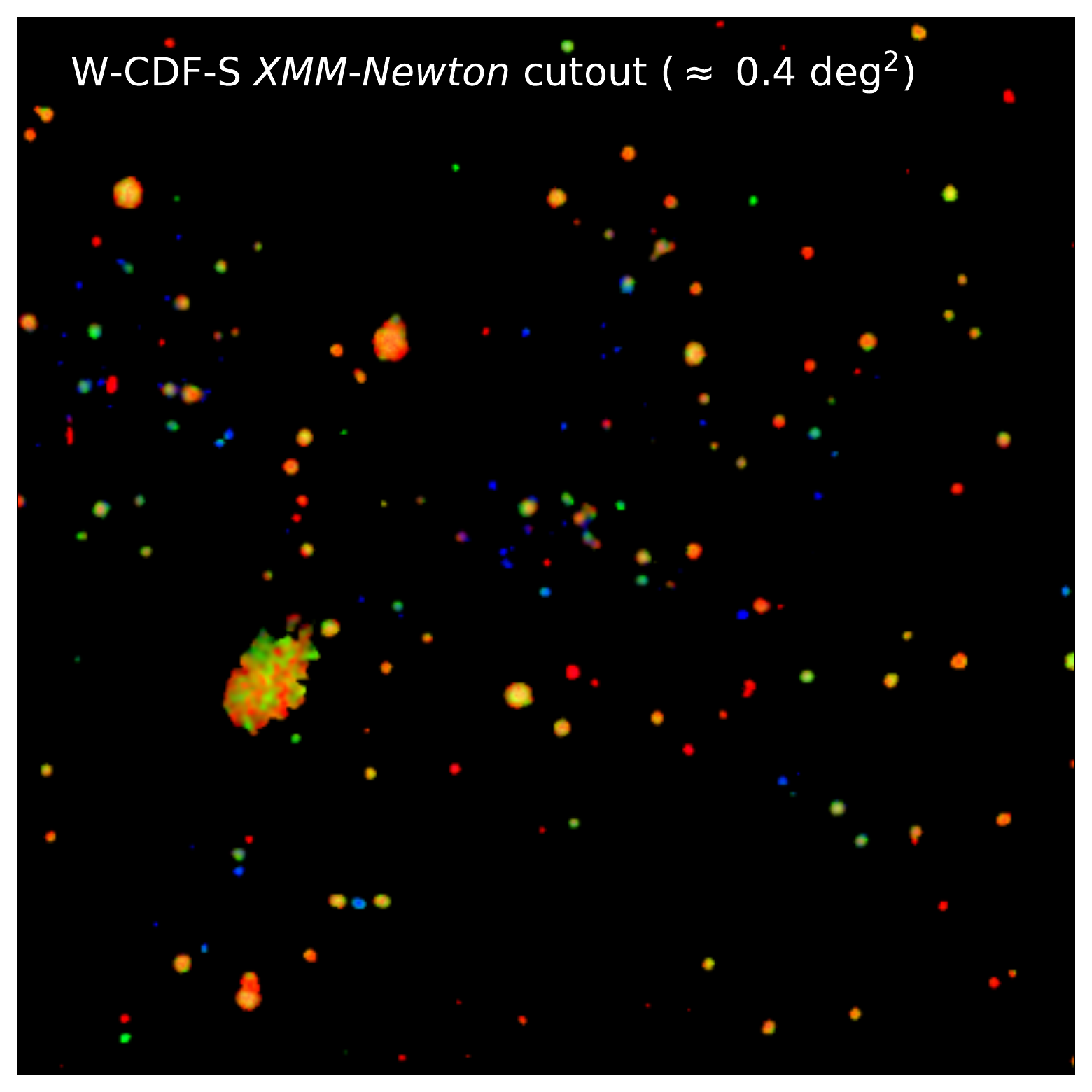}
~
\includegraphics[width = 0.49\textwidth]{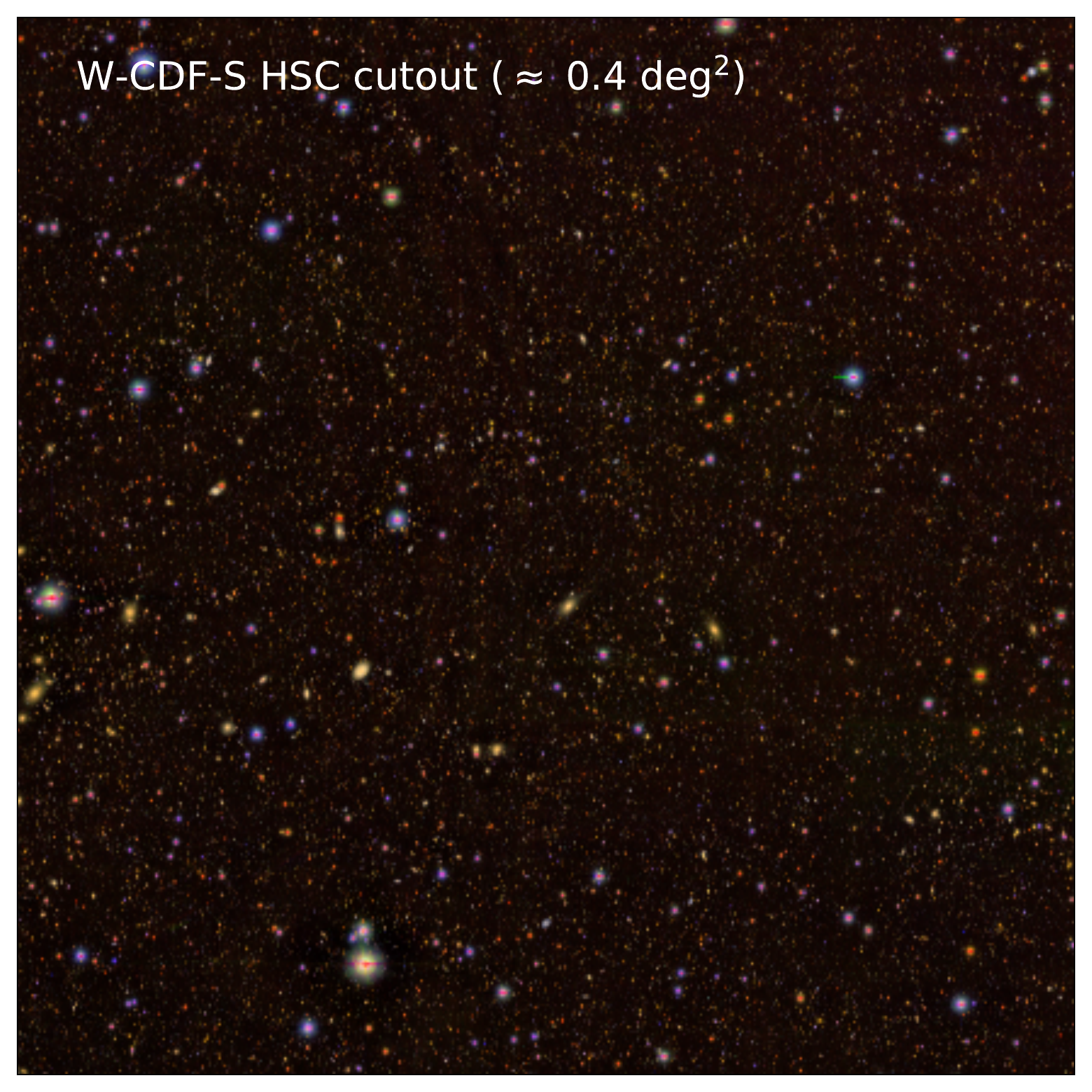}
\caption{\textit{Left:} Example ``false-color'' smoothed \xray\ image cutout of W-CDF-S with field size $\approx 0.4$ deg$^2$ and central RA $= 52.9^\circ$ and DEC $= -27.4^{\circ}$. Band 1+2, band 3, and band 4+5 are represented by the colors red, green, and blue, respectively. Redder sources are softer; bluer sources are harder. An asinh stretch is utilized.
The large extended source toward the lower left is an X-ray cluster at $z = 0.15$.
\textit{Right:} Example ``false-color'' optical image of W-CDF-S in the same sky area as that of the left panel. The $g$/$i$/$z$-band HSC image is represented by the color blue/green/red.}
\label{fig:smoimage}
\end{figure*}

Using the astrometry-corrected event files, we re-create images (see Figure~\ref{fig:smoimagefull} for the smoothed full-field mosaicked \xmm\ images for \wcdfs\ and \es, and Figure~\ref{fig:smoimage} for an example cutout of the smoothed mosaicked image in \wcdfs), exposure maps, detector masks, and background maps in five bands: band 1 (0.2--0.5 keV), band 2 (0.5--1.0 keV), band 3 (1.0--2.0 keV), band 4 (2.0--4.5 keV), and band 5 (4.5--12 keV).
We define the full band as bands 1--5 (0.2--12 keV), soft band as bands 1--3 (0.2--2 keV), and hard band as bands 4--5 (2--12 keV).  Exposure maps and image mosaics are also created for the full/soft/hard band combining all the observations and instruments in the full/soft/hard band.
We then run source detection again with data products from bands 1, 2, 3, 4, and 5, combining all {\it XMM-Newton} observations together.
This five-band detection approach has been widely adopted in \xmm\ catalogs \citep[e.g.,][]{Rosen2016,Traulsen2019,Traulsen2020,Webb2020} since it improves the positional accuracy of sources detected compared to single-band detections.
When detecting sources in the full band (0.2--12 keV), we use bands 1--5 simultaneously; when detecting sources in the soft band (0.2--2 keV), we use bands 1--3 simultaneously; when detecting sources in the hard band (2--12 keV), we use bands 4--5 simultaneously.
As \texttt{emldetect} can only process a limited number of observations, we divide the W-CDF-S/ELAIS-S1 field into a grid when performing the second-pass source detection \citep[e.g.,][]{Chen2018}.
For each cell in the grid, we coadd the images and exposure maps for all observations inside the cell, and run \texttt{ewavelet} with a low detection threshold (4) in the full/soft/hard band.
The source list obtained from \texttt{ewavelet} is then utilized as the input for \texttt{emldetect} (only sources within the celestial coordinate range of the cell plus 1 arcmin ``padding'' on each side of the cell are kept).
The full/soft/hard-band source list from \texttt{emldetect} in each cell is then combined to remove duplications (sources in the ``padding'' area that do not have duplications within 10$''$ are kept).
For each band in each field, we select sources with detection likelihoods ({\sc det\_ml}) larger than the threshold that corresponds to a 1\% spurious fraction according to simulations (see Section~\ref{ss-detmlsim} for details).

\subsection{Simulations to assess catalog reliability} \label{ss-detmlsim}

Similar to \cite{Chen2018}, we perform Monte Carlo simulations of the X-ray observations in \wcdfs\ and \es\ to assess the reliability of the source catalogs.
For each simulation, we generate mock X-ray sources using the \citet{Kim2007} log $N$-log $S$ relations.
The minimum simulated flux is set to be 0.5 dex lower than the minimum detected flux; the maximum flux is set to be $10^{-11}$~erg~cm$^{-2}$~s$^{-1}$.
We then use {\sc CDFS-SIM}\footnote{
\url{https://github.com/piero-ranalli/cdfs-sim}} to convert fluxes to PN/MOS count rates, and place sources at random sky positions, thus creating mock event files.
The images are extracted in the same manner as the real ones.
The background is simulated by re-sampling the original background map according to a Poisson distribution. 
A total of 10 simulations are created for each energy band. 
The same two-stage source-detection procedures are performed on the simulated data; the detected sources are matched to input sources within a $10^{\prime\prime}$ cut-off radius by minimizing the quantity $R^2$:
\begin{equation}
\label{eq:rsq}
R^2 = \big(\frac{\Delta{\rm RA}}{\sigma_{\rm RA}} \big)^2 +  \big(\frac{\Delta{\rm DEC}}{\sigma_{\rm DEC}} \big)^2 + 
 \big(\frac{\Delta{\rm RATE}}{\sigma_{\rm RATE}} \big)^2,
\end{equation}
where $\Delta{\rm RA}$/$\Delta{\rm DEC}$/$\Delta{\rm RATE}$ is the difference between the RA/DEC/count rate of matched detected sources and input sources; 
$\sigma_{\rm RA}$/$\sigma_{\rm DEC}$/$\sigma_{\rm RATE}$ is the uncertainty of the detected sources in RA/DEC/count rate. 
Detected sources without any input sources within $10^{\prime\prime}$ are considered to be spurious detections. 

\begin{figure*}
\centering   
\includegraphics[width=0.47\textwidth]{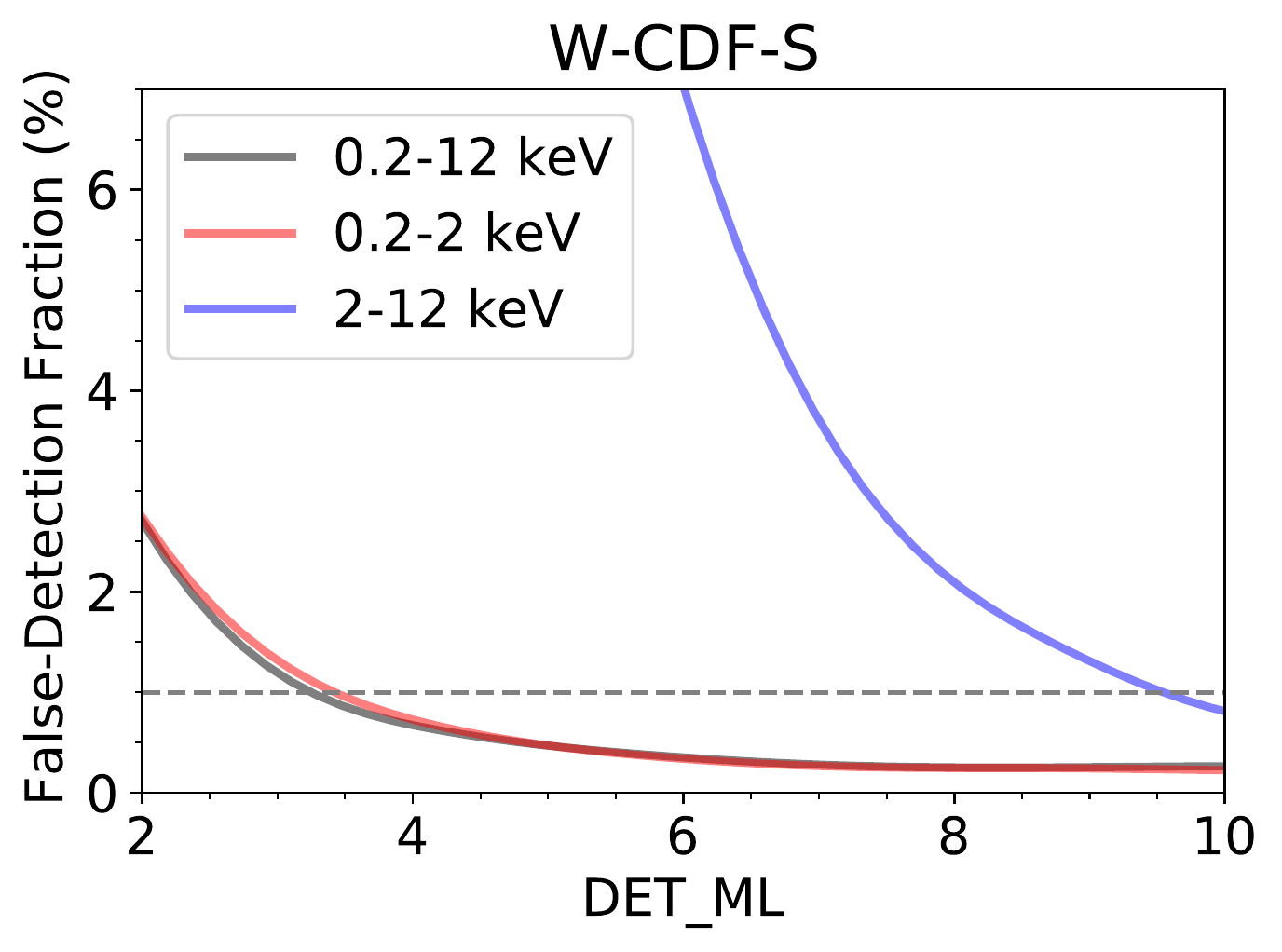}
~
\includegraphics[width=0.48\textwidth]{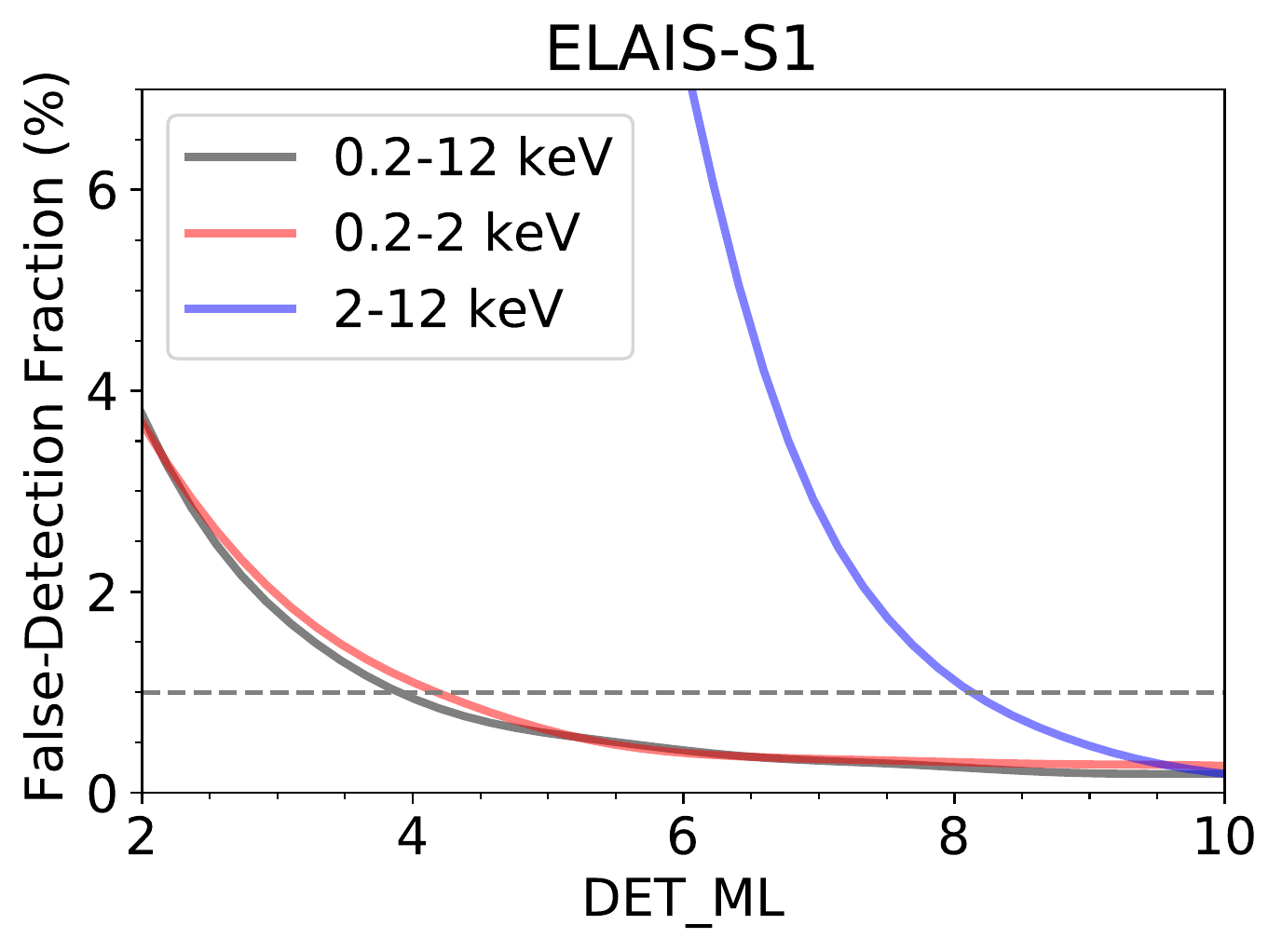}
\caption{
{\it Left}: The fraction of spurious sources as a function of {\sc det\_ml} threshold based on simulations. The horizontal dashed line marks a spurious fraction of 1\%, which determines the {\sc det\_ml} thresholds adopted for \wcdfs.
{\it Right}: Similar to the left panel, but for the \es\ field. 
}
\label{fig:detml}
\end{figure*}

The left/right panel of Figure~\ref{fig:detml} presents the average spurious fraction ($f_{\rm spurious}$) as a function of {\sc det\_ml} in the full/soft/hard band for the \wcdfs/\es\ field obtained from the simulations we ran.
To achieve $f_{\rm spurious}$ $\lesssim 1\%$ for the \wcdfs\ field, a {\sc det\_ml} threshold of $\approx 3.5/3.5/9.5$ is needed for the full/soft/hard band.
For the \es\ field, a {\sc det\_ml} threshold of $\approx 4.0/4.0/8.0$ is required for the full/soft/hard band.
In the soft band, the background levels are similar for the \wcdfs\ and \es\ fields. Thus, due to the slightly larger amount of exposure time in \wcdfs\ than \es, the {\sc det\_ml} threshold in the soft band for the \wcdfs\ field is slightly smaller than that for the \es\ field. In the hard band, the background level for the \wcdfs\ field is higher compared to the \es\ field. Thus, the {\sc det\_ml} threshold in the hard band for the \wcdfs\ field is larger than that for the \es\ field. The source signal in the full band is typically dominated by the source signal from the soft band, so that the {\sc det\_ml} threshold in the full band is close to that in the soft band.

\subsection{Astrometric accuracy}\label{ss:poserr}

\begin{figure}
\centering
\includegraphics[width=0.45\textwidth]{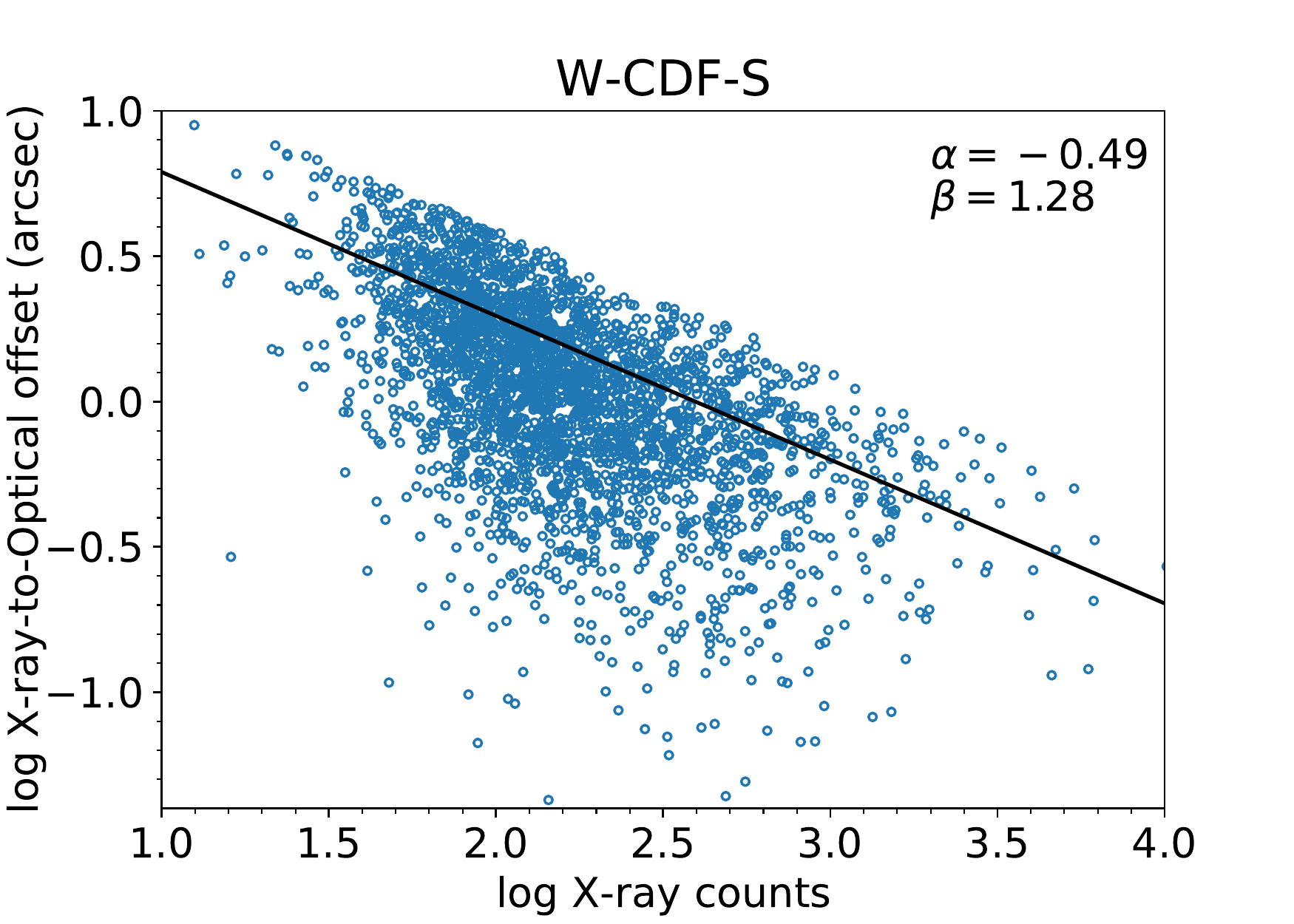}
\includegraphics[width=0.45\textwidth]{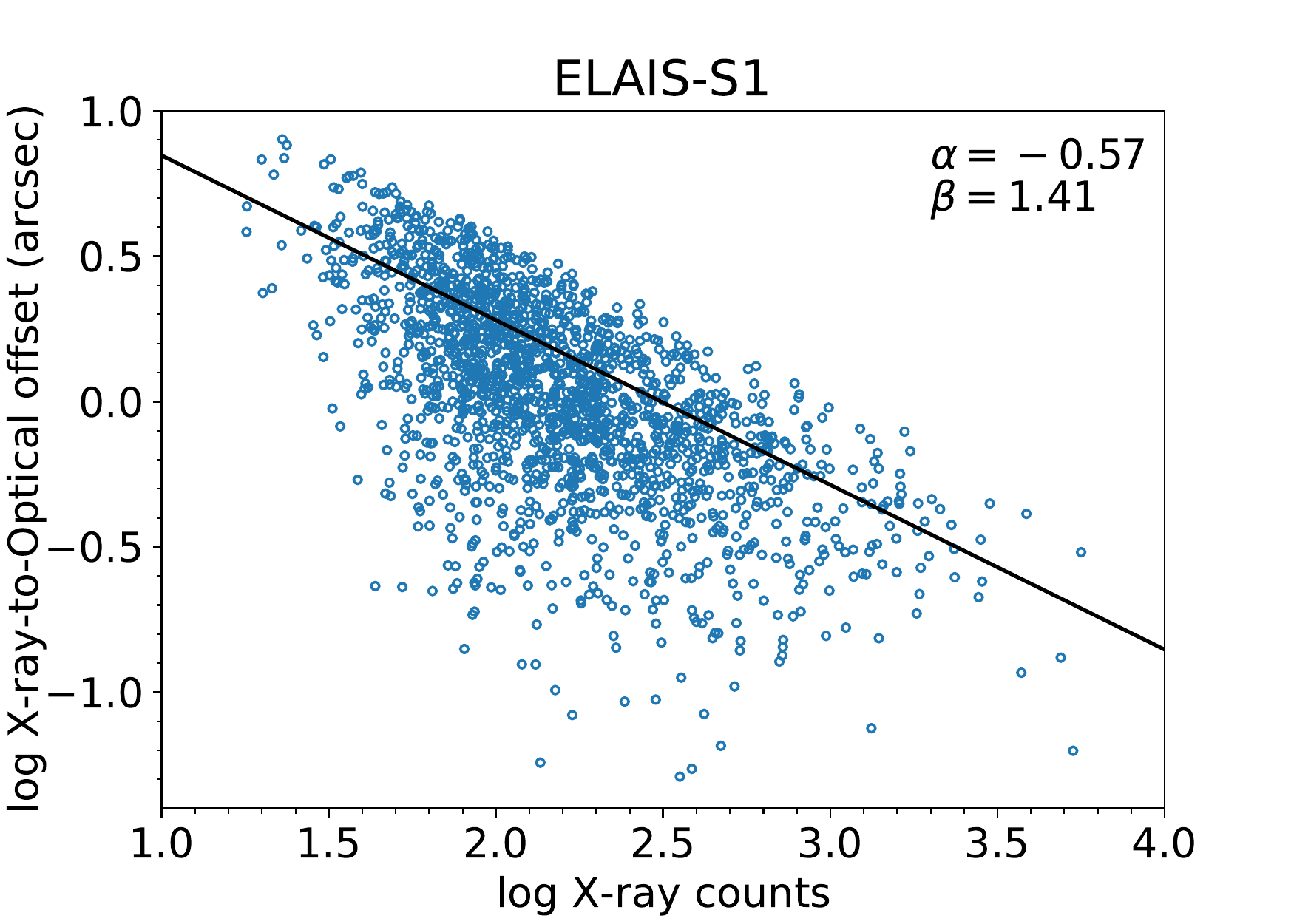}
\caption{The positional offsets between detected \xray\ sources in the full band and their matched optical counterparts vs. the full-band X-ray source counts number ($C$) in \wcdfs\ ({\it top}) and \es\ ({\it bottom}). The derived relation between $\log_{10} r_{68\%}$ and $C$ is marked as the black solid line.}
\label{fig:logr68c}
\end{figure}

To estimate the positional accuracy of the detected \xmm\ sources in the full/soft/hard band, we first matched the sources with optical catalogs.
As described in \citet{Chen2018}, directly matching \xray\ sources to optical counterparts can be associated with a relatively high false-match rate ($\approx 18\%$). 
We therefore chose \texttt{NWAY} (\citealt{Salvato2018}; see Section~\ref{s-mc} for a basic description of \texttt{NWAY}) to match \xmm\ sources with optical/NIR counterparts with priors as described in Section~\ref{s-mc} within 10$''$, using an iterative method.
In the NIR, we use {\it Spitzer} data from the DeepDrill data release \citep{Lacy2021} that includes the SERVS data \citep{Mauduit2012}, and VISTA data from the VIDEO data release in 2020 (M.\ Jarvis et al., private communication) for both the \wcdfs\ and \es\ fields. In the optical, we use HSC data from \citet{Ni2019} for \wcdfs\ and DES DR2 data \citep{Abbott2021} for \es\ (see Table~\ref{tab:wcdfsmw} for the survey descriptions). 
Since a small fraction of \xray\ sources in the \wcdfs\ field lack HSC coverage (see Figure~\ref{fig:loc}), we add DES DR2 sources \citep{Abbott2021} in the \wcdfs\ field that have no HSC counterpart within 1 arcsec to the HSC catalog; this also provides optical coverage in the saturated regions of the HSC image.
In the first iteration, we adopt the quadrature combination of the positional uncertainty derived from \texttt{emldetect} ($\sigma_{\rm eml}$) and a constant 0.5 arcsec systematic uncertainty as the positional uncertainty of \xmm\ sources ($\sigma_x$).
The positional uncertainties adopted for optical/NIR sources are listed in Table~\ref{tab:matching}.
We then select all the X-ray sources in \wcdfs/\es\ with HSC/DES counterparts that have \pany\ $> 0.1$ (which is the threshold adopted in this work, corresponding to a false-match rate of $\sim 5\%$; see Section~\ref{ss-mmfr} and Figure~\ref{fig:mmsim}).\footnote{\pany\ is a parameter in the \texttt{NWAY} output, representing the probability for the source to have any counterpart.}
We also exclude $\approx 4\%$ X-ray sources and their matched optical counterparts that have positional offsets greater than 3$\sigma_x$ from the analysis.
We fit the separations between X-ray sources and optical sources as a linear function of source counts ($C$) in \wcdfs\ and \es, respectively,\footnote{The X-ray positional uncertainty is typically associated with both $C$ and the off-axis angle (see \citealt{Luo2017,Chen2018} for details). For the XMM-SERVS survey, most of the sources are detected in multiple observations, so that their effective average off-axis angles do not vary significantly. Thus, we only associate $\sigma_x$ with $C$ in this work. } and then adjust the intercept so that 68\% of the sources have positional offsets smaller than the expectation from the relation (see Figure~\ref{fig:logr68c} for the obtained relations in the full band).
The intercept and slope are taken as the parameters for the empirical relation between the 68\% positional-uncertainty radius ($r_{68\%}$) and the number of source counts:
\begin{equation}\label{eq:counts}
\log_{10} r_{68\%} = \alpha\times\log_{10} C + \beta.
\end{equation}
Following \citet{Chen2018}, we define $\sigma_x$ to be the same as the uncertainties in RA and DEC ($\sigma_{\rm RA} = \sigma_{\rm DEC} = \sigma_x$), so that $\sigma_x$ = $r_{68\%}$/1.515 (see \citealt{Pineau2017} for details).
With the updated $\sigma_x$, we run \texttt{NWAY} again, iterating until the $\alpha$ and $\beta$ values become stable.

\begin{figure}
\centering
\includegraphics[width=0.45\textwidth]{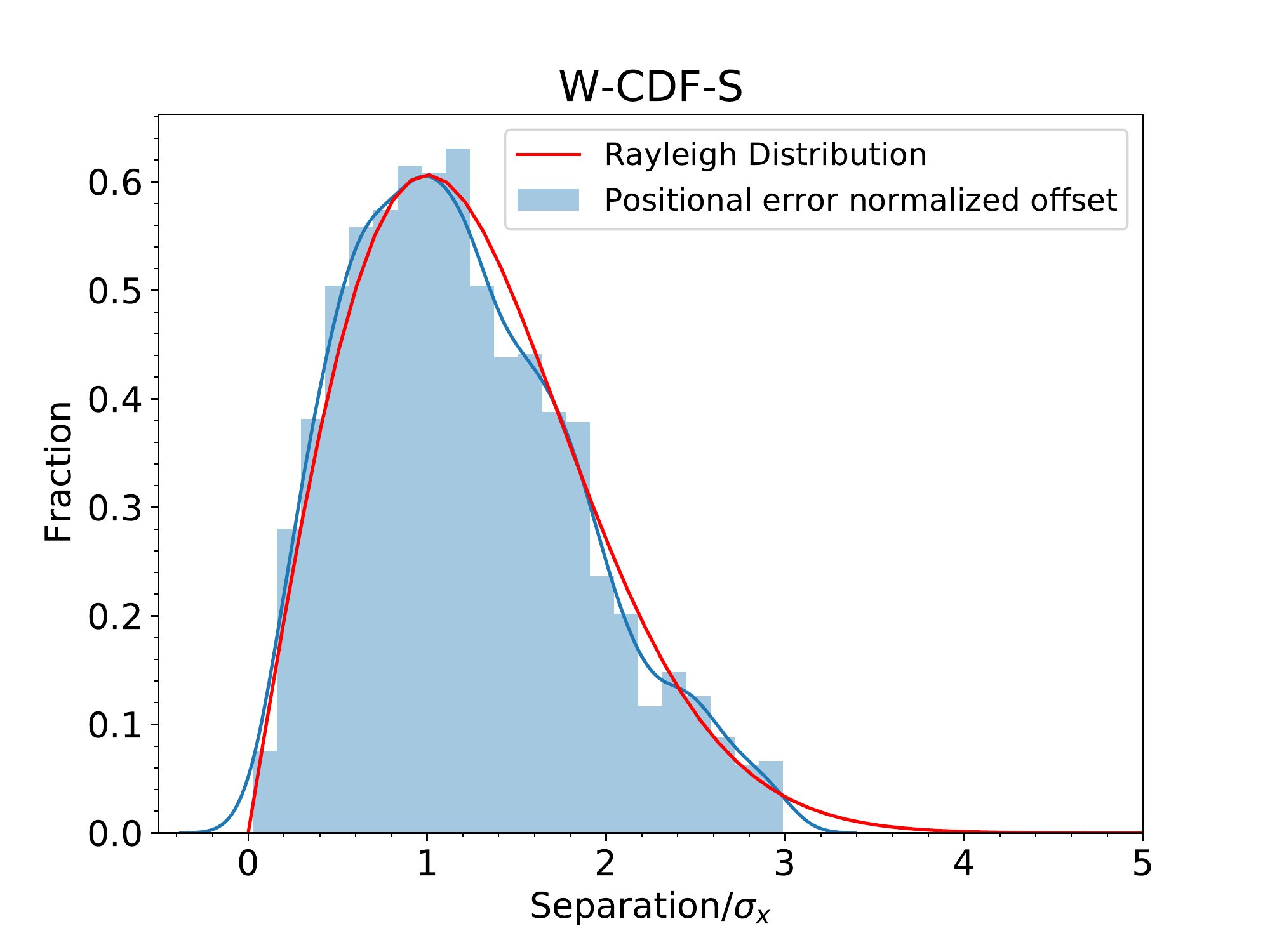}
\includegraphics[width=0.45\textwidth]{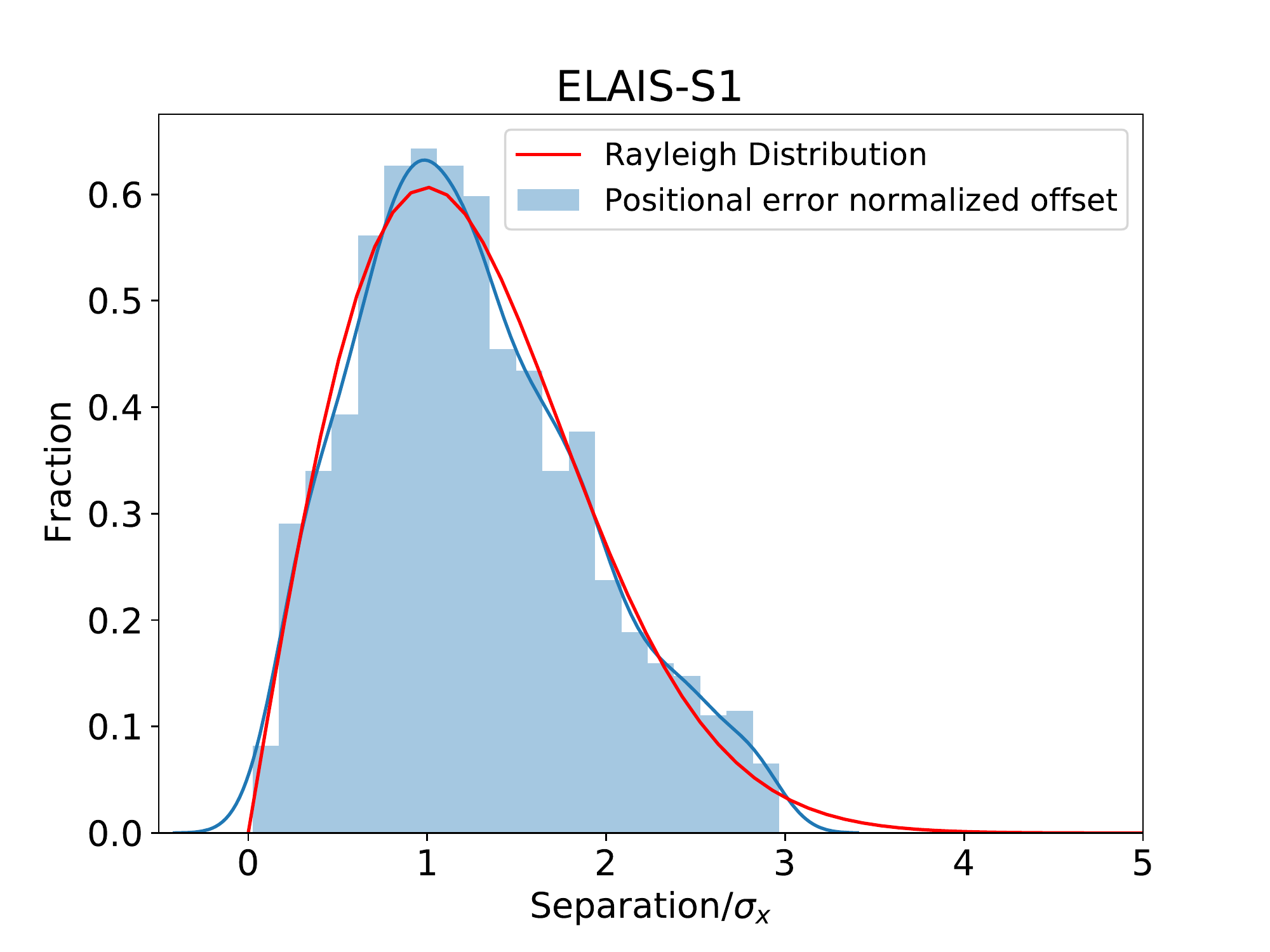}
\caption{Comparison between the distribution of the separations between the full-band \hbox{X-ray} sources and their optical counterparts divided by $\sigma_x$ and the expected Rayleigh distribution (solid red curve) in \wcdfs\ ({\it top}) and \es\ ({\it bottom}). 
The solid blue curve represents the kernel-density estimation of the normalized separation distribution.
The agreement between the two distributions indicates that our empirically derived $\sigma_x$ values are reliable.
}
\label{fig:rayleigh}
\end{figure}

The distribution of $\sigma_x$ can be roughly approximated as a normal distribution.
For the \wcdfs\ field, the average $\sigma_x$ in the full/soft/hard band is 1.15/1.25/1.10 arcsec, with a standard deviation of 0.46/0.51/0.31 arcsec.
For the \es\ field, the average $\sigma_x$ in the full/soft/hard band is 1.15/1.21/1.15 arcsec, with a standard deviation of 0.51/0.55/0.34 arcsec.
Since we assume $\sigma_{\rm RA} = \sigma_{\rm DEC} = \sigma_x$, the separation between X-ray sources and their optical counterparts should follow the Rayleigh distribution (with the scaling parameter $\sigma_x$).
The distribution of the normalized separation (Separation/$\sigma_x$) between the full-band \hbox{X-ray} sources and their optical counterparts is presented in Figure~\ref{fig:rayleigh}, along with the Rayleigh distribution. 
The good agreement between the distribution of separation/$\sigma_x$ and the Rayleigh distribution indicates that our empirically derived $\sigma_x$ values are reliable indicators of the true positional uncertainties.

\subsection{The X-ray source catalogs} \label{ss-sc}
\begin{figure*}  
\includegraphics[width = 0.57 \textwidth]{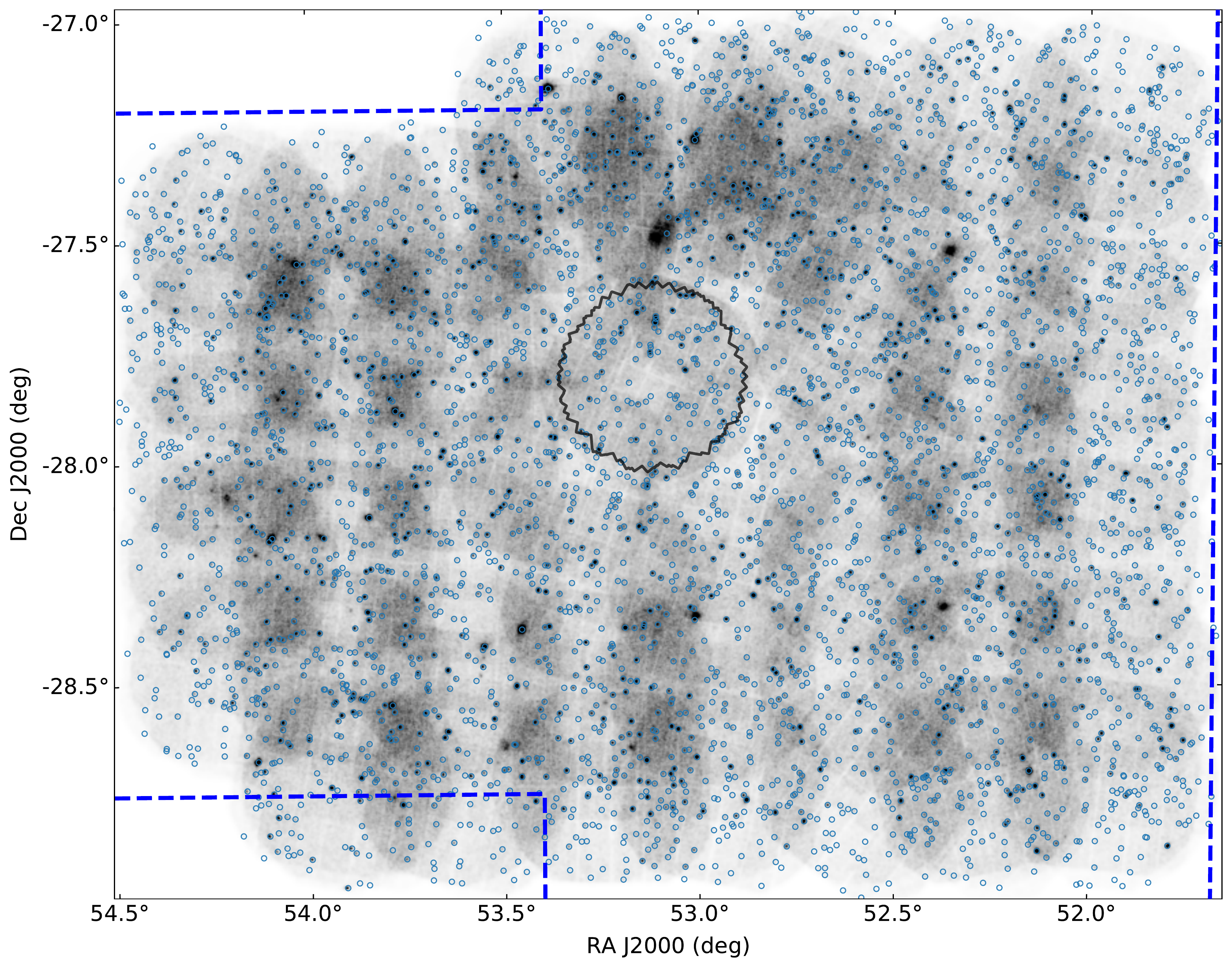}
\includegraphics[width = 0.413 \textwidth]{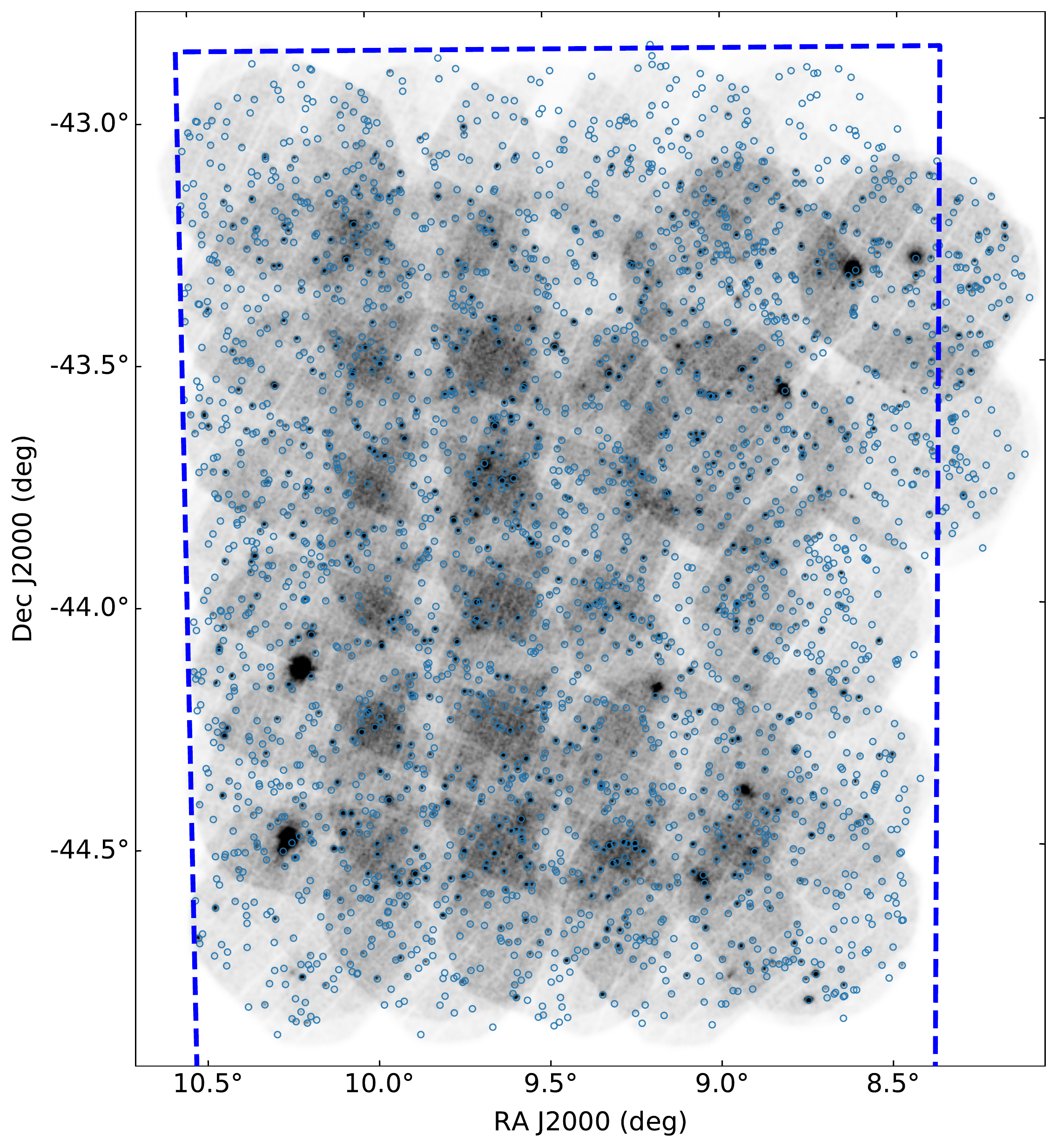}
\vspace{-0.2 cm}
\caption{\textit{Left:} Spatial distribution of the point-like \xray\ detected sources in the \wcdfs\ field (blue circles) projected on the smoothed full-band image. The blue dashed line encloses the region with forced optical-NIR photometry (Nyland et al.\ 2021). The black solid curve indicates the footprint of the 7~Ms CDF-S \citep{Luo2017}. \wcdfs\ is larger than the CDF-S by a factor of $\approx 34$ in solid angle. \textit{Right panel:} Spatial distribution of the point-like \xray\ detected sources in the \es\ field (blue circles) projected on the smoothed full-band image. The blue dashed line encloses the region with forced optical-NIR photometry \citep{Zou2021a}.}
\label{fig:sources}
\end{figure*}

We present the schema of the \xray\ source catalogs for the \wcdfs\ and \es\ fields in Appendix~\ref{a-column}.
With the {\sc det\_ml} thresholds derived in Section~\ref{ss-detmlsim}, we detect 3512/3672/1118 sources in the full/soft/hard band in the \wcdfs\ field, and 2328/2342/884 sources in the full/soft/hard band in the \es\ field.
These numbers only include point-like sources; sources that have $\geqslant 10$ improvements in the detection likelihood when detected as an extended source compared to the likelihood when detected as a point-like source are not included in our X-ray catalogs.
To combine sources detected in the three energy bands, we first need to identify sources that are detected in more than one band.
Two sources detected in different bands are considered to be the same if their angular separation is smaller than 10$^{\prime\prime}$, or the quadratic sum of the $99.73\%$ positional uncertainties from both bands.
Then, we add sources that are only detected in a single band to the source list.
We thus have a catalog of 4053/2630 unique point-like sources (see Figure~\ref{fig:sources} for the spatial distribution of sources) in the \wcdfs/\es\ field.
In the \wcdfs/\es\ field, a total of 2262/1407 sources have more than 100 PN+MOS counts in the full band; 139/78 sources have more than 1000 \hbox{X-ray} counts in the full band (see Figure~\ref{fig:fluxes} for the counts distribution).
For the \wcdfs\ field, $\approx 5/12/1\%$ of the sources are only detected in the full/soft/hard band; for the \es\ field, $\approx 5/10/1\%$ of the sources are only detected in the full/soft/hard band.
We have performed visual examinations to ensure that no obvious sources are missing from the catalogs, and that there are no obvious false matches between different bands.

\begin{figure*}
\centering
\gridline{\fig{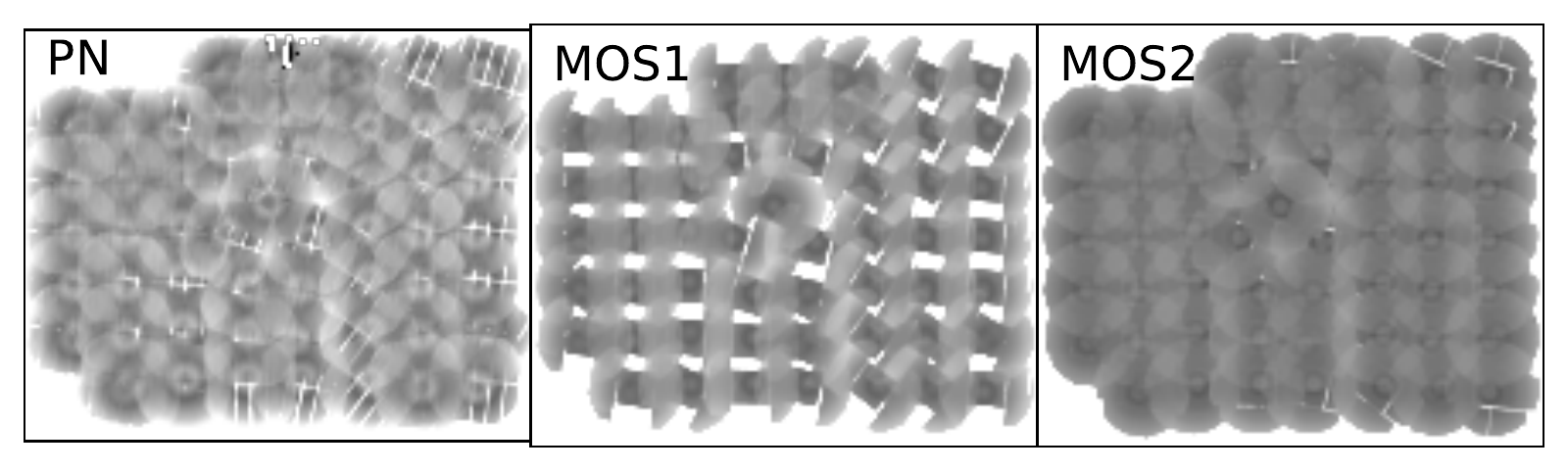}{0.53\textwidth}{}
          \fig{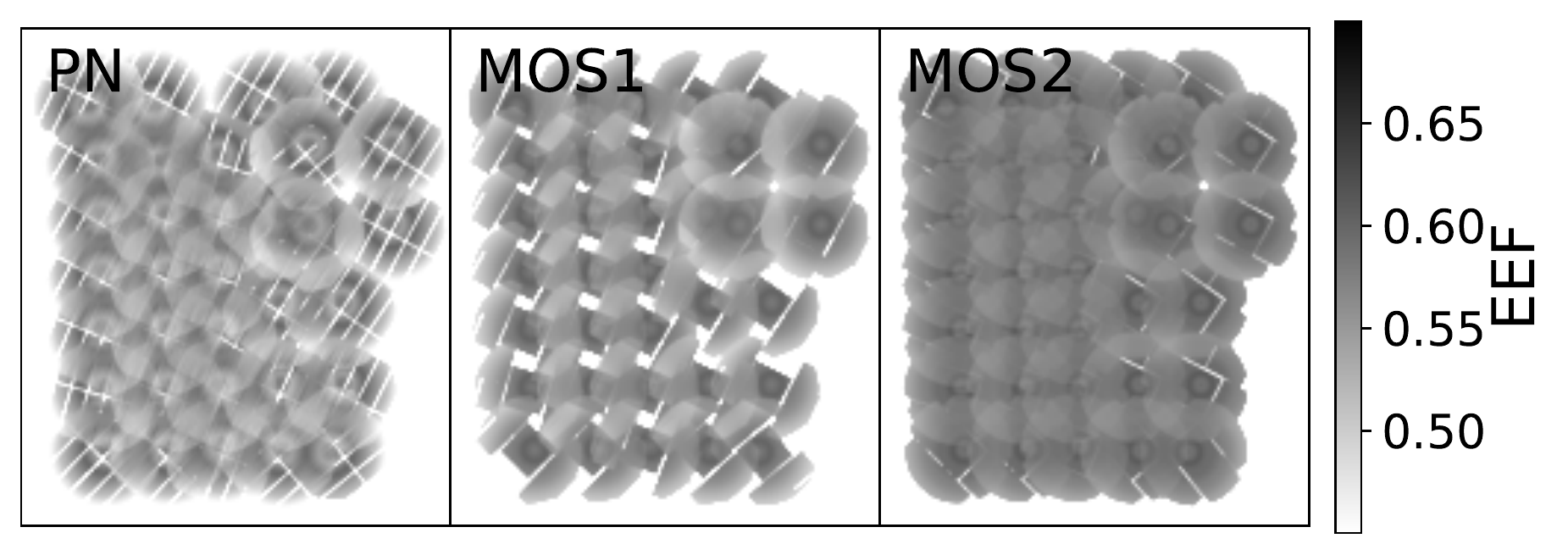}{0.45\textwidth}{}}
\vspace{-0.5 cm}
 \caption{Soft-band encircled energy fraction (in $5 \times\ 5$ pixels) maps for the three EPIC cameras in \wcdfs\ (\textit{left}) and \es\ (\textit{right}). The gaps in the EEF map of MOS1 are due to its lost CCDs.}
\label{fig:eefmap}
\end{figure*}

\begin{figure}   
\includegraphics[width=0.5\textwidth]{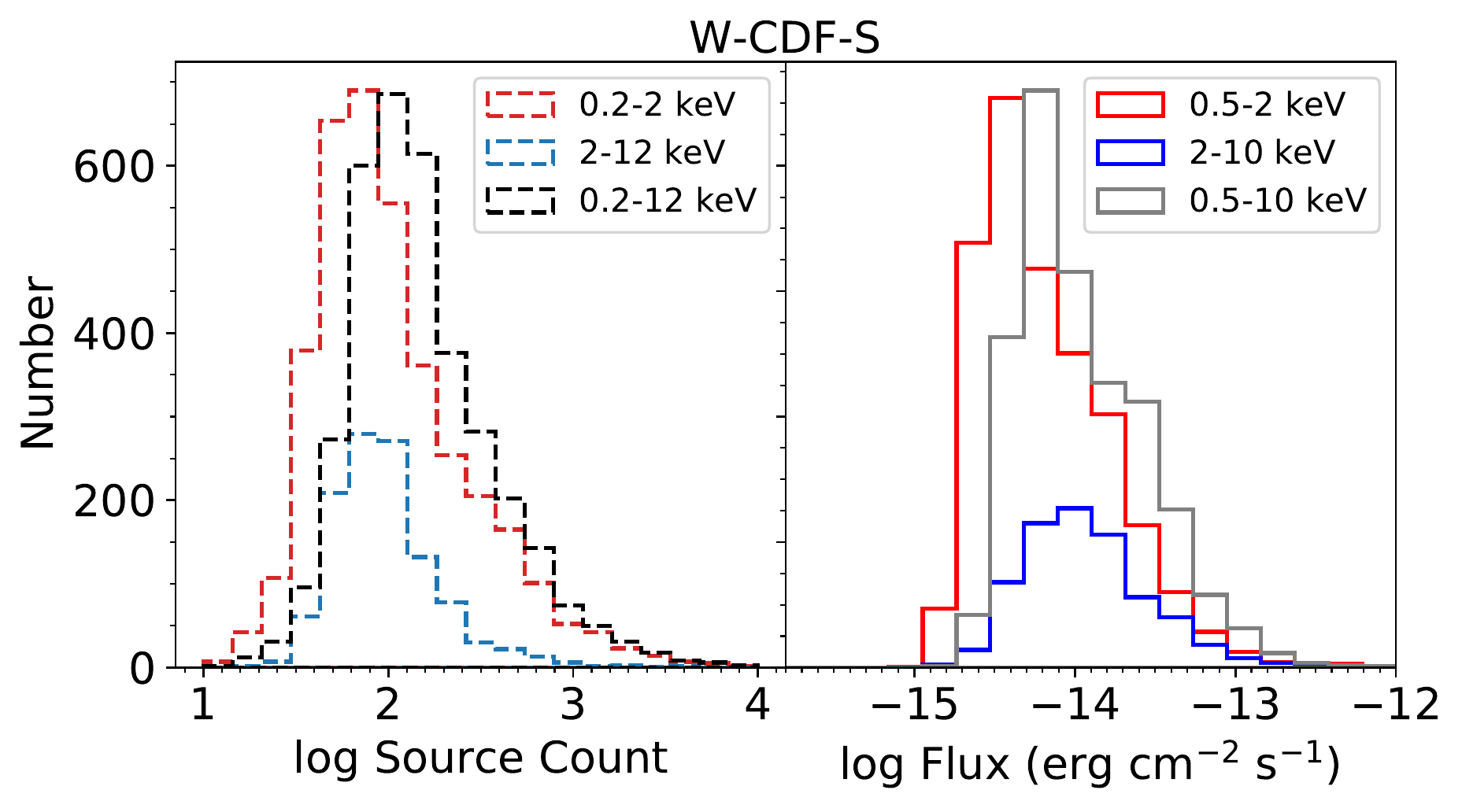}
\includegraphics[width=0.5\textwidth]{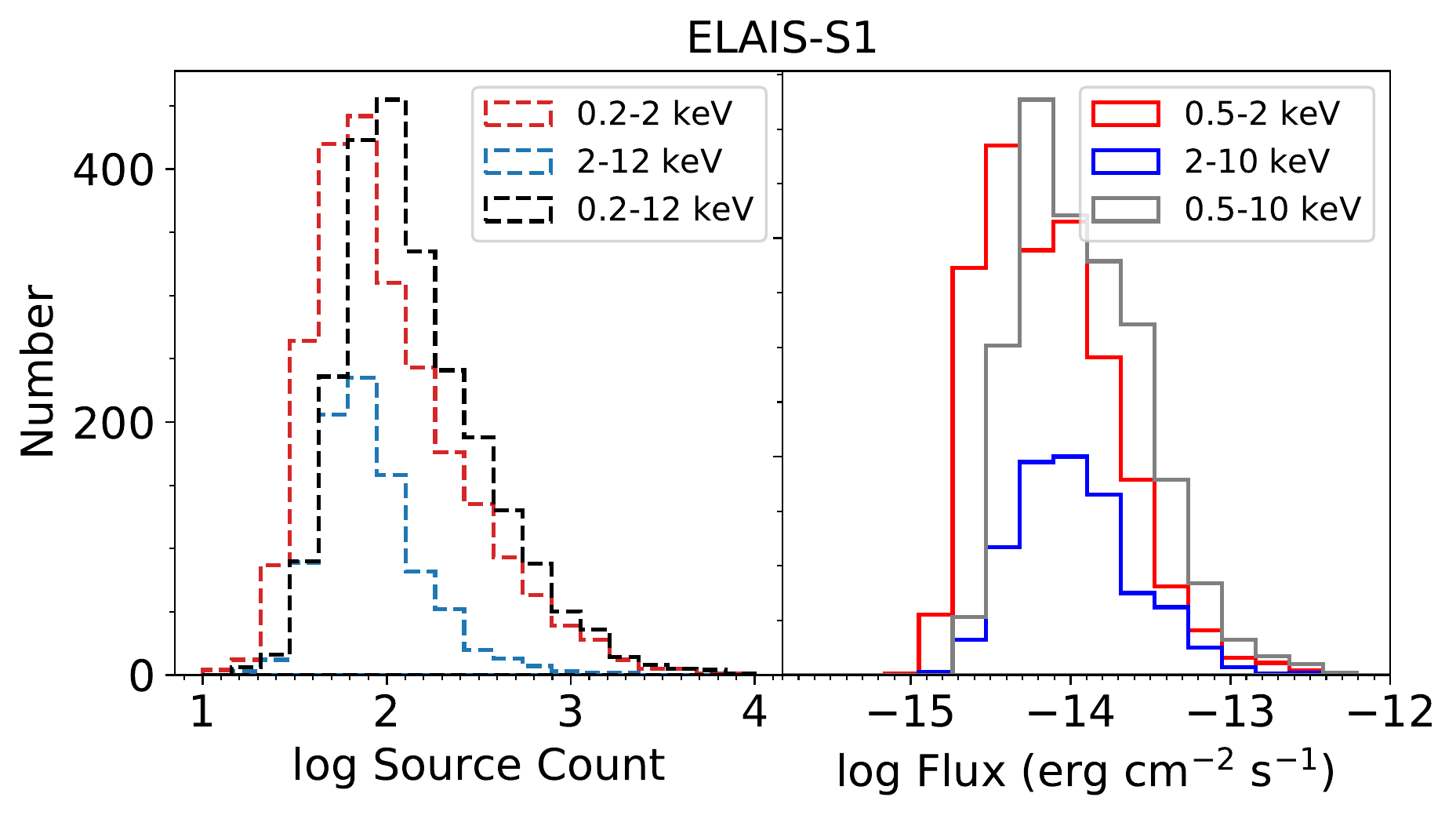}
\caption{
{\it Left panels:} The distributions of source counts in the soft (0.2--2 keV; red dashed), hard (2--12 keV; blue dashed), and full (0.2--12 keV; black dashed) bands.
{\it Right panels:} The distributions of fluxes in the 0.5--2 keV (red), 2--10 keV (blue), and 0.5--10 keV (gray) bands.
}
\label{fig:fluxes}
\end{figure}

When a source is not detected in all the bands, we estimated its count-rate upper limits in bands where the source is undetected.
The minimum required source counts ($m$) for a source to be detected with the \texttt{emldetect} detection threshold ($P_{\rm random}$; \texttt{det\_ml} $=  -~lnP_{\rm random}$) at a given number of background counts ($B$) can be estimated by solving the following regularized upper incomplete $\Gamma$ function \citep{Chen2018}:
\begin{equation}
\label{eq:invg}
P_{\rm Random} = \frac{1}{\Gamma(m)} \int_{B}^{\infty} t^{m-1} e^{-t} dt.
\end{equation}
Here, $B$ is estimated by summing the number of counts in $5 \times 5$ pixels centered at the source position in the mosaicked background map.
We note that the estimated $m$ corresponds to the Poisson detection likelihood of $P_{\rm Random}$, which is not necessarily equal to the detection likelihood from PSF fitting in {\sc emldetect}. However, as the PSF fitting likelihood follows a 1:1 relation with the Poisson likelihood in general \citep{Liu2020}, our estimation roughly holds.
With the estimated $m$, the count-rate upper limit is then calculated with the formula:
\begin{equation}
\label{eq:ctuplim}
{\rm RATE_{upper~limit}} = \frac{m - B}{t_{\rm exp}\times {\rm EEF}},
\end{equation}
where $t_{\rm exp}$ represents the exposure time at the source position, and the encircled energy fraction (EEF) value corresponding to the \hbox{5 $\times$ 5} pixels centered at the source position is obtained from the EEF map.
To derive the EEF map, we use \texttt{psfgen} to generate a series of PSF models for the three EPIC cameras, with different off-axis angles and different energies. 
These PSF models approximate the EEF as a function of the off-axis angle for different EPIC cameras at different energies. 
For each observation, an EEF map is generated for each EPIC camera. 
A mosaicked EEF map for different EPIC instruments at different energies is constructed (see Figure~\ref{fig:eefmap} for the soft-band EEF maps).
The EEF value adopted in Equation~\ref{eq:ctuplim} is the weighted EEF of EEF values at the source position for the three EPIC cameras, with the counts number in the band where the source is detected in each EPIC camera serving as the weight.
Similarly, as exposure times in different EPIC cameras vary, the $t_{\rm exp}$ adopted in Equation~\ref{eq:ctuplim} is the weighted $t_{\rm exp}$ (with the same weights as those utilized to calculate the weighted EEF).

To convert the count rate to flux, we derive the effective power-law photon indices, $\Gamma_{\rm eff}$ (or the upper/lower limits of the indices), for X-ray sources from the hard-to-soft band ratios (or the lower/upper limits of the band ratios), assuming a power law modified by Galactic absorption.
The band ratio is calculated as the ratio between the hard-band count rate and the soft-band count rate. 
The relation between the band ratio and $\Gamma_{\rm eff}$ is derived from the canned response files of EPIC cameras.\footnote{\href{https://www.cosmos.esa.int/web/xmm-Newton/epic-response-files}{https://www.cosmos.esa.int/web/xmm-Newton/epic-response-files}}
The soft/hard/full-band flux of the source is derived from the soft/hard/full-band count rate in each EPIC camera assuming a power-law spectrum with the derived $\Gamma_{\rm eff}$; the weighted mean of fluxes obtained from all available EPIC cameras (the ratio between the count rate and the count-rate error in each camera is utilized as the weight) is reported as the flux of the source.
For sources that are detected in the soft band but not in the hard band in \wcdfs\ and \es, we stack their hard-band counts at the source positions to derive a stacked $\Gamma_{\rm eff}$, which is $\approx 1.9$ in \wcdfs\ and $\approx 2.0$ in \es.
The stacking is performed by summing all the counts in 5 $\times$ 5 pixels of the image centered at the source position, minus all the counts in 5 $\times$ 5 pixels of the background map centered at the source position, and then dividing by the EEF.
Similarly, for sources that are detected in the hard band but not in the soft band, we stack their soft-band counts at the source positions to obtain a stacked $\Gamma_{\rm eff}$, which is $\approx 0.6$ for both \wcdfs\ and \es.
When the stacked $\Gamma_{\rm eff}$ value is consistent with the $\Gamma_{\rm eff}$ limit of a source, the stacked $\Gamma_{\rm eff}$ value is utilized to derive the flux; otherwise, the $\Gamma_{\rm eff}$ limit is utilized to derive the flux.
When a source is only detected in the full band, $\Gamma_{\rm eff} = 1.4$ (which is approximately the slope of the cosmic X-ray background spectrum; e.g., \citealt{Marshall1980}) is assumed to derive the flux.

The distributions of source counts in the soft, hard, and full bands and observed fluxes (i.e., fluxes only corrected for Galactic absorption) of the detected sources in the \hbox{0.5--2} keV, 2--10 keV, and 0.5--10 keV bands are displayed in Figure~\ref{fig:fluxes}; we present the observed fluxes in the \hbox{0.5--2}~keV, 2--10 keV, and 0.5--10 keV bands (calculated with the $\Gamma_{\rm eff}$ values derived in the previous paragraph) to enable direct comparisons with previous \xray\ surveys \citep[e.g.,][]{Cappelluti2009, Chen2018}.
The median observed fluxes of sources in the \wcdfs\ field detected in the \hbox{0.5--2}, \hbox{2--10}, \hbox{0.5--10} keV bands are 
$5.4\times10^{-15}$, $1.1\times10^{-14}$, and $9.0\times10^{-15}$ erg~cm$^{-2}$~s$^{-1}$, respectively.
The median observed fluxes of sources in the \es\ field detected in the \hbox{0.5--2}, \hbox{2--10}, \hbox{0.5--10} keV bands are
$6.6\times10^{-15}$, $1.0\times10^{-14}$, and $1.1\times10^{-14}$ erg~cm$^{-2}$~s$^{-1}$, respectively.

In Table~\ref{tab:compare}, we compare the solid angle and number of detected \xray\ sources for the whole XMM-SERVS survey with several other wide-field \xmm\ surveys, showing the legacy value of XMM-SERVS.

\begin{table}[htbp]
\begin{center}
\footnotesize
\caption{\label{tab:compare}
Comparison of selected wide-field \xmm\ surveys.}
\begin{tabular}{ccccccccccc} 
        \hline
Field   & Area      & Depth & Source & Reference \\
        & (deg$^2$) & (ks)  &  number &    \\
        \hline
XMM-SERVS & 13     &  30   &  11925 & -   \\
SXDS     &  1.14     &  40  &  1245 & \citet{Ueda2008} \\
XMM-COSMOS & 2     &  40   &  1887  & \citet{Cappelluti2009}\\
XMM-XXL-N &  25   &   10   &  14168 & \citet{Chiappetti2018} \\
Stripe 82X & 31.3     &  5   &  6181  & \citet{LaMassa2016} \\
\hline
\end{tabular}
\end{center}
Columns from left to right: survey field, solid-angle coverage, median \xmm\ PN depth  across the field (in ks), number of sources detected, and example reference for the survey.
\end{table}

\subsection{Survey sensitivity, sky coverage, and logN-logS} \label{ss-sens}

\begin{figure}
\includegraphics[width=0.5\textwidth]{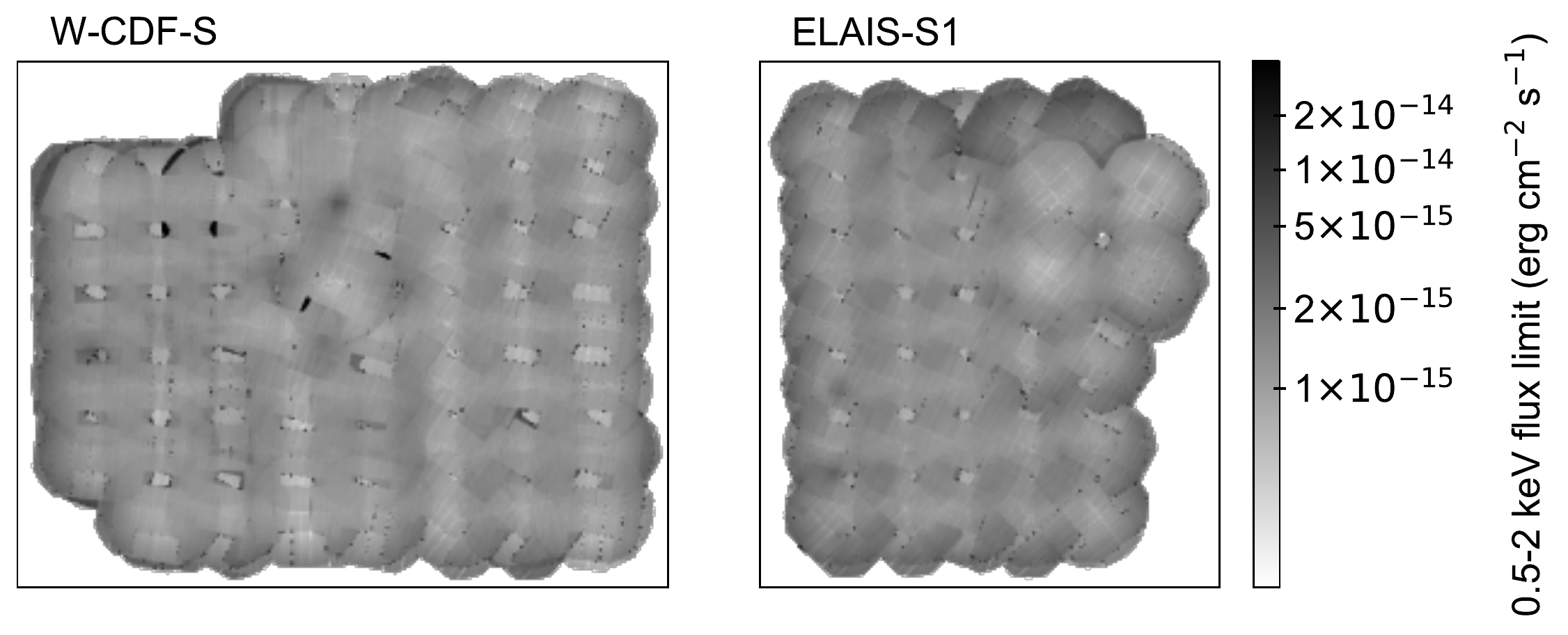}
\vspace{-0.5cm}
 \caption{0.5--2 keV band sensitivity maps in W-CDF-S (\textit{left}) and \es\ (\textit{right}).}
\label{fig:sensmap}
\end{figure}

\begin{figure}
\begin{center}    
\includegraphics[width=0.45\textwidth]{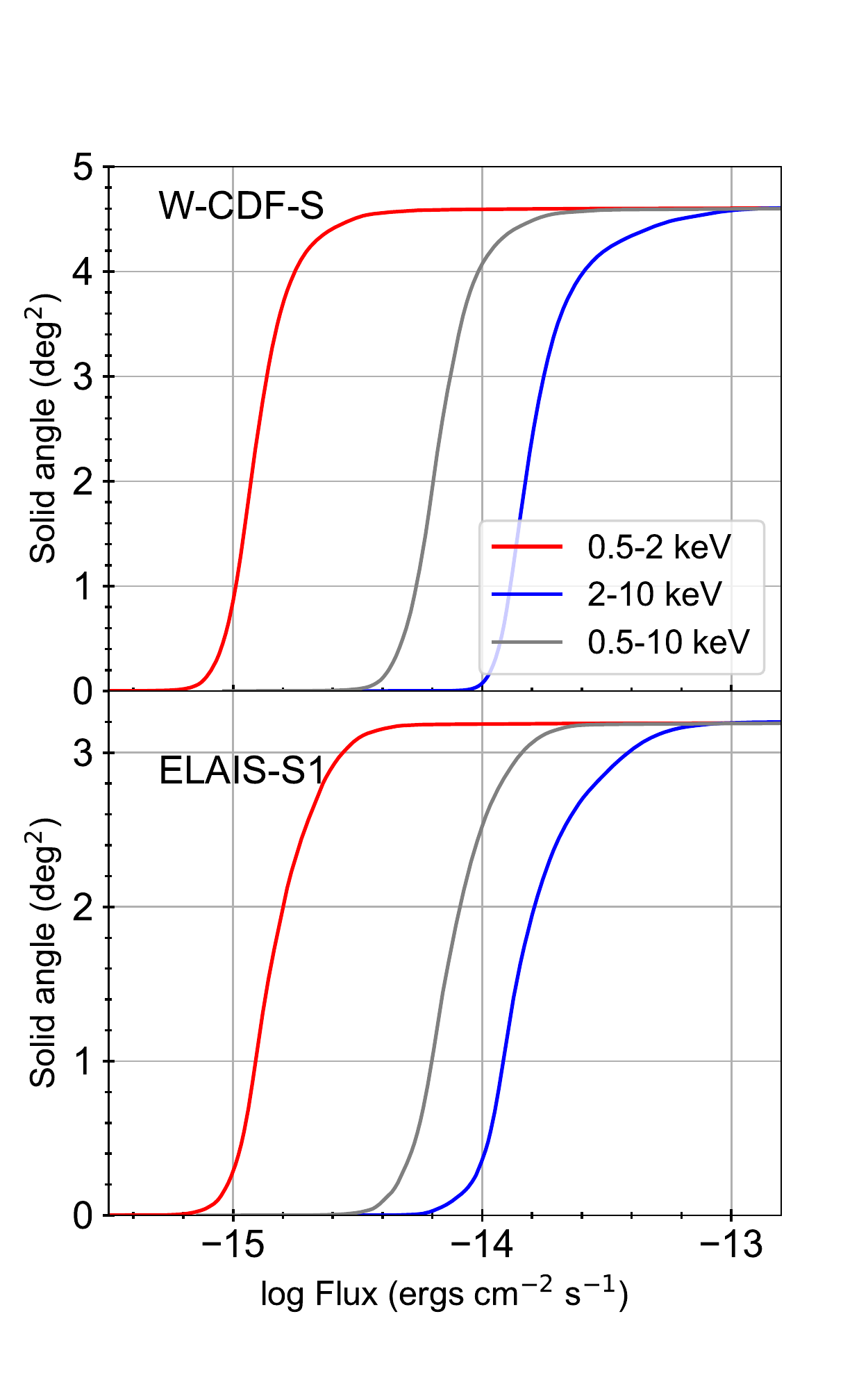}
\caption{
Sensitivity curves in the 0.5--2/2--10/0.5--10 keV band, calculated with the {\sc det\_ml} thresholds in Section~\ref{ss-detmlsim}.
}
\vspace{-0.5cm}
\label{fig:senscurve}
\end{center}   
\end{figure}

We create sensitivity maps in \wcdfs\ and \es\ in the 0.5--2, 2--10, and 0.5--10 keV bands following the methods in section~3.6 of \cite{Chen2018}.
We first bin the mosaicked background and exposure maps in the soft, hard, and full bands for each instrument by 3~$\times$~3 pixels (which is the bin size recommended by the XMM-SAS task \texttt{esensmap}).
For each pixel of the binned background map with a background counts number of $B$, the minimum required source counts ($m$) for a source to be detected with the \texttt{emldetect} detection threshold could be estimated from Equation~\ref{eq:invg}.
The sensitivity is calculated with the formula:
\begin{equation}
\label{eq:sens}
S = \frac{m - B}{t_{\rm exp}\times {\rm EEF} \times {\rm ECF}},
\end{equation}
where energy conversion factors (ECFs) for different bands and different EPIC cameras are derived assuming a power-law spectrum with photon index $\Gamma=1.4$ modified by Galactic absorption.
For \hbox{X-ray} sources in the \wcdfs\ field, the adopted ECF values for PN/MOS1/MOS2 are 8.57/2.27/2.28, 1.10/0.38/0.38, and 3.00/0.86/0.87 counts~s$^{-1}/10^{-11}$erg~cm$^{-2}$~s$^{-1}$, when converting count rates detected in the soft band to fluxes in the 0.5--2 keV band, count rates detected in the hard band to fluxes in the \hbox{2--10} keV band, and count rates detected in the full band to fluxes in the 0.5--10 keV band, respectively.
For \xray\ sources in the \es\ field, the adopted ECF values for PN/MOS1/MOS2 are 8.03/2.21/2.21, 1.10/0.38/0.38, and 2.78/0.82/0.83 counts~s$^{-1}/10^{-11}$erg~cm$^{-2}$~s$^{-1}$.
For each EPIC camera in the soft/hard/full band, we generate a map for the $\frac{1}{t_{\rm exp}\times {\rm EEF} \times {\rm ECF}}$ term in Equation~\ref{eq:sens}, and bin it by 3~$\times$~3 pixels.
As the effective area of PN is $\approx 2.5$ times the effective area of MOS1/MOS2, we combine the $\frac{1}{t_{\rm exp}\times {\rm EEF} \times {\rm ECF}}$ map of PN, MOS1, and MOS2 in each energy band with a weight of 2.5:1:1.
Multiplying this merged map with the $m - B$ value at each pixel, we obtain the sensitivity map at 0.5--2/2--10/0.5--10 keV (see Figure~\ref{fig:sensmap}).
Our survey in the \wcdfs\ field has flux limits of 
$1.9\times10^{-15}$, $2.9\times10^{-14}$, and $1.0\times10^{-14}$~erg~cm$^{-2}$~s$^{-1}$ over 90\% of its area in the 0.5--2, 2--10, and 0.5--10 keV bands, respectively.
Our survey in the \es\ field has flux limits of 
$2.5\times10^{-15}$, $3.2\times10^{-14}$, and $1.3\times10^{-14}$~erg~cm$^{-2}$~s$^{-1}$ over 90\% of its area in the 0.5--2, 2--10, and 0.5--10 keV bands, respectively.
The sensitivity curves corresponding to the {\sc det\_ml} threshold for the 0.5--2/2--10/0.5--10 keV bands are shown in Figure~\ref{fig:senscurve}.

\begin{figure*}
\includegraphics[width=1\textwidth]{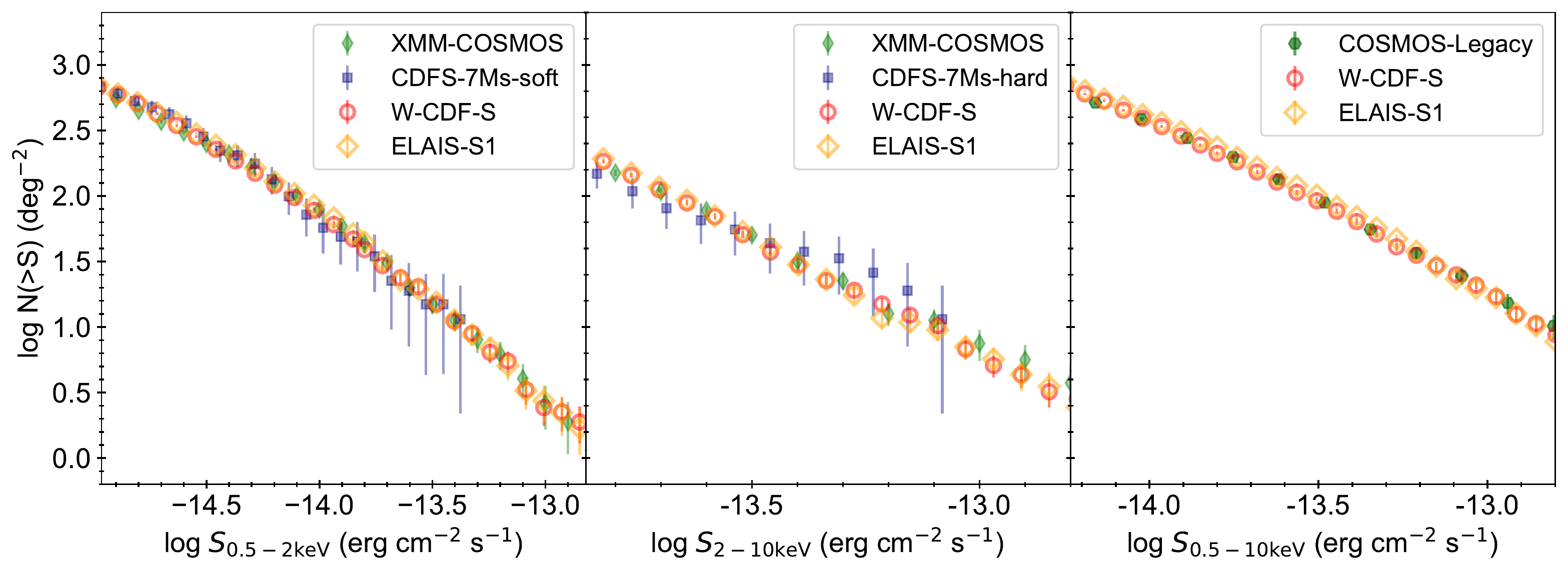}
\caption{
The $\log N - \log S$ relations for our catalogs in the 0.5--2 keV band (left), 2--10 keV band (middle), and 0.5--10 keV band (right). 
For comparison, $\log N - \log S$ relations from other \xray\ surveys are shown (CDF-S 7~Ms, \citealt{Luo2017}; XMM-COSMOS, \citealt{Cappelluti2009}; and COSMOS-Legacy, \citealt{Civano2016}).
The $\log N - \log S$ relations of our survey are generally consistent with those 
of previous studies.}
\label{fig:lognlogs}
\end{figure*}

Utilizing these sensitivity curves, we calculate the $\log N - \log S$ relations for our survey (see Figure~\ref{fig:lognlogs}). 
As can be seen in Figure~\ref{fig:lognlogs}, the $\log N - \log S$ relations in \wcdfs\ and \es\ are, in general, consistent with the relations reported in other studies (CDF-S 7~Ms, \citealt{Luo2017}; XMM-COSMOS, \citealt{Cappelluti2009}; COSMOS-Legacy, \citealt{Civano2016}; and Stripe 82X, \citealt{LaMassa2016}) within the measurement uncertainties. 

\section{Multiwavelength counterparts of X-ray sources} \label{s-mc}

To identify the multiwavelength counterparts for our \xray\ sources, we utilize the Bayesian catalog matching tool \texttt{NWAY} \citep{Salvato2018}, which adopts the distance and magnitude/color priors from multiple catalogs simultaneously to select the most probable counterpart, and allows for the absence of counterparts in some catalogs. \texttt{NWAY} has been widely utilized in matching \xmm\ sources to multiwavelength counterparts \citep[e.g.,][]{Salvato2018, Chen2018, Liu2020}.
Table~3 shows the optical/NIR catalogs utilized in this work. 
In Sections~\ref{ss-chandraprior} and \ref{ss-chooseprior}, we describe the magnitude/color priors utilized. 
In Section~\ref{ss-mmfr}, we present the quality of the matched optical/NIR counterparts.

\subsection{Obtaining priors from \chandra\ sources} \label{ss-chandraprior}

As can be seen from Column 5 of Table 3, it is typical for an \xmm\ source in our catalogs to have multiple optical/NIR sources located within the 99.73\% positional uncertainty ($r_{99\%}$). 
Thus, to compute the magnitude/color priors of the expected counterparts of our \xray\ sources, we make use of the \chandra\ counterparts of our \xmm\ sources within the E-CDF-S, \hbox{CDF-S}, and the original $\approx0.6$~deg$^2$ \es\ regions \citep{Feruglio2008,Xue2016,Luo2017}, along with other sources reported in the \chandra\ Source Catalog (CSC) 2.0 \citep{Evans2010}, as \chandra\ detections have better positional accuracy than \xmm\ detections.
We select \chandra\ sources that are uniquely matched to sources in our X-ray catalogs within the 95\% uncertainties (\chandra\ and \xmm\ positional uncertainties are added in quadrature; the positional uncertainties of the \chandra\ sources are taken from the relevant \chandra\ catalogs). This approach ensures that the \chandra\ sources utilized to obtain the priors have similar flux levels as \xmm\ sources in our catalogs. A total of 264/275 \xmm\ sources are matched to a unique \chandra\ counterpart in \wcdfs/\es. The fluxes and effective power-law indices of matched \chandra\ sources are in agreement with \xmm\ sources. Only a small fraction ($\approx 4\%$) of \xmm\ sources have $>1$ \chandra\ counterpart within the 95\% positional uncertainties.

We search for optical/NIR counterparts within 5$''$ of these \chandra\ sources with \texttt{NWAY}, utilizing the magnitude priors in the $i$, $K_s$, and IRAC 3.6$\mu m$ bands generated from the ``AUTO'' method. 
We select only reliable counterparts with \pany\ $>$ 0.8 (which corresponds to a false-positive fraction of $\approx$  5\% for \chandra\ sources; the false-positive fraction is estimated by matching fake \xray\ sources with optical/NIR counterparts; see Section~\ref{ss-mmfr} for the methods).

As expected from the {spectral energy distributions (SEDs) of AGNs, the matched counterparts of \chandra\ sources occupy a different space in the IRAC [3.6] $-$ [4.5] vs. IRAC [3.6] $+$ [4.5] plane compared with other sources in the DeepDrill catalog (see Figure~\ref{fig:cscservs}). A color and magnitude prior in the NIR has been widely used in the multiwavelength counterpart matching of \xray\ sources (e.g.,\ \citealt{Chen2018}; \citealt{Liu2020}).
Similar to the approach described in \citet{Liu2020}, we pixelate the IRAC [3.6] $-$ [4.5] vs. IRAC [3.6] $+$ [4.5] space into 50 $\times$ 50 pixels, and use a 2D 
Gaussian kernel estimate to generate the prior (``IRAC 2D prior'' hereafter) for the counterparts of X-ray sources in our survey based on the positions of matched \chandra\ sources/other   sources in the DeepDrill catalog on the IRAC [3.6] $-$ [4.5] vs. IRAC [3.6] $+$ [4.5] plane.
We also compute the 1D IRAC [3.6] $-$ [4.5] and IRAC [3.6] + [4.5] priors utilizing a Gaussian kernel estimate; we compute the magnitude prior for the IRAC 3.6$\mu $m band solely as well (see Figure~\ref{fig:cscdistrib}).\footnote{While in Figure~\ref{fig:cscdistrib}, DeepDrill sources with/without \chandra\ counterparts do not seem to have greatly different IRAC [3.6] $-$ [4.5] colors, we note that the peaks of the IRAC [3.6] $-$ [4.5] probability density distributions among these two groups of sources have a difference of $\sim 0.4$ mag, which is roughly consistent with expectation (e.g., see Figure~1 of  \citealt{Stern2005}).}
We do not create VIDEO and HSC (or DES) color priors as done above for the IRAC color, because this action would introduce a bias against type II AGN (just as for most of the VIDEO, HSC, and DES sources that do not have an X-ray counterpart, type II AGNs are typically dominated by host-galaxy light in the optical; e.g., \citealt{Liu2020}). 
We do use a Gaussian kernel estimate to obtain magnitude priors for the HSC/DES $i$ band and VIDEO $K_s$ band (see Figure~\ref{fig:cscdistrib}).

\begin{figure}
\centering
\includegraphics[width=0.45\textwidth]{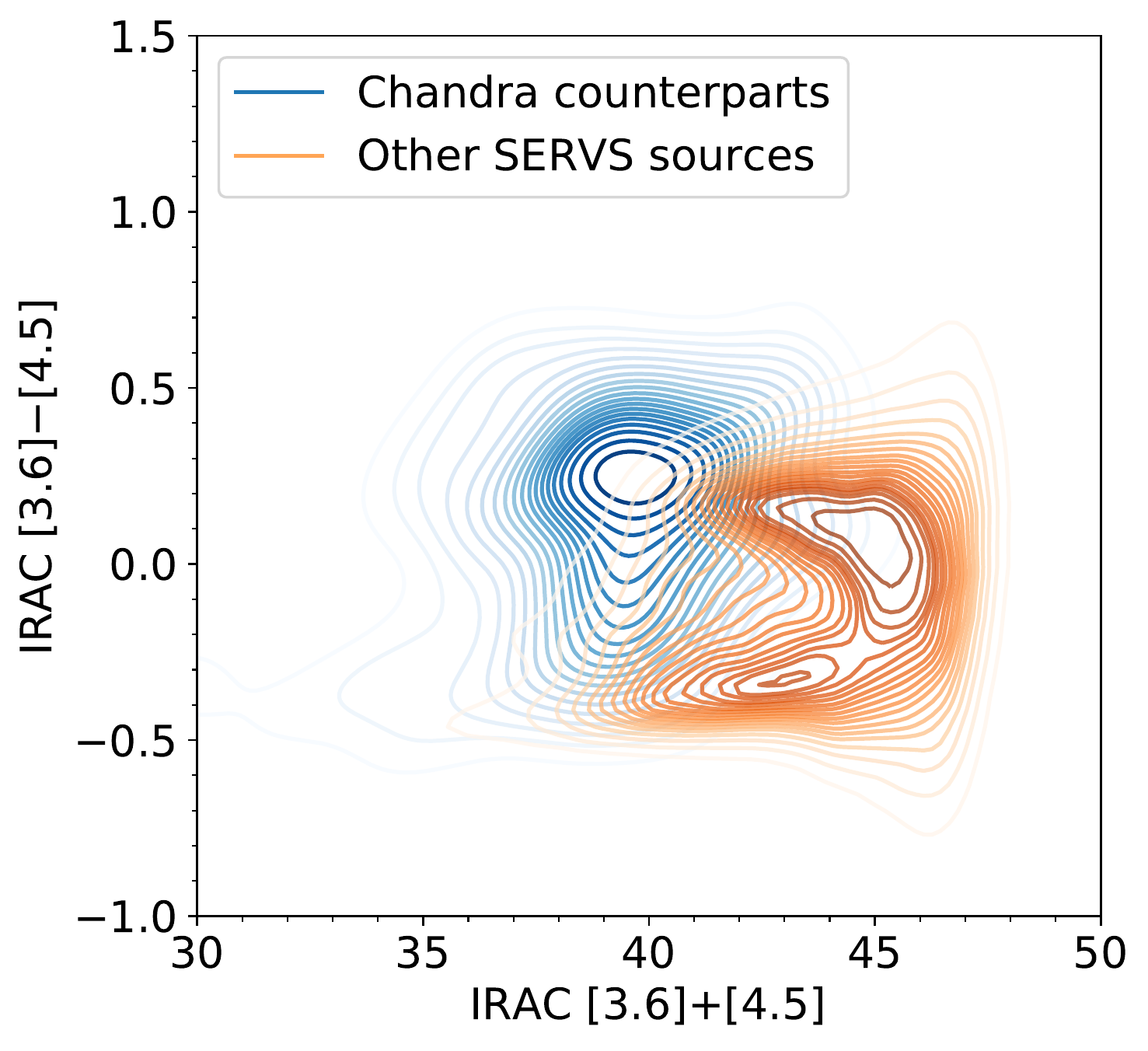}
\caption{
The IRAC [3.6] $-$ [4.5] vs. IRAC [3.6] $+$ [4.5] distribution of sources in the DeepDrill catalog that are matched to \chandra\ sources (blue contours), and the remaining DeepDrill sources (orange contours).
The distribution of DeepDrill sources that have \chandra\ counterparts shows noticeable differences compared to other DeepDrill sources.
}
\label{fig:cscservs}
\end{figure}

\begin{figure*}
\includegraphics[width=1\textwidth]{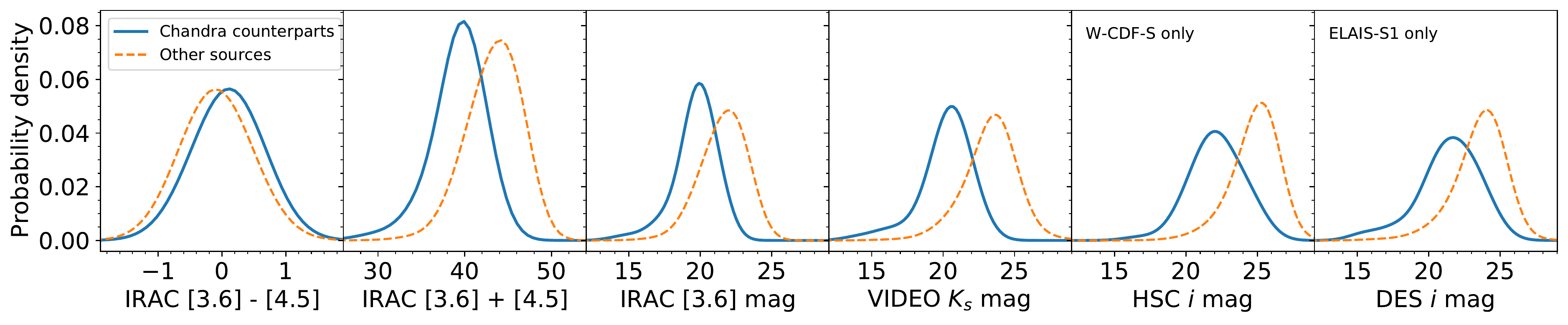}
\caption{
Kernel-density estimations of the IRAC 3.6 $\mu $m $-$ IRAC 4.5 $\mu $m color,  IRAC 3.6 $\mu $m $+$ IRAC 4.5 $\mu $m magnitude, IRAC 3.6~$\mu $m magnitude, VIDEO $K_s$-band magnitude, and HSC/DES $i$-band magnitude distributions of the expected counterparts of X-ray sources in W-CDF-S and/or ELAIS-S1 (blue solid line) and the unmatched optical/NIR sources in the field (orange dashed line).}
\label{fig:cscdistrib}
\end{figure*}

\begin{table*}
\begin{center}
    \caption{\label{tab:matching}
    Summary of multiwavelength counterpart matching results.
    }
    \vspace{-0.3 cm}
    \begin{tabular}{ccccccccccc}
        \hline
        Catalog & 
        Limiting Magnitude & 
        Area &
        $\sigma$ &
        $\overline{N_{99\%}}$ & 
        $f_{\rm matched}$ & 
        $f_{\rm reliable}$ & 
        $f_{\rm FP}$  &
        $f_{\rm AP}$ &
        False Rate &
        Identical Fraction 
       \\
        & 
        & 
        (deg$^2$) & 
        ($''$) &
        & 
        & 
        &
        & (Simulation)
        & (Simulation)
        & ({\it Chandra}) 
       \\
        (1) & (2) & (3) & (4) & (5) & (6) & (7) & (8) & (9) & (10) & (11) 
       \\
        \hline
        \multicolumn{11}{c}{W-CDF-S}\\
        \hline
        DeepDrill                     & $3.6\mu$m $< 23.1$ &  4.6  & 0.5 & 1.2 &
        89.6\% &
        85.2\% &   
        18.4\%  &
        81.0\% & 
        4.8\% &
         96.9\%\\
        VIDEO                     & $K_s < 23.8 $       & 4.5  & 0.3 & 2.0 &
        88.3\%   &
        82.7\% & 

        18.9\%  & 
        78.8\% &   
        5.5\% &
         92.2\%\\
        HSC$^a$                   & $i < 25.8$         &  4.6 & 0.1 &  2.3 &
        95.5\% &  
        86.1\%    & 
        20.8\%  
        &  82.8\% &
        4.6\% & 
        92.6\%\\
        \hline
        Summary & - & - &  -   & -
        & 100\%
        & 88.8\%   
        & 22.3\% & 
          - &
          - &
         91.9\% \\
        \hline
         \multicolumn{11}{c}{ELAIS-S1}\\
        \hline
        DeepDrill                     & $3.6\mu$m $< 23.1$ & 3.2 & 0.5 & 1.2 &
        90.5\% &
        84.0\% &   
        19.2\%  &
        80.4\% & 
        4.9\% &
         97.8\%\\
        VIDEO                     & $K_s < 23.8 $       & 3.0  & 0.3 & 2.1 &
        85.1\%   &
        76.3\% & 

        19.5\%  & 
        71.0\% &   
        7.9\% &
         96.0\%\\
        DES                   & $i < 24.6$         & 3.2  & 0.15 &  1.5 &
        88.4\% &  
        80.8\%    & 
        20.2\%  
        &  76.5\% &
        6.3\% & 
        97.0\%\\
        \hline
        Summary & - & - &  -   & -
        & 100\%
        & 87.0\%   
        & 22.4\% & 
          - &
          - &
         95.2\% \\
        \hline
    \end{tabular}
    \end{center}
    \vspace{0.1cm}
    Col. 1: Catalog name. 
    Col. 2: Magnitude limit. 
    Col. 3: Survey area in the XMM-SERVS survey region (4.6 deg$^2$ for \wcdfs\ and 3.2 deg$^2$ for \es).
    Col. 4: Positional uncertainty adopted for sources in this optical/NIR catalog.
    Col. 5: Average number of sources in this optical/NIR catalog within the 99.73\% positional uncertainty ($r_{99\%}$) of the \hbox{X-ray} sources.
    Col. 6: Percentage of \hbox{X-ray} sources with at least one counterpart in this optical/NIR catalog within the 10\arcsec\ search radius. 
    Col. 7: Percentage of \hbox{X-ray} sources matched with the optical/NIR catalog  that have $p_{\rm any} > 0.1$, which we considered to be reliable matches.
    Col. 8: The fraction of false-positive matches with the optical/NIR catalog among the mock ``isolated population'' with a $p_{\rm any}$ threshold of 0.1.
    Col. 9: The fraction of \hbox{X-ray} sources in the ``associated population'' estimated based on simulations. 
    Col. 10: False-matching rates for \xray\ sources with $p_{\rm any} > 0.1$ estimated from simulations.
    Col. 11: Fraction of the \hbox{X-ray} sources that have identical matching results with the optical/IR catalog when utilizing \chandra\ or \xmm\ positions.\\
In the summary row, col. 6 represents the percentage of \hbox{X-ray} sources that have at least one of the DeepDrill, VIDEO, or HSC counterparts; col. 7 lists the total percentage of \hbox{X-ray} sources that have $p_{\rm any} > 0.1$; col. 8 represents the total fraction of false-positive matches among the mock ``isolated population''; col. 11 contains the fraction of \hbox{X-ray} sources that have identical matched counterparts in all optical and NIR catalogs utilizing \chandra\ or \xmm\ positions.\\
Notes: (a). In a small fraction of the \wcdfs\ area without HSC coverage (see Figure~\ref{fig:loc}), we add DES sources \citep{Abbott2021} to the HSC catalog.
\end{table*}

\subsection{Choosing the priors when performing source matching} \label{ss-chooseprior}

Utilizing different combinations of the priors described above, we run \texttt{NWAY} with a maximum distance of 10$''$ to match detected X-ray sources with the optical/NIR catalogs listed in Table~\ref{tab:matching}.
We also generate mock X-ray sources that are 30$''$ away from any real
X-ray sources with \texttt{NWAY}, thus assessing the false-positive fraction of X-ray sources that should not have counterparts (when different combinations of priors are adopted). 
This false-positive fraction is significantly larger than the expected false rate for the matched counterparts of \xray\ sources in the catalog, as most of the actual \xray\ sources in our catalog are expected to have optical/NIR counterparts (see Section~\ref{ss-mmfr} for details).

The completeness for real X-ray sources, and the false-positive fraction among the mock X-ray sources as a function of adopted $p_{\rm any}$ threshold when different combinations of priors are utilized are presented in Figure~\ref{fig:pany}. 
We also compare the false-positive fraction directly with the completeness when the $p_{\rm any}$ threshold varies.
At a given false-positive fraction, combining the following priors: IRAC 2D, $i$-band mag, and $K_s$-band mag, yields the highest completeness; at a given completeness, adopting these priors produces the lowest false-positive fraction.
Thus, we match \xmm\ sources with these priors.\footnote{We have tested that adding additional magnitude priors from the available optical/NIR bands does not improve the results materially.}
The percentages of \xmm\ sources that are matched to each optical/NIR catalog are listed in Table~\ref{tab:matching}, Column 6.

\begin{figure*}
\centering
\includegraphics[width=0.48\textwidth]{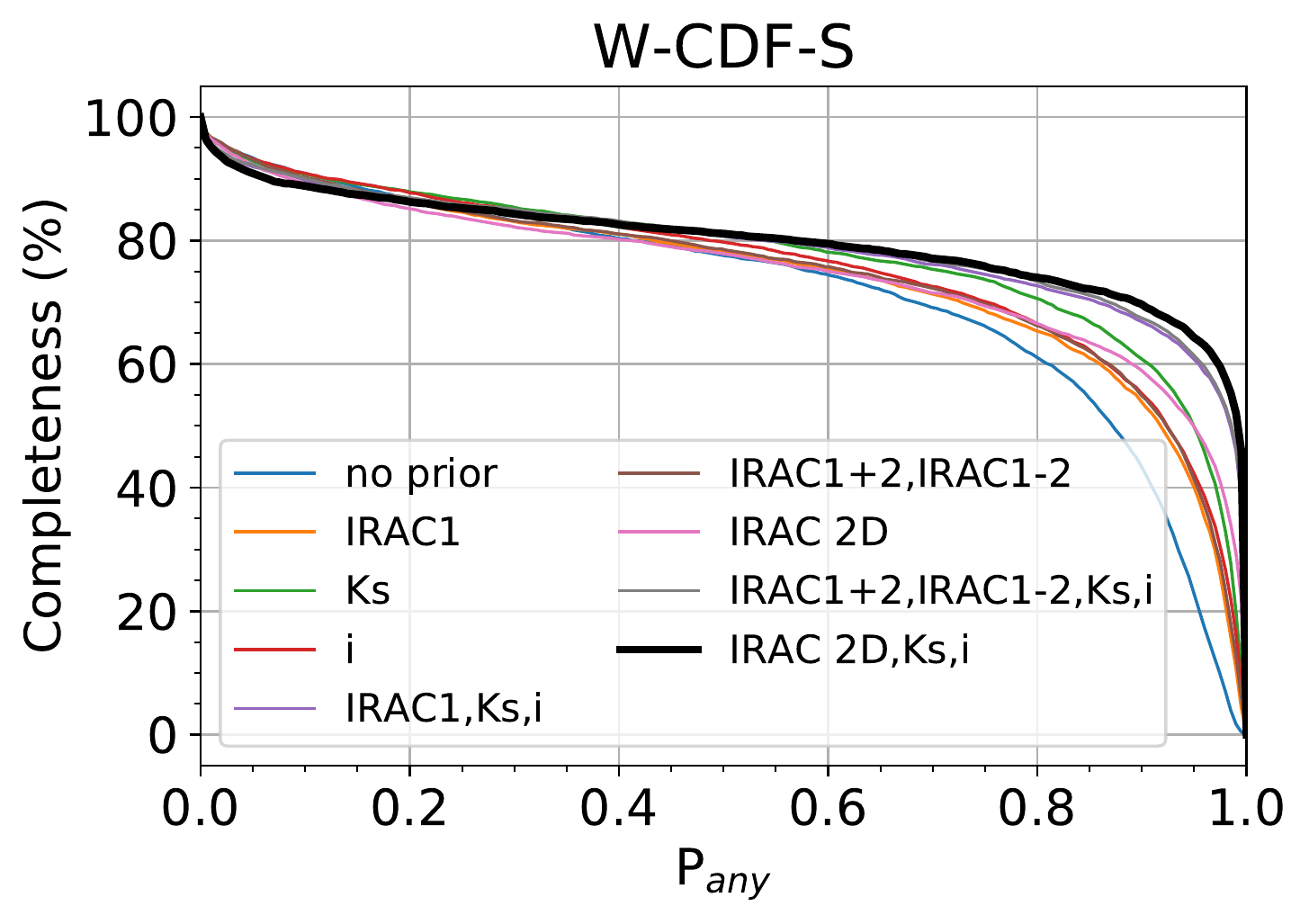}
~
\includegraphics[width=0.48\textwidth]{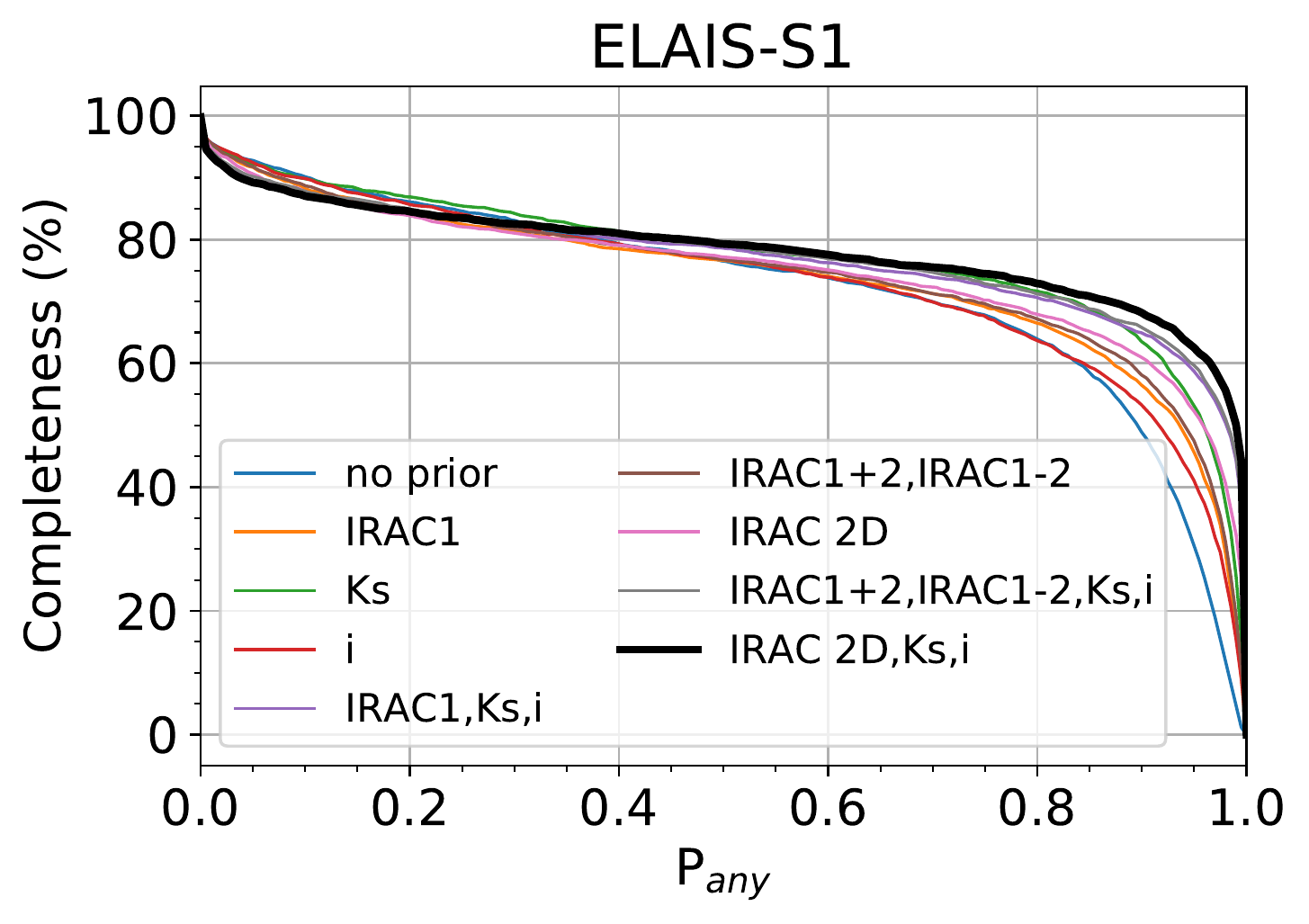}
\includegraphics[width=0.48\textwidth]{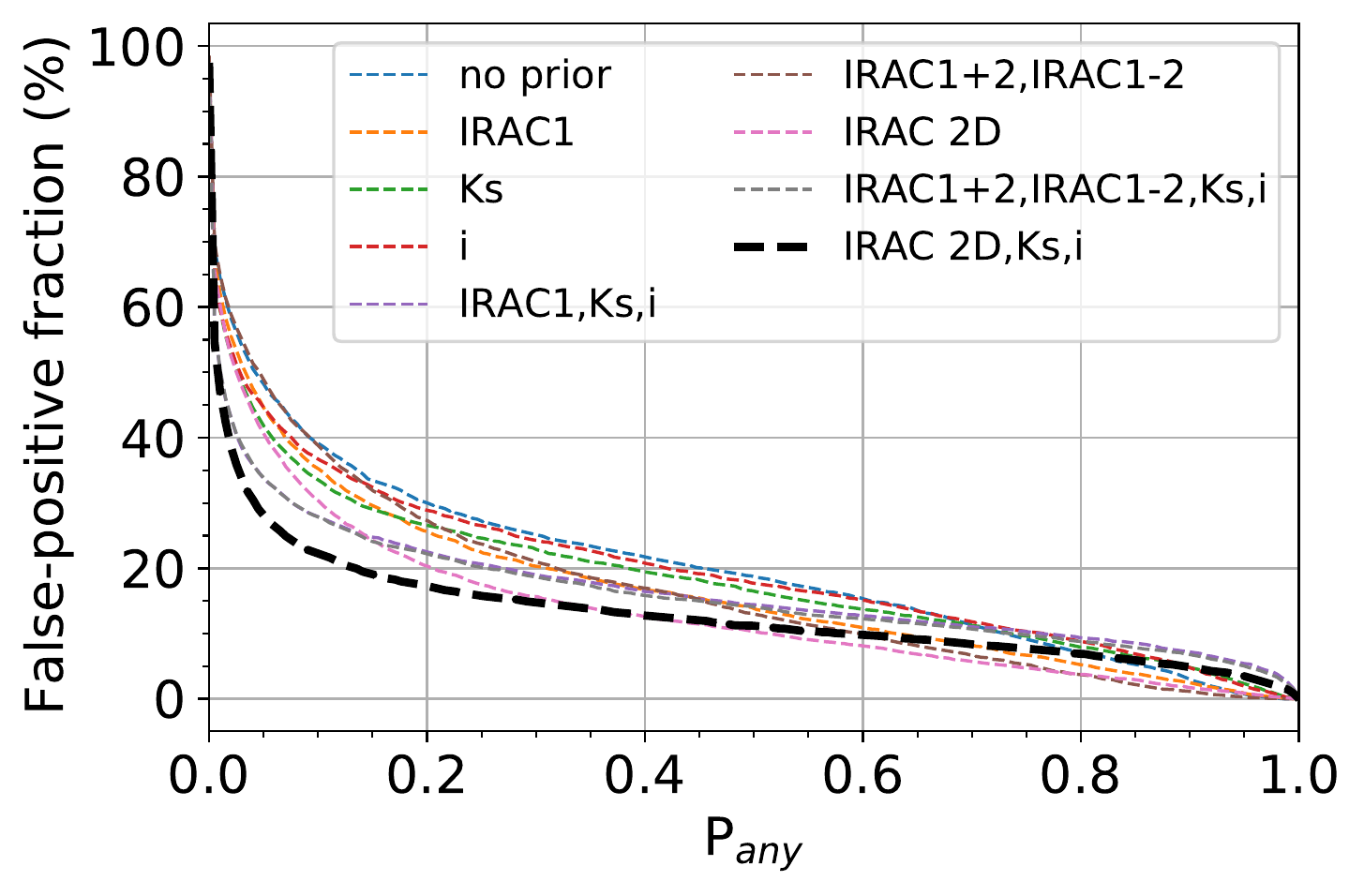}
~
\includegraphics[width=0.48\textwidth]{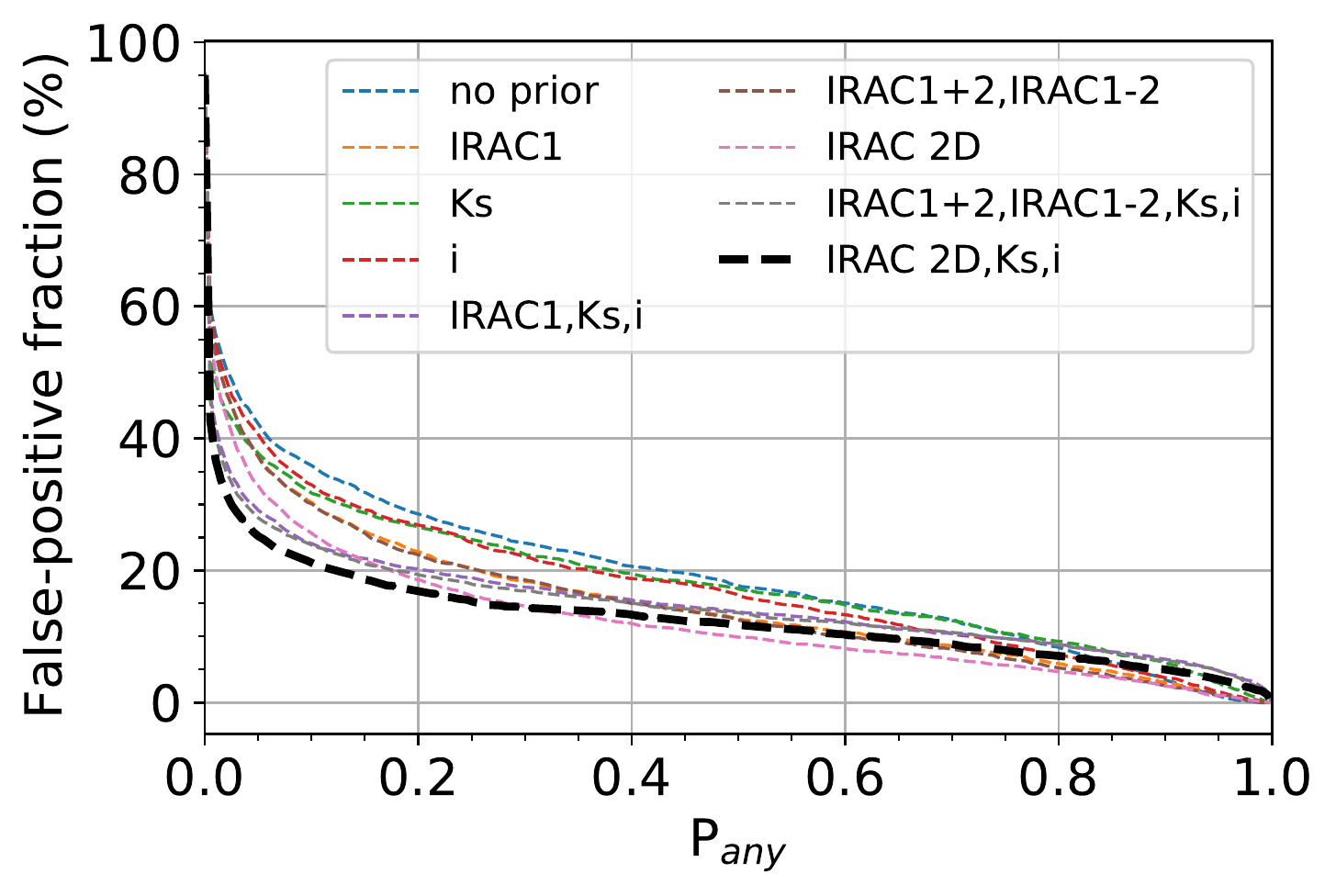}
\includegraphics[width=0.48\textwidth]{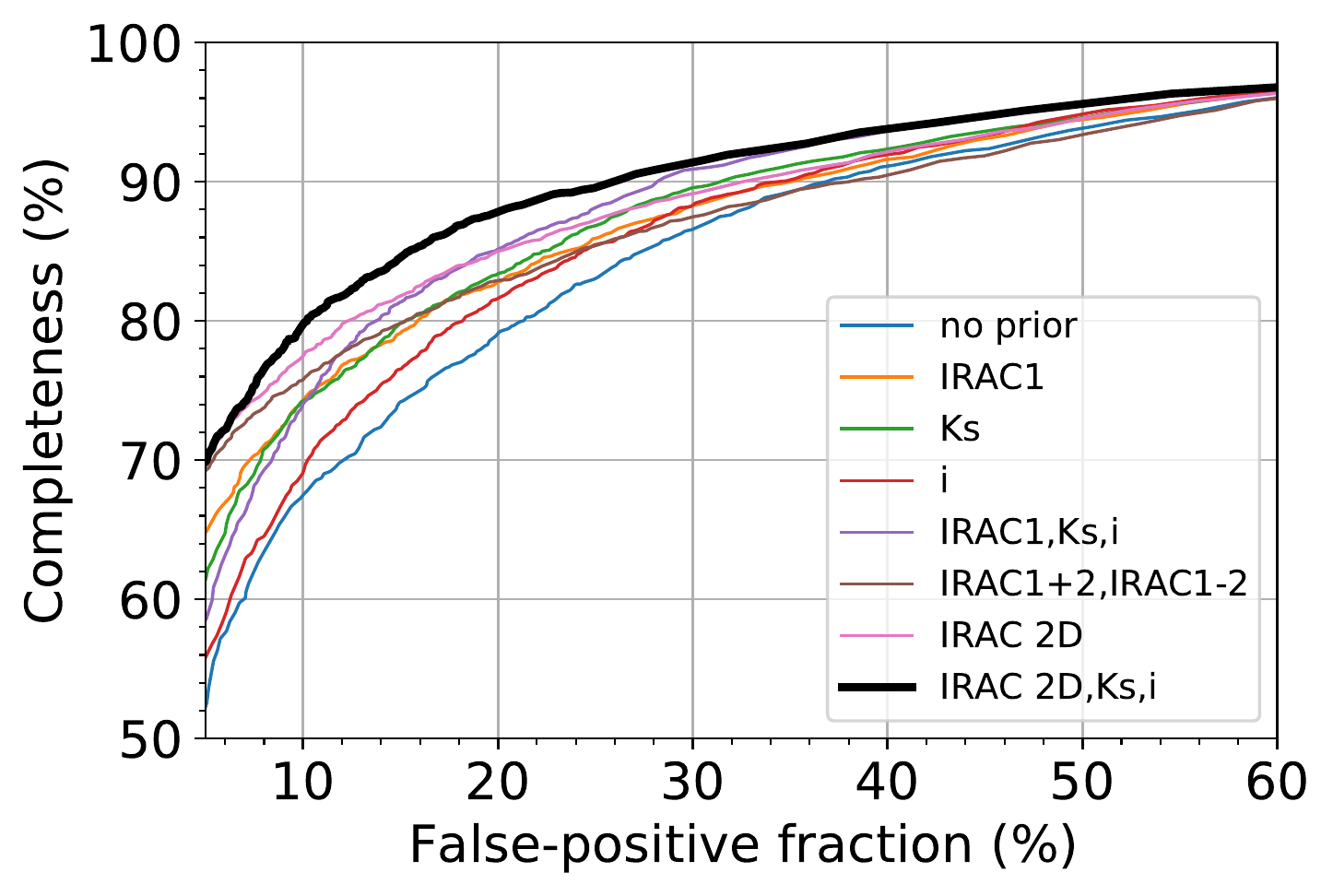}
~
\includegraphics[width=0.48\textwidth]{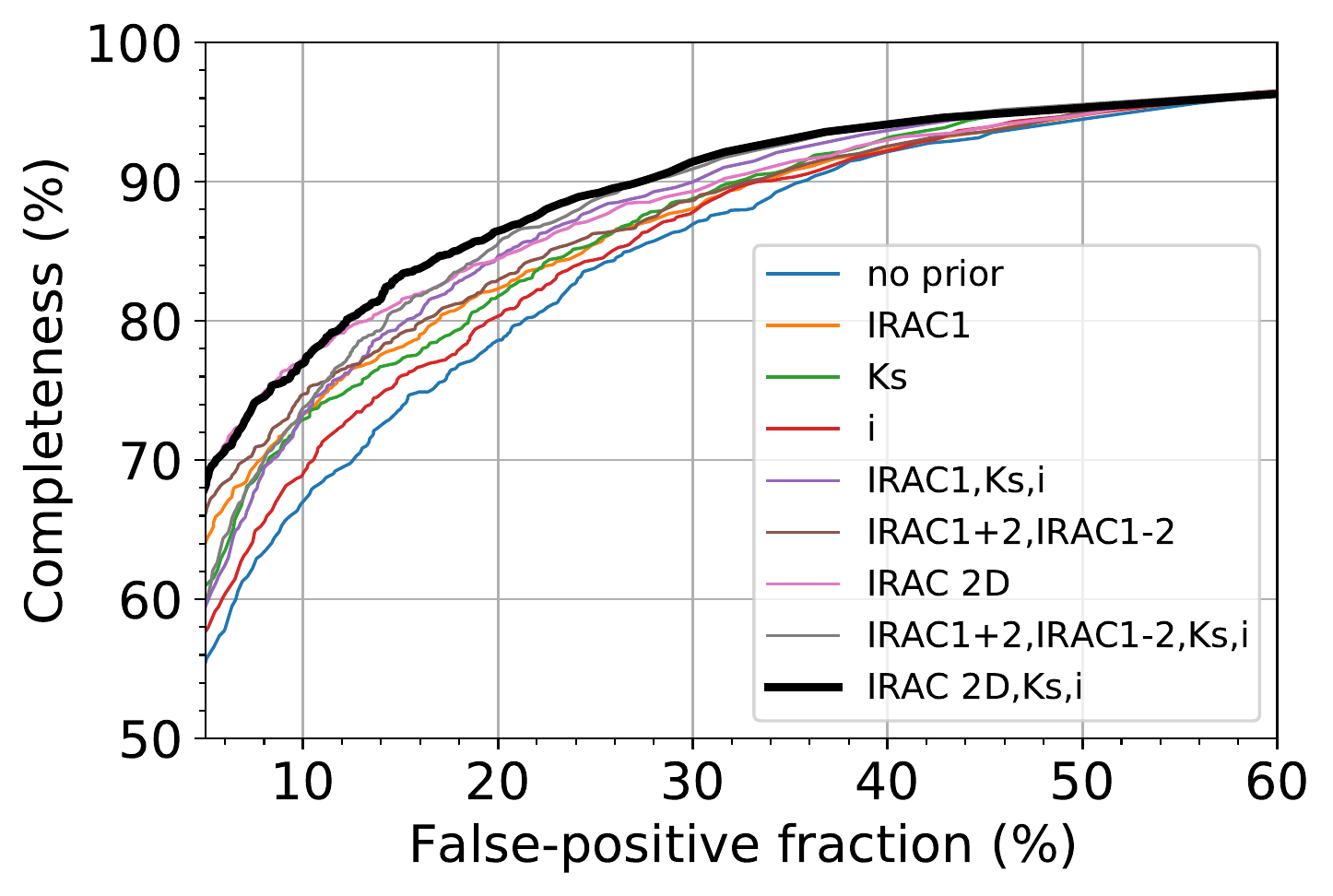}
\caption{
\textit{Top:} Fraction of matched sources above the $p_{\rm any}$ threshold when different priors are adopted.
\textit{Middle:} Fraction of matched mock \xray\  sources above the $p_{\rm any}$ threshold when different priors are adopted.
\textit{Bottom:} Completeness vs. false-positive fraction when different priors are adopted.
}
\label{fig:pany}
\end{figure*}

\subsection{Assessing the matched counterparts} \label{ss-mmfr}
The \texttt{NWAY} matching results can be assessed by investigating the subsample of \xmm\ sources that have matched \chandra\ counterparts.
We compare the matching results of \chandra\ sources with \xmm\ sources. For \wcdfs, the matched DeepDrill counterparts have a $\approx 97\%$ agreement; the matched VIDEO counterparts have a $\approx 92\%$ agreement; the matched HSC counterparts have a $\approx 93\%$ agreement (see Column 11 of  Table~\ref{tab:matching}). 
For \es, the matched DeepDrill counterparts have a $\approx 98\%$ agreement; the matched VIDEO counterparts have a $\approx 96\%$ agreement; the matched DES counterparts have a $\approx 97\%$ agreement.\footnote{The matching results with \chandra\ or \xmm\ positions display a slightly higher level of agreement in \es\ than in the \wcdfs, as the \xmm\ data in the \es\ region with \chandra\ coverage is deeper than the data in the \wcdfs\ region with \chandra\ coverage, leading to better positional accuracy.}
Examples of comparisons between the matching results utilizing \chandra\ and \xmm\ positions are presented in Figure~\ref{fig:chandravsxmm}.

\begin{figure*}
\centering
\includegraphics[width=0.45\textwidth]{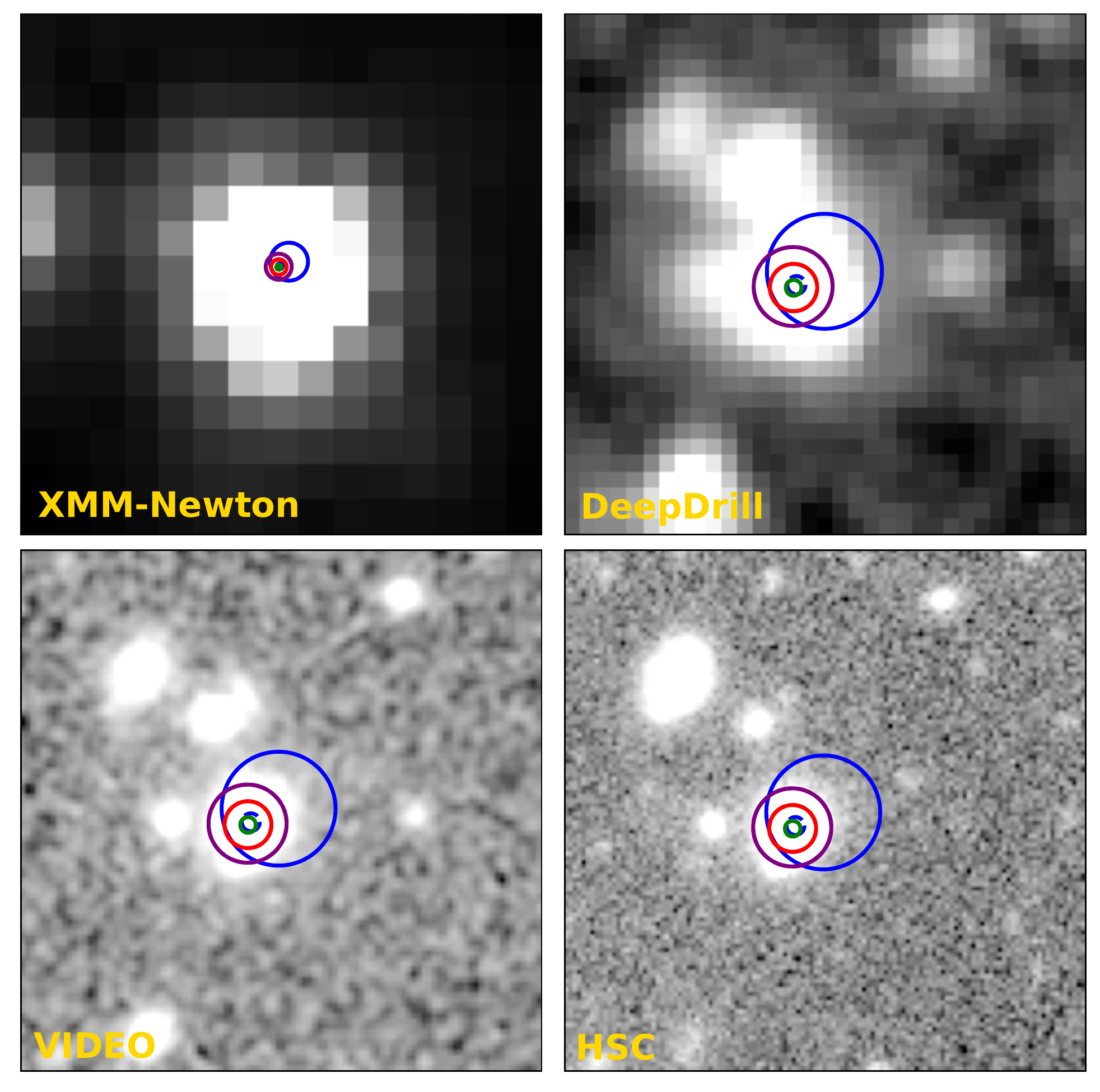}
\includegraphics[width=0.45\textwidth]{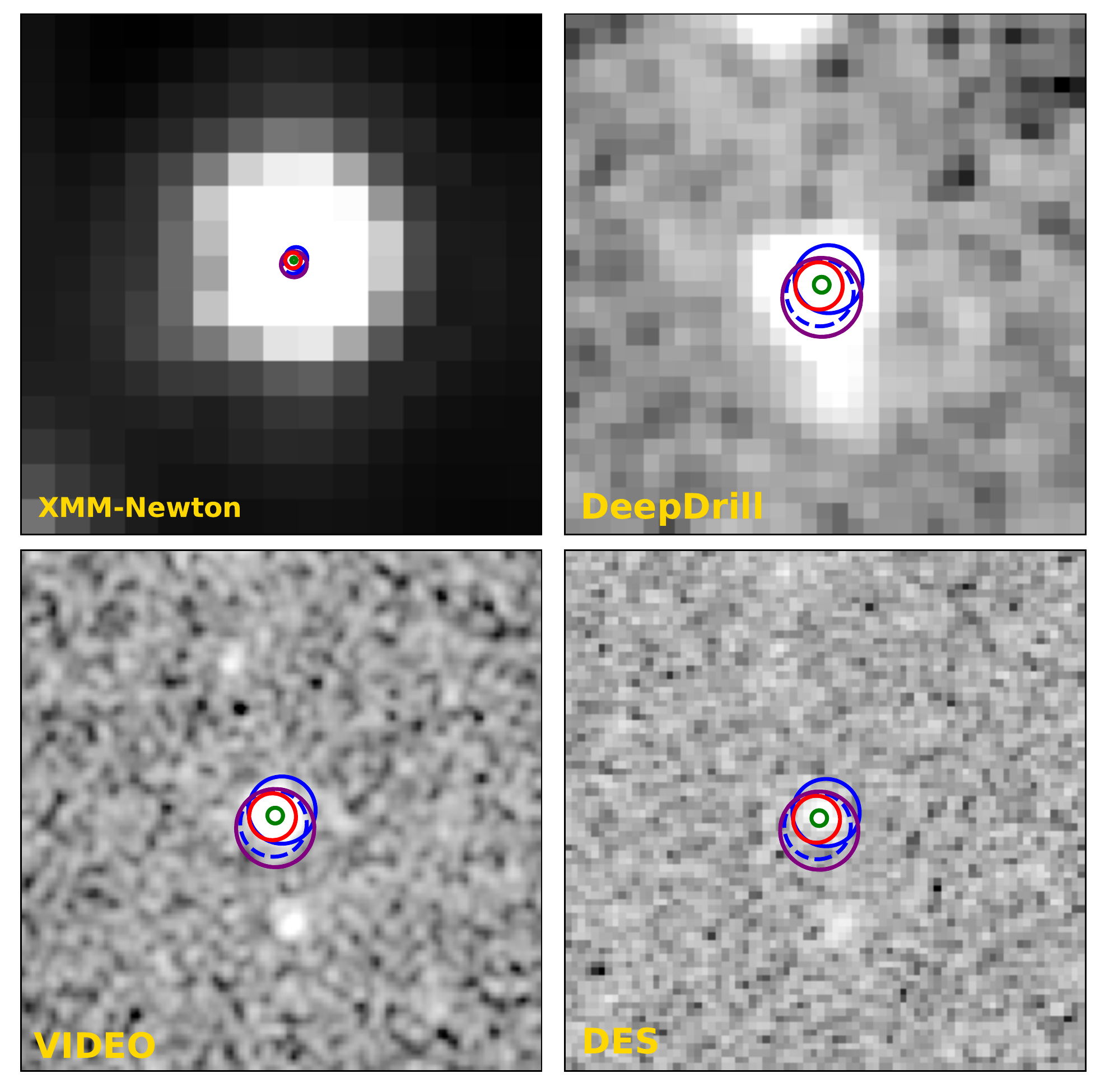}
\includegraphics[width=0.45\textwidth]{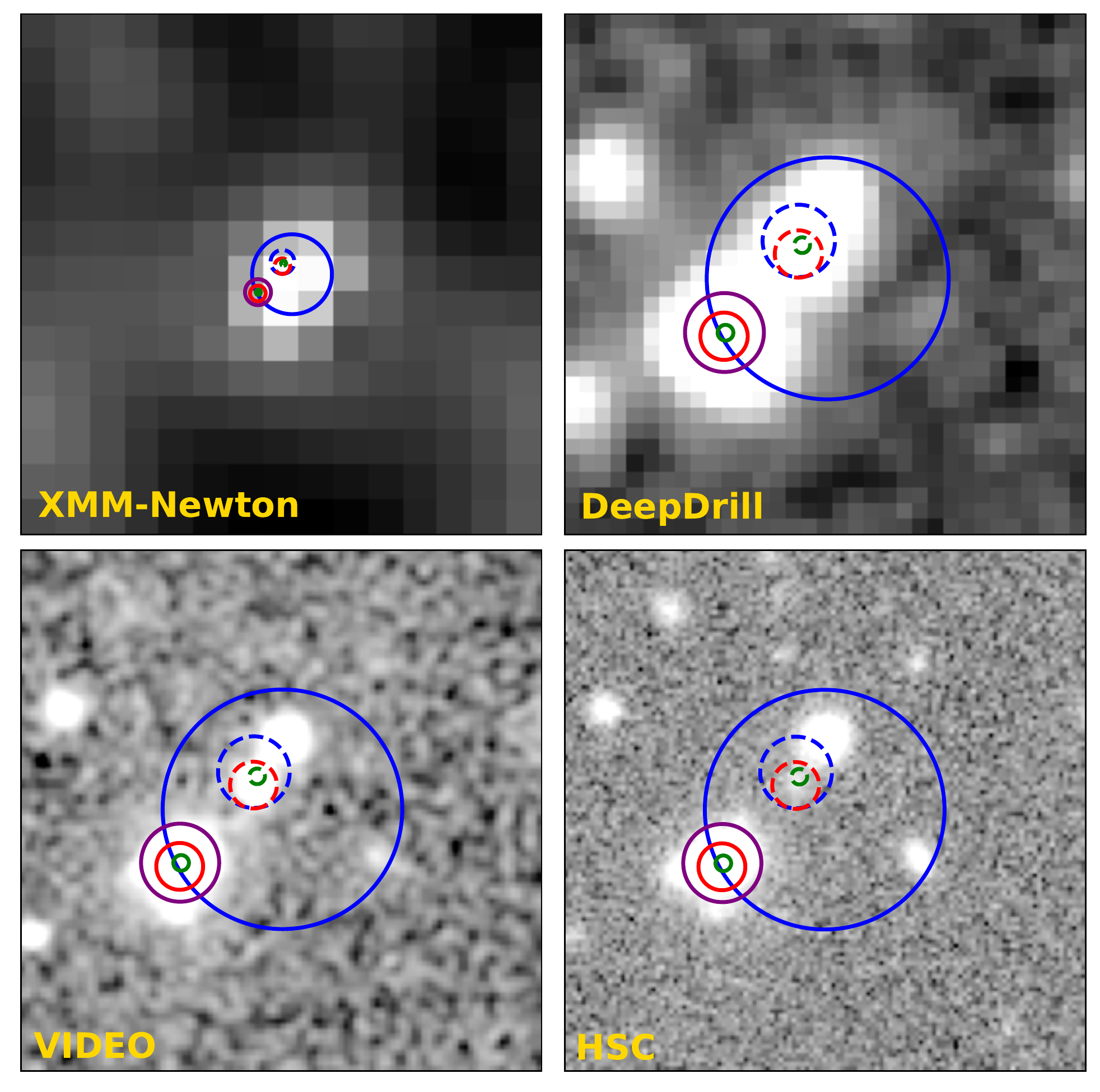}
\includegraphics[width=0.45\textwidth]{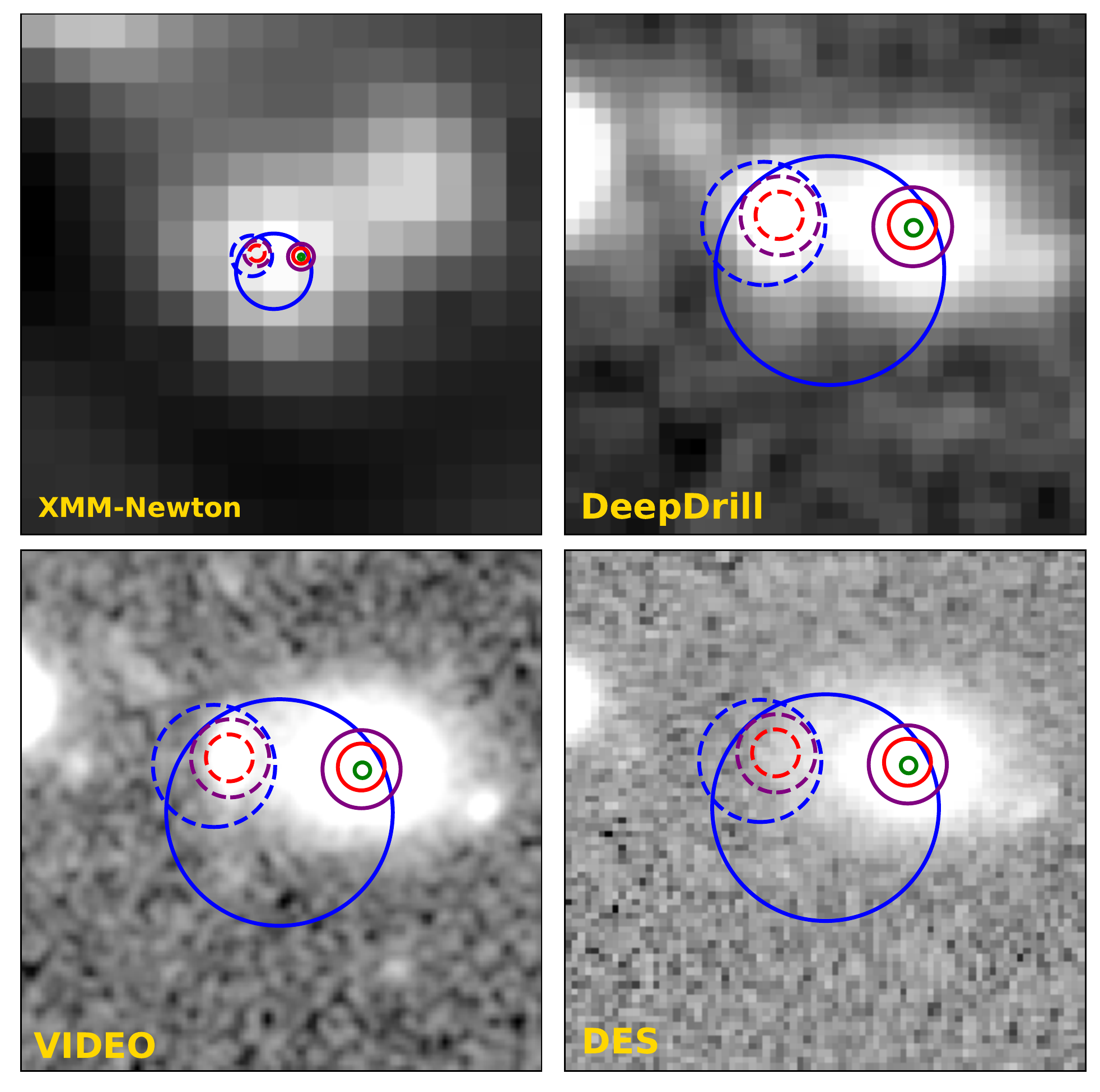}
\caption{Illustrations of the comparison between the matching results using {\it XMM-Newton} positions vs.\ \chandra\ positions. 
Each set of four images shows cutouts from the smoothed \xmm\ \hbox{0.2--12}~keV image (top-left; $60'' \times 60''$), DeepDrill IRAC 3.6$\mu$m band (top-right; $20'' \times 20''$), VIDEO $K_s$-band (bottom-left; $20'' \times 20''$), and HSC $i$-band (for the two \wcdfs\ sets on the left) or DES $i$-band (for the two \es\ sets on the right) (bottom right; $20'' \times 20''$).
\hbox{X-ray} positions are marked as blue circles with a 99.73\% error radius, with the {\it XMM-Newton} positions indicated by solid lines and the \chandra\ positions indicated with dashed lines. 
DeepDrill counterparts matched utilizing \xmm/\chandra\ positions are marked with solid/dashed purple circles; VIDEO counterparts are indicated with solid/dashed red circles; HSC/DES counterparts are identified with solid/dashed green circles.
In most cases, the counterpart-matching results for \xmm\ sources are identical to the results obtained using \chandra\ positions (see the top sets); there is a small fraction of sources where the \xmm\ results do not agree with \chandra\ results (see the bottom sets).}
\label{fig:chandravsxmm}
\end{figure*}

We have also performed simulations in \wcdfs\ and \es, respectively, to assess the results of multiwavelength counterpart matching with \texttt{NWAY}.
Following \cite{Broos2011} and \cite{Chen2018}, we consider our \xray\ sources to have both an ``associated population'' (X-ray sources that do have a real counterpart in the corresponding optical/NIR catalog) and an ``isolated population'' (X-ray sources that do not have a real counterpart in the corresponding optical/NIR catalog).

The fraction of the associated population ($f_{\rm AP}$) can be calculated with the formula:
\begin{equation}
N_{\rm negative} = N_{\rm FN} \times f_{\rm AP} + N_{\rm TN} \times (1 - f_{\rm AP}).
\label{eq:fap}
\end{equation}
$N_{\rm negative}$ is the number of real X-ray sources that do not have a matched counterpart in an optical/NIR catalog; $N_{\rm FN}$ is the number of simulated X-ray sources that belong to the ``associated population'' but do not have a matched counterpart; $N_{\rm TN}$ is the number of mock X-ray sources that belong to the ``isolated population'' and are not matched to a counterpart as expected.
As presented in Section~\ref{ss-chooseprior}, \texttt{NWAY} has a built-in function to simulate the isolated population and obtain $N_{\rm TN}$ with varying $p_{\rm any}$ thresholds. 
To simulate the associated population and calculate $N_{\rm FN}$ with varying $p_{\rm any}$ thresholds, we use a method similar to that in Section~4.2 of \citet{Chen2018}.
For X-ray sources that have \pany\ values above the adopted $p_{\rm any}$ threshold, we remove all their matched optical/NIR counterparts 
in the optical/NIR catalogs, and shift the position of all the remaining optical/NIR sources in the catalog by 1 arcmin in a random direction. 
We then generate fake optical/NIR ``counterparts'' for each X-ray source based on the X-ray and optical/NIR positional uncertainties, with all the priors utilized. When generating the optical/NIR positions for the \wcdfs\ field, we use the positional uncertainty of \xray\ sources and HSC sources to simulate HSC positions from the expected Rayleigh distribution of offsets. The generated HSC positions are utilized to simulate the positions of DeepDrill/VIDEO sources, assuming a Gaussian distribution for the offsets between HSC sources with their DeepDrill/VIDEO counterparts (the standard deviation of the Gaussian distribution is derived from all the matched DeepDrill/VIDEO sources with HSC sources within 1$''$). For the \es\ field, DES sources are simulated instead of HSC sources.
After that, we run \texttt{NWAY} to obtain $N_{\rm FN}$ among the associated population, thus obtaining $f_{\rm AP}$ by solving Equation~\ref{eq:fap}.
With $f_{\rm AP}$, we could obtain the expected false rate ($f_{\rm False}$) of matched counterparts with varying $p_{\rm any}$ thresholds:
\begin{equation}
f_{\rm False} = (N_{\rm IM} \times f_{\rm AP} + N_{\rm FP} \times (1-f_{\rm AP})) / (N_{\rm positive}).
\end{equation}
$N_{\rm IM}$ is the number of incorrect matches among the simulated associated X-ray sources; 
$N_{\rm FP}$ is the number of false positives among the mock isolated X-ray sources.
Figure~\ref{fig:mmsim} presents $f_{\rm False}$ as a function of the $p_{\rm any}$ threshold adopted.
Similar to the finding in \citet{Chen2018}, the matched IRAC counterparts have the smallest $f_{\rm False}$ among all the optical/NIR catalogs.

\begin{figure*}
\centering
\includegraphics[width=0.48\textwidth]{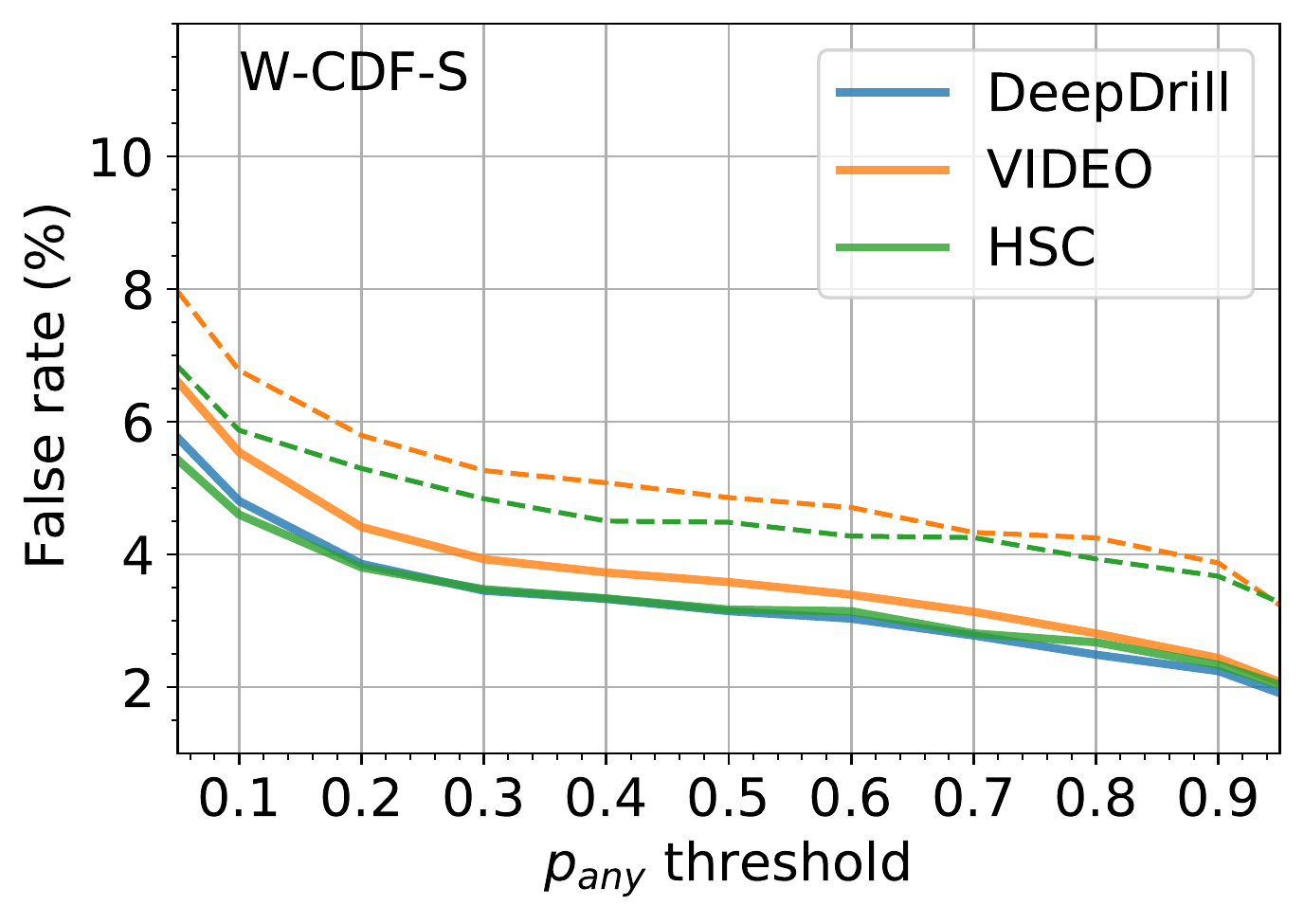}
\includegraphics[width=0.48\textwidth]{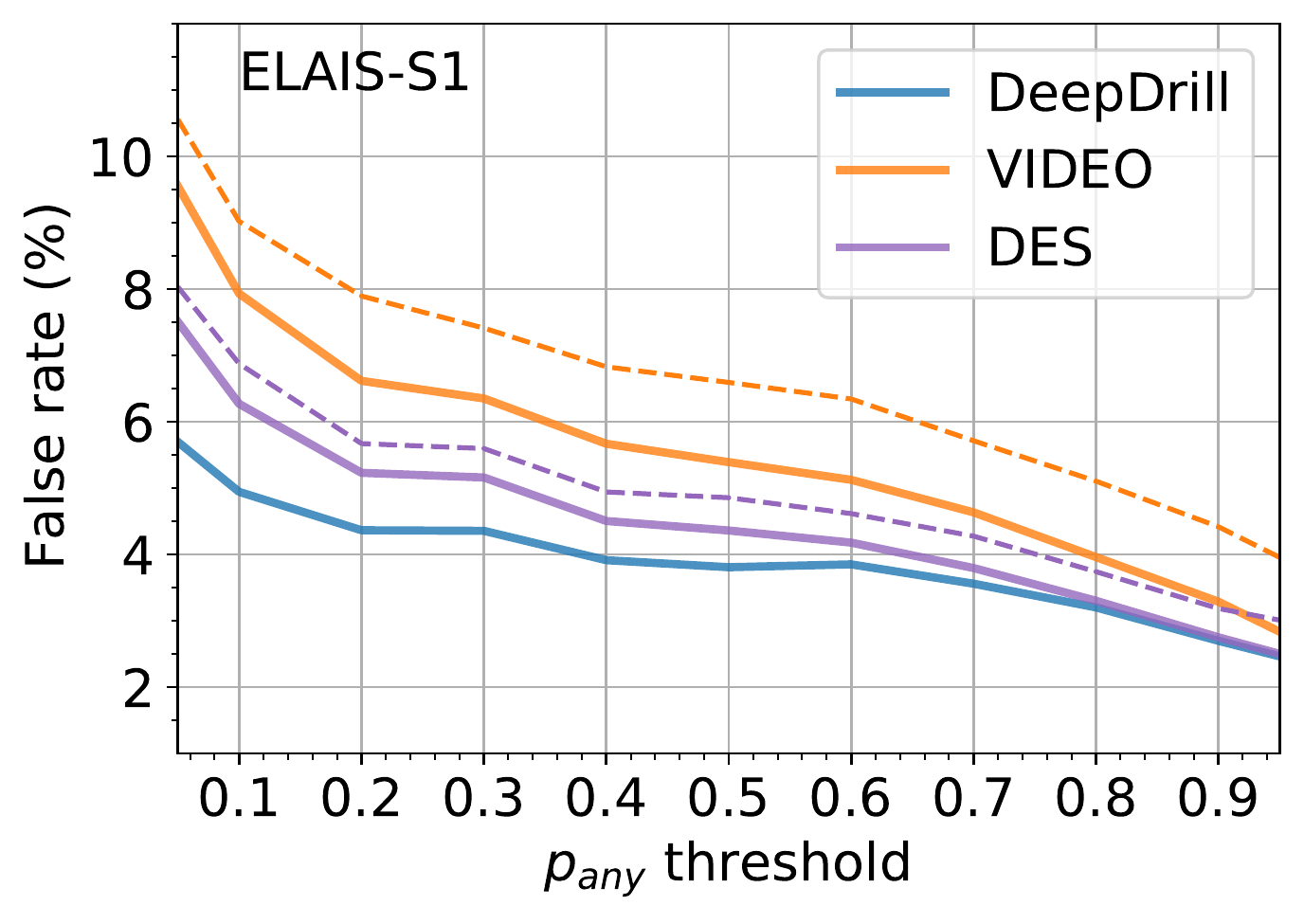}
\caption{\textit{Left:} False rate of matched DeepDrill/VIDEO/HSC counterparts as a function of $p_{\rm any}$ threshold adopted (blue/orange/green solid lines) when matching \xray\ sources in \wcdfs\ to all the optical/IR catalogs simultaneously. The orange/green dashed lines represent the false rates when matching \xray\ sources to VIDEO/HSC sources based on their distances to the matched DeepDrill counterparts (when available). \textit{Right:} False rate of matched DeepDrill/VIDEO/DES counterparts as a function of $p_{\rm any}$ threshold adopted (blue/orange/purple solid lines) when matching \xray\ sources in \es\ to all the optical/IR catalogs simultaneously. The orange/purple dashed lines represent the false rates when matching \xray\ sources to VIDEO/DES sources based on their distances to the matched DeepDrill counterparts (when available).}
\label{fig:mmsim}
\end{figure*}

In Section~4.2 of \citet{Chen2018}, when SERVS counterparts are available for \xray\ sources, other optical/NIR counterparts are selected based on matching them with the matched SERVS counterparts.
We also calculate the $f_{\rm False}$ for VIDEO and HSC (or DES) with the following methodology: when an \xray\ source has a DeepDrill counterpart, we identify VIDEO and HSC (or DES) counterparts purely based on the distance from the DeepDrill counterpart. The results are presented in Figure~\ref{fig:mmsim} as the dashed lines. For VIDEO and HSC (or DES) counterparts, the obtained false rates are slightly higher by $\approx 1$--2\%, revealing the advantages of matching multiple optical/NIR catalogs to \xmm\ sources simultaneously.

In the released catalogs, we do not apply any \pany\ threshold for the identified multiwavelength counterparts with \texttt{match\_flag} = 1 (which indicates that this counterpart is the primary counterpart with the highest likelihood). However, a \pany\ threshold of at least 0.1 is suggested for catalog users so that the false rate of the optical/NIR counterparts is $\sim 5$\% (see Table~\ref{tab:matching}, Column 10).
3600/2288 \xray\ sources in \wcdfs/\es\ have \pany\ $> 0.1$, which is $\approx 89\%$/87\% of the total \xray\ sources detected (see Table~\ref{tab:matching}, Column 7).
For the analyses in Sections~\ref{s-z} and \ref{sc-sp} where the optical/NIR counterparts of \xray\ sources are utilized, we only use \xray\ sources with \pany\ $> 0.1$ counterparts.
Figure~\ref{fig:offset} displays the offsets between X-ray sources (that have \pany\ $> 0.1$) with their optical/NIR counterparts. Following a priority established based on the survey positional uncertainty, we use the HSC (or DES), VIDEO, or DeepDrill positions as the location of optical/NIR counterparts. 
Figure~\ref{fig:offset} also presents histograms of positional offsets when $\sigma_x$ varies, which demonstrates that our estimation of $\sigma_x$ from the empirical relation is reliable in general: since $r_{68\%}$ = $\sigma_x \times$ 1.515 (see Section~\ref{ss:poserr}), we expect the median positional offset in different $\sigma_x$ bins to increase with $\sigma_x$, and roughly 68\% of the sources in a given $\sigma_x$ bin have positional offsets less than the median $r_{68\%}$ in this bin.
Compared to many previous \xmm\ survey catalogs (e.g., \citealt{Chen2018,Liu2020}), this work has substantially reduced the \xray\ positional uncertainty and decreased the offset between X-ray sources and their optical/NIR counterparts.

\begin{figure*}
\centering
\includegraphics[width=0.45\textwidth]{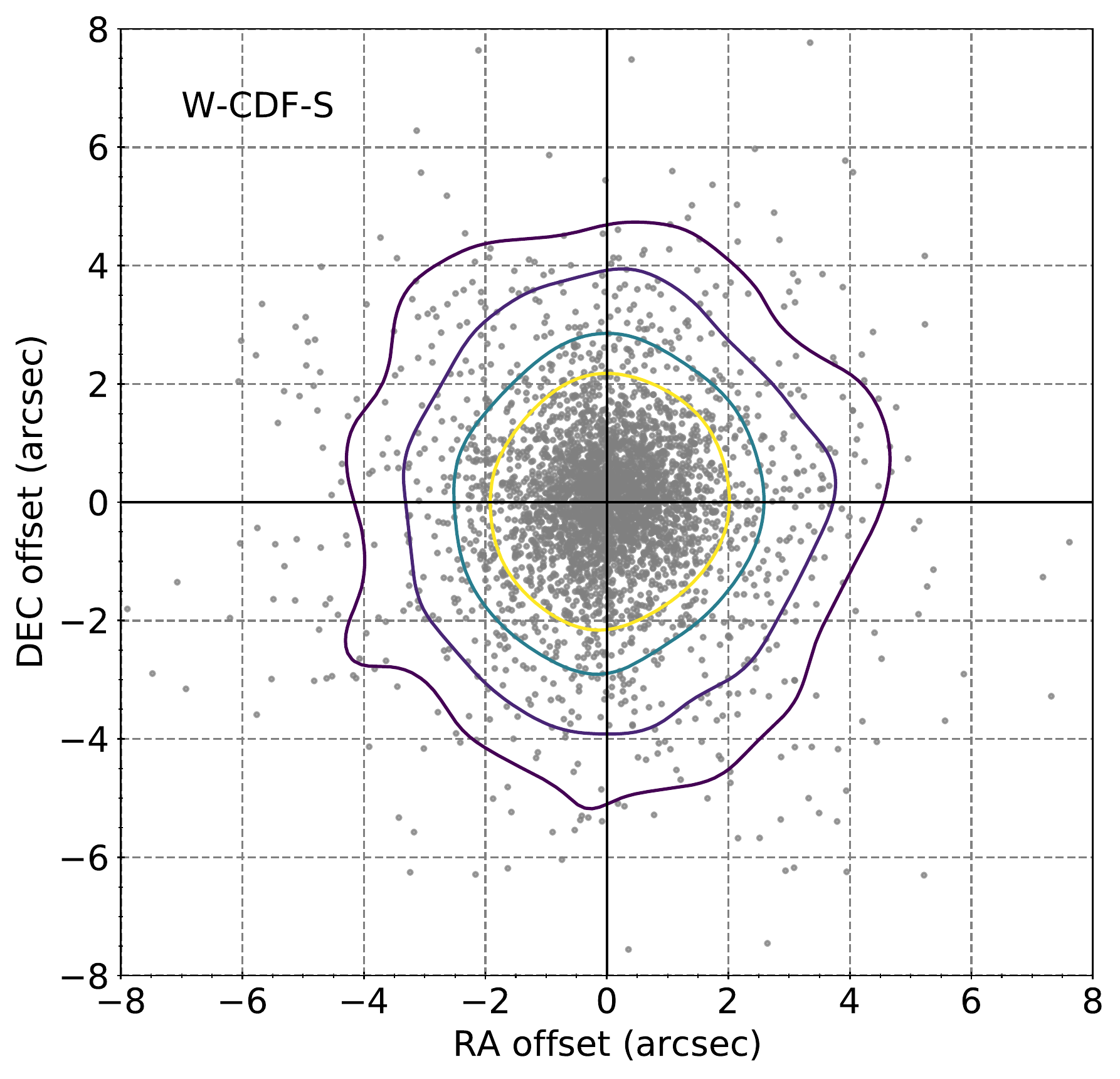}
~
\includegraphics[width=0.45\textwidth]{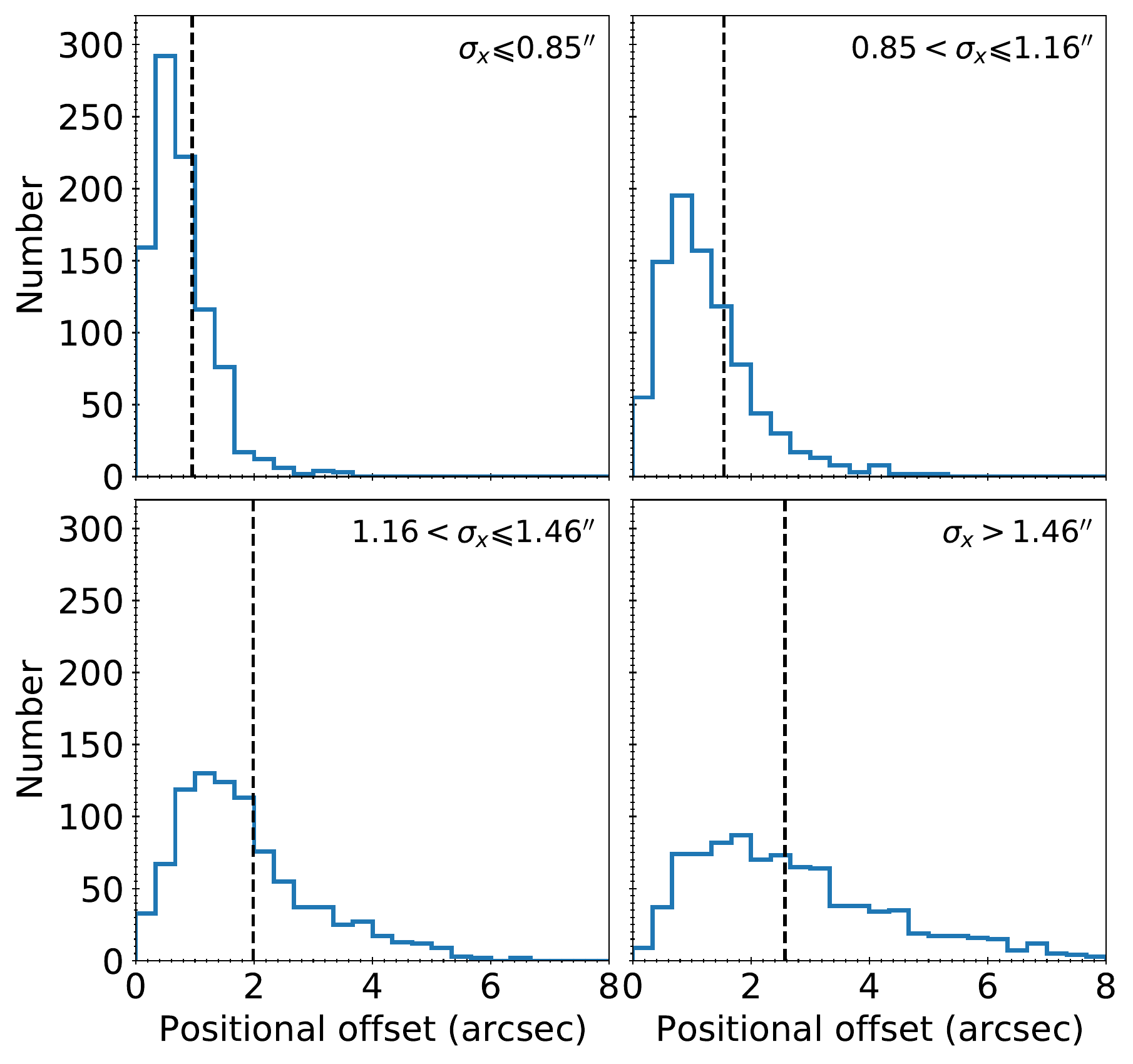}
~
\includegraphics[width=0.45\textwidth]{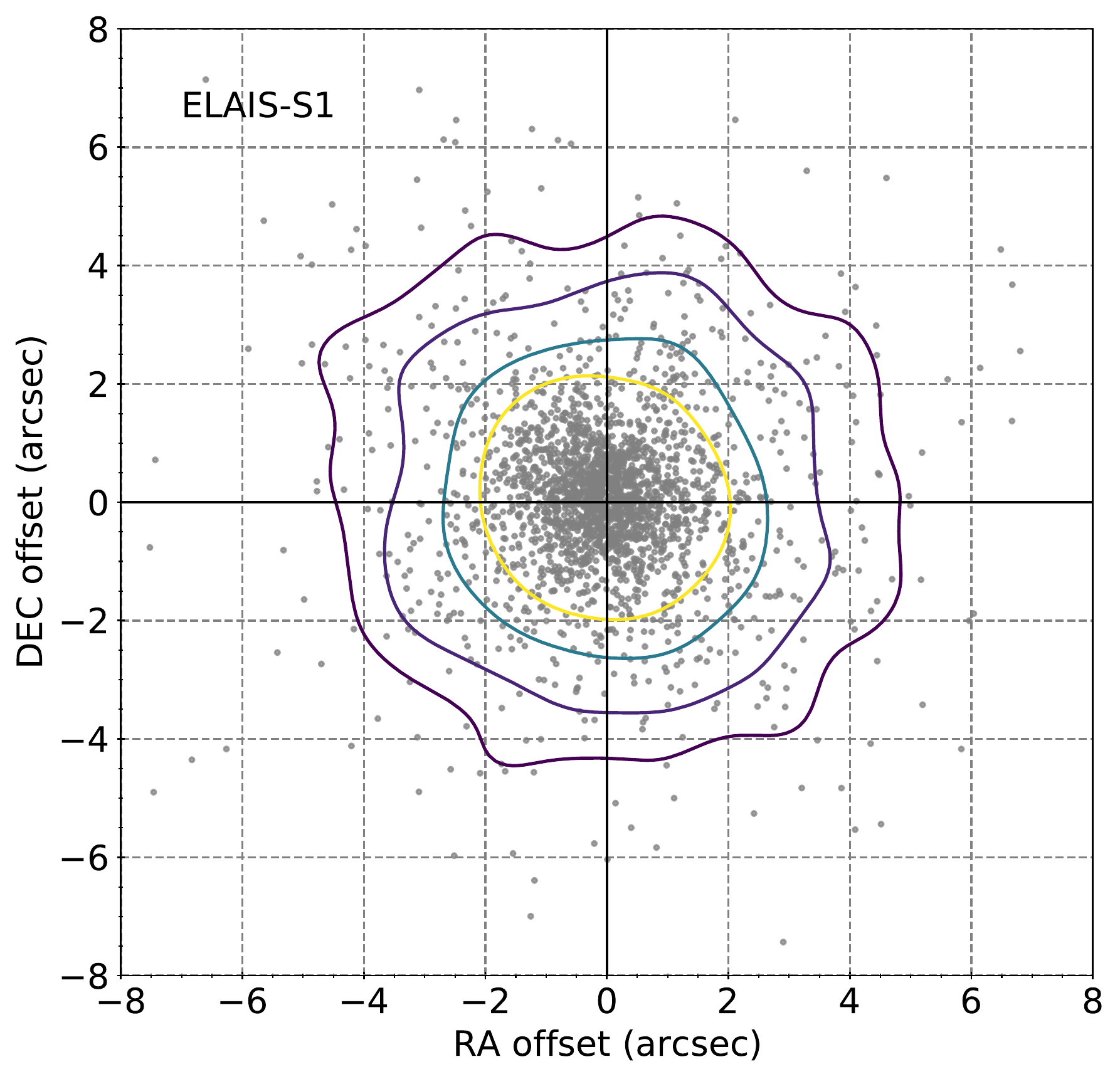}
~
\includegraphics[width=0.45\textwidth]{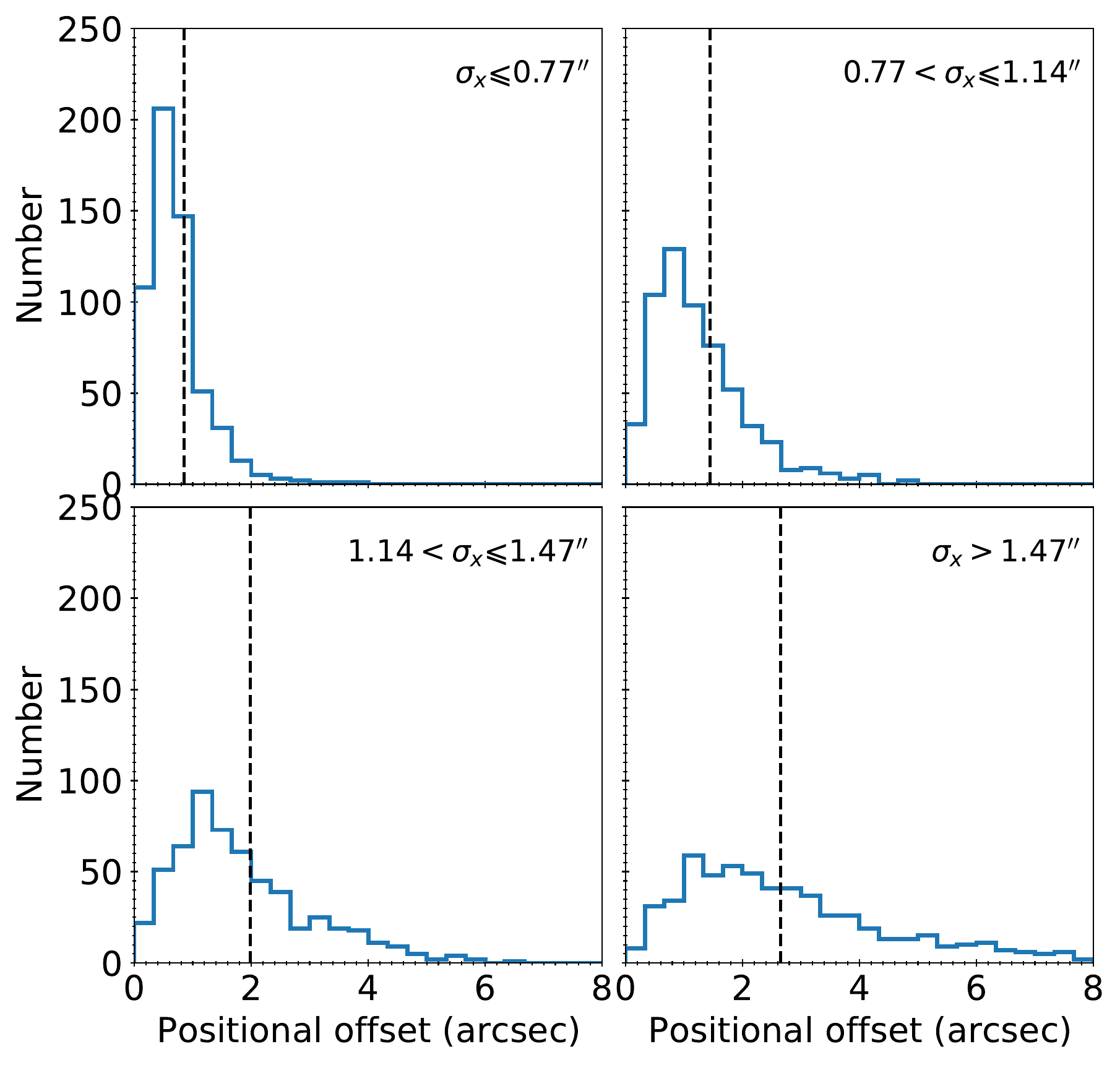}
\caption{{\it Left}: The distribution of DEC offset vs.\ RA offset for X-ray sources in \wcdfs\ ({\it top}) or  \es\ ({\it bottom}) and their optical/NIR counterparts, with contours indicating the isodensity levels that include 68\%, 80\%, 90\%, and 95\% of sources. 
{\it Right}: Histograms of positional offsets between X-ray sources in \wcdfs\ ({\it top}) or  \es\ ({\it bottom}) (that are divided into four bins based on $\sigma_x$) and their matched optical/NIR counterparts. The vertical dashed line in each panel represents the median $r_{68\%}$ ($r_{68\%}$ = 1.515$\sigma_x$; see Section~\ref{ss:poserr}) value in each bin.}
\label{fig:offset}
\end{figure*}

\section{Redshifts} \label{s-z}

\subsection{Spectroscopic redshifts}
 
In addition to the extensive photometric data (see Table~\ref{tab:wcdfsmw}), there are a number of spectroscopic surveys in the \wcdfs/\es\ region (see Table~\ref{tab:redshift}).\footnote{There are spectroscopic surveys in the CDF-S/E-CDF-S region that are not listed in Table~\ref{tab:redshift}, as we mainly focus on the more relevant wide-area surveys.}
We match \xray\ sources to these spectroscopic redshifts (spec-$z$s)  utilizing the positions of matched optical/NIR counterparts: we search for the nearest spectroscopic redshift position that is within 1$''$ of these optical/NIR counterparts.
When an X-ray source is matched to multiple spec-$z$s, we choose redshifts using the priority order in Table~\ref{tab:redshift} (which is ranked based on the spectral resolution, as the accuracy of spec-$z$s is significantly dependent upon the spectral resolution; see Note (a) of Table~\ref{tab:redshift} for more detail). 
Most of the spectroscopic surveys in Table~\ref{tab:redshift} have resolution > 100. As for the low-resolution PRIMUS survey, we only adopt the $Q$ $\geqslant 3$ ($Q$ is the redshift-quality flag provided by \citealt{Coil2011}) objects. Before matching \xray\ sources to the PRIMUS catalog, we utilize the spec-$z$ compilation in the HELP database \citep{Shirley2019}, which provides several additional spec-$z$s.

In the \wcdfs\ catalog, 919 ($\approx 23\%$) \xray\ sources are matched to spec-$z$s ($\approx 750$ of them are outside of the E-CDF-S region), ranging from 0 to 4.56.
In the \es\ catalog, 585 ($\approx 22\%$) \xray\ sources are matched to spec-$z$s, ranging from 0 to 4.04 ($\approx 300$ of them are outside of the original $\approx 0.6$ deg$^2$ \es\ region).
About $84\%$/$98\%$ of the matched spec-$z$ measurements are from catalogs that have spectroscopic classification for AGNs available in \wcdfs/\es.  
Figure~\ref{fig:zhist} shows the distribution of these spec-$z$s.
In the future, there will be more public spectroscopic redshifts from surveys including CSI \citep[e.g.,][]{Kelson2014}, DEVILS \citep[e.g.,][]{Davies2018,Thorne2020}, DESI \citep[e.g.,][]{Levi2019}, MOONS \citep[e.g.,][]{Maiolino2020}, and WAVES \citep[e.g.,][]{Driver2019}.

\begin{table*}
    \caption{\label{tab:redshift}
Spectroscopic-redshift catalogs used in this work, listed with priority from high to low.
}
    \begin{tabular}{cccccccccl} 
        \hline
        Catalog & Instrument & Survey  & Spectral & Targeting & Area & $N_{\rm matched}$ & $N_{\rm assigned}$ & Reference \\
              &     &   Sensitivity & Resolution & Fields & (deg$^2$) & &  & \\
        (1) & (2) & (3) & (4) & (5) & (6) & (7) & (8)  & (9)\\
        \hline 
        \multicolumn{9}{c}{W-CDF-S}\\
        \hline
         OzDES$^*$   & AAOmega    &  $r \lesssim 22.5$                        &  $\sim$1500      & DES-SN C1,C2,C3               & 9              &  406    &  406      &        \citet{Lidman2020}    \\ 
         ATLAS$^*$    &   AAOmega   &   $R < 22$                                & $\sim$1300       &           CDF-S                         &     2.96     &    155  &     97     &    \citet{Mao2012}        \\  
         BLAST$^*$    & AAOmega     &     -                                            &     $\sim$1300   &     GOODS-South                & 3               &    47  &    21      &   \citet{Eales2009}      \\ 
        6dFGS    & UKST            &   $ K \lesssim $ 12.65               & $\sim$1000        &    The Southern Sky             &  17,000   &     13    &      4   &   \citet{Jones2009}    \\ 
        2dFGRS  & AAOmega    &  $\rm b_J < 19.45$                    & $\sim$800        &       SGP strip                        &   2000    &    30    &      5   &       \citet{Colless2011}       \\ 
        ACES     &  IMACS         &  $ R < 24.1$                              &    $\sim$750      &          CDF-S                           &     0.25   &    80   &      61    &   \citet{Cooper2012}   \\  
        -     &  VIMOS/DEIMOS &  R $<$ 25       &    $\sim$180/580      &          
        E-CDF-S                           &     0.33   &    143   &      70   &   \citet{Silverman2010}   \\          
        PRIMUS$^{*a}$ & IMACS         &  $i \lesssim 23.5$                       & $\sim$30         & CDFS-SWIRE,CALIB  & 2.1        &     349 &      252  &    \citet{Coil2011}  \\
         \hline
         \multicolumn{9}{c}{ELAIS-S1}\\
        \hline
         OzDES$^*$   & AAOmega    &  $r \lesssim 22.5$                        &  $\sim$1500      & DES-SN E1,E2              & 6             &  293    &   293     &        \citet{Lidman2020}    \\ 
         ATLAS$^*$    &   AAOmega   &   $R < 22$                                & $\sim$1300       &           ELAIS-S1                         &     4.69     &    46  &     30     &    \citet{Mao2012}        \\  
        6dFGS    & UKST            &   $ K \lesssim $ 12.65               & $\sim$1000        &    The Southern Sky             &  17,000   &     10    &   6     &   \citet{Jones2009}    \\ 
        2dFGRS  & AAOmega    &  $\rm b_J < 19.45$                    & $\sim$800        &       SGP strip                        &   2000    &    5   &      1   &       \citet{Colless2011}       \\ 
        -$^*$    &  EFOSC, FORS2     &     -                                            &   $>$260   &     ELAIS-S1                & 0.6              &    129  &     106      &   \citet{Feruglio2008} \\ 
        -$^*$    & VIMOS     &     $R < 24.2$                                            &     $\sim$210   &     ELAIS-S1                & 0.6              &    134  &    22      &   \citet{Sacchi2009}      \\ 
        PRIMUS$^{*a}$ & IMACS         &  $i \lesssim 23.5$                       & $\sim$30         & ELAIS-S1  & 0.9       &     223 &      123  &    \citet{Coil2011}  \\
         \hline
    \end{tabular}
\raggedright
\\
Col. 1: Redshift survey name. $^*$ marks redshift surveys where spectroscopic classification for AGNs is available (or partially available).
Col. 2: Survey instrument.
Col. 3: Survey sensitivity.
Col. 4: Spectral Resolution.
Col. 5: Targeted fields.
Col. 6: Survey area.
Col. 7: Total number of redshifts matched to the \hbox{X-ray} sources in the catalog.
Col. 8: Total number of redshifts assigned to the \hbox{X-ray} sources in the catalog.
Col. 9: Reference.\\
Notes: (a). The low-resolution PRIMUS survey greatly increases the sample with spectroscopic redshifts, although its measurements are not as accurate as other spectroscopic surveys listed and should be used with appropriate caution. For \xray\ sources in our catalog, when both spec-$z$ measurements from PRIMUS and from other high-resolution spectroscopic surveys are available, $\approx 16\%$ of them have $\mathopen| \Delta z\mathclose|/(1+z_{\rm spec, high-resolution}) > 0.15$.
\end{table*}

\begin{figure}
\includegraphics[width=0.48\textwidth]{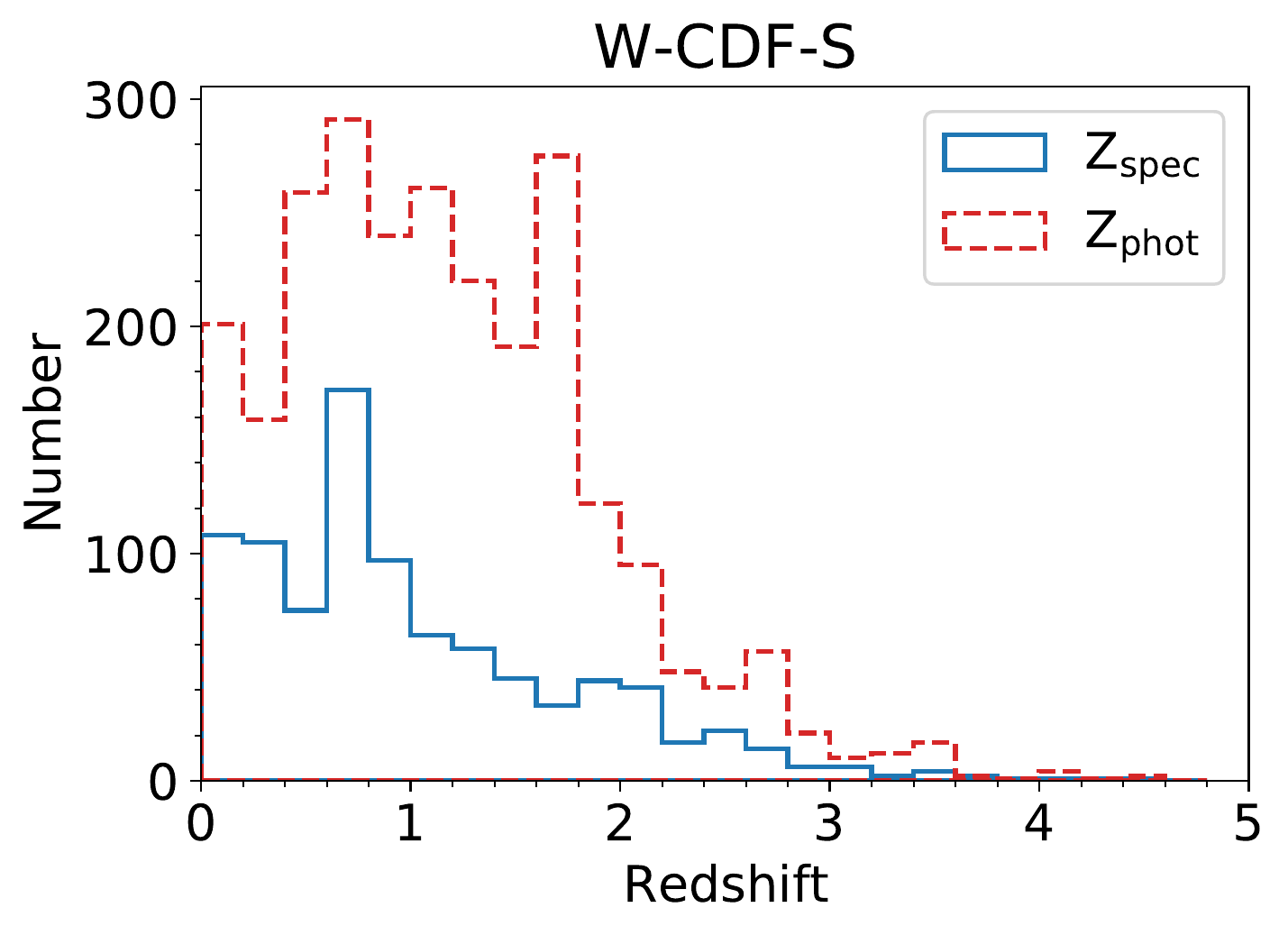}
\includegraphics[width=0.48\textwidth]{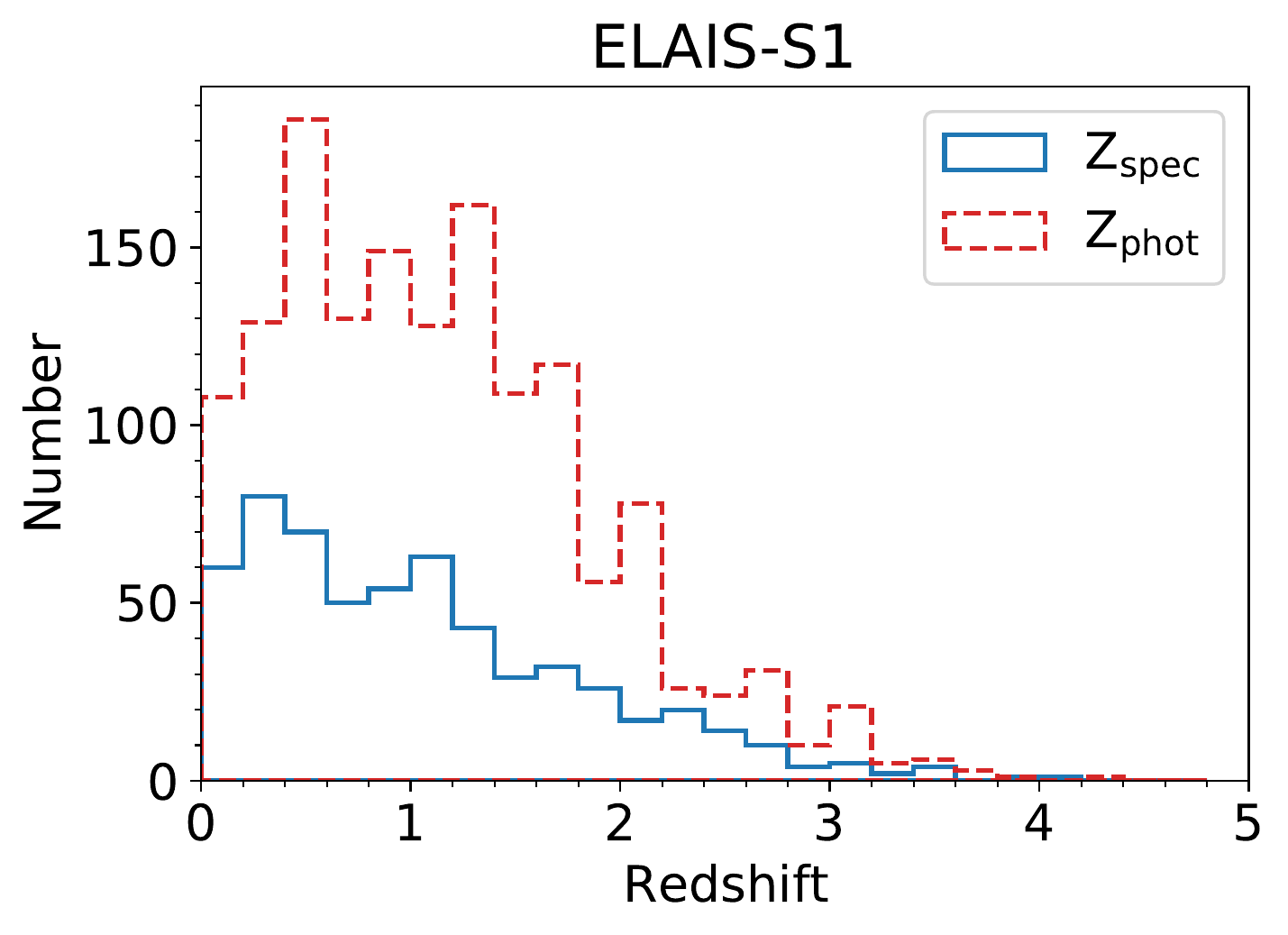}
\caption{Distributions of the spectroscopic/photometric redshifts of \xray\ sources that have spectroscopic/high-quality photometric redshift measurements (see Section~5) in the \wcdfs\ and \es\ fields, represented by the blue/red histograms (plotted respectively).
}
\label{fig:zhist}
\end{figure}

\subsection{Photometric redshifts} \label{ss-photoz}

Photometric redshifts (photo-$z$s) for \xray\ sources in this work are derived from SEDs provided by the forced-photometry catalogs in the $4.5/3 \rm~deg^2$ area covered by VIDEO in \wcdfs/\es\ (Nyland et al.\ 2021; \citealt{Zou2021a}); these catalogs were generated utilizing {\it The Tractor} \citep{Lang2016}. {\it The Tractor} derives consistent flux measurements in all bands with priors of source positions and surface-brightness profiles obtained from a fiducial band. As can be seen in Figure~\ref{fig:sources}, most ($\approx 95\%$) of our \xray\ catalog areas are covered by these catalogs. 
The forced-photometry catalogs are generated following the methods in \cite{Nyland2017}, where prior measurements of source positions and surface-brightness profiles from a fiducial VIDEO band (that has high resolution) are employed to model and fit the fluxes at other bands.
The photometric bands utilized in \wcdfs\ include the $u$, $g$, $r$ and $i$ bands in VOICE; $g$, $r$, $i$ and $z$ bands in HSC; $Z$, $Y$, $J$, $H$, and $K_s$ bands in VIDEO; and 3.6$\mu m$ and 4.5$\mu m$ bands in DeepDrill (see Table~\ref{tab:wcdfsmw} for the survey information). 
In total, 3319 \xray\ sources in \wcdfs\ have forced-photometry measurements.
For the HSC bands, we only utilize ``clean'' HSC photometry (see \citealt{Ni2019} for details). For a band that is included in two surveys ($g$/$r$/$i$), the scatter in two sets of photometry is small ($\approx 0.2$ dex), and both detections are utilized in the photo-$z$ calculation.
The photometric bands utilized in \es\ include the $g$, $r$, $i$, $z$, and $Y$ bands in DES; $B$, $V$, and $R$ bands in ESIS; $u$ band in VOICE; $Z$, $Y$, $J$, $H$, and $K_s$ bands in VIDEO; and 3.6$\mu m$ and 4.5$\mu m$ bands in DeepDrill \citep{Zou2021a}.
In total, 2001 \xray\ sources in \es\ have forced-photometry measurements.
When matching X-ray sources to the forced-photometry catalog, we utilize the position of their matched VIDEO counterparts. Galactic extinction corrections are applied to the photometry utilized (see \citealt{Zou2021b} for details).

Photo-$z$s for \xray\ sources that are BL AGNs or non-BL AGNs are derived separately in our work. Here, BL AGNs are identified via classifications from spectroscopic surveys or the \texttt{SED\_BLAGN\_FLAG} $= 1$ flag in this work (see Appendix~\ref{a-blagn} for details of selecting BL AGN candidates through observed-frame SEDs; sources marked with \texttt{SED\_BLAGN\_FLAG} $= 1$ are likely to be BL AGNs).
Photo-$z$s of \xray\ sources that are not BL AGNs are provided in \citet{Zou2021b} for both the \wcdfs\ and \es\ fields, which use the default templates of the SED-fitting code \texttt{EAZY} \citep{Brammer2008}. In \wcdfs, 1792 of the matched photo-$z$s ($\approx 68\%$) have $Q_z < 1$ ($Q_z$ evaluates the quality of the photo-$z$; see equation 8 of \citealt{Brammer2008}), which are considered to be of high quality.
There are 455 sources with both spec-$z$ and $Q_z < 1$ photo-$z$ measurements, which are utilized to assess the reliability of the photo-$z$s.
The normalized median absolute deviation (NMAD) is $\sigma_{\rm NMAD} = 0.04$, and the outlier fraction ($f_{\rm outlier}$; defined as $ \Delta z/(1+z_{\rm spec}) > 0.15$, see \citealt{Zou2021b}) is $6.8\%$; these numbers are similar to the photometric-redshift reliability in \citet{Chen2018}.
The distribution of $\mathopen| \Delta z\mathclose|/(1+z_{\rm spec})$ is given in Figure~\ref{fig:photoz}; the distribution of phot-$z$ vs. spec-$z$ is also presented in Figure~\ref{fig:photoz}.
In \es, 1020 ($\approx 65\%$) of the photo-$z$s have $Q_z < 1$.
Among 230 sources with both spec-$z$ and $Q_z < 1$ photo-$z$ measurements, 
the comparison between spec-$z$ and photo-$z$ produces $\sigma_{\rm NMAD} = 0.03$ and $f_{\rm outlier} = 5.2\%$ (see Figure~\ref{fig:photoz}).\footnote{As stated in \citet{Zou2021b}, due to the deeper spectroscopic coverage in \wcdfs\ compared to \es, the photo-$z$ qualities in \wcdfs\ and \es\ listed here are not directly comparable.}

\begin{figure*}
\centering
\includegraphics[width=0.52\textwidth]{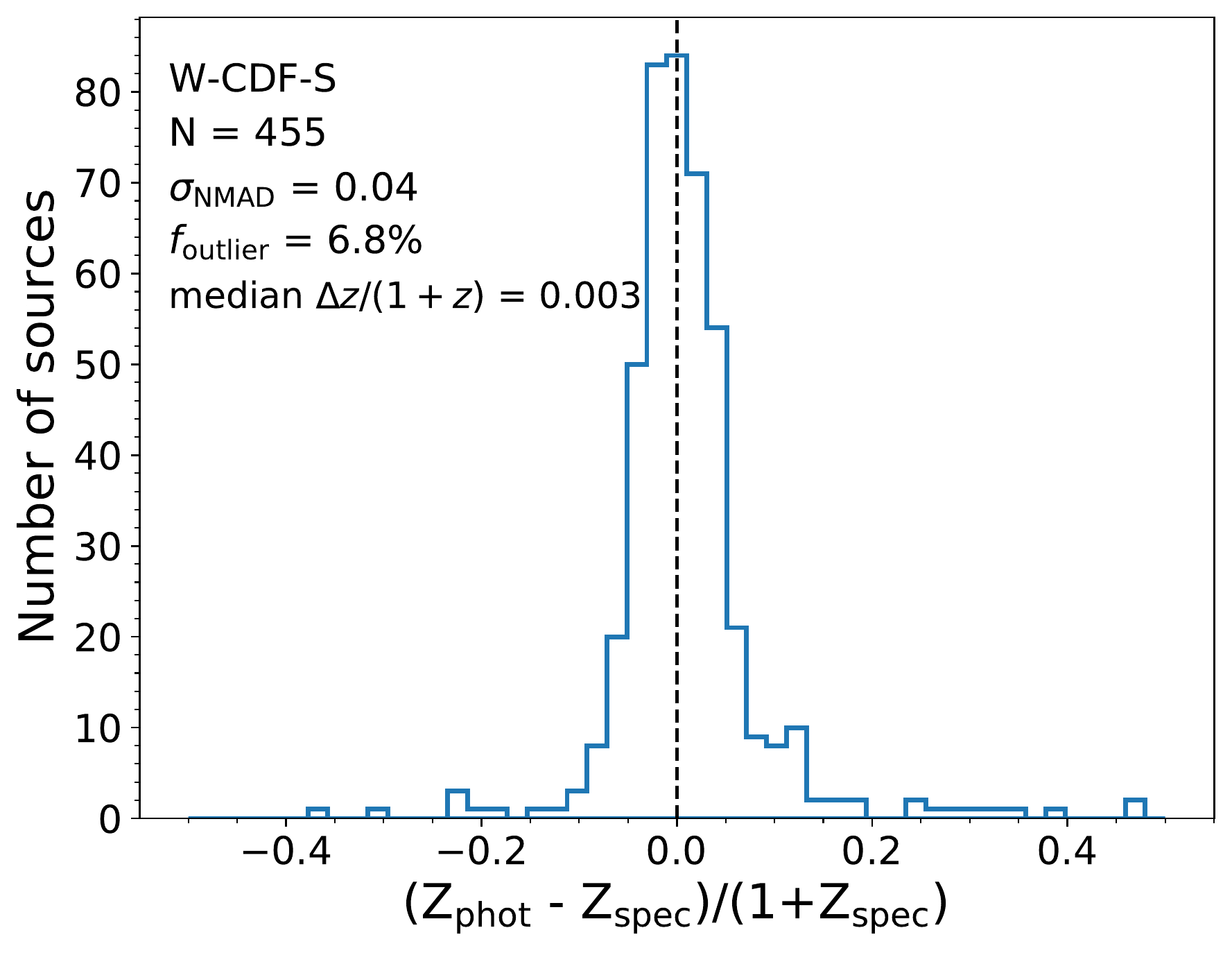}
\includegraphics[width=0.44\textwidth]{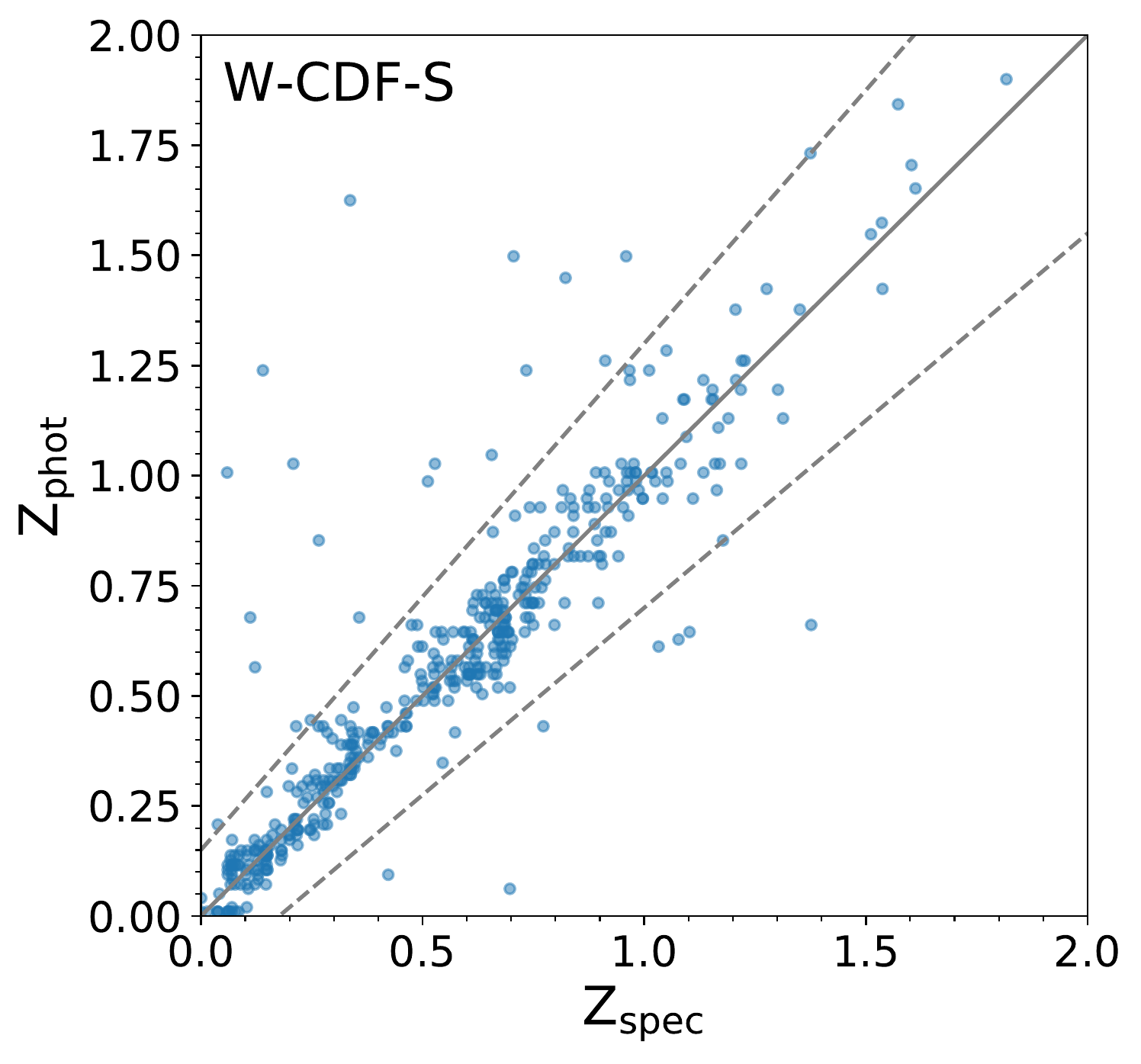}
\includegraphics[width=0.52\textwidth]{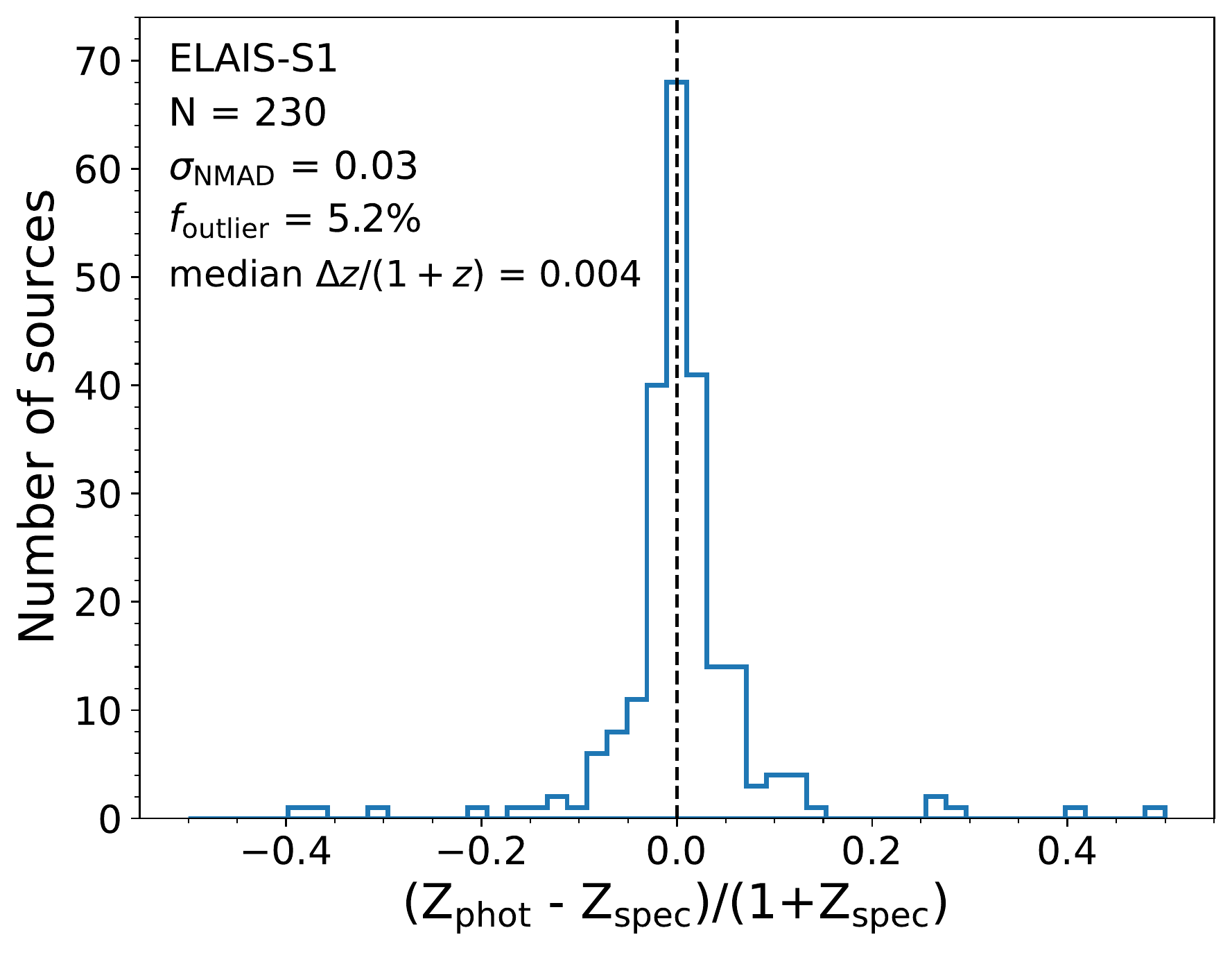}
\includegraphics[width=0.44\textwidth]{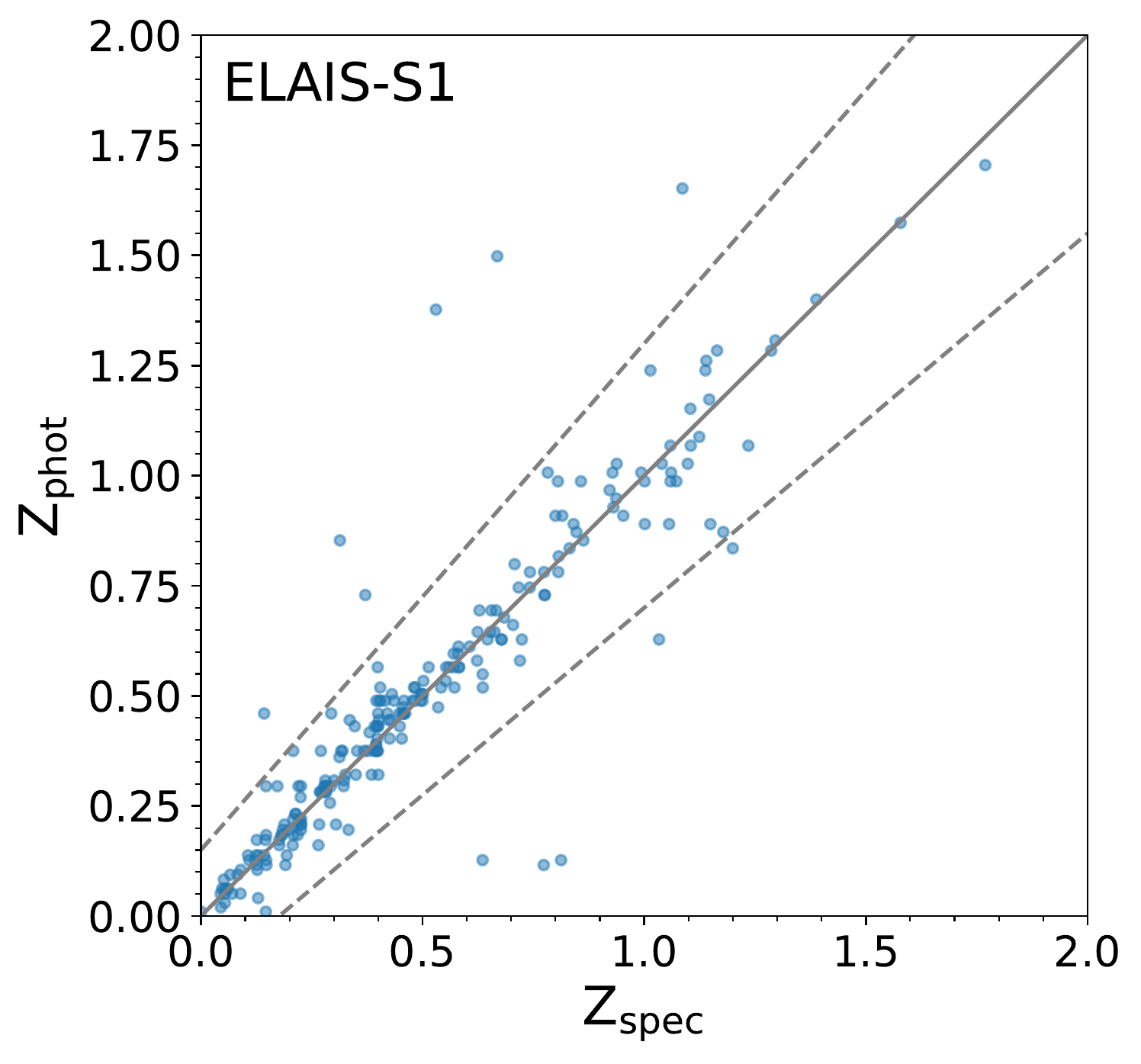}
\caption{
{\it Left:} Histogram of the fractional difference between the \texttt{EAZY} high-quality photo-$z$s and the spec-$z$s.
{\it Right:} Comparison between the \texttt{EAZY} high-quality photo-$z$s and the spec-$z$s. The gray solid line represents the $z_{\rm spec} = z_{\rm phot}$ relation; the gray dashed lines represent the $\mathopen|\Delta z\mathclose|/(1+z_{\rm spec}) = 0.15$ boundary. 
}
\label{fig:photoz}
\end{figure*}

For the $\approx 760$/430 \texttt{SED\_BLAGN\_FLAG} $= 1$ objects and the $Q_z \geqslant 1$ \texttt{SED\_BLAGN\_FLAG} $= 0.5$ objects (sources marked with \texttt{SED\_BLAGN\_FLAG} $= 0.5$ are possibly BL AGNs; see Appendix~\ref{a-blagn}) in \wcdfs/\es, we utilized an SED library designed for fitting AGN-dominated sources \citep{Salvato2009,Salvato2011} with 30 templates in total to estimate the photo-$z$ of these BL AGN candidates with {\sc LePhare} \citep{Arnouts1999,Ilbert2006}.
As the characterization of AGN-dominated sources can be substantially improved when the Lyman break is detected (the optical-to-NIR SED of BL AGNs roughly follows a featureless power law, which may produce large errors for photometric redshifts derived from the template fitting), we match the positions of the optical/NIR counterparts of X-ray sources to the {\it GALEX} catalog \citep{Martin2005} with a matching radius of 1$''$, and utilize the NUV and FUV fluxes when available. This approach allows the Lyman break to be detected at redshifts as low as $z = 0.7$ (when the FUV flux is available) or $z = 1.5$ (when the NUV flux is available).
$\chi_{\rm red}^{2} < 2$ and band number $> 10$ are utilized to select high-quality photo-$z$ estimates ($\approx 74\%$ of them have high-quality photo-$z$).
BL AGNs identified in spectroscopic surveys that have high-resolution ($> 100$) spec-$z$ measurements are utilized to assess the {\sc LePhare} photo-$z$ quality.
Among these 174/138 sources in \wcdfs/\es, 130/102 have high-quality {\sc LePhare} photo-$z$ measurements utilizing the method above. 
A comparison between these spec-$z$ and photo-$z$ measurements produces $\sigma_{\rm NMAD} \approx 0.07$ and $f_{\rm outlier} \approx 18\%$ for \wcdfs, and $\sigma_{\rm NMAD} \approx 0.06$ and $f_{\rm outlier} \approx 20\%$ for \es\ (see Figure~\ref{fig:type1photoz}). 
However, as noted in \citet{Salvato2009}, the photo-$z$ performance deteriorates when a source is fainter in the optical. For the BL AGNs with spec-$z$ measurements, the median $i$-band mag is $\approx 20$; this brightness is $\approx 22$ for BL AGN candidates without spec-$z$ measurements. 
In addition, only $\approx 40\%$ of the \texttt{SED\_BLAGN\_FLAG} $= 1$ objects and the $Q_z \geqslant 1$ \texttt{SED\_BLAGN\_FLAG} $= 0.5$ objects are matched to {\it GALEX} sources; this number is $\approx 85\%$ for spectroscopically confirmed BL AGNs.
Thus, caution is advised when using {\sc LePhare} photo-$z$ measurements for \texttt{SED\_BLAGN\_FLAG} $= 0.5$ or $1$ sources.

\begin{figure*}
\centering 
\includegraphics[width=0.45\textwidth]{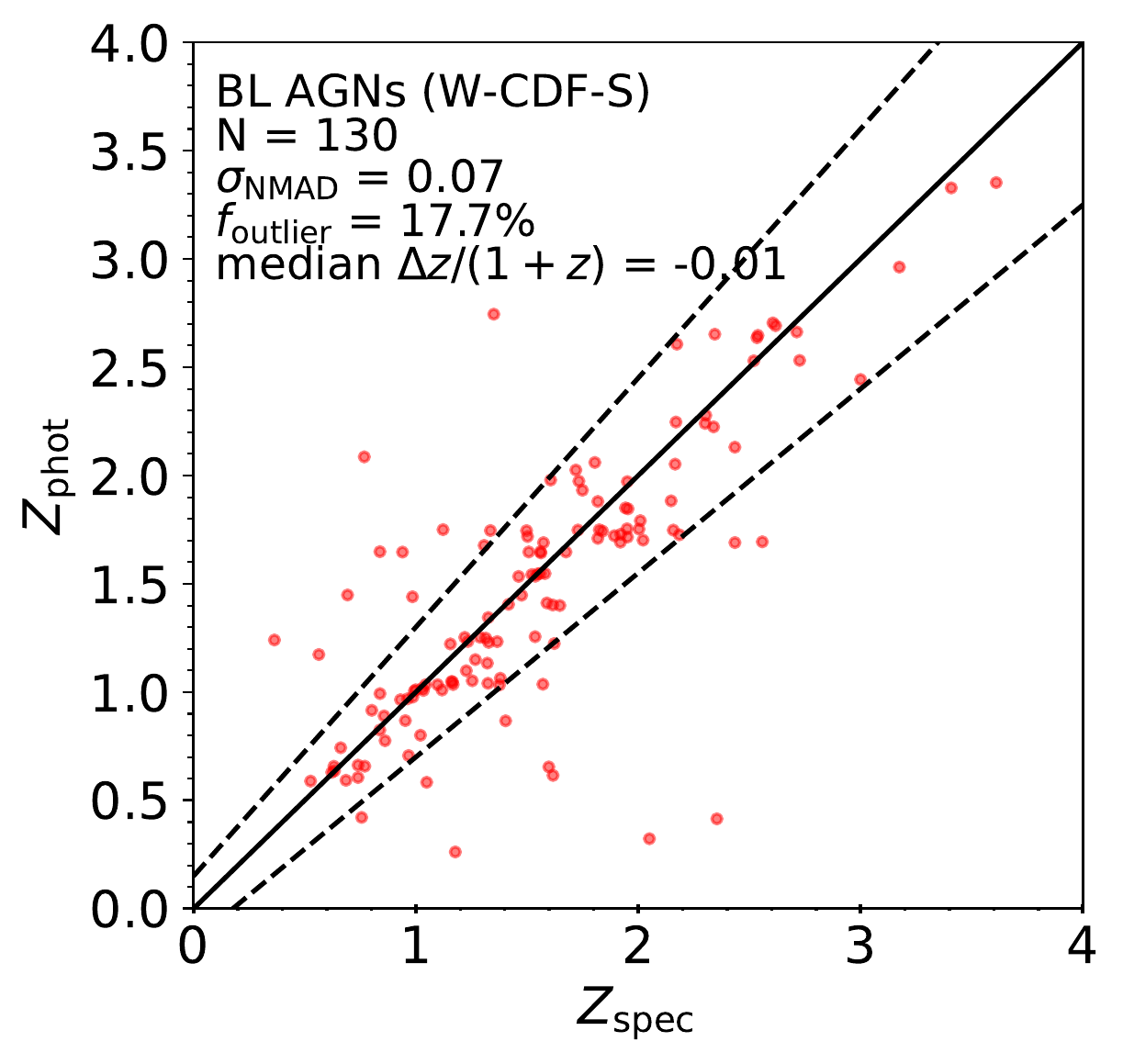}
\includegraphics[width=0.45\textwidth]{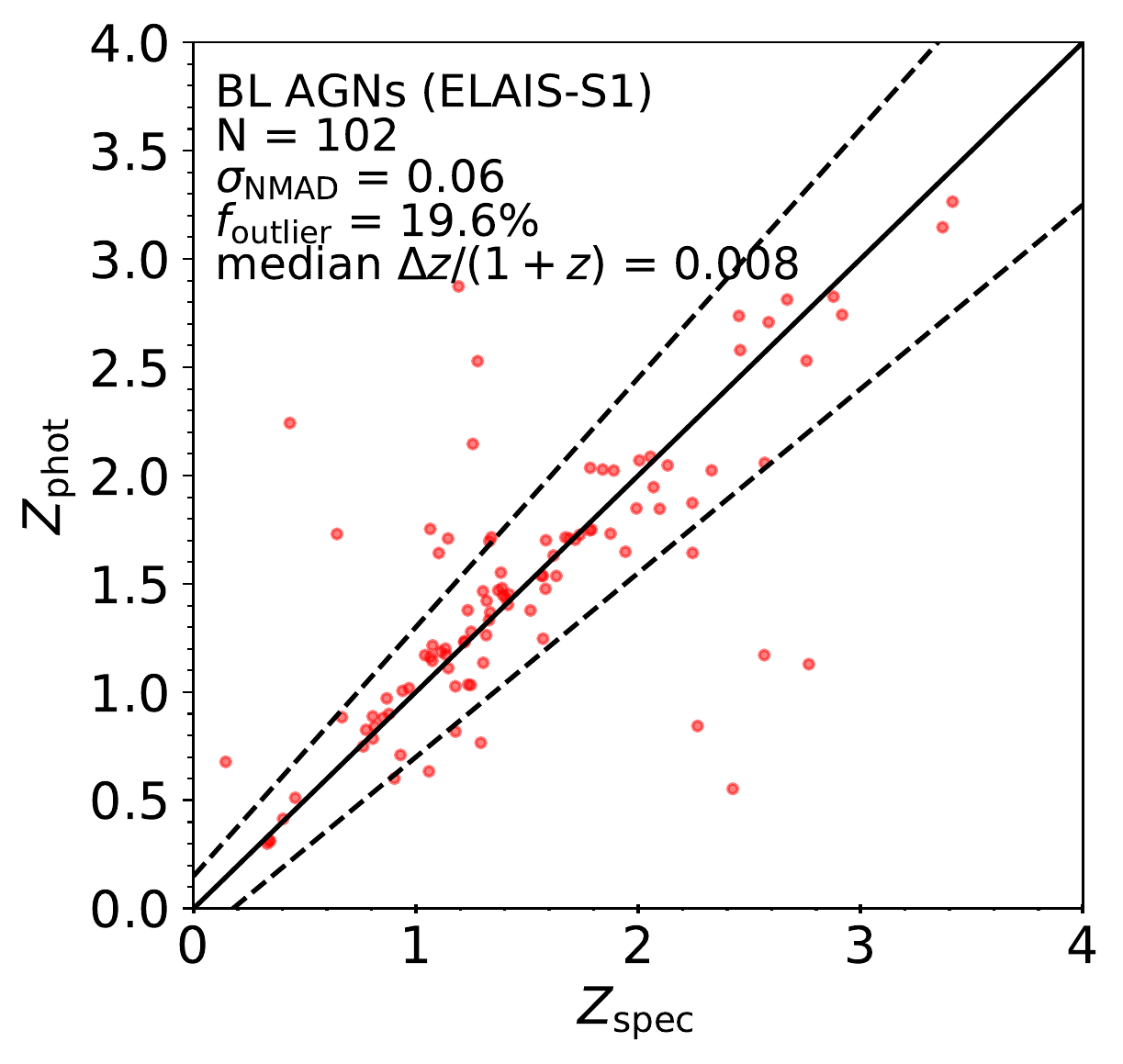}
\caption{The comparison between the photo-$z$s and the spec-$z$s of broad-line AGNs that have both high-quality photo-$z$ from {\sc LePhare} and resolution $> 100$ spec-$z$ measurements. The black solid line represents the $z_{\rm spec} = z_{\rm phot}$ relation; the black dashed lines represent the $\mathopen|\Delta z\mathclose|/(1+z_{\rm spec}) = 0.15$ boundary.}
\label{fig:type1photoz}
\end{figure*}

Combining all the information above, we report the high-quality photo-$z$ measurements in the column \texttt{PHOTOZ\_BEST} (see Appendix~\ref{a-column}):
$Q_z < 1$ \texttt{EAZY} photo-$z$ measurements are adopted for sources that have  \texttt{SED\_BLAGN\_FLAG} $< 1$ and are not identified as BL AGNs in spectroscopic surveys (1792 in \wcdfs\ and 1020 in \es); high-quality {\sc LePhare} photo-$z$ measurements are adopted for spectroscopically identified BL AGNs, \texttt{SED\_BLAGN\_FLAG} $= 1$ objects, and \texttt{SED\_BLAGN\_FLAG} $= 0.5$ objects without $Q_z < 1$ \texttt{EAZY} photo-$z$ measurements (738 in \wcdfs\ and 460 in \es).
The catalog has high-quality photo-$z$ measurements for 1833/1117 \xray\ sources in \wcdfs/\es\ without spec-$z$ measurements.
The cumulative histogram of the $i$-band magnitude of \xray\ sources with either spec-$z$ measurements or high-quality photo-$z$ measurements is presented in Figure~\ref{fig:zcumu}.
 
\begin{figure*}
\centering
\includegraphics[width=0.48\textwidth]{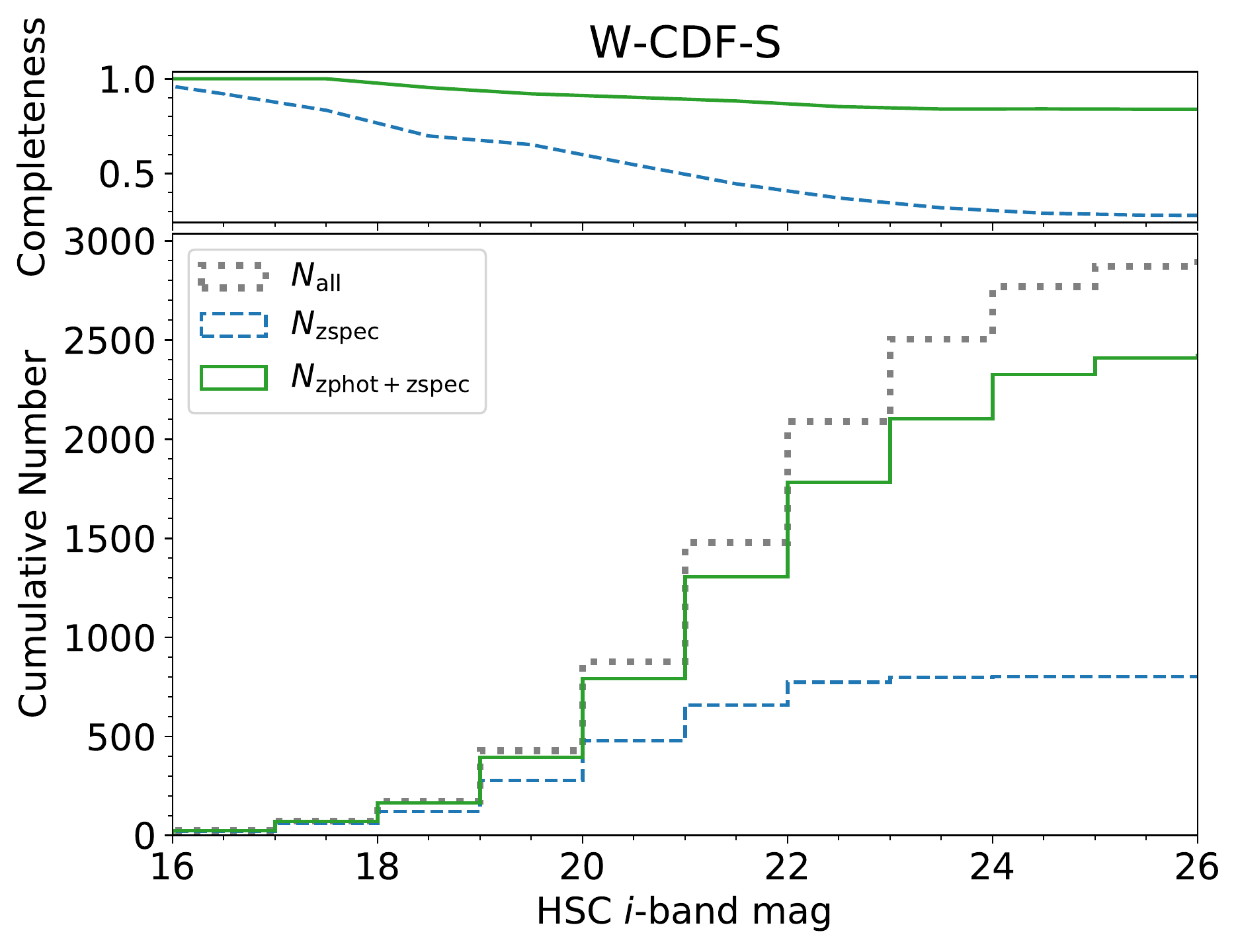}
\includegraphics[width=0.48\textwidth]{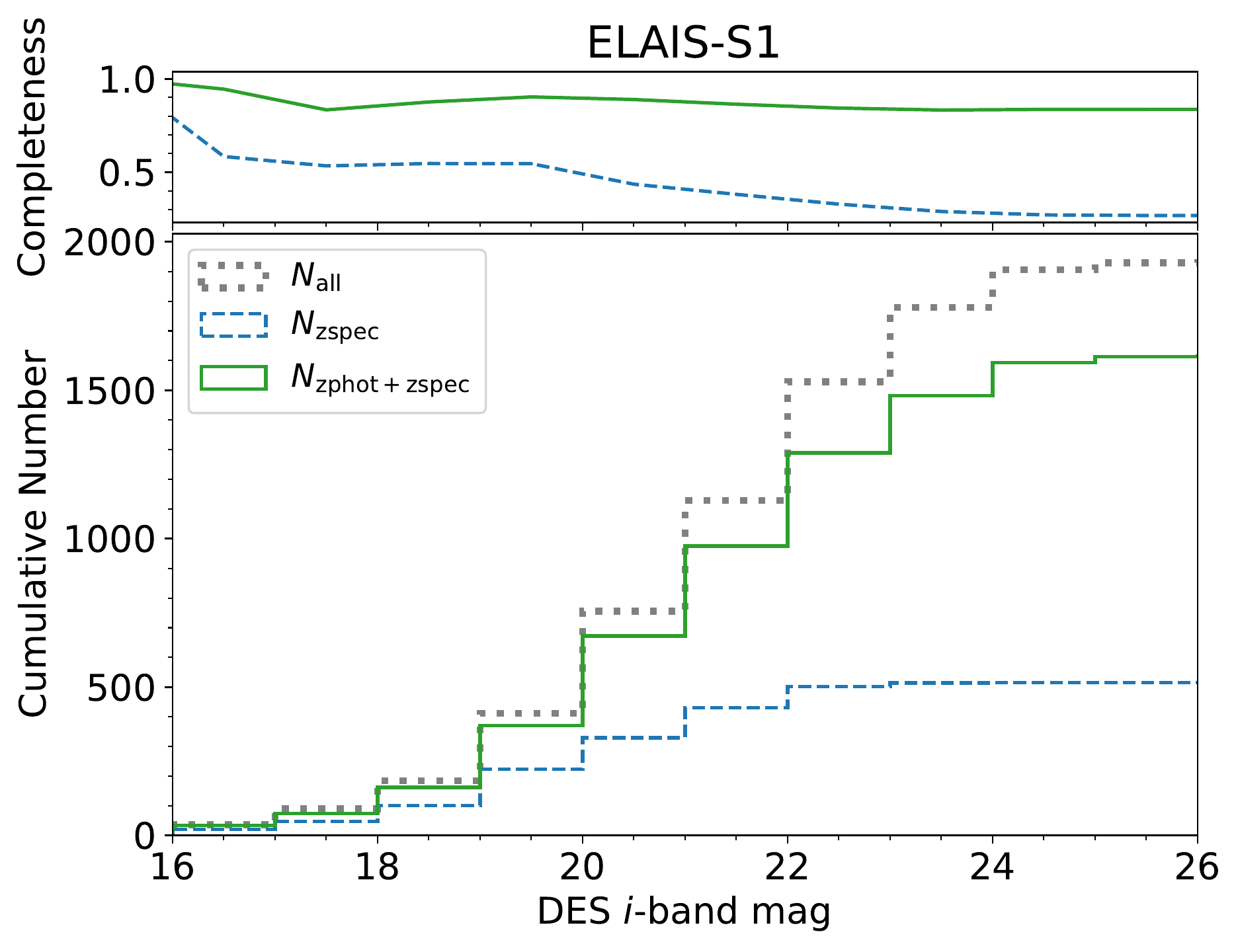}
\caption{The cumulative distributions of the HSC/DES $i$-band magnitudes for $i$-band detected \xray\ sources in \wcdfs/\es\ located within the region with forced photometry (gray dotted histogram), \xray\ sources that have spec-$z$ measurements (blue dashed histogram), and \xray\ sources that have either spec-$z$ or high-quality photo-$z$ measurements (green solid histogram). The blue dashed/green solid curve in the top panels is the fraction of \xray\ sources in the region with forced photometry with spec-$z$/high-quality photo-$z$ or spec-$z$ measurements as a function of $i$-band magnitude. Objects without spec-$z$ or high-quality photo-$z$ measurements are \texttt{SED\_BLAGN\_FLAG} $= 0$ objects with \texttt{EAZY} $Q_z > 1$ or \texttt{SED\_BLAGN\_FLAG} $> 0$ objects with {\sc LePhare} $\chi^2_{\rm red} \geqslant 2$ or band number $\leqslant 10$.}
\label{fig:zcumu}
\end{figure*}

\section{Source properties and classification} \label{sc-sp}

For the 919/585 \xray\ sources with spec-$z$ measurements and 1833/1117 \xray\ sources with high-quality \texttt{EAZY} or {\sc LePhare} photo-$z$ measurements (but lacking spec-$z$ measurements) in \wcdfs/\es, 
we estimate their \xray\ luminosity at rest-frame 2--10 keV ($L_{2-10 \rm~keV}$) assuming a power-law spectrum with $\Gamma_{\rm eff} = 1.8$ (which is a typical power-law photon index for AGNs; e.g., \citealt{Lanzuisi2013,Yang2016,Liu2017}) modified by Galactic absorption, utilizing source count rates in the priority order of hard band, full band, and soft band.
This prioritization minimizes \xray\ absorption effects.
Figure~\ref{fig:lxdistrib} displays the distribution of 
$L_{2-10 \rm~keV}$, as well as the $L_{2-10 \rm~keV}$ vs. $z$ distribution.
In Figure~\ref{fig:lxz_compare}, we show the $L_{2-10 \rm~keV}$ vs. $z$ distribution for the whole XMM-SERVS survey, and compare it with distributions from selected deep pencil-beam \xray\ surveys (CDF-S; \citealt{Luo2017}; \hbox{CDF-N}; \citealt{Xue2016}) and shallower X-ray surveys over wider areas (XMM-XXL North; e.g., \citealt{Menzel2016}; Stripe 82X; e.g., \citealt{Ananna2017,Lamassa2019}).\footnote{We note that for X-ray sources in CDF-S, \hbox{CDF-N}, and Stripe 82X, both spectroscopic redshifts and high-quality photometric redshifts are available, so the sources included in our comparison are those with either spec-$z$ or photo-$z$; for the XMM-XXL North survey, the sources included are only those with spec-$z$ measurements, due to the lack of available photo-$z$ measurements in this area, currently.}
While deep pencil-beam surveys can detect less-luminous X-ray sources, the AGN sample size provided by the XMM-SERVS survey is substantially larger than the sample size these deep surveys could provide.
When compared to shallower X-ray surveys over wider areas, we can see that the XMM-SERVS survey detects a significantly larger number of moderate-luminosity AGNs at log $L_{\rm X}$ $\sim 42$--44; also, due to the superb multiwavelength coverage of XMM-SERVS, the overall number of detected X-ray sources with reliable redshift measurements is larger at all redshifts.
The $L_{2-10 \rm~keV}$ vs. $z$ coverage of XMM-SERVS is similar to that of the \chandra\ \textit{COSMOS-Legacy} survey \citep[e.g.,][]{Marchesi2016}, though the \chandra\ \textit{COSMOS-Legacy} survey is somewhat deeper: the peak of the log $L_{2-10 \rm~keV}$ distribution of \xray\ sources in the \chandra\ \textit{COSMOS-Legacy} survey is $\approx 0.5$ dex smaller than that of X-ray sources in XMM-SERVS. At the same time, the sample size of AGNs with reliable $L_{\rm X}$ estimation provided by XMM-SERVS is $\approx 3$ times that of the \chandra\ \textit{COSMOS-Legacy} survey.

\begin{figure*}
\includegraphics[width=0.45\textwidth]{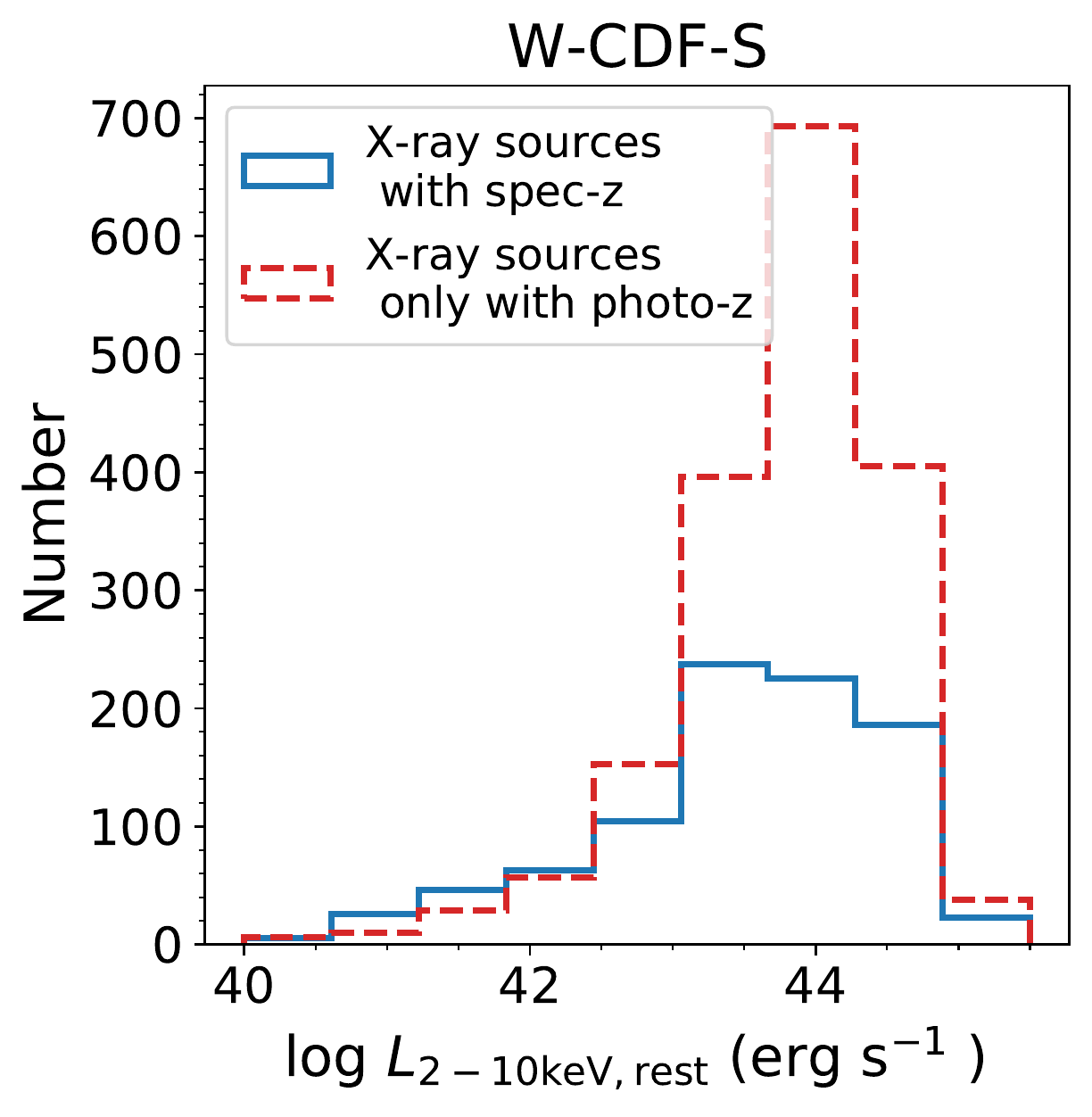}
\includegraphics[width=0.45\textwidth]{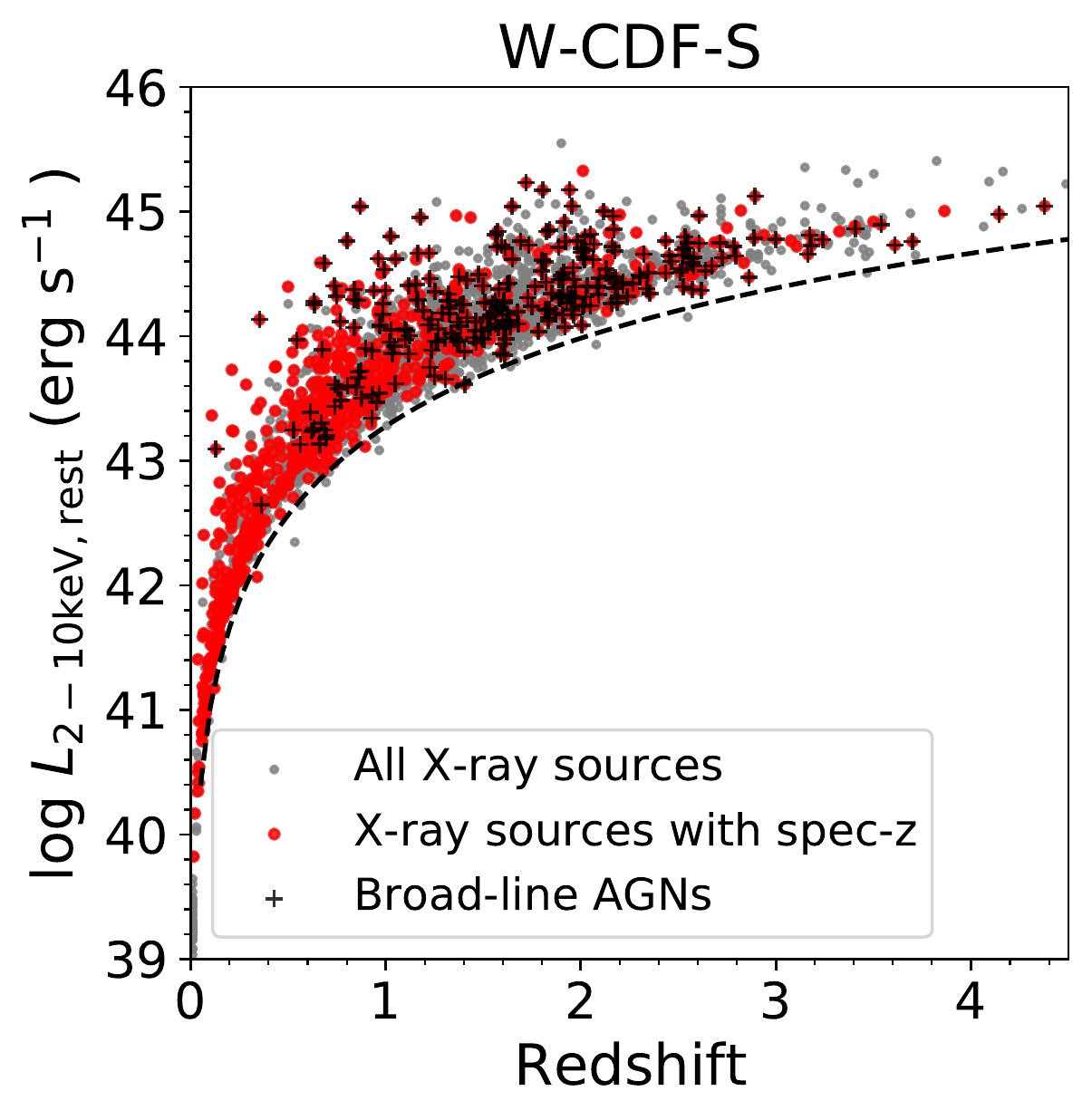}
\newline
\includegraphics[width=0.45\textwidth]{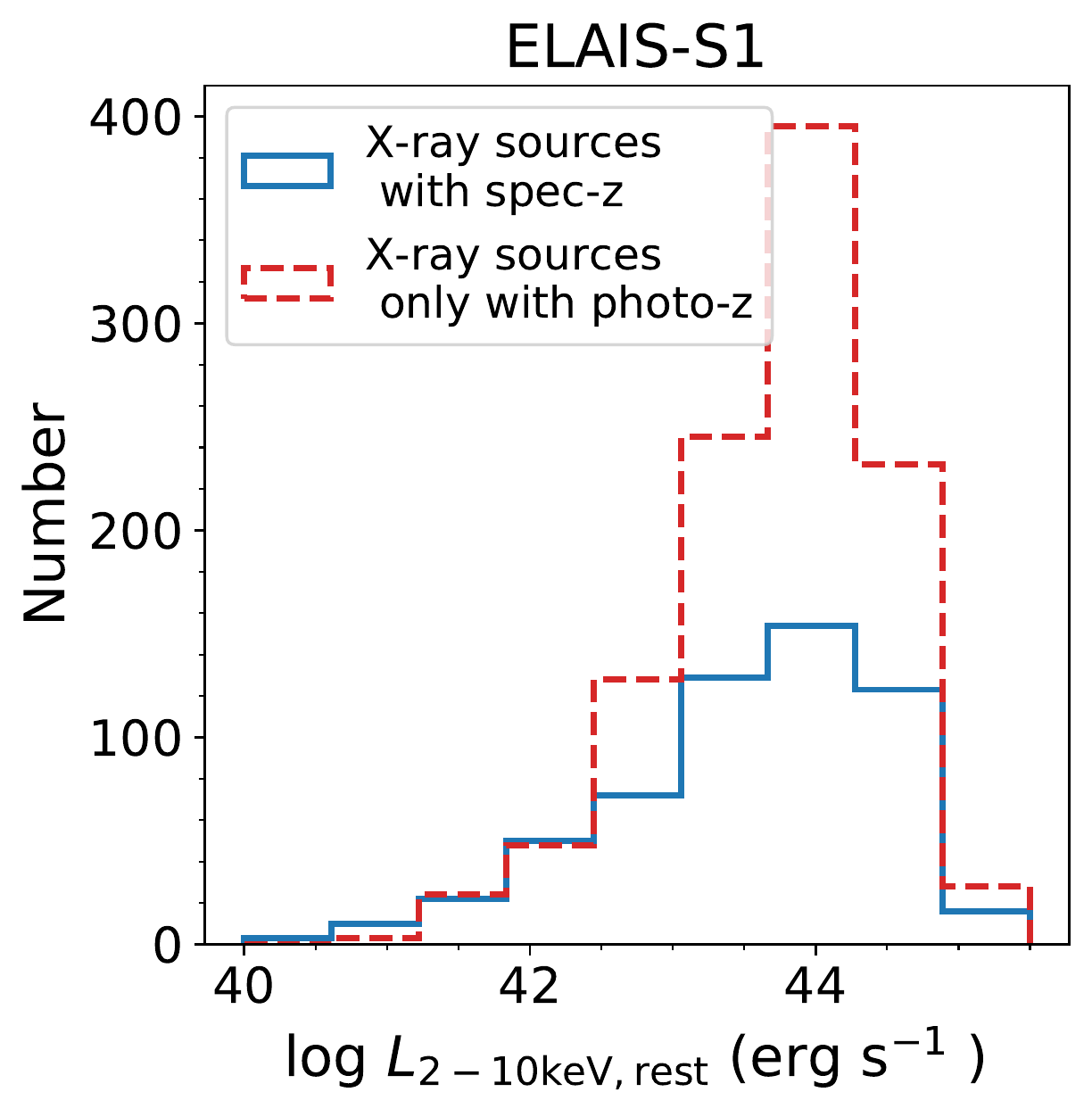}
\includegraphics[width=0.45\textwidth]{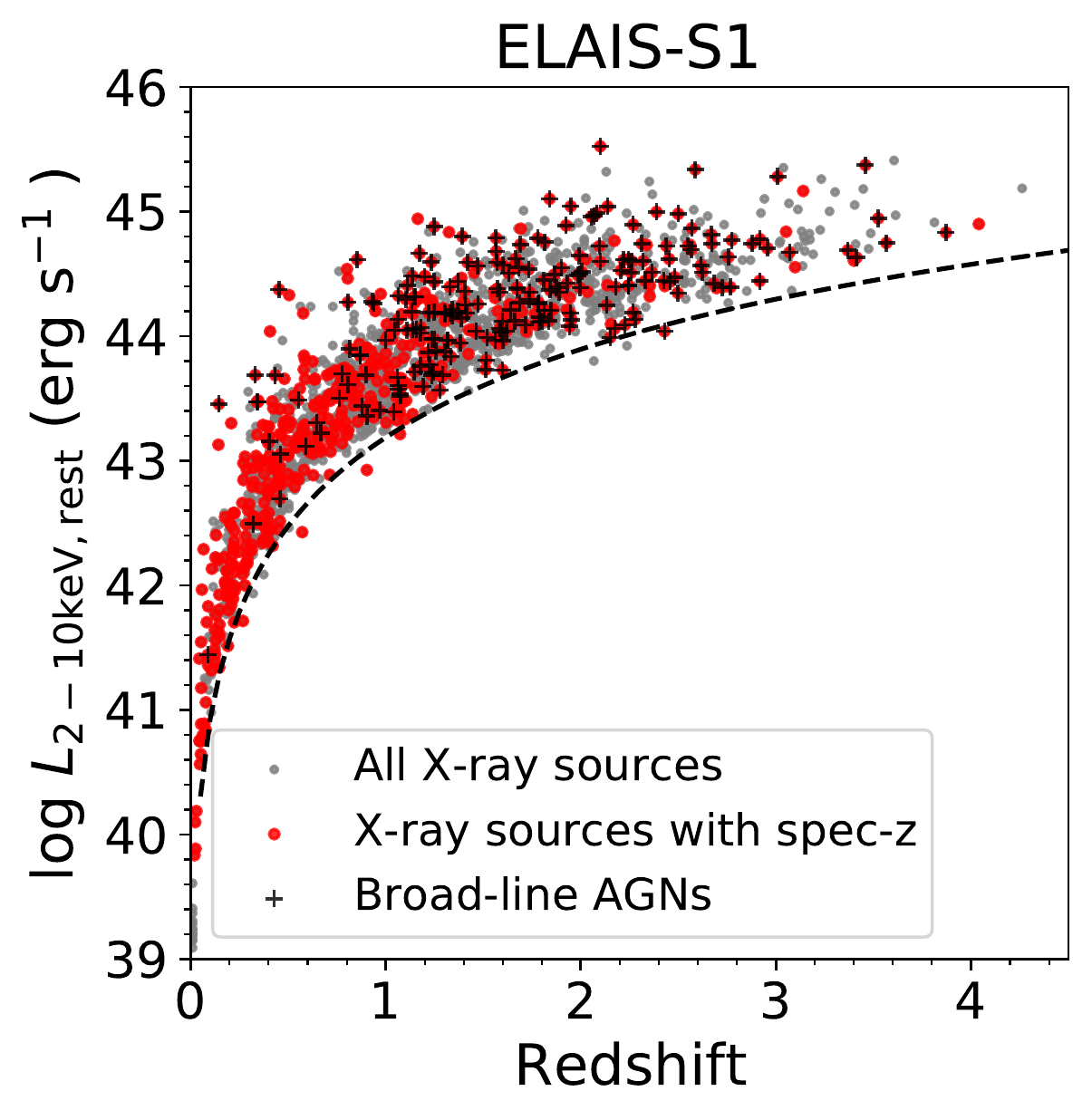}
\caption{
{\it Left:} The distribution of rest-frame 2--10 keV X-ray luminosity for X-ray sources with spec-$z$ measurements (blue solid histogram) and \xray\ sources without spec-$z$ measurement but having high-quality photo-$z$ measurements (red dashed histogram), respectively.
{\it Right:} The $L_{2-10 \rm~keV}$ vs. $z$ distribution of all X-ray sources with spec-$z$ or high-quality photo-$z$ measurements (gray dots). X-ray sources with spec-$z$ measurements are marked by the red circles; among these sources, broad-line AGNs are marked as black pluses. The full-band sensitivity limit generated assuming $\Gamma_{\rm eff} = 1.8$ is represented by the black dashed line.
}
\label{fig:lxdistrib}
\end{figure*}

\begin{figure}
\includegraphics[width=0.45\textwidth]{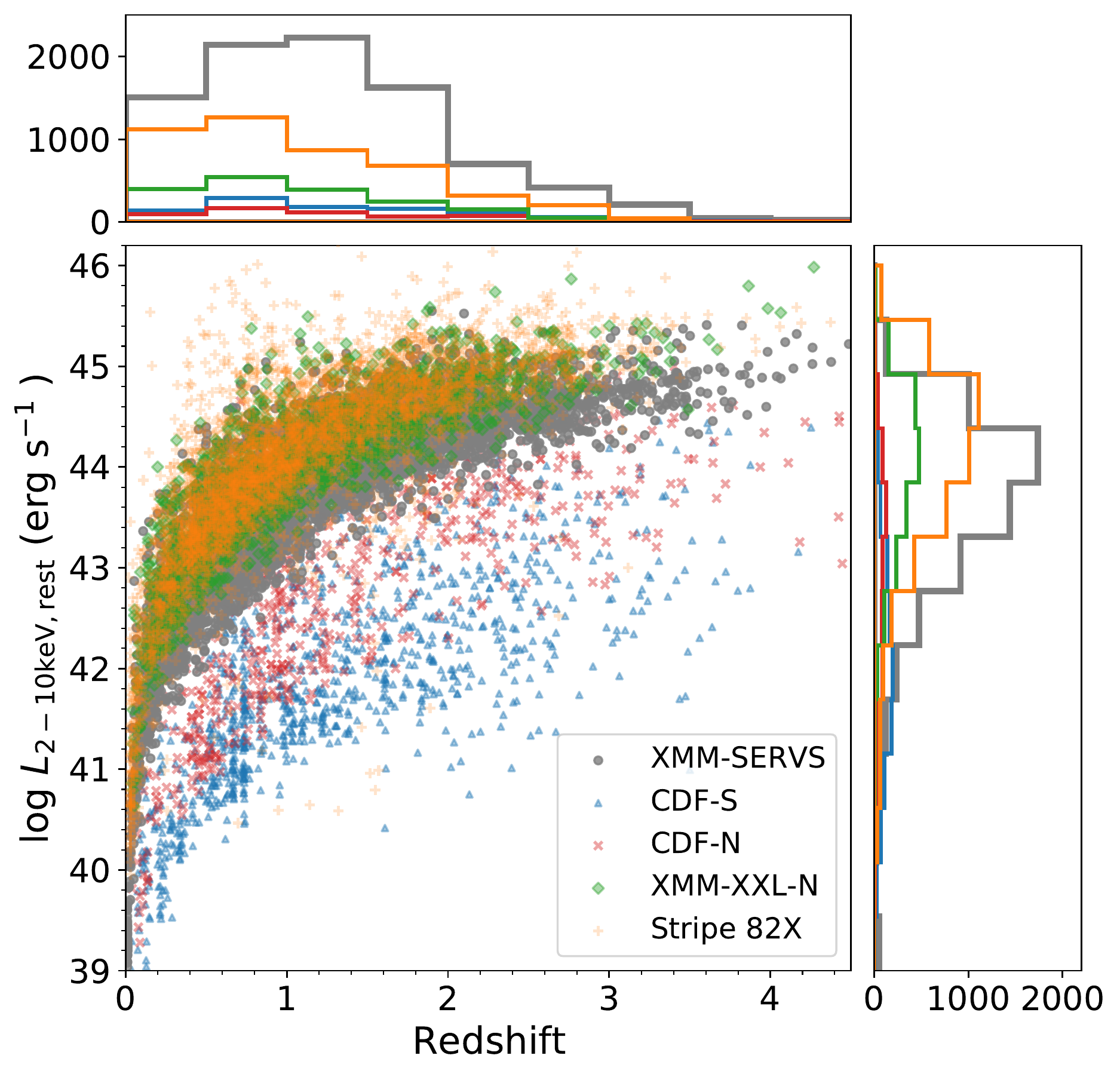}
\caption{
The $L_{2-10 \rm~keV}$ vs.\ $z$ distribution of X-ray sources with spec-$z$ or high-quality photo-$z$ measurements in the full XMM-SERVS survey (gray dots). 
For comparison, the $L_{2-10 \rm~keV}$ vs. $z$ distributions of X-ray sources in \hbox{CDF-S} \citep{Luo2017}, \hbox{CDF-N} \citep{Xue2016}, \hbox{XMM-XXL North} \citep[e.g.,][]{Menzel2016}, and \hbox{Stripe 82X} \citep[e.g.,][]{Ananna2017,Lamassa2019} are shown as blue triangles, red crosses, green diamonds, and orange pluses, respectively.
The distributions of $z$ for X-ray sources in the survey fields mentioned above are shown in the top sub-panel, with colors the same as those in the legend; the distributions of $L_{2-10 \rm~keV}$ are shown in the right sub-panel, with colors the same as those in the legend.}
\label{fig:lxz_compare}
\end{figure}

We also perform basic AGN selection for X-ray sources in our catalogs following criteria from section~2.3 of \citet{Brandt2015} and references therein. The specific criteria utilized are the following:

\begin{enumerate}
  \item Identified as broad-line AGNs in spectroscopic surveys (280 AGNs in \wcdfs; 208 AGNs in \es).
  \item Has observed $L_{2-10 \rm keV} > 3 \times 10^{42}~\rm erg~s^{-1}$ (in the rest frame) when spec-$z$ measurements or high-quality \texttt{EAZY} or {\sc LePhare} photo-$z$ measurements are available (2459 AGNs in \wcdfs; 1511 AGNs in \es).
  \item Has a power-law effective photon index $\Gamma  \leqslant 1$ (412 AGNs in \wcdfs; 314 AGNs in \es; see Figure~\ref{fig:hardness}). This criterion helps select hard X-ray sources that are heavily obscured, which are likely AGNs rather than  X-ray binary populations \citep[e.g.,][]{Alexander2005,Brandt2015}.
  \item Identified as AGNs when utilizing \texttt{X-CIGALE} to perform SED template-fitting in Appendix~\ref{a-blagn} (2717 AGNs in \wcdfs; 1684 AGNs in \es). AGNs selected via this SED-based selection method already include AGNs selected from empirical methods that use large X-ray-to-optical or X-ray-to-NIR flux ratios: $\log f_x/f_{i} > -1$ or $\log f_x/f_{Ks} > -1.2$ (see Figure~\ref{fig:fxfoir}). For the small fraction of \xray\ sources that are not located within the VIDEO footprint that has forced photometry from optical to NIR available, we adopt large X-ray-to-optical flux ratios ($\log f_x/f_{i} > -1$) to identify AGNs (126 AGNs in \wcdfs; 109 AGNs in \es).
  \item Have red MIR colors (obtained from the four-band IRAC data) that meet the AGN criteria in \citet{Lacy2004}, \citet{Stern2005}, or \cite{Donley2012}. Among 1897/1042 \xray\ sources in \wcdfs/\es\ that are detected in four IRAC bands (as reported in \textit{Spitzer} Data Fusion DR1; \citealt{Vaccari2015}), 1441 objects in \wcdfs\ ($\approx 76\%$)  and 810 objects in \es\ ($\approx 78\%$) are MIR-selected AGNs. Only 52/38 of these objects in \wcdfs/\es\ are not already identified as AGNs with the first four methods (see Figure~\ref{fig:iraccolor}). If we only adopt the conservative selection criteria from \cite{Donley2012}, only 24/16 additional AGNs are identified in \wcdfs/\es\ via the MIR approach.
  \item Have ATLAS radio counterparts and SWIRE 24$\mu$m counterparts and satisfy the $q_{24} < 0$ radio AGN selection criterion in \citet{Donley2005}, where $q_{24}$ is defined as ${\rm log}(S_{\rm 24,obs}/S_{\rm 1.4~GHz,obs})$ ($S_{\rm 24,obs}$ is the SWIRE 24~$\mu$m flux density, and $S_{\rm 1.4~GHz,obs}$ is the 1.4~GHz flux density). Among 213 objects in \wcdfs\ and 86 objects in \es\ that have both 24~$\mu$m and 1.4~GHz counterparts detected, 49/15 objects in \wcdfs/\es\ are identified as AGNs. 11/0 of these objects in \wcdfs/\es\ are not already identified as AGNs via the first four methods.
\end{enumerate}

The combination of all these methods identifies 3186 AGNs in \wcdfs\ and 1985 AGNs in \es, which is $\approx 89\%$/87\% of X-ray sources matched to multiwavelength counterparts with \pany\ $> 0.1$. The non-AGN \xray\ sources could be attributed to stars, bright galaxies (which can contain \xray\ binaries and/or low-luminosity AGNs), and other source classes (see Appendix~\ref{a-star}).

\begin{figure*}
\includegraphics[width=0.48\textwidth]{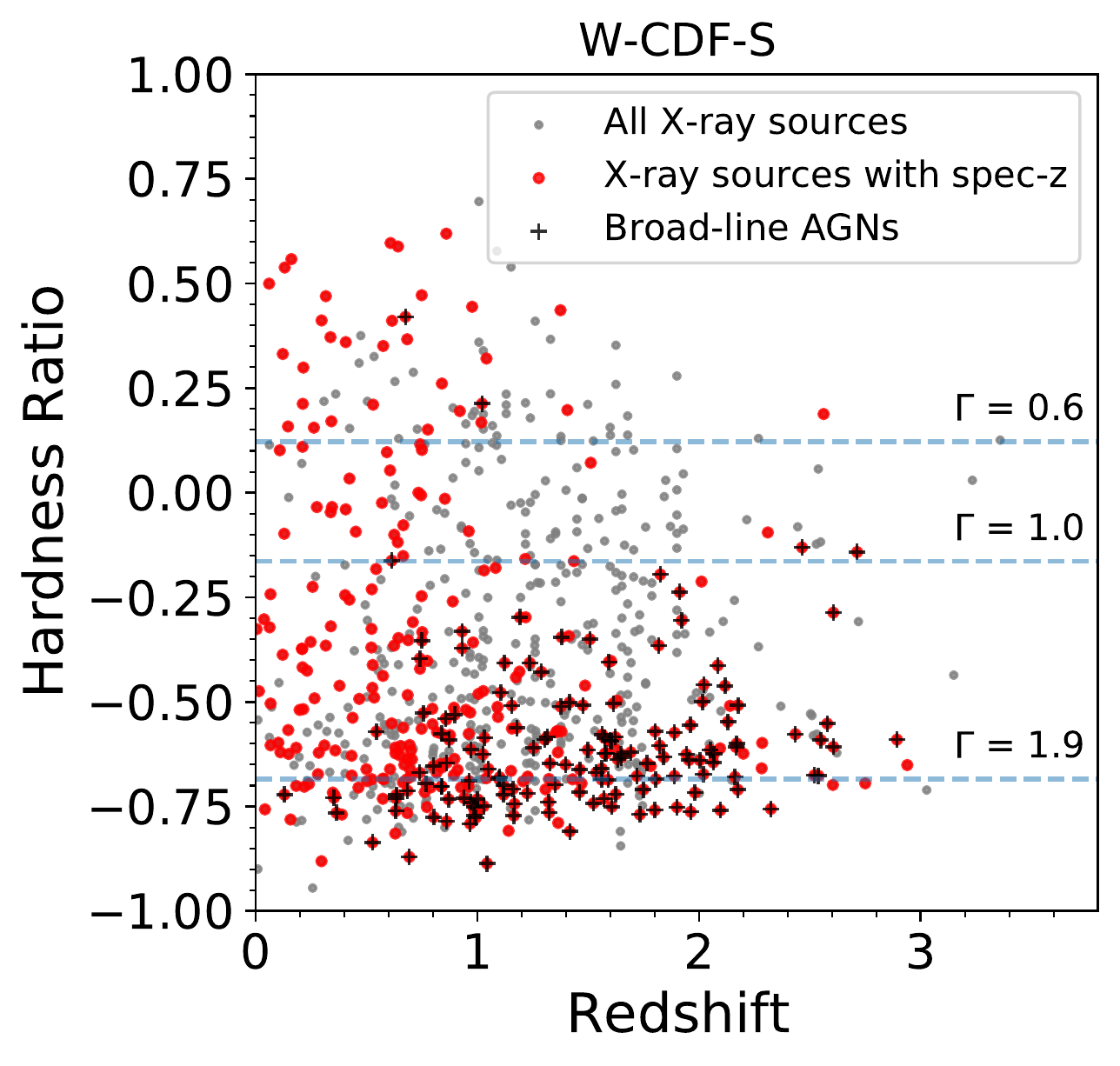}
\includegraphics[width=0.48\textwidth]{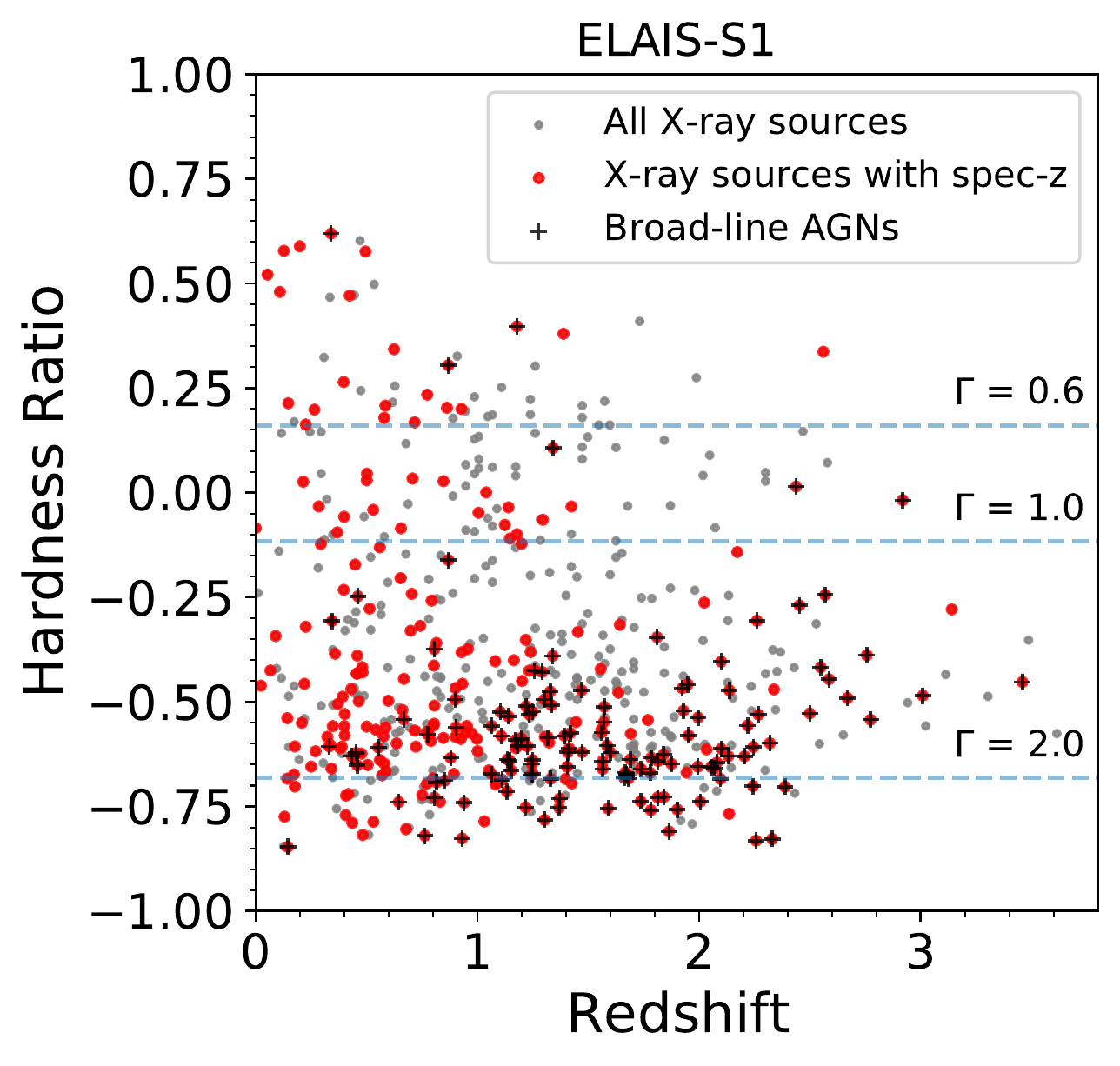}
\caption{Hardness ratio vs.\ redshift for X-ray sources that are detected in both the soft and hard bands, and have reliable redshift measurements (grey dots) in \wcdfs\ (\textit{left}) and \es\ (\textit{right}). Assuming a power-law spectrum modified by Galactic absorption, effective power-law photon indices can be derived from hardness ratios, which are utilized in AGN selection (objects with $\Gamma_{\rm eff}  \leqslant 1$ are classified as AGNs).
X-ray sources with spec-$z$ measurements are marked by the red circles; among these sources, broad-line AGNs are marked as black pluses.
}
\label{fig:hardness}
\end{figure*}

\begin{figure*}
\centering
\includegraphics[width=0.48\textwidth]{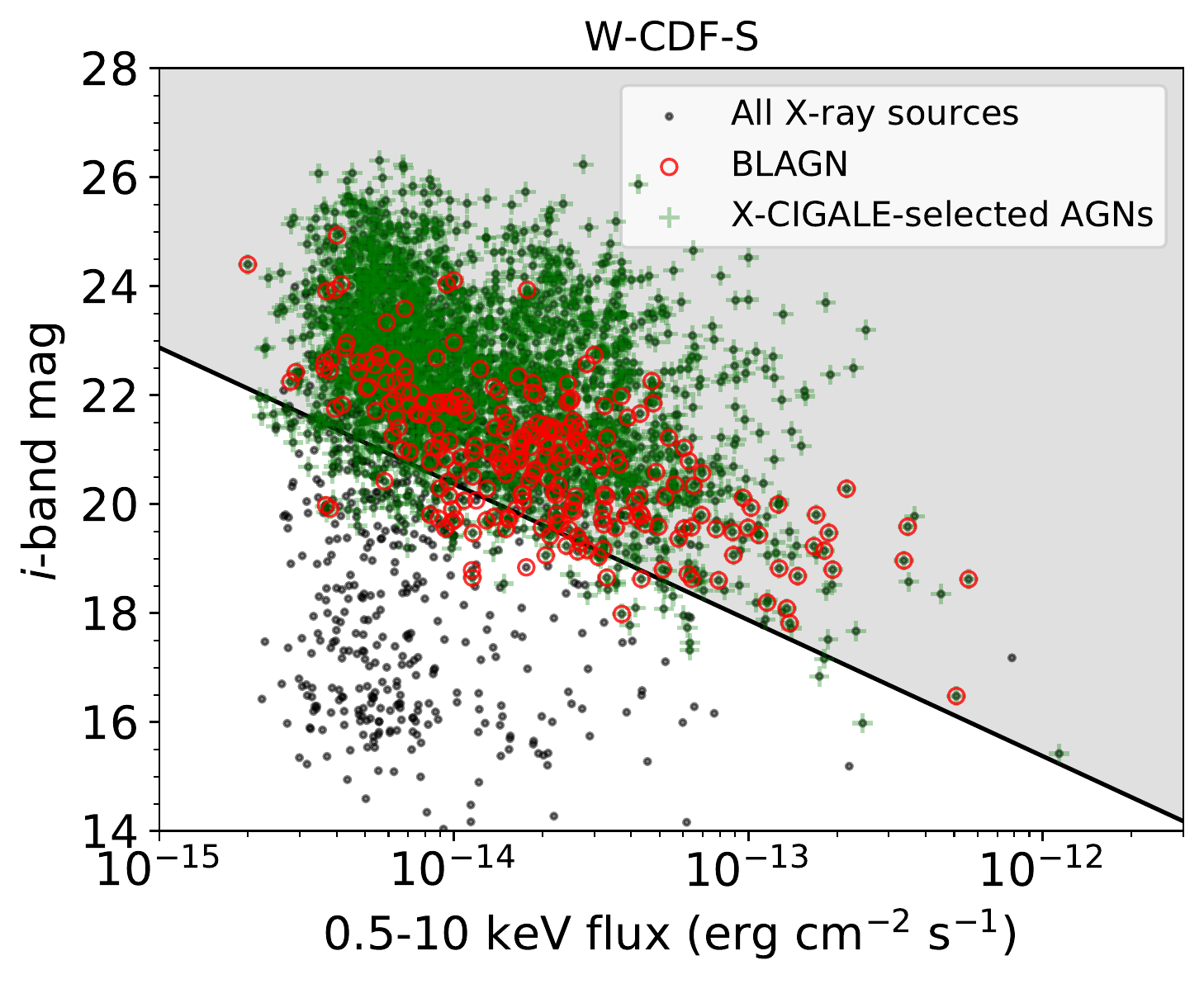}
\includegraphics[width=0.48\textwidth]{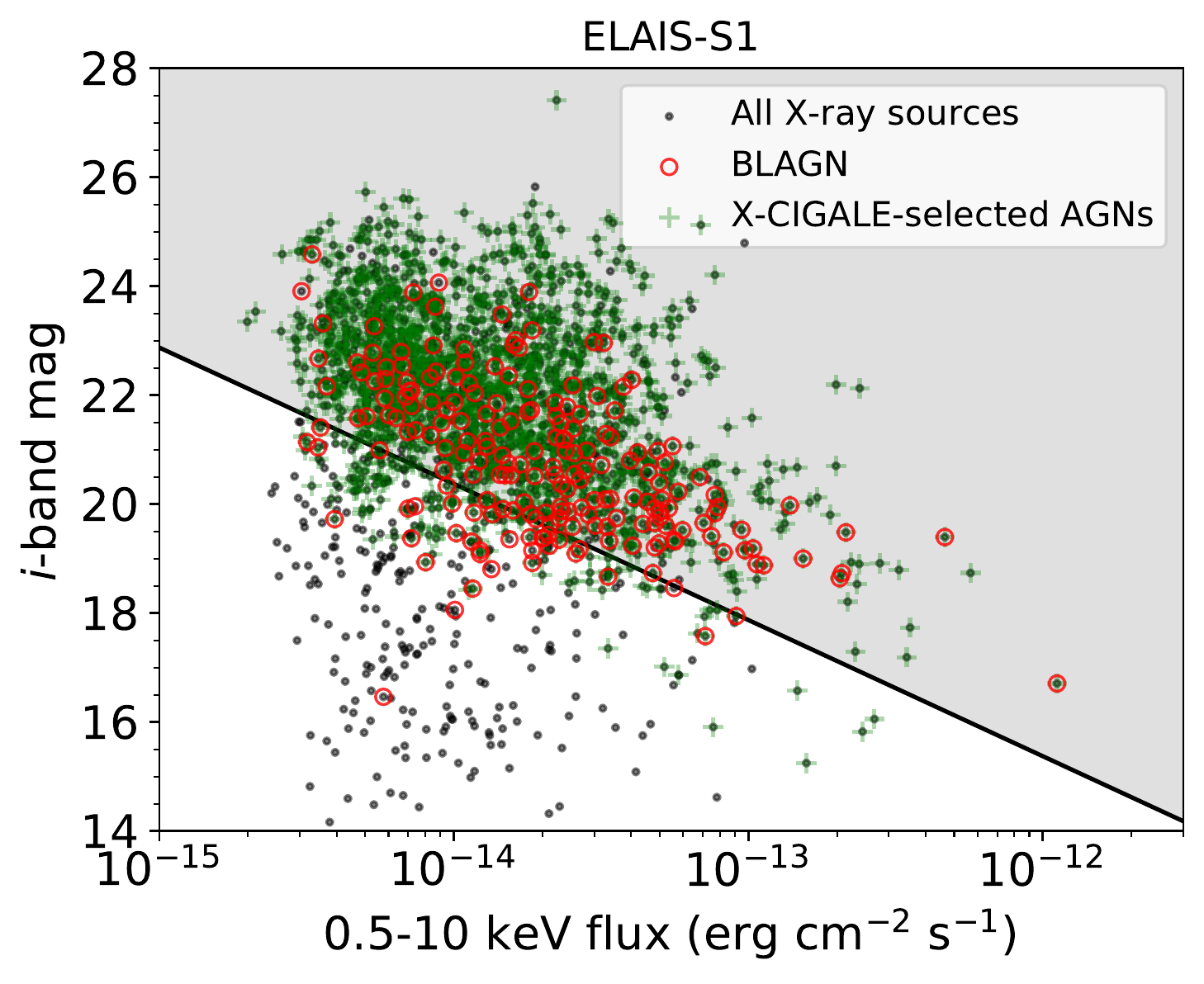}
\includegraphics[width=0.48\textwidth]{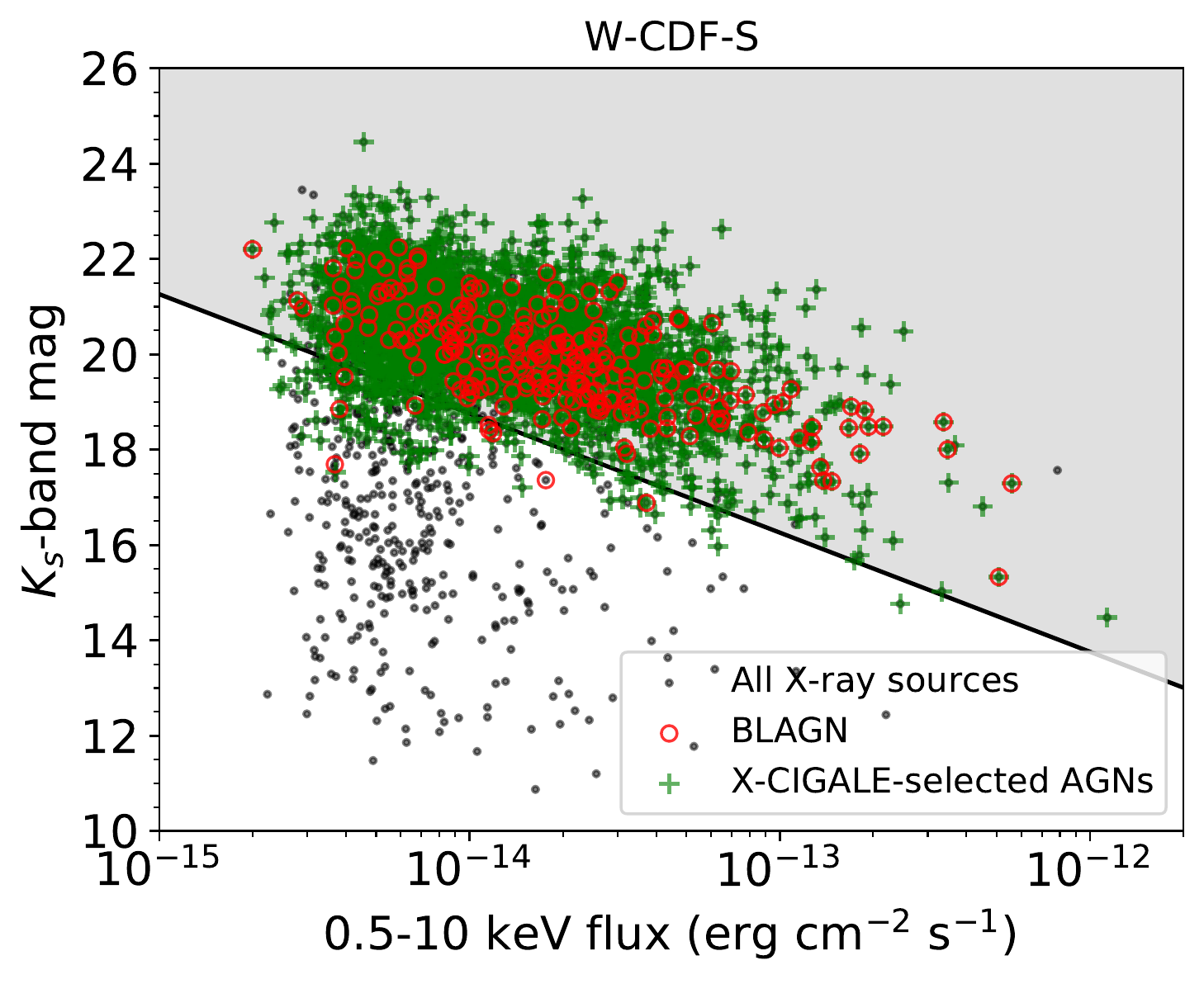}
\includegraphics[width=0.48\textwidth]{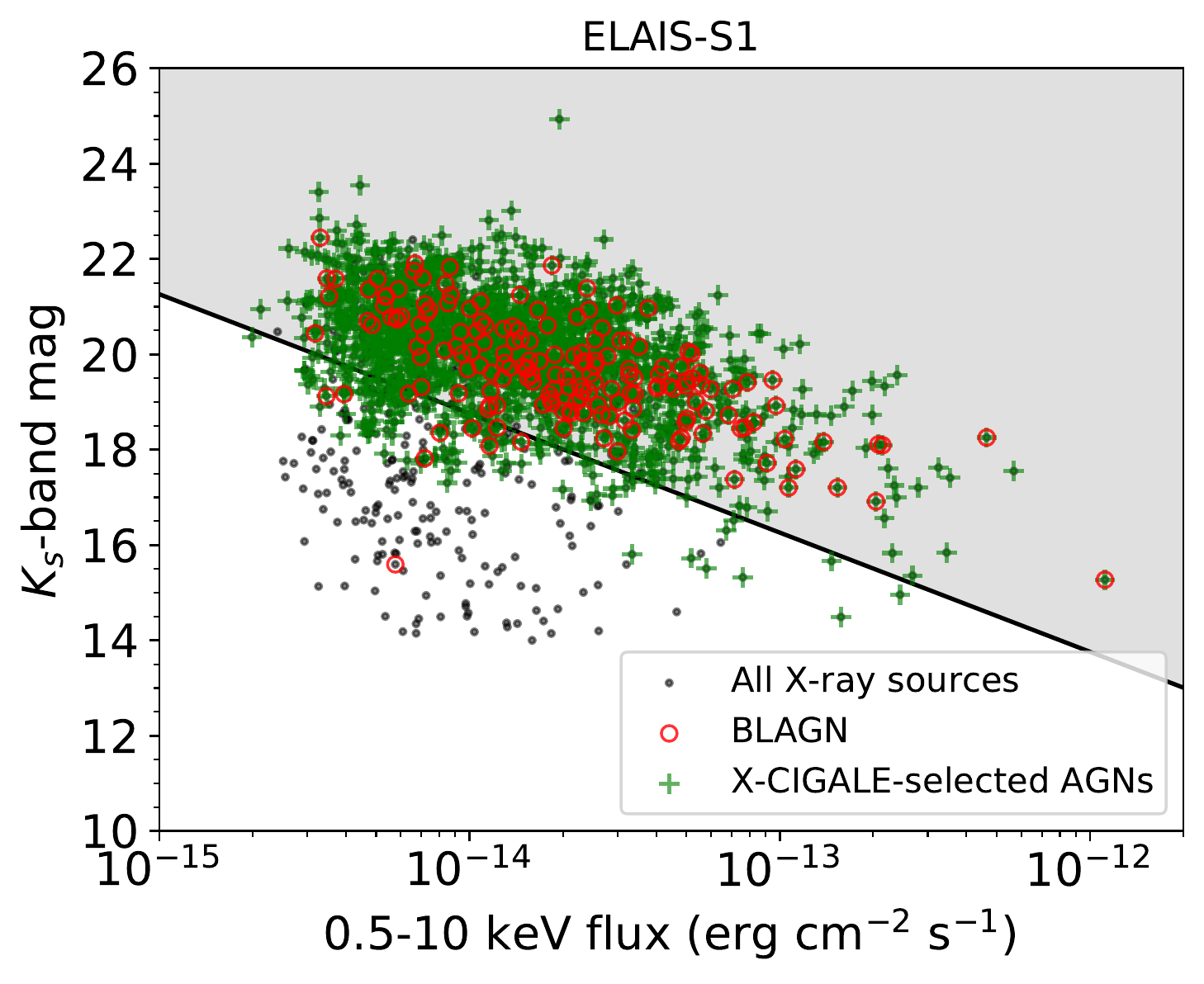}
\caption{
{\it Left/Right:} HSC (or DES) $i$-band/VIDEO $K_s$-band magnitude vs. \xray\ flux in the full band for all the \xray\ sources detected in the $i$/$K_s$-band (black dots), $i$/$K_s$-band detected broad-line AGNs (red circles), $i$/$K_s$-band detected \xray\ sources that are classified as AGNs from {\sc X-CIGALE} SED fitting (green pluses) in \wcdfs\ (\textit{top panels}) and \es\ (\textit{bottom panels}). The shaded region marks the ``AGN region'' defined by the  $\log_{10} f_x/f_i > -1$ or $\log_{10} f_x/f_{K_s} > -1.2$ threshold (represented by the black solid line).  
}
\label{fig:fxfoir}
\end{figure*}

\begin{figure*}
\centering
\includegraphics[width=0.48\textwidth]{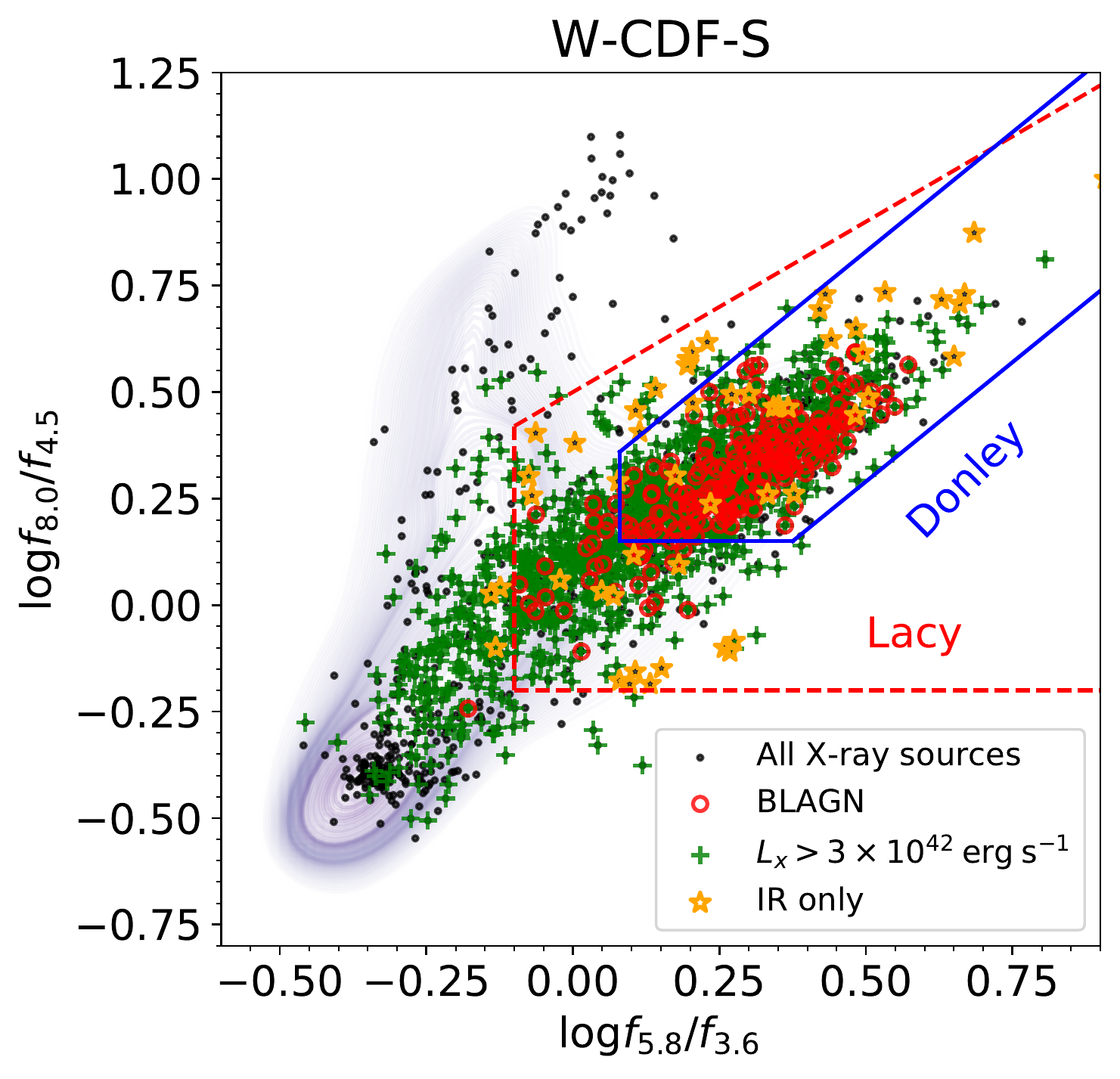}
\includegraphics[width=0.48\textwidth]{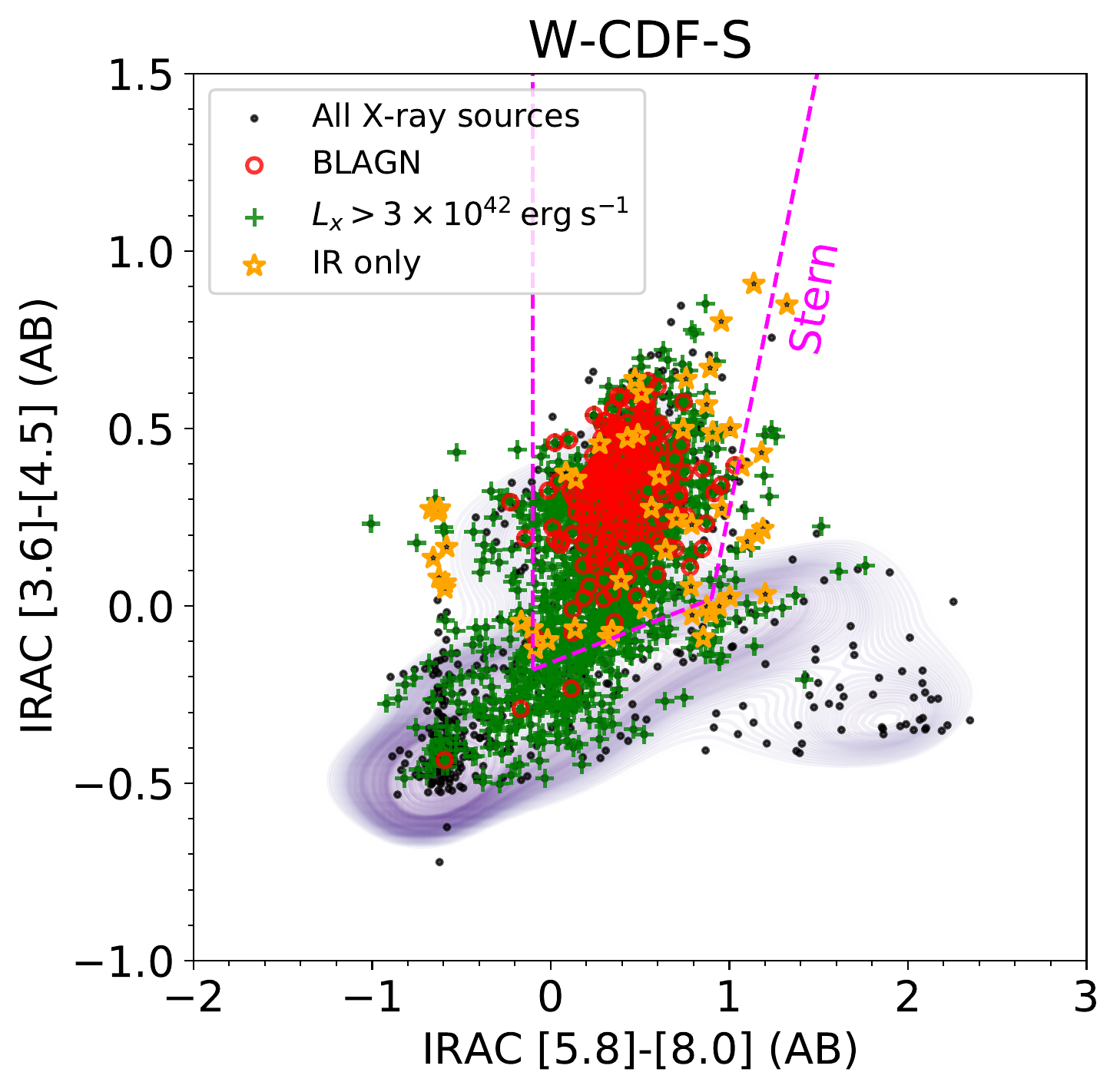}
\includegraphics[width=0.48\textwidth]{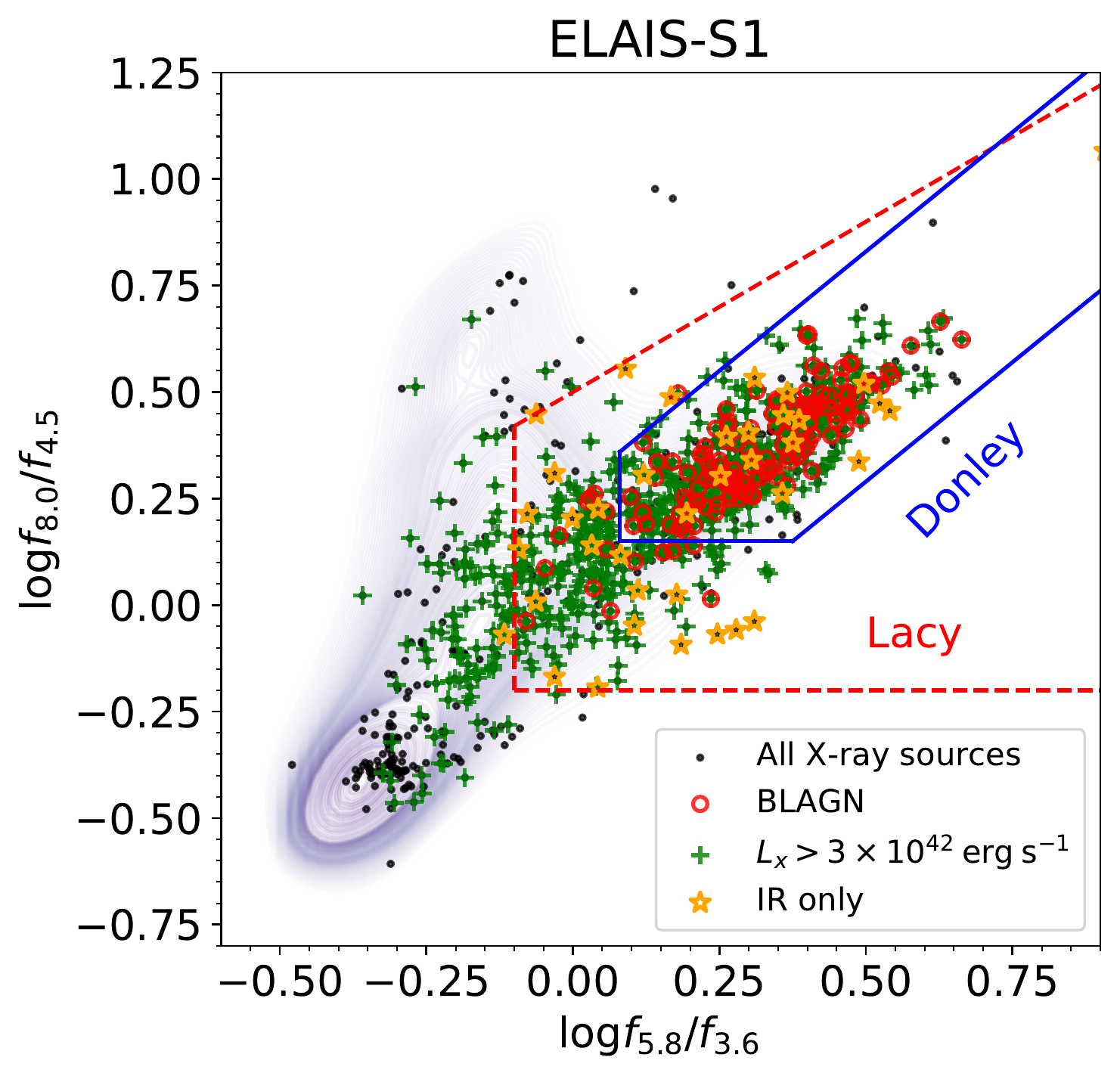}
\includegraphics[width=0.48\textwidth]{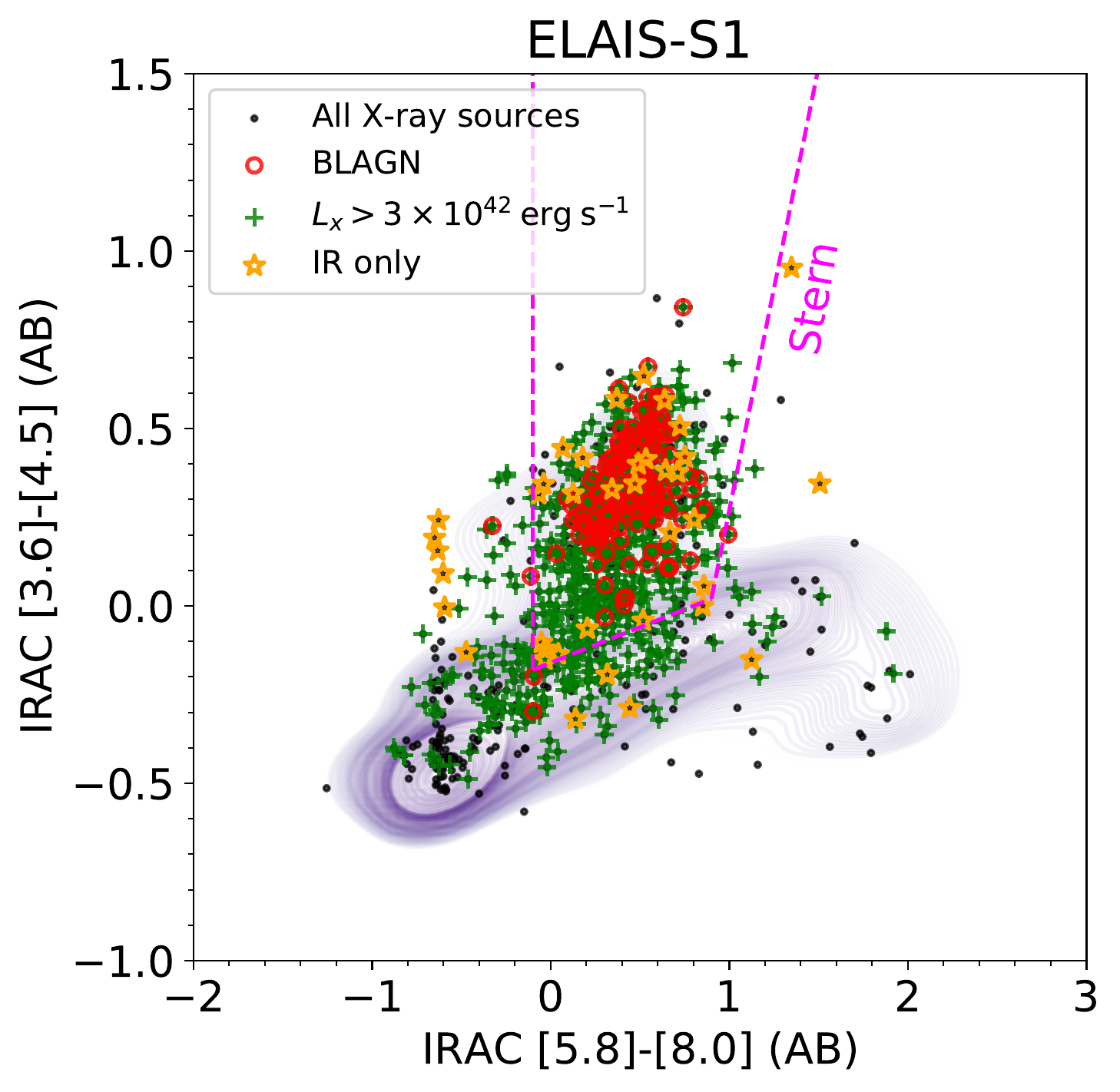}
\caption{
{\it Left panels:} The distribution of $\log f_{5.8}/f_{3.6}$ vs. $\log f_{8.0}/f_{4.5}$ for \hbox{X-ray} sources that are detected in four IRAC bands (black dots), with the 2D kernel-density plot of all SWIRE sources detected in four IRAC bands in the background in \wcdfs\ (\textit{top}) and \es\ (\textit{bottom}).
Among these \xray\ sources, broad-line AGNs are marked by the red circles; AGNs identified with high $L_X$ values are marked by green pluses; and X-ray sources that are only identified as AGN via MIR colors are marked by orange stars.
The blue lines denote the Donley wedge \citep{Donley2012}; the red dashed lines denote the Lacy wedge \citep{Lacy2004}.
{\it Right panels:} The distribution of IRAC$[5.8] - [8.0]$ vs. $[3.6] - [4.5]$ (AB) for all \hbox{X-ray} sources that are detected in four IRAC bands (black dots), with the 2D kernel-density plot of all SWIRE sources detected in four IRAC bands in the background in \wcdfs\ (\textit{top}) and \es\ (\textit{bottom}). Symbols represent the similar objects as the left panel.
The magenta dashed lines denote the Stern wedge \citep{Stern2005}.}
\label{fig:iraccolor}
\end{figure*}

\section{Summary and Future Work} \label{s-sum}
We have presented the \xray\ point-source catalogs for two of the XMM-SERVS fields, \wcdfs\ and \es, in this work. 
These are the final two fields of the $\approx 30$ ks depth \xmm\ survey, XMM-SERVS ($\approx 13~{\rm deg}^2$ in total). The main results are the following:

\begin{enumerate}
  \item 2.3 Ms and 1.0 Ms of \xmm\ observations were performed in the $\approx 4.6$ deg$^2$ \wcdfs\ field and the $\approx 3.2$ deg$^2$ \es\ field, respectively. After background filtering, the median cleaned PN+MOS1+MOS2 exposure time is $\approx 84$ ks in \wcdfs\ and $\approx 80$ ks in \es\ (see Section~\ref{s-xo}). Our survey in \wcdfs/\es\ has a flux limit of $1.0\times10^{-14}$/ $1.3\times10^{-14}$ \flux\ over 90\% of its area in the 0.5--10 keV band (see Section~\ref{ss-sens}).
  \item We compiled the X-ray point-source catalogs in \wcdfs\ and \es\ with the SAS task {\sc emldetect}. Adopting detection likelihoods that correspond to a spurious fraction of $\approx 1\%$ (obtained through simulations; see Section~\ref{ss-detmlsim}), 4053 point sources are detected in \wcdfs, and 2630 point sources are detected in \es. These \xray\ sources have a median positional uncertainty of $\approx 1.2$ arcsec (see Section~\ref{s-xcat}). 
  \item Utilizing optical-to-NIR data from DES, HSC, VOICE, VIDEO, and DeepDrill, we use \texttt{NWAY} to identify multiwavelength counterparts for \xray\ sources in the catalogs. 3600 ($\approx 89\%$) \xray\ sources in \wcdfs\ and 2288 ($\approx 87\%$) \xray\ sources in \es\ are matched to reliable optical and/or NIR counterparts (see Section~\ref{s-mc}).
  \item Photometric redshifts are estimated for 3319/2001 \xray\ sources in \wcdfs/\es\ with optical-to-NIR forced photometry available; type 1 AGNs are identified and fit separately with a suitable SED library. 2752 \xray\  sources in \wcdfs\ and 1702 \xray\ sources in \es\ have either spectroscopic or high-quality ($\sigma_{\rm NMAD} \approx 0.03$--0.04 for non-BL AGNs and $\sigma_{\rm NMAD} \approx 0.06$--0.07 for BL AGNs when compared to spec-$z$s) photometric redshifts (see Section~\ref{s-z}).
  \item We identify 3186 X-ray sources in \wcdfs\ and 1985 X-ray sources in \es\ as AGNs based on their optical spectroscopic properties, \xray\ luminosity and/or spectral shape, and \xray-to-NIR SED template fitting results. MIR color and radio luminosity are also utilized to select AGNs when available (see Section~\ref{sc-sp}).

  \end{enumerate}

The \xray\ point-source catalogs provided in this work will have great legacy value for studies of AGNs across the full range of cosmic environments, and will enable large-scale studies of SMBH growth in the multi-dimensional space of galaxy parameters. 
We note that all the XMM-SERVS fields, including \wcdfs\ and \es, are selected LSST deep-drilling fields, which will have $\approx 900$ epoch $ugrizy$ coverage with coadded depth reaching $i \approx 28$; the robustly identified \xray\ AGNs will be useful for calibrating LSST AGN selection in the deep-drilling fields and the main survey.
Future deep radio coverage from MIGHTEE \citep[e.g.,][]{Jarvis2016b}, sub-millimeter coverage from LMT and ALMA, and spectroscopic data from DEVILS, MOONS, and WAVES \citep[e.g.,][]{Davies2018,Driver2019,Maiolino2020} will also contribute to the legacy value of the \wcdfs\ and \es\ fields.
The SDSS-V Black Hole Mapper Program \citep{Kollmeier2017} and the 4MOST TiDES Program \citep{Swann2019} will provide direct SMBH masses for hundreds of the AGNs in these fields.
Together with this superior multiwavelength coverage, the \xray\ catalogs presented in this work will enable outstanding studies of the $\approx$ 5200 AGNs reported.
We leave detailed characterization of extended \xray\ sources in the XMM-SERVS fields for future work, which will contribute to the studies of \xray\ groups and clusters \citep[e.g.,][]{Pierre2016}.

\section*{Acknowledgements}
We thank the anonymous referee for constructive feedback.
We thank Roberto Assef and Teng Liu for helpful discussions.
We thank Pedro Rodriguez, Norbert Schartel, and the \xmm\ Science Operations Centre for kind help with scheduling these \xmm\ observations.
QN, WNB, and FZ acknowledge support from NASA grant 80NSSC19K0961 and the V.M. Willaman Endowment.
BL acknowledges financial support from the NSFC grant 11991053 and
National Key R\&D Program of China grant 2016YFA0400702.
KN acknowledges basic research in radio astronomy at the U.S. Naval Research Laboratory is supported by 6.1 Base Funding.
JA acknowledges support from a UKRI Future Leaders Fellowship (grant code: MR/T020989/1).
DMA acknowledges support from the Science and Technology Facilities Council through grant ST/T000244/1.
FEB acknowledges support from ANID-Chile Basal AFB-170002, FONDECYT Regular 1200495 and 1190818, and Millennium Science Initiative Program - ICN12\_009.
ADS and IT acknowledge the support of SSC work at AIP by Deutsches Zentrum f\"{u}r Luftund Raumfahrt (DLR) through grant 50 OX 1901.
MV acknowledges support from the Italian Ministry of Foreign Affairs and
International Cooperation (MAECI Grant Number ZA18GR02) and the South African
Department of Science and Technology's National Research Foundation (DST-NRF
Grant Number 113121) as part of the ISARP RADIOSKY2020 Joint Research Scheme.
YQX acknowledges support from the National Natural Science Foundation of China (NSFC-12025303, 11890693, 11421303), the CAS Frontier Science Key Research Program (QYZDJ-SSW-SLH006), and the K.C. Wong Education Foundation.
RG acknowledges support from the agreement ASI-INAF n. 2017-14-H.O.
MP acknowledges financial contribution from the agreement ASI-INAF n.2017-14-H.O.
The National Radio Astronomy Observatory is a facility of the National Science Foundation operated under cooperative agreement by Associated Universities, Inc.

\appendix
\section{Main X-ray source catalog description} \label{a-column}
The descriptions of the columns included in our main \hbox{X-ray} source catalogs in \wcdfs\ and \es\ (see Tables~\ref{tab:mainxtab} and \ref{tab:mainxtab_es}) are presented below in a format similar to that of \citet{Chen2018}. Throughout the table, null values are set to $-99$.
All celestial coordinates are given in equinox J2000.

\begin{table}
\begin{center}
\caption{\label{tab:mainxtab}
The main \xray\ source catalog in \wcdfs\ with a selection of columns.}
\vspace{-0.3 cm}
\begin{tabular}{ccccccccccccc}
    \hline
	XID &  RA &  DEC &  XPOSERR  &  FB\_EXP &  FB\_BKG &  FB\_SCTS &  0p5\_10\_FLUX &   SPECZ  &  AGN\_FLAG \\
    (1) & (2) & (3) & (4)   &  (19)    &  (31)    &  (43)    &  (104)  &  (189) &  (206)
   \\
    \hline
WCDFS0000 & 52.152070 & $-$28.698755 & 0.14 & 90110.6 &  3.78 & 8505.6 & 1.13635 & 0.10870 & 1  \\
WCDFS0001 & 52.168228 & $-$28.669922 & 0.29 & 118105.5  & 4.87 & 1948.1 & 0.05208 & 0.77247 & 1 \\
WCDFS0002 & 52.130722 & $-$28.880466 & 0.30 & 64600.6  &  3.95 & 1917.3  & 0.11816 & 1.04995	& 1 \\
WCDFS0003 & 52.136161 & $-$28.733398 & 0.33 & 100678.1  & 4.05 & 1555.2 & 0.04719  & $-$99  & 1 \\
WCDFS0004 & 51.876881 & $-$28.512461 & 0.34 & 99457.5  & 3.23 & 1430.5 & 0.05912 & 0.38893 & 1 \\
... & ... & ... &  ... & ... & ... &  ... & ... & ... &  ...  \\
\hline
\end{tabular} 
\end{center}
\vspace{-0.2cm}
A detailed description of each column is presented in Appendix~\ref{a-column}. This table is available in its entirety in machine-readable form online.
\end{table}

\begin{table}
\caption{\label{tab:mainxtab_es}
The main \xray\ source catalog in \es\ with a selection of columns.}
\centering
\begin{tabular}{cccccccccc}
    \hline
XID &  RA &  DEC &  XPOSERR  &  FB\_EXP &  FB\_BKG &  FB\_SCTS &  0p5\_10\_FLUX &   SPECZ  &  AGN\_FLAG \\
(1) & (2) & (3) & (4)   &  (19)    &  (31)    &  (43)    &  (104)  &  (180) &  (197)
   \\
    \hline
ES0000 & 8.747565 & $-$44.824939 & 0.13 & 42850.9 &  1.96 & 5539.8 & 0.56817   & -99.0 & 1  \\
ES0001 & 8.726280 & $-$44.771419 & 0.16 & 52141.0  & 2.00 & 3727.6 & 0.32483 & 0.40723 & 1 \\
ES0002 & 9.324548 & $-$44.503995 & 0.21 & 118323.1  & 5.71 & 2356.2  &  0.11067 & 1.32429	& 1 \\
ES0003 & 9.087287 & $-$44.144419 & 0.23 & 58336.4 & 2.29 & 2011.9 & 0.23091 & 0.20828  & 1 \\
ES0004 & 8.787302 & $-$44.310709 & 0.25 & 70255.8  & 1.81 & 1770.0 & 0.13339 & 0.34187 & 1 \\
... & ... & ... &  ... & ... & ... &  ... & ... & ... &  ...  \\
\hline \\
\end{tabular} 
\vspace{-0.2 cm}
\\
A detailed description of each column is presented in Appendix~\ref{a-column}. This table is available in its entirety in machine-readable form online.
\end{table}

\hfill\\
\noindent
\textbf{Table~\ref{tab:mainxtab}: \xray\ source catalog in \wcdfs} \\
\vspace{0.2 cm}
\textbf{\textit{X-ray properties}}\\
Columns 1--110 list  the \hbox{X-ray} properties of our sources. 
Columns for the soft/hard/full-band results are marked with the ``{\sc SB\_}''/``{\sc HB\_}''/``{\sc FB\_}'' prefix. 

\begin{itemize}
    \setlength{\itemindent}{-1ex}
    \item Column 1, \texttt{XID}: The source ID of each \hbox{X-ray} source.
    \item Columns 2--3, \texttt{RA, DEC}: RA and DEC (in degrees) of the \hbox{X-ray} source. Based on availability, we use the positions from, in priority order, the full band, soft band, and hard band as the primary position. Band-specific positions are listed in Columns 8--13.
    \item Column 4, \texttt{XPOSERR}: \hbox{X-ray} positional uncertainty ($\sigma_x$) in arcsec (reported with the same priority order as that of positions).
    \item Columns 5--6, \texttt{R68, R99}: 68\% and 99.73\% \hbox{X-ray} positional uncertainties in arcsec based on the Rayleigh distribution (see Section~\ref{ss:poserr}). 
    \item Column 7, \texttt{EMLERR}: Positional uncertainties calculated by {\sc emldetect}, $\sigma_{\rm eml}$, in arcsec (with the same priority order as that of positions).
    \item Columns 8--13, \texttt{SB\_RA, SB\_DEC, HB\_RA, HB\_DEC, FB\_RA, FB\_DEC}: RA and DEC (in degrees) of the source in the soft, hard, and full bands, respectively.
    \item Columns 14--16, \texttt{SB\_DET\_ML, HB\_DET\_ML, FB\_DET\_ML}: The {\sc emldetect} source-detection likelihood in each band.
    \item Columns 17--19, \texttt{SB\_EXP, HB\_EXP, FB\_EXP}: Total (PN + MOS1 + MOS2) exposure time in seconds in each band.
    \item Columns 20--28, \texttt{SB\_EXPPN, SB\_EXPM1, SB\_EXPM2, HB\_EXPPN, HB\_EXPM1, HB\_EXPM2, FB\_EXPPN, FB\_EXPM1, FB\_EXPM2}: PN, MOS1, and MOS2 exposure times in seconds in each band.
    \item Columns 29--31, \texttt{SB\_BKG, HB\_BKG, FB\_BKG}: Total background-map values (PN + MOS1 + MOS2) in counts per pixel in each band. 
    \item Columns 32--40, \texttt{SB\_BKGPN, SB\_BKGM1, SB\_BKGM2, HB\_BKGPN, HB\_BKGM1, HB\_BKGM2, FB\_BKGPN, FB\_BKGM1, FB\_BKGM2}: PN, MOS1, and MOS2 background-map values in counts per pixel in each band. 
    \item Columns 41--43, \texttt{SB\_SCTS, HB\_SCTS, FB\_SCTS}: Total (PN + MOS1 + MOS2) net counts in each band.
    \item Columns 44--52, \texttt{SB\_SCTSPN, SB\_SCTSM1, SB\_SCTSM2, HB\_SCTSPN, HB\_SCTSM1, HB\_SCTSM2, FB\_SCTSPN, FB\_SCTSM1, FB\_SCTSM2}: PN, MOS1, and MOS2 net counts in each band.
    \item Columns 53--64, \texttt{SB\_SCTS\_ERR, HB\_SCTS\_ERR, FB\_SCTS\_ERR, SB\_SCTSPN\_ERR, SB\_SCTSM1\_ERR, SB\_SCTSM2\_ERR, HB\_SCTSPN\_ERR, HB\_SCTSM1\_ERR, HB\_SCTSM2\_ERR, FB\_SCTSPN\_ERR, FB\_SCTSM1\_ERR, FB\_SCTSM2\_ERR}: Uncertainties of total, PN, MOS1, and MOS2 net counts in each band reported in {\sc emldetect}.
    \item Columns 65--76, \texttt{SB\_RATE, HB\_RATE, FB\_RATE, SB\_RATEPN, SB\_RATEM1, SB\_RATEM2, HB\_RATEPN, HB\_RATEM1, HB\_RATEM2, FB\_RATEPN, FB\_RATEM1, FB\_RATEM2}: Total, PN, MOS1, and MOS2 net count rates in each band, in count s$^{-1}$.
    \item Columns 77--88, \texttt{SB\_RATE\_ERR, HB\_RATE\_ERR, FB\_RATE\_ERR, SB\_RATEPN\_ERR, SB\_RATEM1\_ERR, SB\_RATEM2\_ERR, HB\_RATEPN\_ERR, HB\_RATEM1\_ERR, HB\_RATEM2\_ERR, FB\_RATEPN\_ERR, FB\_RATEM1\_ERR, FB\_RATEM2\_ERR}: Uncertainties of total, PN, MOS1, and MOS2 net count rates in each band, in count s$^{-1}$.
    \item Column 89--96, \texttt{BR, BR\_ERR, BRPN, BRPN\_ERR, BRM1, BRM1\_ERR, BRM2, BRM2\_ERR}: Total hard-to-soft band ratio and its uncertainty, and the hard-to-soft band ratio and its uncertainty for each EPIC detector.
    \item Column 97--98, \texttt{HR, HR\_ERR}: Hardness ratio and its uncertainty. 
    \item Column 99, \texttt{GAMMA}: The effective power-law photon index, $\Gamma_{\rm eff}$, derived for each source based on the hard-to-soft band ratio. 
    \item Columns 100--105, \texttt{0p5\_2\_FLUX, 0p5\_2\_FLUX\_ERR, 2\_10\_FLUX, 2\_10\_FLUX\_ERR, 0p5\_10\_FLUX, 0p5\_10\_FLUX\_ERR}: Observed flux and flux uncertainty in the 0.5--2, 2--10, and 0.5--10 keV bands, in $ 10^{-12}$ erg~cm$^{-2}$~s$^{-1}$, after correcting for Galactic absorption.
    The fluxes and uncertainties reported here are the error-weighted average of all EPIC detectors.
    \item Column 106, \texttt{LX}: Logarithm of rest-frame observed \hbox{2--10}~keV \xray\ luminosity (in erg~s$^{-1}$) after correcting for Galactic absorption.
    \item Column 107, \texttt{CHANDRA\_SOURCE}: The catalog origin of the nearest \chandra\ source within 10$''$. An entry of ``1'' stands for the CDF-S catalog \citep{Luo2017}, ``2'' stands for the E-CDF-S catalog \citep{Xue2016}, and ``3'' stands for the CSC 2.0 catalog.
    \item Column 108, \texttt{CHANDRA\_ID}: \chandra\ source ID.
    \item Column 109--110, \texttt{CHANDRA\_RA, CHANDRA\_DEC}: RA and DEC (in degrees) of the matched \chandra\ counterpart.
\end{itemize}

\hfill\\
\noindent
\textbf{\textit{Multiwavelength properties}}\\
Columns 111--207 provide the multiwavelength properties of the matched counterparts with {\sc match\_flag} $= 1$ utilizing \texttt{NWAY}.

\begin{itemize}
    \setlength{\itemindent}{-1ex}
        \item Column 111, \texttt{P\_ANY}: The posterior probability of the \hbox{X-ray} source having any correct counterparts (\pany). 
    \item Column 112, \texttt{P\_I}: The relative probability ($p\_i$) of the reported {\sc match\_flag} $= 1$ counterpart to be the correct match. 
    \item Column 113, \texttt{FLAG\_SECOND}: Warning flag for sources where a second possible counterpart is indicated by \texttt{NWAY}.
    \item Columns 114--121, \texttt{IRAC\_RA, IRAC\_DEC, VIDEO\_RA, VIDEO\_DEC, HSC\_RA, HSC\_DEC, DES\_RA, DES\_DEC}: RA and DEC of the counterpart in the DeepDrill/VIDEO/HSC/DES catalog in degrees. Note that DES counterparts are only reported in areas lacking HSC coverage.
    \item Columns 122--125, \texttt{SEP\_IRAC, SEP\_VIDEO, SEP\_HSC, SEP\_DES}: Separation of the \hbox{X-ray} position from the DeepDrill/VIDEO/HSC/DES counterpart in arcseconds.
    \item Columns 126--129, \texttt{IRAC\_1\_MAG, IRAC\_1\_MAG\_ERR, IRAC\_2\_MAG, IRAC\_2\_MAG\_ERR}: $1.9''$ aperture photometry and uncertainties in the IRAC $3.6\mu$m and $4.5\mu$m bands reported in the DeepDrill catalog.
    \item Columns 130--139, \texttt{VIDEO\_Z\_MAG, VIDEO\_Z\_MAG\_ERR, VIDEO\_Y\_MAG, VIDEO\_Y\_MAG\_ERR, VIDEO\_J\_MAG, VIDEO\_J\_MAG\_ERR, VIDEO\_H\_MAG, VIDEO\_H\_MAG\_ERR, VIDEO\_KS\_MAG, VIDEO\_KS\_MAG\_ERR}: VIDEO 2$''$ aperture photometry and uncertainties in the $Z$, $Y$, $J$, $H$, and $K_s$ bands.
    \item Columns 140--147, \texttt{HSC\_G\_MAG, HSC\_G\_MAG\_ERR, HSC\_R\_MAG, HSC\_R\_MAG\_ERR, HSC\_I\_MAG, HSC\_I\_MAG\_ERR, HSC\_Z\_MAG, HSC\_Z\_MAG\_ERR}: HSC CModel photometry and uncertainties in the $g$, $r$, $i$, and $z$ bands.
     \item Columns 148--157, \texttt{DES\_G\_MAG, DES\_G\_MAG\_ERR, DES\_R\_MAG, DES\_R\_MAG\_ERR, DES\_I\_MAG, DES\_I\_MAG\_ERR, DES\_Z\_MAG, DES\_Z\_MAG\_ERR, DES\_Y\_MAG, DES\_Y\_MAG\_ERR}: DES Kron Magnitude and uncertainties in the $g$, $r$, $i$, $z$, and $Y$ bands.
    \item Column 158, \texttt{TRACTOR\_ID}: The object ID of the VIDEO counterpart in the forced-photometry catalog (Nyland et al.\ 2021).
    \item Columns 159--188, \texttt{IRAC\_1\_FP\_MAG, IRAC\_1\_FP\_MAG\_ERR, IRAC\_2\_FP\_MAG, IRAC\_2\_FP\_MAG\_ERR, VIDEO\_Z\_FP\_MAG, VIDEO\_Z\_FP\_MAG\_ERR, VIDEO\_Y\_FP\_MAG,VIDEO\_Y\_FP\_MAG\_ERR, VIDEO\_J\_FP\_MAG, VIDEO\_J\_FP\_MAG\_ERR, VIDEO\_H\_FP\_MAG, VIDEO\_H\_FP\_MAG\_ERR, VIDEO\_KS\_FP\_MAG, VIDEO\_KS\_FP\_MAG\_ERR, HSC\_G\_FP\_MAG, HSC\_G\_FP\_MAG\_ERR, HSC\_R\_FP\_MAG, HSC\_R\_FP\_MAG\_ERR, HSC\_I\_FP\_MAG, HSC\_I\_FP\_MAG\_ERR, HSC\_Z\_FP\_MAG, HSC\_Z\_FP\_MAG\_ERR, VOICE\_U\_FP\_MAG, VOICE\_U\_FP\_MAG\_ERR, VOICE\_G\_FP\_MAG, \\VOICE\_G\_FP\_MAG\_ERR, VOICE\_R\_FP\_MAG, VOICE\_R\_FP\_MAG\_ERR, VOICE\_I\_FP\_MAG, VOICE\_I\_FP\_MAG\_ERR}: Forced photometry and uncertainties of DeepDrill $3.6\mu$m and $4.5\mu$m bands, VIDEO $ZYJHK_s$ bands, HSC $griz$ bands, and VOICE $ugri$ bands reported in the forced-photometry catalog (Nyland et al.\ 2021).
    \item Column 189, \texttt{SPECZ}: Spectroscopic redshift adopted for the \xray\ source.
    \item Column 190, \texttt{SPECZ\_CLASS}: Spectroscopic classification of the source. ``1'' stands for BL AGNs; ``0''  stands for galaxies or non-BL AGNs; ``$-1$'' stands for stars. 
    \item Column 191, \texttt{SPECZ\_Q}: Spectroscopic quality flag of the source reported in the original catalog.
    \item Columns 192--193, \texttt{SPECZ\_RA, SPECZ\_DEC}: RA and DEC (in degrees) of the spec-$z$. 
    \item Column 194, \texttt{SPECZ\_SOURCE}: The spectroscopic catalog listed in Table~\ref{tab:redshift} that provides the spec-$z$.
    \item Column 195, \texttt{SED\_BLAGN\_FLAG}: Flag for BL AGN candidates identified in Appendix~\ref{a-blagn}. An entry of ``1'' stands for sources that are classified as BL AGN candidates by two different methods; ``0.5'' stands for sources identified as BL AGN candidates using one method but not the other; ``0'' indicates sources identified as non-BL AGNs by both methods.
    \item Columns 196--197, \texttt{PHOTOZ\_RA, PHOTOZ\_DEC}: RA and DEC (in degrees) of the source in the forced-photometry catalog (Nyland et al.\ in 2021), which includes forced photometry from DeepDrill, VIDEO, HSC, and VOICE that is utilized to compute photo-$z$s.
    \item Column 198, \texttt{PHOTOZ\_BEST}: Photometric redshift adopted for the source. \texttt{PHOTOZ\_EAZY} values are adopted for sources that have \texttt{SED\_BLAGN\_FLAG} $<$ 1 and are not identified as BL AGNs in spectroscopic surveys, when \texttt{PHOTOZ\_EAZY\_Q} $< 1$; \texttt{PHOTOZ\_LEPHARE} values are adopted for spectroscopically identified BL AGNs, \texttt{SED\_BLAGN\_FLAG} = 1 objects, and \texttt{SED\_BLAGN\_FLAG} $= 0.5$ objects with \texttt{PHOTOZ\_EAZY\_Q} $\geqslant 1$ (see Section~\ref{ss-photoz} for details).
    \item Columns 199--202, \texttt{PHOTOZ\_EAZY, PHOTOZ\_EAZY\_UERR, PHOTOZ\_EAZY\_LERR, PHOTOZ\_EAZY\_Q}: Photometric redshift computed by \texttt{EAZY}, the associated upper and lower uncertainties, and the photometric-redshift quality parameter ($Q_z$). 
    \item Columns 203--205, \texttt{PHOTOZ\_LEPHARE, PHOTOZ\_LEPHARE\_UERR, PHOTOZ\_LEPHARE\_LERR}: Photometric redshift computed by {\sc LePhare} and the associated upper and lower uncertainties. We only report {\sc LePhare} photo-$z$s with $\chi_{\rm red}^{2} < 2$ and band number $> 10$ (see Section~\ref{ss-photoz} for details).
    \item Column 206, \texttt{AGN\_FLAG}: Flag for AGNs identified in Section~\ref{sc-sp}.
    \item  Column 207, \texttt{STAR\_FLAG}: Flag for stars identified in Appendix~\ref{a-star}.
\end{itemize}

\hfill\\
\noindent
\textbf{Table~\ref{tab:mainxtab_es}: \xray\ source catalog in \es} \\
\vspace{0.2 cm}
\textbf{\textit{X-ray properties}}\\
Columns 1--110 give the \hbox{X-ray} properties of our sources in the same format as that of Table~~\ref{tab:mainxtab}.

\hfill\\
\noindent
\textbf{\textit{Multiwavelength properties}}\\
Columns 111--198 provide the multiwavelength properties of the matched counterparts with {\sc match\_flag} $= 1$ utilizing \texttt{NWAY}.
\begin{itemize}
    \setlength{\itemindent}{-1ex}
        \item Columns 111--113, \texttt{P\_ANY, P\_I, FLAG\_SECOND}: see Columns 111--113 of Table~\ref{tab:mainxtab}.
    \item Columns 114--119, \texttt{IRAC\_RA, IRAC\_DEC, VIDEO\_RA, VIDEO\_DEC, DES\_RA, DES\_DEC}: RA and DEC of the counterpart in the DeepDrill/VIDEO/DES catalog in degrees.
    \item Columns 120--122, \texttt{SEP\_IRAC, SEP\_VIDEO, SEP\_DES}: Separation of the \hbox{X-ray} position from the DeepDrill/VIDEO/DES counterpart in arcseconds.
    \item Columns 123--126, \texttt{IRAC\_1\_MAG, IRAC\_1\_MAG\_ERR, IRAC\_2\_MAG, IRAC\_2\_MAG\_ERR}: $1.9''$ aperture photometry and uncertainties in the IRAC $3.6\mu$m and $4.5\mu$m bands reported in the DeepDrill catalog.
    \item Columns 127--136, \texttt{VIDEO\_Z\_MAG, VIDEO\_Z\_MAG\_ERR, VIDEO\_Y\_MAG, VIDEO\_Y\_MAG\_ERR, VIDEO\_J\_MAG, VIDEO\_J\_MAG\_ERR, VIDEO\_H\_MAG, VIDEO\_H\_MAG\_ERR, VIDEO\_KS\_MAG, VIDEO\_KS\_MAG\_ERR}: VIDEO 2$''$ aperture photometry and uncertainties in the $Z$, $Y$, $J$, $H$, and $K_s$ bands.
    \item Columns 137--146, \texttt{DES\_G\_MAG, DES\_G\_MAG\_ERR, DES\_R\_MAG, DES\_R\_MAG\_ERR, DES\_I\_MAG, DES\_I\_MAG\_ERR, DES\_Z\_MAG, DES\_Z\_MAG\_ERR, DES\_Y\_MAG, DES\_Y\_MAG\_ERR}: DES Kron Magnitude and uncertainties in the $g$, $r$, $i$, $z$, and $Y$ bands.
    \item Column 147, \texttt{TRACTOR\_ID}: The object ID of the VIDEO counterpart in the forced-photometry catalog \citep{Zou2021a}.
    \item Columns 148--179, \texttt{IRAC\_1\_FP\_MAG, IRAC\_1\_FP\_MAG\_ERR, IRAC\_2\_FP\_MAG, IRAC\_2\_FP\_MAG\_ERR, VIDEO\_Z\_FP\_MAG, VIDEO\_Z\_FP\_MAG\_ERR, VIDEO\_Y\_FP\_MAG, VIDEO\_Y\_FP\_MAG\_ERR, VIDEO\_J\_FP\_MAG, VIDEO\_J\_FP\_MAG\_ERR, VIDEO\_H\_FP\_MAG, VIDEO\_H\_FP\_MAG\_ERR, VIDEO\_KS\_FP\_MAG, VIDEO\_KS\_FP\_MAG\_ERR, DES\_G\_FP\_MAG, DES\_G\_FP\_MAG\_ERR, DES\_R\_FP\_MAG, DES\_R\_FP\_MAG\_ERR, DES\_I\_FP\_MAG, DES\_I\_FP\_MAG\_ERR, DES\_Z\_FP\_MAG, DES\_Z\_FP\_MAG\_ERR, DES\_Y\_FP\_MAG, DES\_Y\_FP\_MAG\_ERR, ESIS\_B\_FP\_MAG, \\ESIS\_B\_FP\_MAG\_ERR, ESIS\_V\_FP\_MAG, ESIS\_V\_FP\_MAG\_ERR, ESIS\_R\_FP\_MAG, ESIS\_R\_FP\_MAG\_ERR, VOICE\_U\_FP\_MAG, VOICE\_U\_FP\_MAG\_ERR}: Forced photometry and uncertainties of DeepDrill $3.6\mu$m and $4.5\mu$m bands, VIDEO $ZYJHK_s$ bands, DES $grizY$ bands, ESIS $BVR$ bands, and VOICE $u$ band reported in the forced-photometry catalog \citep{Zou2021a}.
    \item Columns 180--198, \texttt{SPECZ, SPECZ\_CLASS, SPECZ\_Q, SPECZ\_RA, SPECZ\_DEC, SPECZ\_SOURCE, SED\_BLAGN\_FLAG, PHOTOZ\_RA, PHOTOZ\_DEC, PHOTOZ\_BEST, PHOTOZ\_EAZY,PHOTOZ\_EAZY\_UERR, PHOTOZ\_EAZY\_LERR, PHOTOZ\_EAZY\_Q, PHOTOZ\_LEPHARE, PHOTOZ\_LEPHARE\_UERR, PHOTOZ\_LEPHARE\_LERR, AGN\_FLAG, STAR\_FLAG}: See Columns 188--206 of Table~\ref{tab:mainxtab}.
\end{itemize}

\section{Identifying BL AGN candidates} \label{a-blagn}

Considering its \xray\ sensitivity limit, the XMM-SERVS survey would be able to detect $\approx 95\%$ of the spectroscopically identified BL AGNs in the COSMOS field \citep[e.g.,][]{Marchesi2016}, which is a survey field with rich spectroscopic observations.\footnote{As estimated in Section 2.1.3 of \citet{Yang2018}, the fraction of BL AGNs missed by spectroscopic campaigns in the COSMOS field is likely less than $\approx 18\%$.}
As the sky density of spectroscopically confirmed BL AGNs ($\approx 60$ deg$^{-2}$ for both \wcdfs\ and \es) in our study is much less than that of COSMOS ($\approx 290$ deg$^{-2}$; \citealt{Marchesi2016}), we also identify BL AGN candidates in our X-ray catalogs that do not have spectroscopic classifications utilizing their SEDs via two independent methods: one method is based on machine learning, and the other is based on SED template fitting. As each of these methods has its own advantages and disadvantages, combining the results from both methods provides more reliable predictions for AGNs in our catalogs.

\subsection{Machine-learning-based classification} 
For X-ray sources in \wcdfs\ and \es\ that have both spectroscopic classifications and forced photometry, we use machine learning to assess the differences between the optical-to-IR SEDs of X-ray sources that are classified as BL AGNs (260 in \wcdfs\ and 179 in \es) and X-ray sources that are not BL AGNs (470 in \wcdfs\ and 333 in \es).
Utilizing all the available photometric data points, we normalized the SEDs so that all the data points have a maximum value of 0 in log space, and use the \texttt{interp1d} function in \texttt{scipy} to interpolate the log-space SED shape. We extract 16 data points at common observed-frame wavelengths from the interpolated SEDs (see Figure~\ref{fig:normalizedsed}) to feed a 1D convolutional neutral network (CNN). Approximately $60\%$ of the objects are used as the training set; $\approx 20\%$ of the objects are used as the validation set; the remaining $\approx 20\%$ of the objects are used as the test set. 
After training the network and selecting the best model utilizing the validation set, we could correctly predict $\approx 86\%$ of the BL AGNs in the test set and $\approx 91\%$ of the sources that are not BL AGNs in \wcdfs\ (see panel~(a) of Figure~\ref{fig:cm} for the confusion matrix). 
When we apply the trained model to the remaining X-ray sources in the \wcdfs\ forced-photometry catalog with signal-to-noise ratio (SNR) > 3 detections in more than 5 bands, $\approx 790$ ($\approx 30\%$) of the sources are classified as BL AGN candidates. Combining these objects with the spectroscopically identified BL AGNs, the BL AGN density reaches $\approx 230$ deg$^{-2}$.
Considering that only $\approx 80\%$ of the detected \xray\ sources in \wcdfs\ have reliably matched counterparts in the forced-photometry catalog (Nyland et al. 2021) with SNR > 3 detections in more than 5 bands, this number is roughly consistent with the expectation from the COSMOS field.
Similarly, after training the network in \es, we could correctly predict $\approx 80\%$ of the BL AGNs and $\approx 85\%$ of the sources that are not BL AGNs (see panel~(c) of Figure~\ref{fig:cm}). 
About $450$ ($\approx 30\%$) of the remaining X-ray sources in the \es\ forced-photometry catalog with SNR > 3 detections in more than 5 bands are classified as BL AGN candidates. The BL AGN density reaches $\approx 200$ deg$^{-2}$. Considering that only $\approx 70\%$ of the \xray\ sources detected in \es\ have SNR > 3 detections in more than 5 bands in the \citet{Zou2021a} forced-photometry catalog (while this number is 80\% for \wcdfs), this relatively low BL AGN density is also acceptable.

\begin{figure*}    
\includegraphics[width=0.48\textwidth]{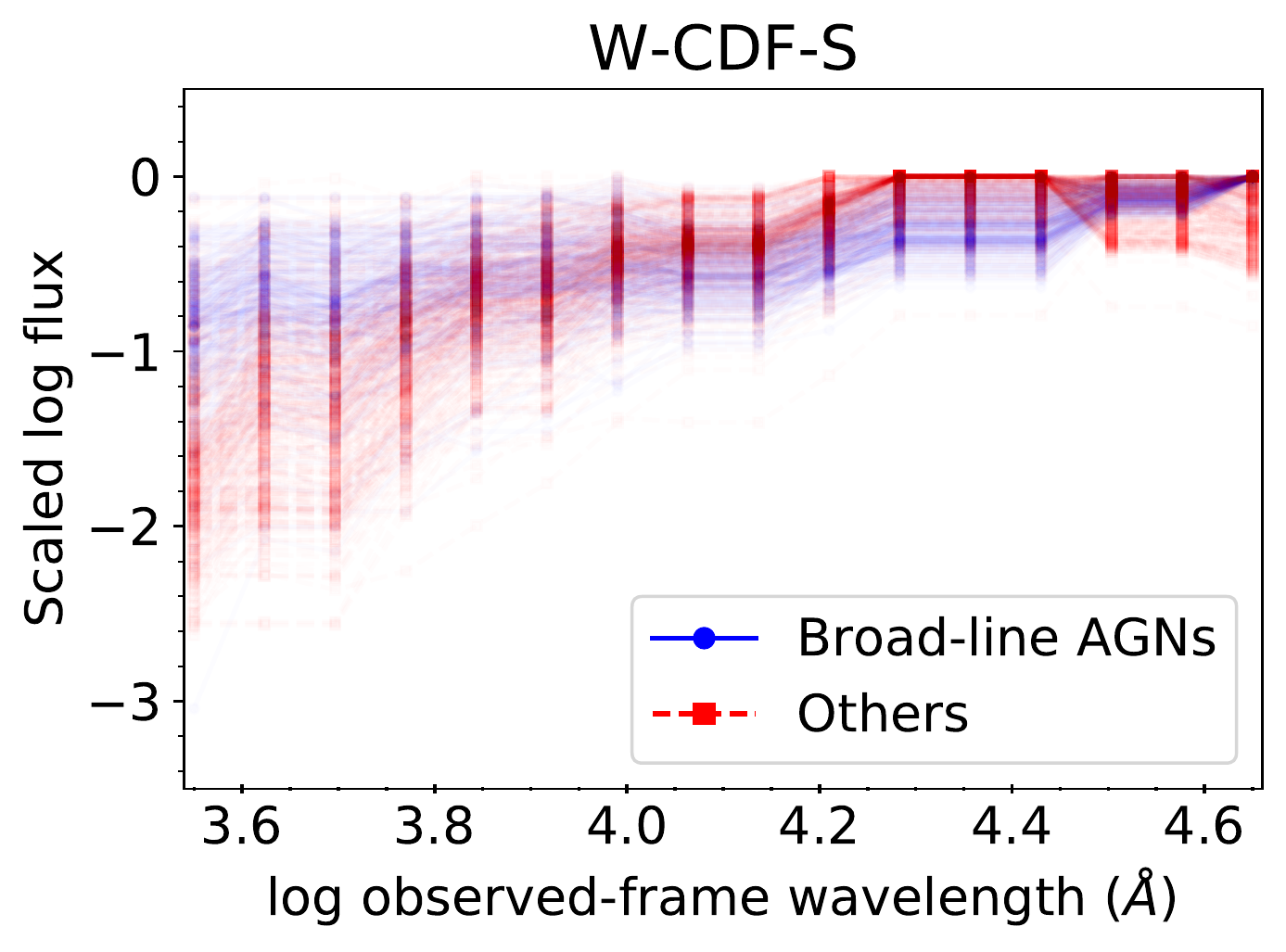}
\includegraphics[width=0.48\textwidth]{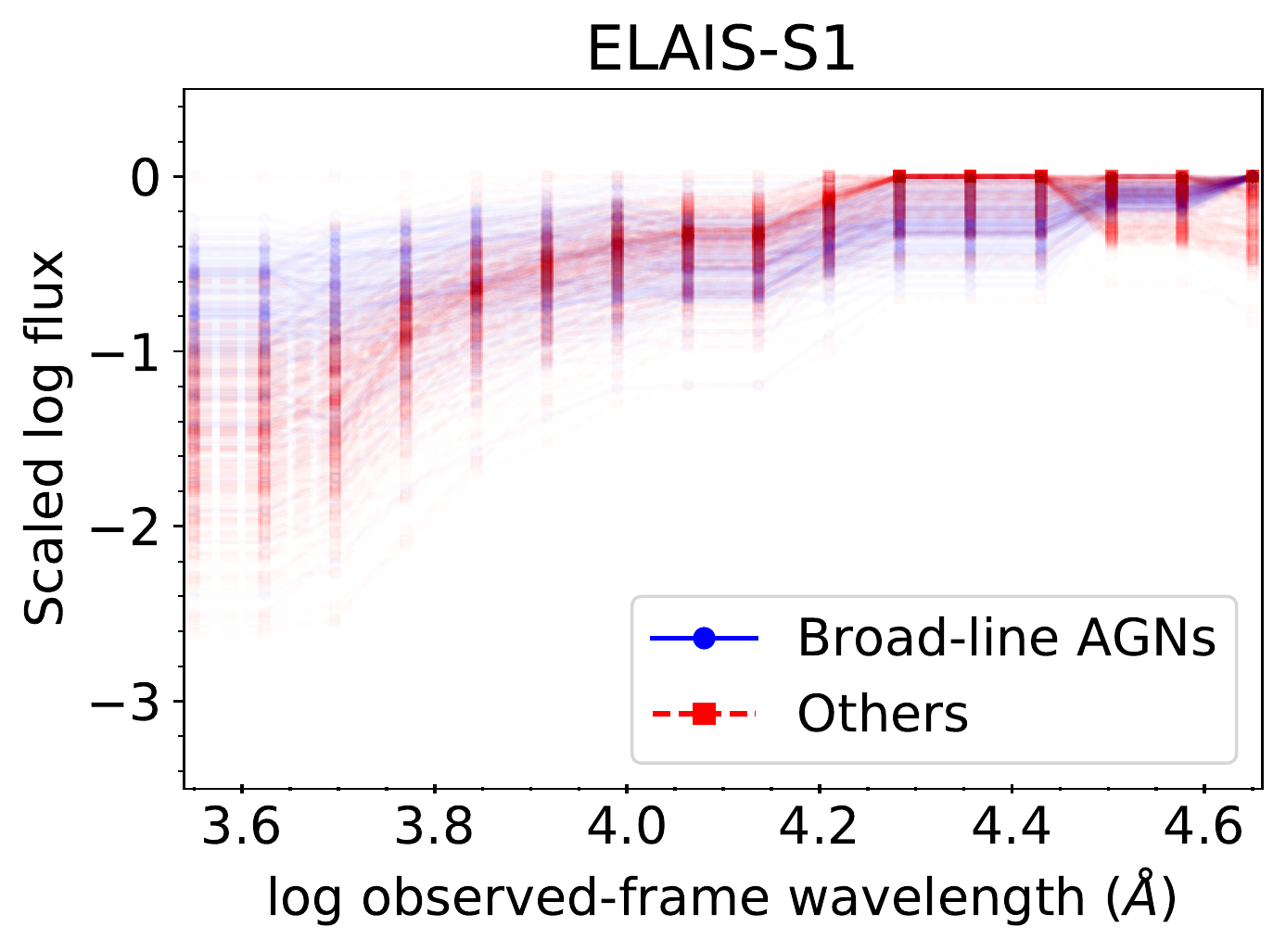}
\caption{Extracted data points from the interpolated SEDs of broad-line AGNs identified in spectroscopic surveys (blue) compared with other \xray\ sources that have spectroscopic classifications (red) in 
\wcdfs\ (\textit{left}) and \es\ (\textit{right}).}
\label{fig:normalizedsed}
\end{figure*}

\begin{figure} 
\centering
\gridline{\fig{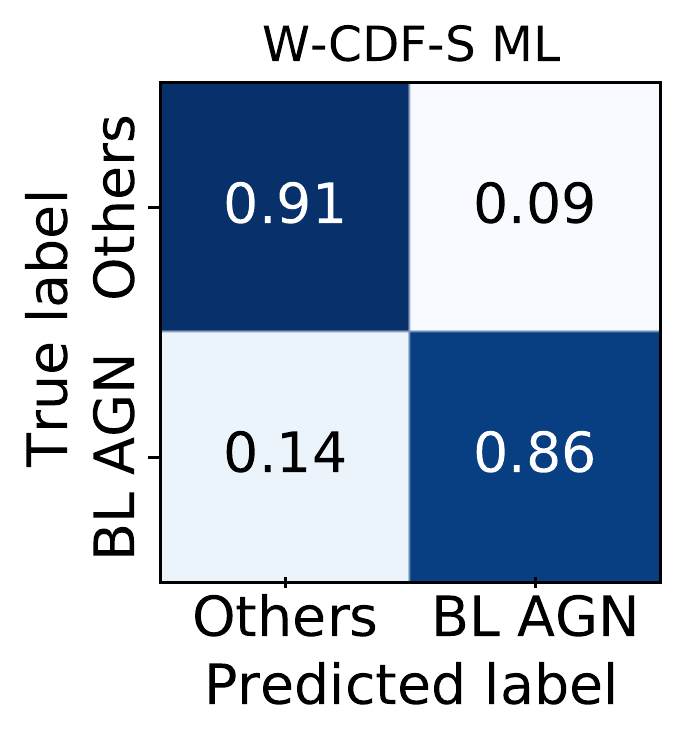}{0.24\textwidth}{(a)}
           \fig{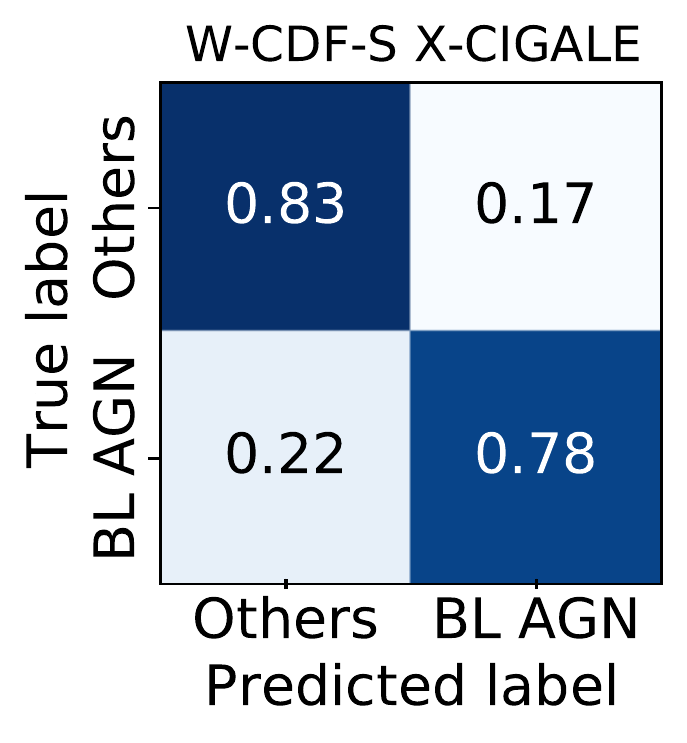}{0.24\textwidth}{(b)}
           \fig{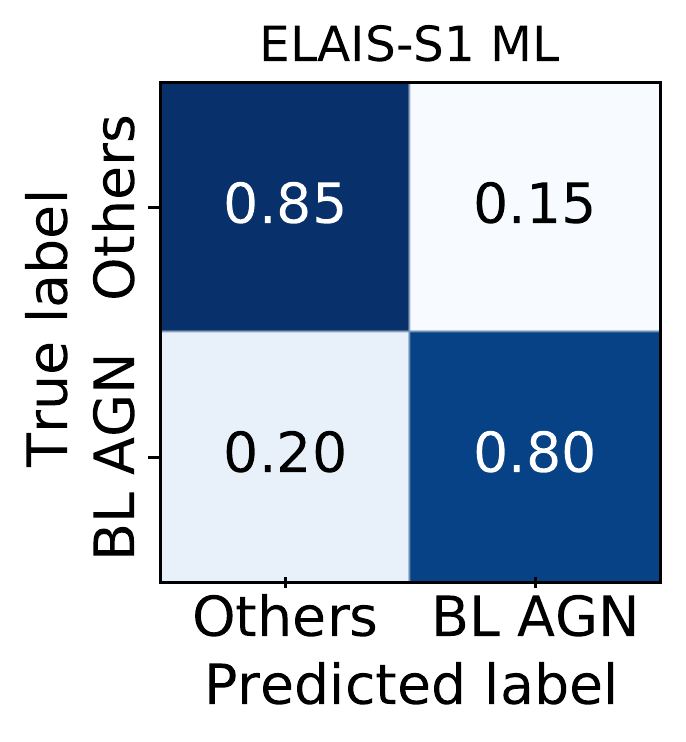}{0.24\textwidth}{(c)}
           \fig{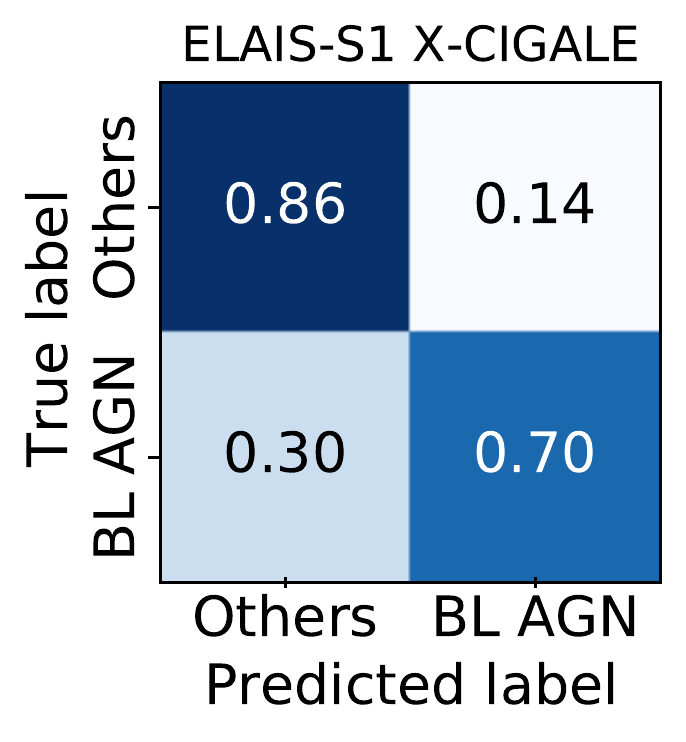}{0.24\textwidth}{(d)}
           }
\caption{(a). The confusion matrix for the machine-learning based classification of \xray\ sources in \wcdfs. (b). The confusion matrix for the \texttt{X-CIGALE}-based classification of \xray\ sources in \wcdfs. (c). The confusion matrix for the machine-learning based classification of \xray\ sources in \es. (d). The confusion matrix for the \texttt{X-CIGALE}-based classification of \xray\ sources in \es.}
\label{fig:cm}
\end{figure}

\subsection{SED-template-fitting-based classification}
We also utilize {\sc X-CIGALE} \citep{Yang2020} to identify BL AGNs from their optical-to-IR SEDs in combination with the \xray\ flux level. We do not provide redshift information to \texttt{X-CIGALE}, and allow \texttt{X-CIGALE} fit redshift as a free parameter.
We adopt a delayed exponentially declining star formation history, a Chabrier initial mass function \citep{Chabrier2003}, the extinction law from \citet{Calzetti2000}, the dust-emission template from \citet{Dale2014}, the AGN component SKIRTOR (that is established based on \citealt{Stalevski2012,Stalevski2016}), and the X-ray module following \citet{Yang2020}. Details of the fitting parameters are given in Table~\ref{tab:xcigalep}. 
As \texttt{X-CIGALE} requires intrinsic \xray\ fluxes, we convert the observed \xray\ flux derived in Section~\ref{ss-sc} to intrinsic absorption-corrected \xray\ flux following the method in Section~4.4 of \citet{Luo2017}. Basically, we assume that all \xray\ sources with $\Gamma_{\rm eff} < 1.8$ suffer from some level of intrinsic absorption, and their intrinsic spectra have power-law shapes with a fixed photon index of 1.8.
To identify AGNs, we utilized the ratio between the Bayesian estimation of the AGN 2--10~keV luminosity and the sum of the Bayesian estimation of the 2--10~keV LMXB luminosity and HMXB luminosity in the \texttt{X-CIGALE} output: if this ratio is greater than 10, we identify the source as an AGN.
As BL AGNs generally do not suffer from high levels of extinction of the AGN emission, if the Bayesian estimation for the $E(B-V)$ parameter of the AGN component is smaller than 0.2, we classify the AGN as a BL AGN candidate.
We tested the accuracy of the template-fitting-based classification utilizing X-ray sources that have spectroscopic classifications available; the confusion matrix can be seen in the relevant panels of Figure~\ref{fig:cm}.
We correctly predict $\approx 78\%$ of the BL AGNs and $\approx 83\%$ of the sources that are not BL AGNs in \wcdfs; we correctly predict $\approx 70\%$ of the BL AGNs and $\approx 86\%$ of the sources that are not BL AGNs in \es.
When we fit the remaining X-ray sources in the forced-photometry catalog that do not have spectroscopic classifications available, $\approx 690$/350 of the sources in \wcdfs/\es\ are classified as BL AGN candidates.  As template fitting strongly relies on the number of photometric points available, the resulting BL AGN sky density ($\approx 210/180$ deg$^{-2}$ in \wcdfs/\es) is roughly consistent with the expectation from the COSMOS field: only $\approx 65\%$ of the \xray\ sources in \wcdfs\ and \es\ without spectroscopic redshifts have SNR > 3 detections in at least 10 bands.

\begin{table*}
 \begin{center}
 \caption{Utilized {\sc X-CIGALE} modules with fitting parameters. Default values are adopted for parameters not listed.}
 \vspace{-0.3 cm}
  \begin{tabular}{ccccccccccc}
  \hline\hline
Module  & Parameters  & Values \\
\hline
Star formation history: {\it sfhdelayed} & $\tau$ (Myr) & 100, 300, 500, 1000, 5000, 8000 \\
                                         & $t$ (Myr)    &  100, 300, 500, 1000, 2000, 3000, 5000, 7000, 10000 \\ 
Stellar population synthesis model: {\it bc03} & Initial mass function &   \citet{Chabrier2003} \\
Dust attenuation: {\it dustatt\_calzetti}  & $E(B-V)$ &     0.2--1.0 with steps of 0.1 \\
Dust emission: {\it dale2014}   & $\alpha$ in d$M_{\rm dust} \propto U^{- \alpha}{\rm d}U$ &  2.0 \\
AGN emission: {\it skirtor2016}  & Torus optical depth at 9.7$\mu$m &   7 \\
                                                & Viewing angle ($^{\circ}$) &  30 \\
                                               & AGN fraction of total IR luminosity ($\rm frac_{AGN}$)& 0--0.9 with steps of 0.1, and 0.99\\
                                               & $E(B-V)$ of AGN polar dust & 0--0.6 with steps of 0.1\\
 X-ray               & $\Gamma$& 1.8 \\  
                                & max$\vert \Delta \alpha_{\rm OX}\vert$& 0.2 \\       
\hline \hline
  \end{tabular}
  \end{center}
  \label{tab:xcigalep}
\end{table*}

\begin{figure}
\centering    
\includegraphics[width=0.4\textwidth]{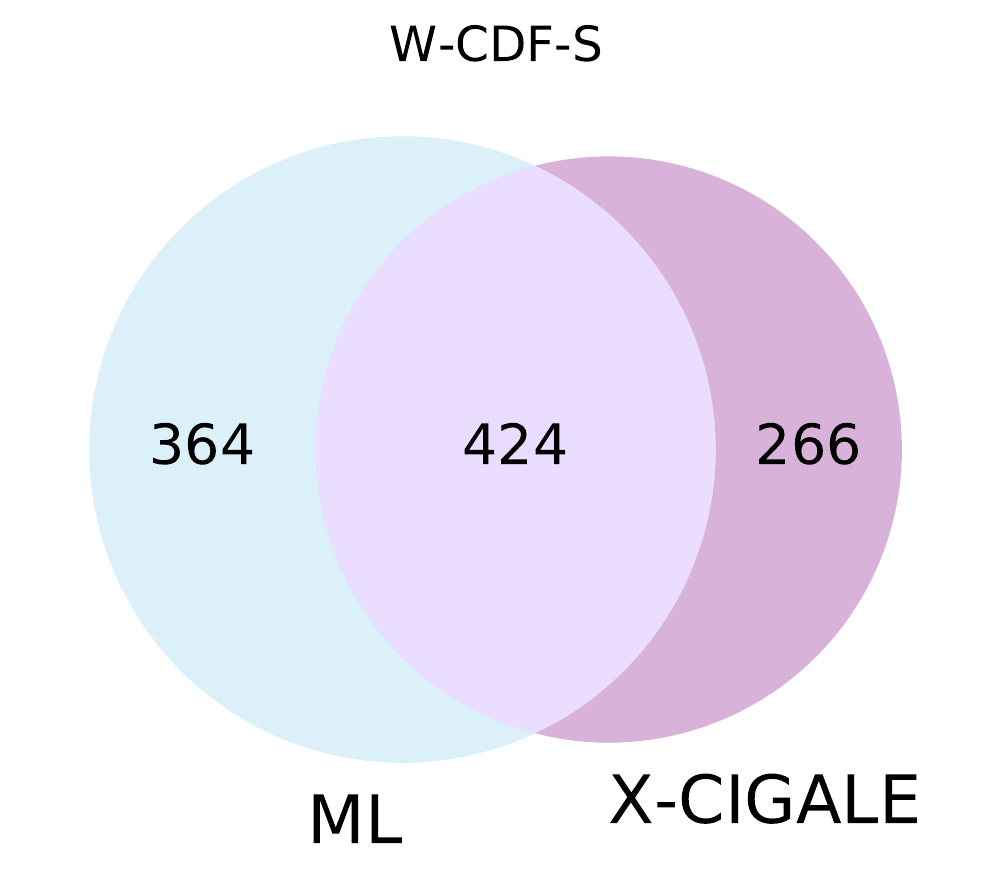}
\includegraphics[width=0.4\textwidth]{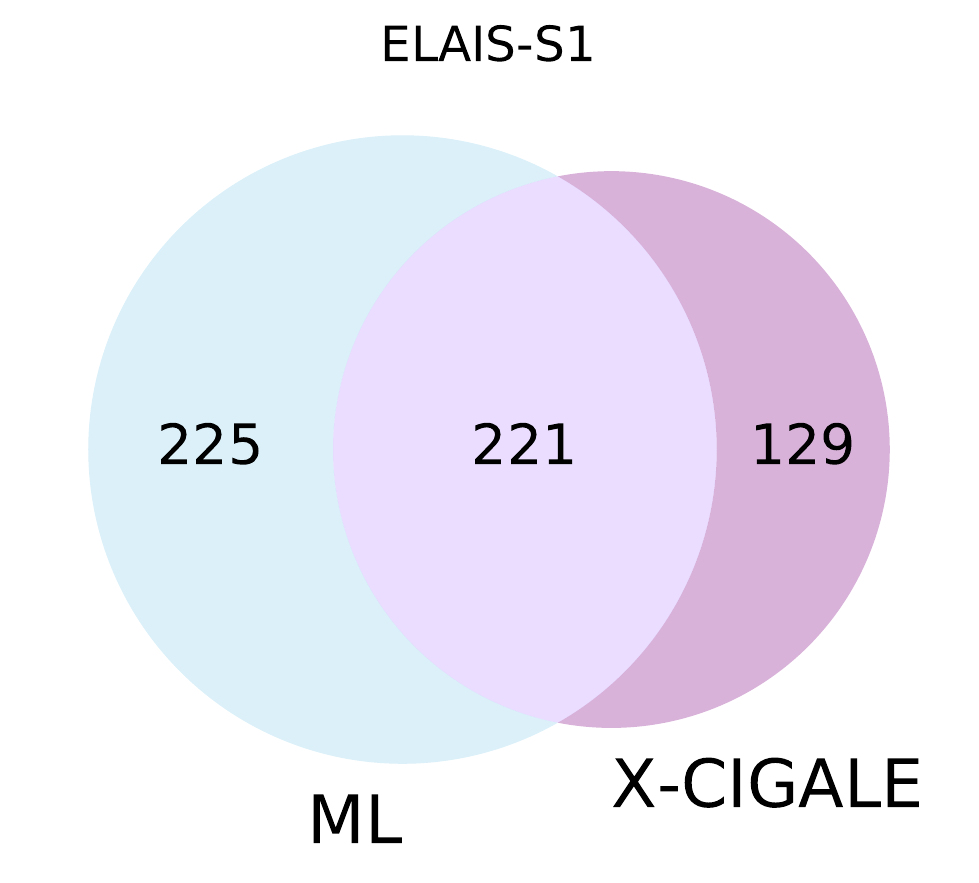}
\caption{The Venn diagram for BL AGNs identified with the machine-learning-based approach and the \texttt{X-CIGALE} template-fitting-based approach.}
\label{fig:venn2}
\end{figure}

\subsection{Assessing the reliability of BL AGN candidates identified}

The above two methods each have advantages and disadvantages. The machine-learning-based method achieves a higher apparent level of accuracy; at the same time, as the model is trained based upon BL AGNs with spectroscopic classifications (which are brighter compared with sources not identified in spectroscopic surveys), there might be biases associated with the predictions. 
While the template-fitting-based method does not suffer from the potential bias introduced from the training set, its accuracy is lower.
Also, for both methods, the prediction accuracy declines when the available number of photometric data points is smaller.
Thus, we create a flag, \texttt{SED\_BLAGN\_FLAG}, for the BL AGN candidates identified. \texttt{SED\_BLAGN\_FLAG} $= 1$ is assigned to $\approx 420$/220 X-ray sources in \wcdfs/\es\ that are identified as BL AGN candidates by both methods; \texttt{SED\_BLAGN\_FLAG} $= 0.5$ is assigned to $\approx 630$/350 X-ray sources in \wcdfs/\es\ that are identified as BL AGN candidates with one method but not the other (see Figure~\ref{fig:venn2} for the Venn diagram).

To assess the reliability of the BL AGN candidates identified, we check the morphology and optical variability of BL AGN candidates identified in \wcdfs.
As luminous BL AGNs often appear to be point-like sources when they dominate over host-galaxy starlight, morphological information has been adopted in selecting BL AGN candidates in some literature works \citep[e.g.,][]{Salvato2009,Salvato2011}. 
In the HSC catalog presented in \citet{Ni2019}, the \texttt{sdss\_pointlike} flag selects point-like sources with the SDSS algorithm, \texttt{psfMag - CmodelMag} < 0.145, in the reference band (the band in which the source is detected with the highest SNR).
Utilizing this \texttt{sdss\_pointlike} column, 1346 \xray\ sources in \wcdfs\ are considered to be point-like through HSC morphology.
Among 731 \xray\ sources where spectroscopic classifications are available, HSC morphology could correctly classify $\approx 93\%$ of the BL AGNs and $\approx 82\%$ of the sources that are not BL AGNs. However, as HSC morphology becomes less accurate at higher redshift and fainter magnitudes, we only utilize it to assess the reliability of objects marked with \texttt{SED\_BLAGN\_FLAG} $= 1$ or $= 0.5$: although not all point-like sources identified from HSC morphology are BL AGNs, we do expect a substantial fraction of the BL AGNs identified via SEDs to have \texttt{sdss\_pointlike} $= 1$.
Figure~\ref{fig:venn3} shows that $\approx 64\%$ of the X-ray sources marked with \texttt{SED\_BLAGN\_FLAG} $= 1$ or $= 0.5$ in \wcdfs\ are identified as point-like sources utilizing HSC morphology. This fraction is $\approx 87\%$ for \texttt{SED\_BLAGN\_FLAG} $= 1$ sources, and $\approx 48\%$ for \texttt{SED\_BLAGN\_FLAG} $= 0.5$ sources, suggesting that \texttt{SED\_BLAGN\_FLAG} $= 1$ is more reliable in identifying BL AGNs, as expected.

The catalogs of optically variable sources in \wcdfs\ from \citet{Falocco2015} and \citet{Poulain2020} are also utilized to assess the quality of the BL AGN candidates selected from SEDs. 333 of our \xray\ sources in \wcdfs\ are identified as potential AGNs in these catalogs ({\sc class} $\geqslant$ 0), and these sources are likely to be BL AGNs. Approximately 160 of them have spectroscopic classification: $\approx 70\%$ of them are real BL AGNs. Although the sample of BL AGN candidates identified via optical variability is incomplete and may have contamination (e.g., from supernovae or stars whose observed fluxes vary due to internal or external reasons), it remains a useful sample for testing the completeness of BL AGNs identified via SEDs.
As can be seen in Figure~\ref{fig:venn3}, \texttt{SED\_BLAGN\_FLAG} $= 1$ or $= 0.5$ objects in \wcdfs\ include $\approx 81\%$ of the optically variable sources; \texttt{SED\_BLAGN\_FLAG} $= 1$ objects alone only include $\approx 58\%$ of the optically variable sources. Thus, the utilization of \texttt{SED\_BLAGN\_FLAG} $= 1$ alone will likely lead to a relatively incomplete BL AGN identification.

For the purposes of this work, we would like to provide reliable photo-$z$ estimations for X-ray sources. 
About $70\%$/$60\%$ of the \texttt{SED\_BLAGN\_FLAG} $= 1/0.5$ objects in \wcdfs\ and \es\ do not have high-quality ($Q_z < 1$) \texttt{EAZY} photo-$z$ measurements utilizing galaxy and obscured AGN templates (see Section~\ref{ss-photoz}).
Thus, we use the AGN-dominated SED templates to fit all the \texttt{SED\_BLAGN\_FLAG} $= 1$ objects, and all the \texttt{SED\_BLAGN\_FLAG} $= 0.5$ sources that cannot be characterized well (i.e., with $Q_z < 1$) by galaxy/obscured AGN templates (see Section~\ref{ss-photoz}).

\begin{figure}
\centering    
\includegraphics[width=0.33\textwidth]{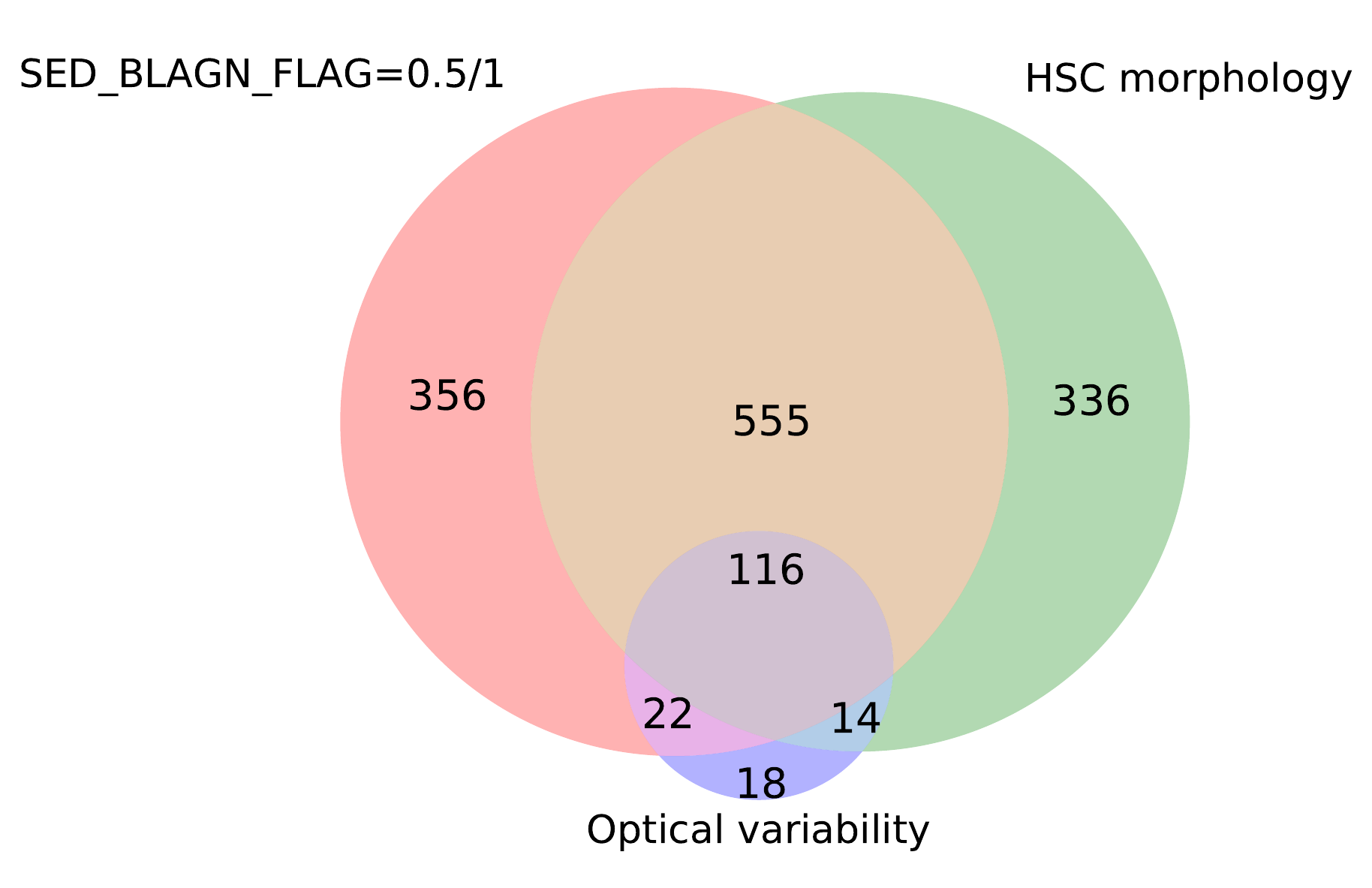}
\includegraphics[width=0.33\textwidth]{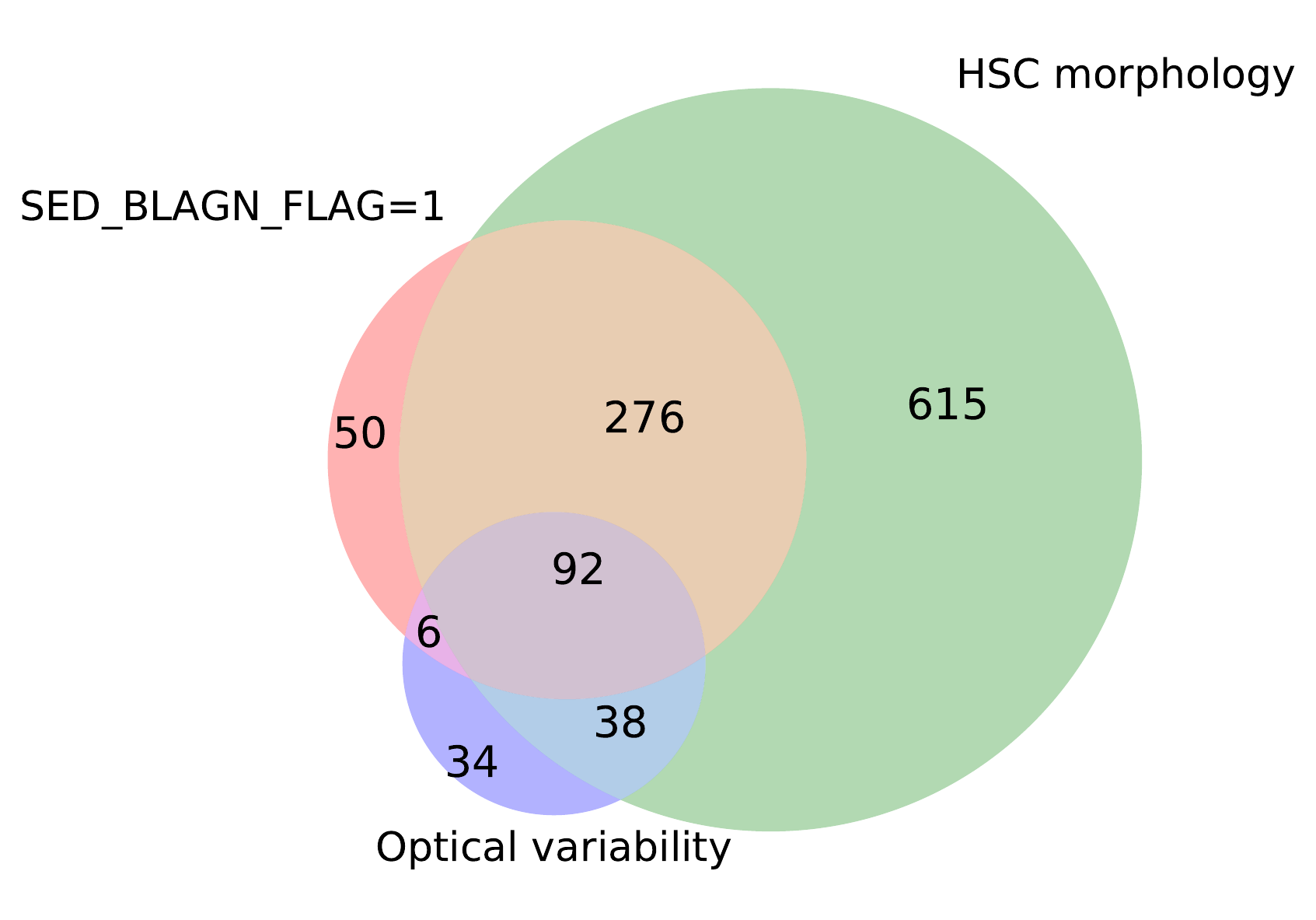}
\includegraphics[width=0.33\textwidth]{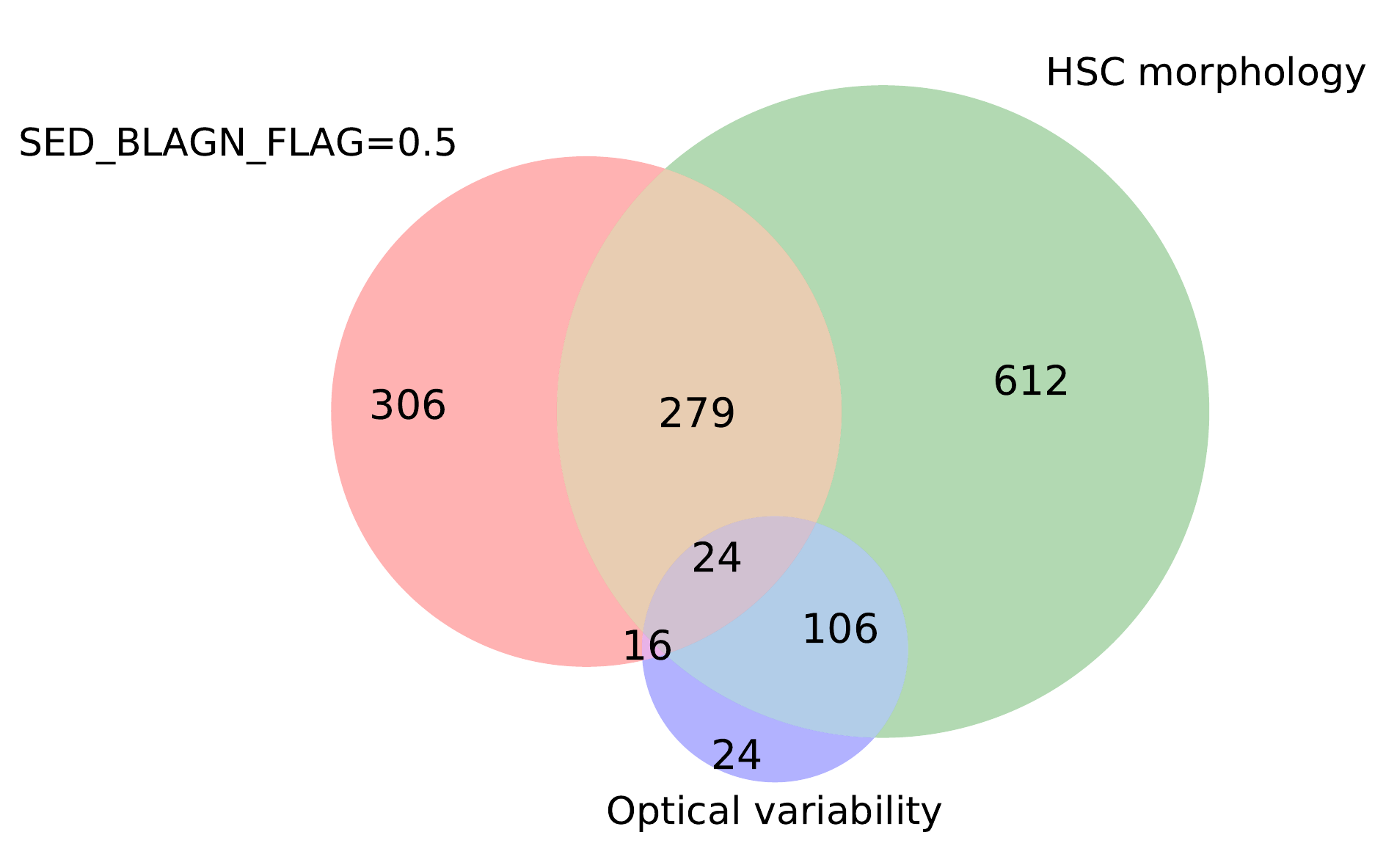}
\caption{{\it Left:} The Venn diagram for the \texttt{SED\_BLAGN\_FLAG} $= 1$ or 0.5 sources, point-like sources identified via HSC morphology, and reported optically variable sources in the \wcdfs\ field. {\it Middle:} Similar to the left panel, but for the \texttt{SED\_BLAGN\_FLAG} $= 1$ sources. {\it Right:} Similar to the left panel, but for the \texttt{SED\_BLAGN\_FLAG} $= 0.5$ sources.}
\label{fig:venn3}
\end{figure}

\section{Identifying non-AGN X-ray sources} \label{a-star}
We have identified Galactic stars among non-AGN X-ray sources that are associated with reliable multiwavelength counterparts (see Section~\ref{sc-sp}) based on the $g-z$ (or $B-z$) vs. $z-K_s$ diagram in \wcdfs\ (or \es); sources falling below the dashed line in Figure~\ref{fig:stargzk} are classified as stars. 
Figure~\ref{fig:stargzk} demonstrates that this criterion successfully identifies almost all of the spectroscopically-confirmed stars.
We also match the non-AGN \xray\ sources to \textit{Gaia} sources \citep[e.g.,][]{Gaia2018} with a matching radius of 1$''$, and classify sources with significant proper motions as stars.
We visually examined the optical imaging to remove contaminating galaxies with obviously extended morphology.
In total, 163 out of the 429 non-AGN X-ray sources in \wcdfs\ are classified as stars; 
89 out of the 317 non-AGN X-ray sources in \es\ are classified as stars.
For the remaining non-AGN X-ray sources, most are bright and large foreground galaxies (identified via visual examination) that contain a population of \xray\ binaries and/or a low-luminosity AGN (see Figure~\ref{fig:xstargal} for example cutouts).

\begin{figure*}
\centering    
\includegraphics[width=0.49\textwidth]{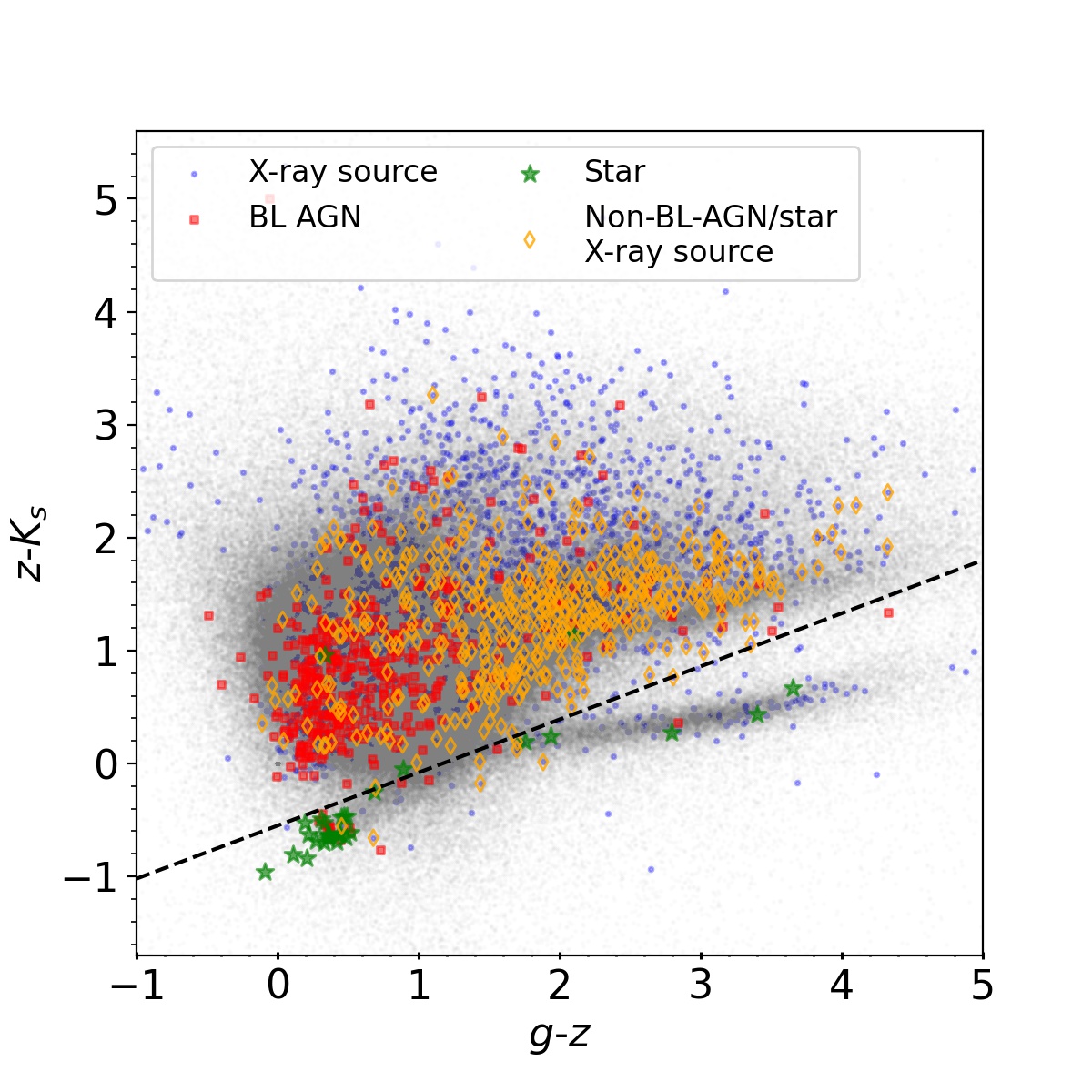}
\includegraphics[width=0.49\textwidth]{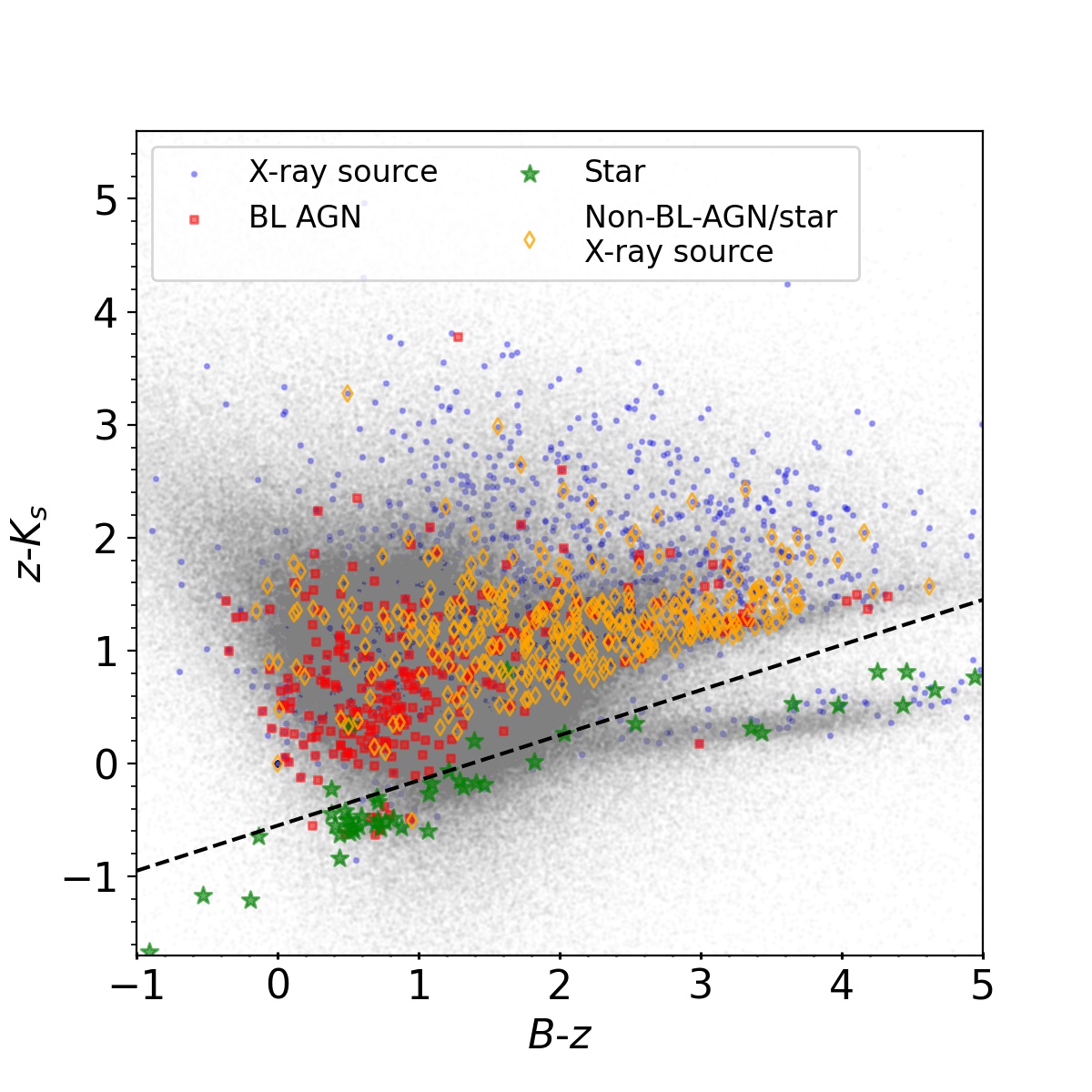}
\caption{\textit{Left:} The $g-z$ vs. $z-K_s$ diagram for sources in the \wcdfs\ forced-photometry catalog (Nyland et al.\ 2021) represented by the grey dots, which can be utilized to separate stars from other \xray\ sources. X-ray sources are marked as blue circles; spectroscopically confirmed BL AGNs are indicated by red squares; spectroscopically identified stars are marked as green stars; X-ray sources with spectroscopic classification that are not identified as BL AGNs nor stars are represented by orange diamonds. The black dashed line is utilized to identify stars among non-AGN sources. \textit{Right:} The $B-z$ vs. $z-K_s$ diagram for sources in the \es\ forced-photometry catalog \citep{Zou2021a}. Symbols are similar to the top panel.}
\label{fig:stargzk}
\end{figure*}

\begin{figure}
\centering
\gridline{\fig{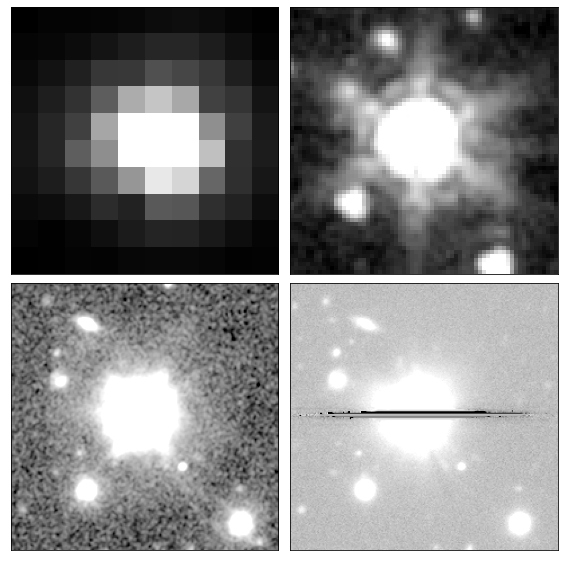}{0.24\textwidth}{(a)}
           \fig{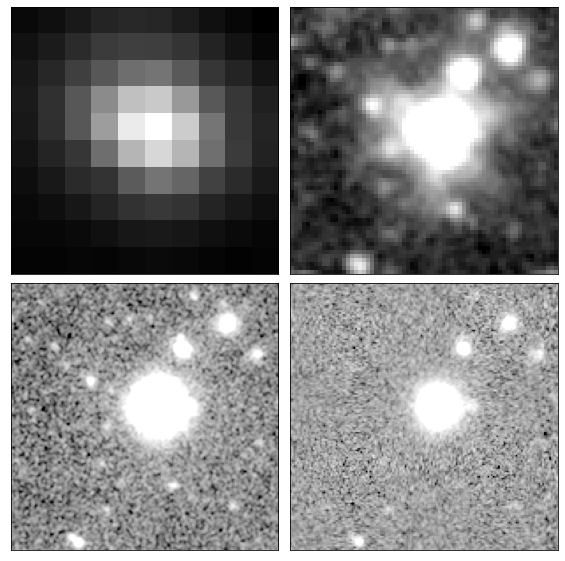}{0.24\textwidth}{(b)}
           ~~
           \fig{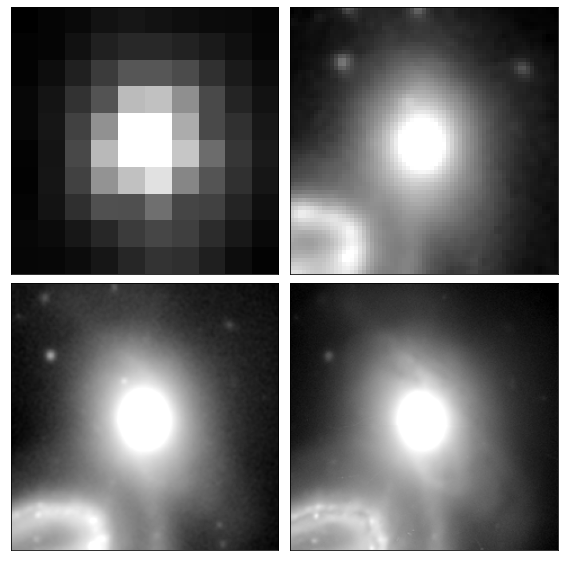}{0.24\textwidth}{(c)}
           \fig{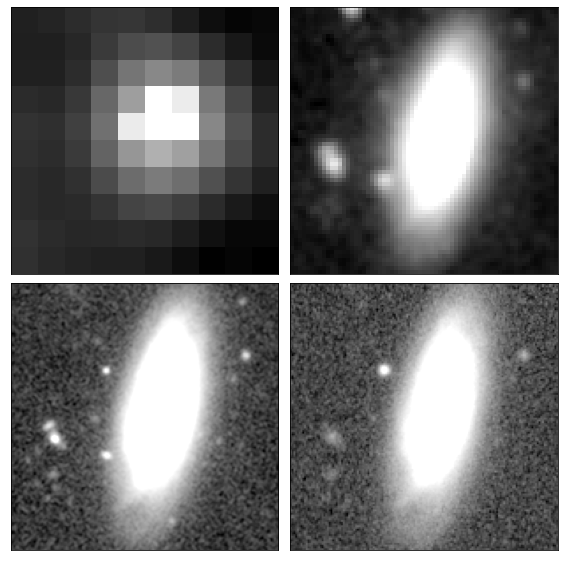}{0.24\textwidth}{(d)}
           }
\caption{Example multiwavelength cutouts of X-ray detected stars (a/b) and \xray\ detected galaxies (c/d) in \wcdfs/\es.
Each panel shows four $40'' \times 40''$ cutouts: \xmm\ \hbox{0.2--12}~keV (top-left), DeepDrill 3.6$\mu$m band (top-right), VIDEO $K_s$-band (bottom-left), and HSC (or DES) $i$-band (bottom-right).}
\label{fig:xstargal}
\end{figure}

\bibliography{xmmservs.bib}
\bibliographystyle{aasjournal}

\end{document}